\documentclass[12pt,a4paper,twoside, headsepline=true]{scrartcl}

\usepackage{lmodern}
\usepackage{geometry} 
\usepackage[automark]{scrlayer-scrpage}
\usepackage[USenglish,ngerman]{babel}
\usepackage[utf8]{inputenc}
\usepackage[T1]{fontenc}
\usepackage{color}
\usepackage{graphicx}
\usepackage{tikz}
\usetikzlibrary{shadings}
\usepackage[outercaption]{sidecap}
\usepackage{tablefootnote}
\usepackage{subcaption}
\usepackage{epstopdf}
\usepackage{floatflt}
\usepackage[numbers,sort&compress]{natbib}
\newcommand*{\doi}[1]{\href{http://dx.doi.org/#1}{\nolinkurl{#1}}}
\usepackage{pdfpages}
\usepackage{wrapfig}	
\usepackage{fancybox} 
\usepackage{amsmath}
\usepackage{MnSymbol}
\usepackage{mathtools}
\usepackage{array}
\usepackage{bbold}
\DeclarePairedDelimiter{\abs}{\lvert}{\rvert}
\usepackage{verbatim}
\usepackage{fancyvrb} 
\usepackage{listings}
\usepackage{enumitem}
\usepackage{setspace}
\usepackage{bookmark}
\usepackage{hyperref}
\hypersetup{
    bookmarks=true,
    unicode=false,
    breaklinks=true,
    pdftoolbar=false,
    pdfpagelayout=TwoPageRight,
    pdfstartpage=2,
    pdfstartview={Fit},
    pdftitle={f-electron charge densities probed using core level non-resonant inelastic x-ray scattering},
	pdfdisplaydoctitle={true},
    pdfauthor={Martin Sundermann},
    pdfsubject={Thesis},
    pdfcreator={pdfLaTeX},
	pdflang=en,
    pdfproducer={},
    pdfkeywords={},
    pdfnewwindow=true,
	hyperfootnotes=false,
    colorlinks=true,
    linkcolor=black,
    citecolor=black,
    filecolor=black,
    menucolor=black,
    urlcolor=blue
}
\usepackage{cancel}
\ihead[]{\thepage}
\ohead[]{\rightmark}
\ifoot[]{}
\ofoot[]{}

\renewcommand{\sectionmark}[1]{\markright{#1}{}}
\renewcommand{\subsectionmark}[1]{\markright{#1}}

\renewcommand{\theequation}{\arabic{section}.\arabic{equation}}
\renewcommand{\thetable}{\arabic{section}.\arabic{table}}
\renewcommand{\thefigure}{\arabic{section}.\arabic{figure}}
\newcommand{\sectioff}[1]{\section*{#1}\addcontentsline{toc}{section}{\protect\numberline{}#1}\sectionmark{#1}}
\numberwithin{equation}{section}	
\numberwithin{table}{section}
\numberwithin{figure}{section}
\usepackage[perpage,hang]{footmisc}
\renewcommand{\thefootnote}{\ifcase\value{footnote}\or*\or
**\or***\fi}
\deffootnote[1.3em]{2em}{0em}{\textsuperscript{\thefootnote}}
\newcommand{\monthword}[1]{\ifcase#1\or Jan\or Feb\or Mar\or Apr\or May\or Jun\or Jul\or Aug\or Sep\or Oct\or Nov\or Dec\fi}
\newcommand{\Month}[1]{\ifthenelse{#1<10}{0#1}{#1}}
\newcommand{\Day}[1]{\ifthenelse{#1<10}{0#1}{#1}}

\usepackage{multirow}
\usepackage{color}
\usepackage{chngcntr}
\usepackage{placeins}
\usepackage{pdfcomment}

\begin{document}

\newcommand{\note}[1]{\textcolor{red}{#1}}
\renewcommand{\emph}[1]{\textit{\textbf{#1}}}

\renewcommand{\vec}[1]{\text{\(\boldsymbol{#1}\)}}
\newcommand{\op}[1]{\hat{#1}}
\newcommand{\bra}[1]{\text{\(\langle\)#1|}}
\newcommand{\ket}[1]{\text{|#1\(\rangle\)}}

\newcommand{\textsub}[1]{\textsubscript{#1}}
\newcommand{\textsup}[1]{\textsuperscript{#1}}
\newcommand{\textfrac}[2]{\raisebox{2pt}{\scalebox{1}{\(#1\)}} \scalebox{1.3}{\(/\)} \raisebox{-2pt}{\scalebox{1}{\(#2\)}}\,}
\newcommand{\etal}{\textit{et al.\@}}

\newcommand{\chap}[2][]{\hyperref[#2]{Ch.\,\ref*{#2}#1}}
\newcommand{\kap}[2][]{\hyperref[#2]{Kap.\,\ref*{#2}#1}}
\newcommand{\Chap}[2][]{\hyperref[#2]{Chapter~\ref*{#2}#1}}
\newcommand{\app}[2][]{\hyperref[#2]{App.\,\ref*{#2}#1}}
\newcommand{\App}[2][]{\hyperref[#2]{Appendix~\ref*{#2}#1}}
\newcommand{\fig}[2][]{\hyperref[#2]{Fig.\,\ref*{#2}#1}}
\newcommand{\Fig}[2][]{\hyperref[#2]{Figure~\ref*{#2}#1}}
\newcommand{\tab}[2][]{\hyperref[#2]{Tab.\,\ref*{#2}#1}}
\newcommand{\Tab}[2][]{\hyperref[#2]{Table~\ref*{#2}#1}}
\newcommand{\equ}[2][]{\hyperref[#2]{Eq.\,\ref*{#2}#1}}
\newcommand{\Equ}[2][]{\hyperref[#2]{Equation~\ref*{#2}#1}}

\def\CeB{\hyperref[sec:CeB6]{\text{CeB\textsub{6}}}}
\def\SmB{\hyperref[sec:SmB6]{\text{SmB\textsub{6}}}}
\def\CeRuSn{\hyperref[sec:CeRu4Sn6]{\text{CeRu\textsub{4}Sn\textsub{6}}}}
\def\UO{\hyperref[sec:UO2]{\text{UO\textsub{2}}}}
\def\URuSi{\hyperref[sec:URu2Si2]{\text{URu\textsub{2}Si\textsub{2}}}}

\newcommand{\newabbr}[3]{\expandafter\newcommand\csname #1\endcsname{\pdftooltip{#3}{#2}}}

\newcommand{\opcr}[1]{\text{\(\pdftooltip{\op{\textsl{a}}^\dag}{creation operator}_\text{#1}\)}}
\newcommand{\opan}[1]{\text{\(\pdftooltip{\op{\textsl{a}}}{annihilation operator}_\text{#1}\)}}

\newabbr{AIM}{Anderson impurity model}{\text{AIM}}
\newabbr{AFM}{antiferromagnetic}{\text{AFM}}
\newabbr{CF}{crystal field}{\text{CF}}
\newabbr{CIM}{configuration interaction model}{\text{CIM}}
\newabbr{CI}{Coulomb repulsion}{\text{CR}}
\newabbr{DCM}{double crystal monochromator}{\text{DCM}}
\newabbr{DOS}{density of states}{\text{DOS}}
\newabbr{Ef}{$f$-electron binding energy}{\text{\textit{E\textsub{f}}}}
\newabbr{EFermi}{Fermi energy}{\text{\(\varepsilon\)\textsub{F}}}
\newabbr{echarge}{elementary charge}{\text{\(e^\text{-}\)}}
\newabbr{emass}{free electron mass}{\text{\(m_{e^\text{-}}\)}}
\newabbr{FWHM}{full width at half maximum}{\text{FWHM}}
\newabbr{FWHMG}{Gaussian full width at half maximum}{\text{FWHM\textsub{G}}}
\newabbr{FWHML}{Lorentzian full width at half maximum}{\text{FWHM\textsub{L}}}
\newabbr{HO}{hidden order}{\text{HO}}
\newabbr{isingindex}{Ising index}{\text{Z\textsub{2}}}
\newabbr{kB}{Boltzmann constant}{\text{\textit{k}\textsub{B}}}
\newabbr{kFermi}{Fermi wave vector}{\text{\textit{\textbf{k}}\textsub{F}}}
\newabbr{LMAF}{large moment antiferromagnetic}{\text{\textit{LMAF}}}
\newabbr{opT}{transition operator}{\text{\textit{T}}}
\newabbr{QCP}{quantum critical point}{\text{QCP}}
\newabbr{rBohr}{Bohr radius}{\text{\textit{a}\textsub{0}}}
\newabbr{redG}{additional scaling of the core-valence dipole Coulomb interaction}{\text{red\textsub{G\textsup{1}}}}
\newabbr{redfc}{scaling of the core-valence Coulomb interaction}{\text{red\(_{fc}\)}}
\newabbr{redff}{scaling of the valence-valence Coulomb interaction}{\text{red\(_{ff}\)}}
\newabbr{redZc}{scaling of the core spin-orbit coupling}{\text{red\(_{\zeta_{c}}\)}}
\newabbr{redZv}{scaling of the valence spin-orbit coupling}{\text{red\(_{\zeta_{v}}\)}}
\newabbr{Res}{resistivity}{\text{\textit{R}}}
\newabbr{Rigc}{ideal gas constant}{\text{\textsl{R}}}
\newabbr{ROI}{region of interest}{\text{ROI}}
\newabbr{SOC}{spin-orbit coupling}{\text{SOC}}
\newabbr{sol}{speed of light}{\text{\text{\textsl{c}}}}
\newabbr{T}{temperature}{\text{\textit{T}}}
\newabbr{TK}{Kondo temperature}{\text{\textit{T}\textsub{K}}}
\newabbr{TKC}{coherence temperature}{\text{\textit{T}*}}

\newabbr{Veff}{effective hybridization}{\text{\(V_\text{eff}\)}}
\newabbr{Deltaf}{effective \textit{f}-electron binding energy}{\text{\(\Delta_f\)}}
\newabbr{bandwidth}{bandwidth}{\text{\textit{D}}}
\newabbr{onsiteU}{onsite Coulomb repulsion}{\text{\textit{U\textsub{ff}}}}

\newabbr{NIXS}{non-resonant inelastic x-ray scattering}{\text{NIXS}}
\newabbr{HAXPES}{hard x-ray photoelectron spectroscopy}{\text{HAXPES}}
\newabbr{PES}{photoelectron spectroscopy}{\text{PES}}
\newabbr{RIXS}{resonant inelastic x-ray scattering}{\text{RIXS}}
\newabbr{XAS}{x-ray absorption spectroscopy}{\text{XAS}}
\newabbr{PFY}{partial fluorescence yield}{\text{PFY}}
\newabbr{TEY}{total electron yield}{\text{TEY}}
\newabbr{TFY}{total fluorescence yield}{\text{TFY}}
\newabbr{LD}{linear dichroism}{\text{LD}}

\newabbr{kin}{wave vector of the incoming photon}{\text{\(\vec{k}_\text{in}\)}}
\newabbr{kout}{wave vector of the outgoing photon}{\text{\(\vec{k}_\text{out}\)}}
\newabbr{oin}{angular frequency of the incoming photon}{\text{\(\omega_\text{in}\)}}
\newabbr{oout}{angular frequency of the outgoing photon}{\text{\(\omega_\text{out}\)}}
\newabbr{otr}{energy transfer}{\text{\(\omega\)}}
\newabbr{Ein}{energy of the incoming photon}{\text{\(E_\text{in}\)}}
\newabbr{Eout}{energy of the outgoing photon}{\text{\(E_\text{out}\)}}
\newabbr{Etr}{energy transfer}{\text{\(\Delta E\)}}

\newabbr{thS}{scattering angle}{\text{\(\theta\)}}
\newabbr{thB}{Bragg angle}{\text{\(\theta_\text{B}\)}}

\newcommand{\Ylm}[2]{\text{\(\pdftooltip{Y}{spherical harmonic}_\text{#1}^\text{#2}\)}}
\newcommand{\Clm}[2]{\text{\(\pdftooltip{C}{renormalized spherical harmonic}_\text{#1}^\text{#2}\)}}
\newcommand{\Zlm}[2]{\text{\(\pdftooltip{Z}{tesseral harmonic}_\text{#1}^\text{#2}\)}}
\newcommand{\Dlm}[2]{\text{\(\pdftooltip{D}{renormalized tesseral harmonic}_\text{#1}^\text{#2}\)}}
\newcommand{\Klm}[2]{\text{\(\pdftooltip{K}{cubic harmonic}_\text{#1}^\text{#2}\)}}
\newcommand{\Elm}[2]{\text{\(\pdftooltip{E}{renormalized cubic harmonic}_\text{#1}^\text{#2}\)}}
\newcommand{\Rnl}[2]{\text{\(\pdftooltip{R}{radial wave function}_\text{#1#2}^\text{}\)}}
\newcommand{\Pnl}[2]{\text{\(\pdftooltip{P}{radial expansions (r*R)}_\text{#1#2}^\text{}\)}}

\newcommand{\opClm}[2]{\text{\(\op{C}_\text{#1}^\text{#2}\)}}

\newabbr{euler}{Euler's number}{\text{e}}
\renewcommand{\exp}[1]{\text{\euler\(^{#1}\)}}

\newcommand{\Quanty}{\text{\textit{Quanty}\,\cite{Quanty, Haverkort2012, Haverkort2016}}}
\newcommand{\termsymbol}[3]{\pdftooltip{\text{\textsup{#1}#2\textsub{#3}}}{term symbol: (2S+1)L(J)}}
\newcommand{\Legendre}[2]{\pdftooltip{\text{\(\mathcal{L}_{#1}(#2)\)}}{Legendre polynomial}}
\newcommand{\edge}[3][\!]{\text{\pdftooltip{#1\,\textit{#2}\textsub{#3}}{absorption edge}}}

\newcommand*\del{\raisebox{0pt}{\scalebox{1.25}{\(\delta\)}}}
\newcommand*\diff{\mathrm{d}}
\newcommand{\order}[1]{\scalebox{1.25}{\(\mathcal{O}\)}\hspace{-3pt}\left(#1\right)}

\newabbr{csJ}{coupling strength}{\text{\(\mathcal{J}\)}}
\newabbr{pfZ}{partition function}{\text{\(\mathcal{Z}\)}}

\newabbr{qnQ}{multipolar order}{\text{\textsl{k}}}
\newabbr{qnZ}{atomic number}{\text{\textsl{Z}}}
\newabbr{qnn}{principal quantum number}{\text{\textsl{n}}}
\newabbr{rotn}{order of the rotational symmetry}{\text{\textit{n}}}

\newabbr{opj}{momentum operator}{\text{\textbf{j}}}
\newabbr{qnj}{momentum}{\text{\textsl{j}}}
\newabbr{qnmj}{projection of the momentum}{\text{\textsl{m\textsub{j}}}}
\newabbr{opJ}{total momentum operator}{\text{\textbf{J}}}
\newabbr{qnJ}{total momentum}{\text{\textsl{J}}}
\newabbr{qnJz}{projection of the total momentum}{\text{\textsl{J\textsub{z}}}}

\newabbr{opl}{angular momentum operator}{\text{\textbf{l}}}
\newabbr{qnl}{angular momentum}{\text{\textsl{l}}}
\newabbr{qnml}{projection of the angular momentum}{\text{\textsl{m}}}
\newabbr{opL}{total angular momentum operator}{\text{\textbf{L}}}
\newabbr{qnL}{total angular momentum}{\text{\textsl{L}}}
\newabbr{qnLz}{projection of the total angular momentum}{\text{\textsl{L\textsub{z}}}}

\newabbr{ops}{spin operator}{\text{\textbf{s}}}
\newabbr{qns}{spin}{\text{\textsl{s}}}
\newabbr{qnms}{projection of the spin}{\text{\(\sigma\)}}
\newabbr{opS}{total spin operator}{\text{\textbf{S}}}
\newabbr{qnS}{total spin}{\text{\textsl{S}}}
\newabbr{qnSz}{projection of the total spin}{\text{\textsl{S\textsub{z}}}}

\newcommand{\xyz}[1]{\pdftooltip{\text{[#1]}}{direct lattice vector}}
\newabbr{vecq}{momentum transfer}{\text{\(\vec{q}\)}}
\newcommand{\absq}{\text{|\vecq|}}
\newcommand{\qp}[1]{\text{\vecq\,||\,\xyz{#1}}}
\newabbr{sqw}{dynamic structure factor}{\text{S(\(\vec{q}\),\,\(\omega\))}}

\selectlanguage{USenglish}

\setcounter{page}{0}
\pdfbookmark[section]{Title}{toc}
\thispagestyle{empty}
\vspace*{\fill}

\begin{center}
{\setstretch{1.5}
\scalebox{1.5}{\textbf{\textit{f}-electron charge densities}}\\
\scalebox{1.25}{\textbf{probed using}}\\
\scalebox{1.5}{\textbf{core level non-resonant inelastic x-ray scattering}}\\\vspace{1.5cm}
\scalebox{1.5}{D\,I\,S\,S\,E\,R\,T\,A\,T\,I\,O\,N}\\
\scalebox{1.25}{\--- {\"u}berarbeitete Fassung \---}\\\vspace{1.5cm}
\scalebox{1.25}{von}\\
\scalebox{1.5}{Martin Sundermann}\\\vspace{3cm}
}
\end{center}

\begin{tabular}{l}
Vorgelegt der Fakult{\"a}t Mathematik und Naturwissenschaften\\
\phantom{Vorgelegt} der Technischen Universit{\"a}t Dresden\\ \\
\end{tabular}\\
\begin{tabular}{ll}
Eingereicht am: & 15.11.2018 \\
Tag der Verteidigung: & 05.06.2019\\ \\
\end{tabular}\\
\begin{tabular}{l}
Gutachter:\\
\quad Prof. Dr. L. H. Tjeng \\
\quad\quad Max-Planck-Institut f{\"u}r Chemische Physik fester Stoffe, Dresden \\
\quad Prof. Dr. J. Geck \\
\quad\quad Technische Universit{\"a}t Dresden \\
\quad Prof. Dr. M. W. Haverkort \\
\quad\quad Universität Heidelberg
\end{tabular}\vspace{-3cm}

\cleardoublepage
\pagenumbering{Roman}
\setcounter{page}{1}
\sectionmark{Table of contents}
\setlength{\parskip}{-.4pt}
\vspace{0 pt}
\setcounter{tocdepth}{3}
\pdfbookmark[section]{Table of contents}{Contents}
\makeatletter\@starttoc{toc}\makeatother
\setlength{\parskip}{1em} 
\cleardoublestandardpage

\pagenumbering{arabic}
\setcounter{page}{1}
\setlength{\parskip}{1ex}
\renewcommand{\subsectionmark}[1]{}

\sectioff{Abstract}
Strongly correlated materials are characterized by the presence of electron-electron interactions in their electronic structure.
They often have remarkable properties and transitions between competing phases of very different electronic and magnetic order.
This thesis focuses on \textit{strongly correlated f-electron compounds} containing Ce, Sm, and U.
These materials exhibit a so-called \textit{heavy-fermion} or \textit{Kondo-lattice} behavior.
They can become insulating due to hybridization effects (\textit{Kondo-insulator}) or develop \textit{multipolar} (\textit{hidden}) \textit{order}.
Kondo insulators have recently been discussed in the context of \textit{strongly correlated topological insulators}.
This new aspect caused an enormous activity in the field of Kondo insulators, theoretically as well as experimentally. 

Multipolar order as well as the formation of a Kondo insulating state strongly depend on the symmetry of the $f$ states involved.
Also the character of the surface states in a topological insulator is determined by the properties of the bulk states.
Therefore the scope of this thesis has been to unveil the underlying symmetries of the bulk $f$ states.
Here the compounds CeB$_6$, UO$_2$, and URu$_2$Si$_2$, which exhibit multipolar order, as well as the Kondo insulators (semimetals) SmB$_6$ and CeRu$_4$Sn$_6$ have been studied.

Non-resonant inelastic x-ray scattering (\NIXS{}) has been established  as a bulk sensitive technique for determining ground-state wave functions.
\NIXS{} at large momentum transfer measures a higher than dipole scattering signal, so that anisotropies of cubic point symmetries are accessible, and it is the only technique that shows multiplet structures in metallic uranium compounds.

CeB$_6$ (\chap{sec:CeB6}) and UO$_2$ (\chap{sec:UO2}) were measured because they can serve as benchmark compounds; their ground states were known from other studies, so that the validity of the \NIXS{} experiment and its analysis could be tested.
For SmB$_6$ (\chap{sec:SmB6}), CeRu$_4$Sn$_6$ (\chap{sec:CeRu4Sn6}), and URu$_2$Si$_2$ (\chap{sec:URu2Si2}) the ground states were determined.
For SmB$_6$ \NIXS{} was needed because SmB$_6$ is cubic, for CeRu$_4$Sn$_6$ because \NIXS{} is bulk sensitive, and for URu$_2$Si$_2$ because the multipolar \NIXS{} signal is more excitonic.

Finally, the linear independent directions of the multipole expansion of the transition operator in \NIXS{} have been investigated (\chap{ch:vecqdependence}) and a tensor notation has been developed (\chap{sec:scatteringtensor}). This is useful for a better understanding of the directional dependence of the momentum transfer of the single crystal \NIXS{} spectra.

\clearpage
\sectioff{Kurzzusammenfassung}
\begin{otherlanguage}{ngerman}

Stark korrelierte Systeme werden stark von der Elektron-Elektron-Wechselwirkung gepr\"agt.
Dies f\"uhrt h\"aufig zu bemerkenswerten Eigenschaften und \"Uberg\"angen zwischen konkurrierenden Phasen sehr unterschiedlicher elektronischer oder magnetischer Ordnung.
Diese Dissertation behandelt \textit{stark korrelierte f-Elektronen-Systeme} die Ce, Sm und U beinhalten.
Diese weisen sogenanntes \textit{schwerfermionisches} oder \textit{Kondo-Gitter}-Verhalten auf.
Sie k\"onnen durch Hybridisierung isolierend werden (\textit{Kondo-Isolator}) oder eine \textit{multipolare} (\textit{verborgene}) \textit{Ordnung} ausbilden.
Seit Kurzem werden Kondo-Isolatoren als m\"ogliche \textit{stark korrelierte topologische Isolatoren} diskutiert.
Dieser neue Aspekt rief eine Vielzahl an T\"atigkeiten im Bereich der Kondo-Isolatoren hervor, sowohl theoretisch als auch experimentell.

Sowohl multipolare Ordnung als auch die Formation des Kondo-isolierenden Zustandes h\"angen stark von der Symmetrie der \textit{f}-Elektronen ab.
Auch die Eigenschaften der Oberfl\"achenzust\"ande topologischer Isolatoren werden von den Zust\"anden im Inneren bestimmt.
Daher befasst sich diese Dissertation mit der Identifizierung der Symmetrie der inneren \textit{f}-Elekronen-Zust\"ande.
Es wurden CeB$_6$, UO$_2$ und URu$_2$Si$_2$, welche multipolare Ordnung aufweisen, sowie die Kondo-Isolatoren/\hyphenation{Semi-metalle}Semi-metalle SmB$_6$ und CeRu$_4$Sn$_6$ untersucht.

Nicht-resonante inelastische R\"ontgenstreuung (\NIXS{}) wurde als bulksensitive Methode zur Bestimmung der Grundzustandswellenfunktion etabliert.
F\"ur große Impuls\"ubertr\"age weist NIXS{} Signale jenseits der Dipol-Auswahlregeln auf, sodass Anisotropien auch in kubischer Punktsymmetrie feststellbar sind, und zeigt als einzige Methode klare exzitonische Multiplettstrukturen f\"ur metallische Uranverbindungen.

CeB$_6$ (\kap{sec:CeB6}) und UO$_2$ (\kap{sec:UO2}), deren Grundzustand bereits aus anderen Studien bekannt sind, wurden als Referenzmaterialien untersucht um die G\"ultigkeit der Methode und der Auswertung zu \"uberpr\"ufen.
F\"ur SmB$_6$ (\kap{sec:SmB6}), CeRu$_4$Sn$_6$ (\kap{sec:CeRu4Sn6}) und URu$_2$Si$_2$ (\kap{sec:URu2Si2}) wurden die Grundzust\"ande bestimmt.
Im Fall von SmB$_6$ wurde \NIXS{} ben\"otigt da SmB$_6$ kubisch ist, f\"ur CeRu$_4$Sn$_6$ da \NIXS{} bulksensitiv ist und f\"ur URu$_2$Si$_2$ wegen des exzitonischen \NIXS{} Signals.

Weiter wurden die linear unabh\"angigen Richtungen der multipolaren \"Uberg\"ange des \"Ubergangsmatrixelements von \NIXS{} untersucht (\kap{ch:vecqdependence}) und eine Tensorschreibweise entwickelt (\kap{sec:scatteringtensor}).
Dies erlaubt ein besseres Verst\"andnis der Richtungsabh\"angigkeit des Impuls\"ubertrages der \NIXS{} Spektren von Einkristallen.

\end{otherlanguage}

\clearpage
\section{Correlated \textit{f}-electron physics}
\FloatBarrier

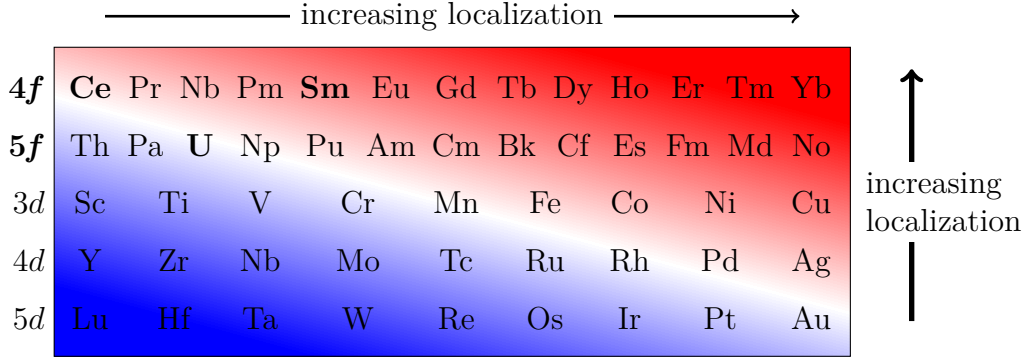
\begin{figure}
  \centering
  \def\arraystretch{1.5}
  \setlength{\tabcolsep}{3pt}
  \pgfdeclareverticalshading{bwr}{100bp}{color(0bp)=(red); color(25bp)=(red); color(50bp)=(white); color(75bp)=(blue); color(100bp)=(blue)}
  \begin{tikzpicture}
    \draw[->, line width=1pt] (-130pt,70pt)--(130pt,70pt);
    \node[rectangle, fill=white, align=left] at (0,70pt){increasing localization};
    \node[rectangle , shading=bwr, shading angle=145, draw=black] at (0,0){
      \begin{tabular}{rccccccccccccc}
       \hspace{-24pt}\textbf{4\textit{f}}\hspace{3pt} & \textbf{Ce} & Pr & Nb & Pm & \textbf{Sm} & Eu & Gd & Tb & Dy & Ho & Er & Tm & Yb \\
       \hspace{-24pt}\textbf{5\textit{f}}\hspace{3pt} & Th & Pa & \textbf{U} & Np & Pu & Am & Cm & Bk & Cf & Es & Fm & Md & No \\
       \hspace{-24pt}3$d$\hspace{3pt} & Sc & \multicolumn{2}{c}{Ti} & V  & \multicolumn{2}{c}{Cr} & Mn & \multicolumn{2}{c}{Fe} & Co & \multicolumn{2}{c}{Ni} & Cu \\
       \hspace{-24pt}4$d$\hspace{3pt} & Y  & \multicolumn{2}{c}{Zr} & Nb & \multicolumn{2}{c}{Mo} & Tc & \multicolumn{2}{c}{Ru} & Rh & \multicolumn{2}{c}{Pd} & Ag \\
       \hspace{-24pt}5$d$\hspace{3pt} & Lu & \multicolumn{2}{c}{Hf} & Ta & \multicolumn{2}{c}{W}  & Re &  \multicolumn{2}{c}{Os} & Ir & \multicolumn{2}{c}{Pt} & Au \\
    \end{tabular}};
    \draw[->, line width=2pt] (172pt,-45pt)--(172pt,50pt);
    \node[rectangle, fill=white, align=left] at (184pt,0){increasing\\localization};
  \end{tikzpicture}\hspace{-24pt}
  \caption{Kmetko-Smith diagram adapted from Ref.\,\cite{Coleman2015}. It shows a rearrangement of the $d$ and $f$ elements of the periodic table to visualize the trend of increasing localization in these ions, i.e.\ $5d$\,<\,$4d$\,<\,$3d$\,<\,$5f$\,<\,$4f$.}
  \label{fig:KmetkoSmith}
\end{figure}

New phenomena occur when magnetic ions are placed inside a metallic host.
But magnetic states can survive only when local Coulomb interactions suppress the charge fluctuation, i.e.\ when the valence electrons are localized.
\Fig{fig:KmetkoSmith} shows the so-called \textit{Kmetko-Smith diagram}, which sorts the $d$ and $f$ elements of the periodic table according to the localization of the valence shell.
This arrangement is particularly useful, as it relates also to the energy scaling of the important interactions of correlated materials:
The Coulomb interaction, which increases with increasing localization, the strength of the crystal field (\CF{}), which decreases with increasing localization, and the spin-orbit coupling (\SOC{}), which increases strongly with the principal quantum number (\qnn{}) and further with the atomic number (\qnZ{}).
The increasing localization from the bottom left to the top right goes along with a crossover from itinerant electrons to localized magnetic moments and particularly interesting phenomena occur in the crossover regime.
In the following several phenomena will be described, which are mostly relevant to the energy scales of the rare earths (4$f$) and actinides (5$f$) materials investigated in the framework of this thesis, summarizing Ref.\,\cite{Kroha2017,Khomskii2010,Coleman2015}.

The rare earth and, to a lesser extent, the actinides stand out by their strong localization.
This leads to strong Coulomb interactions and weak \CF{} splittings in comparison to the transition metals, and the Hund's rule ground state remains mostly unaffected by crystalline \CF{} potentials.
A more detailed overview of the atomic interactions of the $f$ electrons will be given in \chap{sec:calcH}.
In the following it will be discussed, what happens when the rare earth or actinide ions are placed inside a metallic host.

\subsection{Kondo effect}

The Kondo effect\,\cite{Kondo1964} describes what happens when a single local moment is put inside a metallic host.
Here the single ion is referred to as \textit{impurity}.

The magnetic moment of the impurity acts as a scattering center for the conduction electrons.
The scattering amplitude for the spin flip case behaves logarithmic in \T{}\textsup{-1} so that the resistivity \Res{} increases at low temperature \T{} with \( \Res{}(\T{}) \propto J\rho(\EFermi{}) \ln(\EFermi{} / \kB{}\T{}) \).
Here \csJ{} represents the antiferromagnetic coupling strength between conduction electrons and impurity and $\rho(\varepsilon_\text{F})$, the density of states (\DOS{}) at the Fermi energy (\EFermi{}).
This causes the famous increase in the resistivity at low temperature, and the minimum in the resistivity first observed in \textit{dirty} Au in 1934\,\cite{deHaas1934} and other metals with magnetic impurities thereafter\,\cite{vdBerg1965}.
 
Below the so-called Kondo temperature
\begin{align}
\kB{}\TK{} &= \varepsilon_\text{F} ~ e^{-1/J\rho(\varepsilon_\text{F})}, \label{eq:Tkondo}
\end{align}
perturbation theory breaks down.
The resistivity does not diverge completely and remains constant for \T{}\,$\to$\,0\,K.

The magnetic coupling between the impurity and the conduction electrons lowers the energy of the conduction electrons with opposite spin in the vicinity of the impurity.
These electrons condensate around the impurity when the temperature decreases below a characteristic temperature \TK{}.
As a result, the magnetic moment of the impurity gets screened by the surrounding conduction electron cloud and a strongly enhanced Pauli paramagnetism becomes prominent.
The new Pauli paramagnetic state is referred to as Kondo singlet state because of the \qnS{}\,=\,0 singlet character.
The remaining conduction electrons no longer see a magnetic scattering center, thus preventing the divergence of the resistivity.
The \textit{bound} conduction electrons condensate at the Fermi energy leading also to an increase of the \DOS{} and giving rise to a peak with the width of the characteristic temperature \TK{}, the so-called Kondo resonance peak.

\begin{figure}
  \centering
  \includegraphics[width=\textwidth]{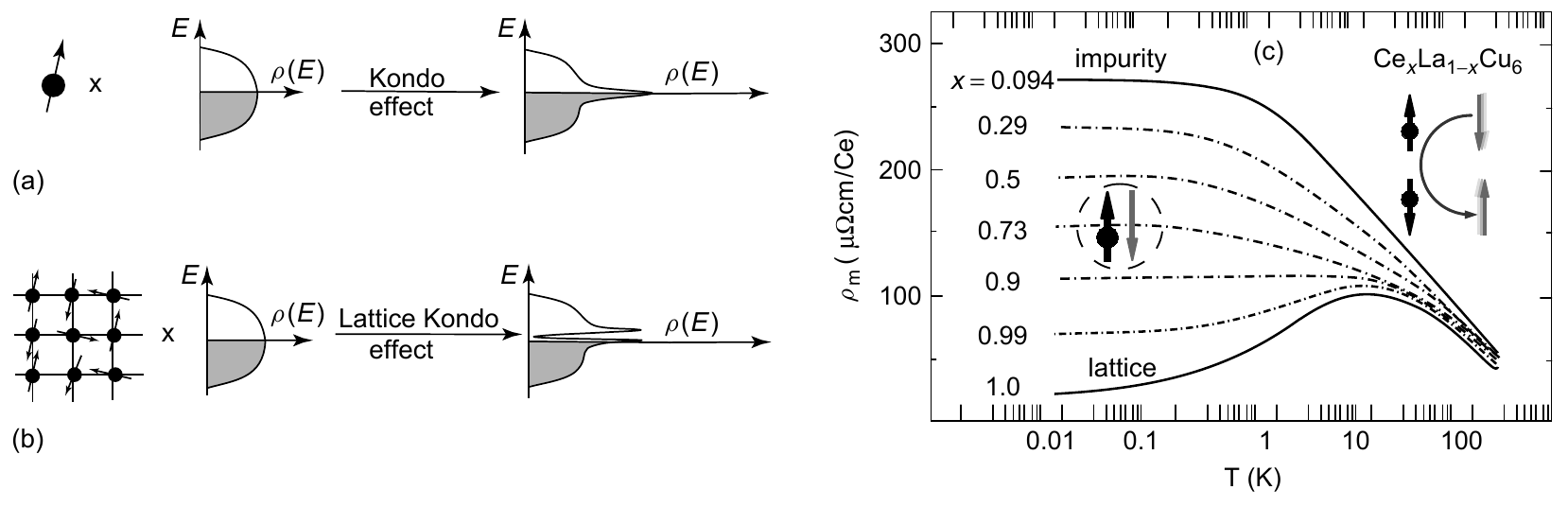}
  \caption{Schematic density of states in the presence of the Kondo effect for a single magnetic impurity (a) and a lattice with periodic arranges ions (b). (c) Temperature dependent resistivity data of Ce$_x$La$_{1-x}$Cu$_6$. Figure adapted from Ref.\,\cite{Coleman2007}.}
  \label{fig:Kondo}
\end{figure}

\Fig[a]{fig:Kondo} visualizes the formation of the Kondo peak for a single impurity and the resistivity of Ce$_x$La$_{1-x}$Cu$_6$ is shown in \fig[c]{fig:Kondo}.
When small amounts (see $x$\,=\,0.094) of the non-magnetic La is replaced by the magnetic Ce inside the Cu host, the typical low temperature Kondo behavior is shown (logarithmic increase with \T{}\textsup{-1} and the flattening below a certain temperature).
The insets in \fig[c]{fig:Kondo} show graphically the microscopic interpretation of the scattering process and the Kondo singlet state.
For higher Ce concentrations $x$ the resistivity decreases after passing through a maximum when \T{} decreases further.
This is a new phenomenon that happens when going to a dense \textit{Kondo lattice system}.

\paragraph{Heavy fermions}

The $f$-electron physics becomes even more fascinating and richer when going to systems with a high density of impurities; in particular when the \textit{impurity} becomes part of the crystal unit cell and forms a periodic lattice, the so-called \textit{Kondo lattice} (see \fig[b]{fig:Kondo}).
Then the physics of the \textit{impurities} can no longer be treated independently.

When the temperature goes below \TK{} the Kondo singlet state forms at every impurity.
The quasi-particles which screen the local moments will more and more overlap and interact with increasing concentration of the magnetic ions.
At this point, they start to form new bands, which have highly renormalized properties, which explains the greatly enhanced specific heat first observed in the dense Kondo lattice system CeAl$_3$\,\cite{Andres1975}.
This renormalization can be expressed in terms of an \textit{effective mass}, which can be several orders of magnitude larger than the free electron mass (\emass{}), thus giving the name \textit{heavy fermions} to this material class.
There is a double peak at the Fermi level in \fig[b]{fig:Kondo}), which can be understood as an electron or hole doping of the Kondo singlet state\,\cite{Dzero2016}.

The charge carriers behave no longer independently, as in a Fermi gas due to the electron-electron correlations of the quasi-particles.
Their correlations lead to a Fermi liquid behavior, which leads to the $\T{}^2$ behavior of the resistivity observed in the pure CeCu$_6$ ($x$\,=\,1 in \fig[c]{fig:Kondo}).

The bands form and the Fermi liquid behavior sets in below the so-called coherence temperature (\TKC{}).
Due to the continuous crossover between the regime of the Kondo effect and the Fermi liquid state, it is difficult to distinguish between \TK{} and \TKC{} and in the framework of this thesis, both scales will be referred to as \TK{}.

While the Kondo effect tends to screen the localized moments, there is another interaction that tends to form long range magnetic order.
This so-called RKKY interaction is described below.

\subsection{RKKY interaction}

The Ruderman-Kittel-Kasuya-Yosida (RKKY) interaction\,\cite{Ruderman1954, Kasuya1956, Yosida1957} describes the interaction of the impurity moments via the conduction electrons.

The moment of the impurity causes a spin polarization of the conduction electrons which couples by the same exchange interaction \csJ{} as for the Kondo effect.
Assuming, that the conduction electrons remain mobile, they will travel through the lattice with the spin density oscillating as described by Friedel oscillations.
In a dense system the spin density will interact with other impurities, which leads to the RKKY exchange interaction
\begin{align}
H_\text{RKKY} &= \alpha \csJ{}^2 \vec{S}_i \vec{S}_j \cos\left(2\kFermi{}\vec{r}_{ij}+\phi\right) \abs{\vec{r}_{ij}}^{-3}. \label{eq:Hrkky}
\end{align}
The prefactor $\alpha$ and the Fermi wave vector (\kFermi{}) depend greatly on the electronic structure of the system.
In an ordered Kondo lattice system, the RKKY interaction may lead to magnetic order and even complex magnetic structures can form, e.g.\ with incommensurate propagation vectors or multiple magnetic phases in a single compound.

The two effects, RKKY interaction, favoring a magnetically ordered ground state, and Kondo effect, favoring a non-magnetic singlet state, compete with each other.
This is pictured by the so-called Doniach phase diagram.

\paragraph{Doniach phase diagram}

\begin{figure}
  \centering
  \includegraphics[width=0.75\textwidth]{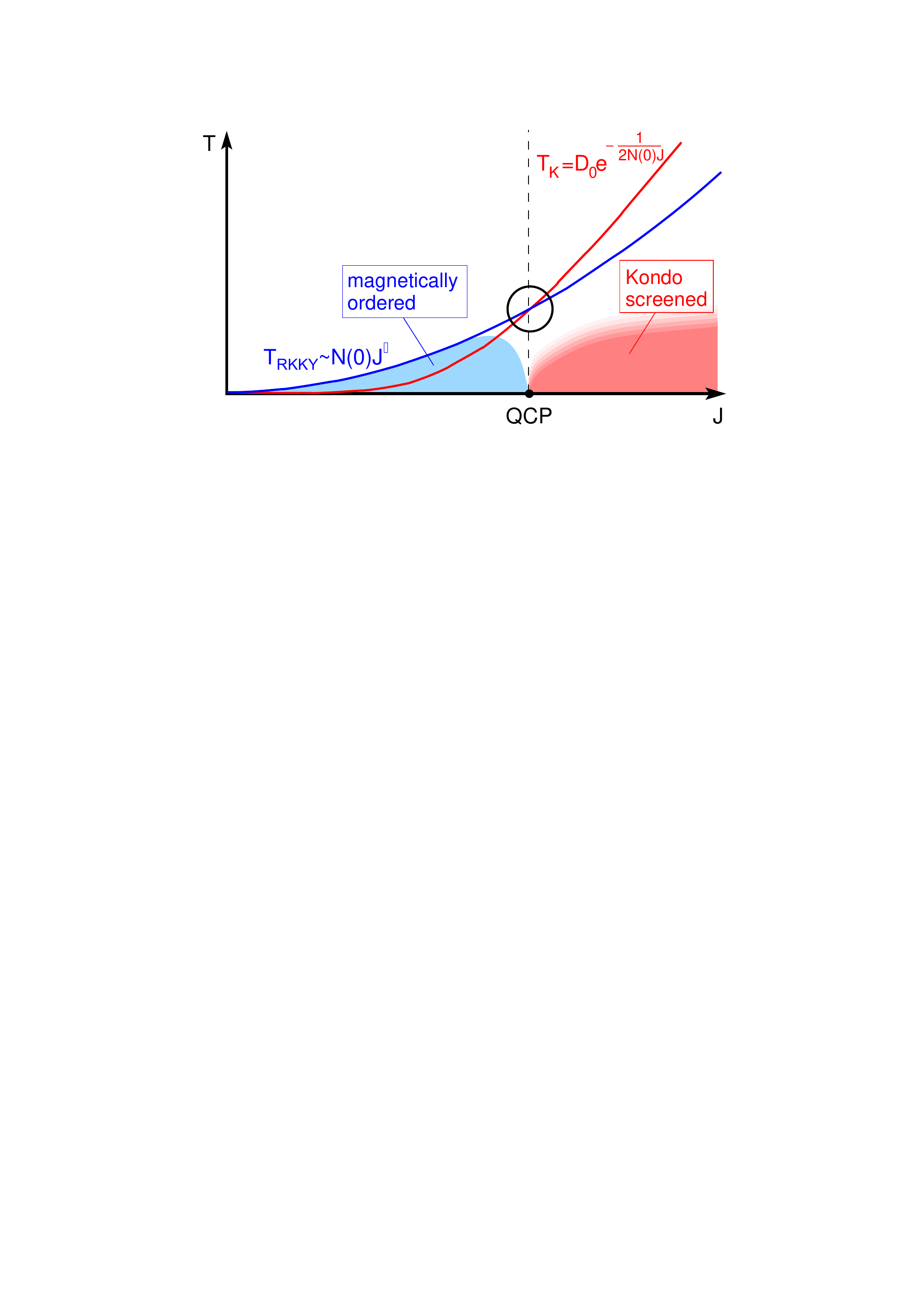}
  \caption{The Doniach phase diagram, which summarizes the interplay of the Kondo effect and the RKKY interaction adapted from Ref.\,\cite{Kroha2017}.}
  \label{fig:DoniachPD}
\end{figure}

The Doniach phase diagram (\fig{fig:DoniachPD}) describes phenomenologically the competition of the Kondo effect and the RKKY interaction as a function of the coupling strength \csJ{} between conduction electrons and local moments.
The energy scale of the Kondo interaction is proportional to \exp{-1/\csJ{}\rho(\EFermi{})} (\equ{eq:Tkondo}) and to \csJ{}\textsup{2} for the RKKY interaction (\equ{eq:Hrkky}).
For weak coupling \csJ{}, the RKKY energy scale is preponderant and characterizes the low temperature properties of the system, leading to a magnetically ordered ground state.
When the coupling \csJ{} becomes stronger, the Kondo energy scale catches up with the RKKY energy scale.
At first the ordered moments become reduced due a partial screening of the Kondo effect and the magnetic transition temperature decreases and at some point goes to zero.
Beyond that point the low temperature properties of the system are determined by the Kondo effect, which yields the non-magnetic Fermi liquid state.
Since the transition temperature goes to 0\,K, i.e.\ in the energy range where there are no thermal fluctuations, the crossover between the two regimes infers the existence of a so-called quantum critical point (\QCP{}) which was later observed in CeCu$_{6-x}$Au$_x$\cite{vLoehneysen1994}.
In the vicinity of this point the fluctuations are of purely quantum nature and can trigger new phenomena like unconventional superconductivity in the presence of magnetic ions with critical temperatures $T_\text{c}\lesssim 1$\,K; first observed in CeCu$_2$Si$_2$\,\cite{Steglich1979}.
Even the coexistence of superconductivity and a magnetically ordered phase have been observed\,\cite{Machida1984}.

\subsection{Anderson Impurity Model}\label{subsec:AIM}

The prior models were assuming that the local moments remain fully localized, i.e.\ only the spin interacts with the conduction electrons, but the charge remains unaffected.

The Anderson impurity model (\AIM{})\,\cite{Anderson1961} describes the system including the coupling of the charges of the $f$ electrons to the conduction electrons.
The \AIM{} Hamiltonian consists of three contributions, one describing the conduction electrons, one the impurity, and one their interaction expressed by a so-called hybridization $V_{\vec{k}}$.

The hybridization can be expressed by annihilation of the $f$ electron and creation of a conduction electron.
The configurations with different occupations of the impurity ion are coupled in the \AIM{}.
For example the Ce$^{3+}$ state with one $f$ electron \ket{$f^{1}\underline{L}$} ($f^1$) hybridizes with the states with zero \ket{$f^{0}$} ($f^0$) or two \ket{$f^{2}\underline{\underline{L}}$} ($f^2$) $f$ electrons, with $\underline{L}$ and $\underline{\underline{L}}$ denoting one and two holes in the ligand states, which form the conduction band.
The presence of strong Coulomb interactions will make states with double electron occupation to be high in energy.
The \AIM{} Hamiltonian can then be written as
\begin{align}
H'_{\AIM{}} &= \left( \begin{array}{ccc}
  -\Deltaf{} & \Veff{} &       \Veff{} \\
    \Veff{}  &       0 &             0 \\
    \Veff{}  &       0 & -2\Deltaf{}+\onsiteU{} \\
\end{array} \right). \label{eq:Hexc}
\end{align}
In this representation it is referred to an effective, $\vec{k}$-independent hybridization \Veff{}, the conduction electrons are approximated by one discrete energy level $\varepsilon_{\vec{k}}$\,=\,0, and the $f$ electrons have an effective binding energy \Deltaf{}\,=\,$\varepsilon_{\vec{k}}$$-$$\varepsilon_f$.
The basis of $H'_{\AIM{}}$ is given by the three configurations as (\ket{$f^{1}\underline{L}$}, \ket{$f^{0}$}, \ket{$f^{2}\underline{\underline{L}}$}) assuming only a single state for each configuration for simplicity.
At this point, it should be pointed out that the eigenstates of the Hamiltonian are admixtures of the above basis configurations due to the off-diagonal hybridization.
Consequently, the Ce ion will have a mixed valence ground state, which can be expressed as
\begin{align}
 \ket{GS} &= c_{0} \ket{$f^{0}$} + c_{1} \ket{$f^{1}\underline{L}$} + c_{2} \ket{$f^{2}\underline{\underline{L}}$}, \label{eq:mixnf}
\end{align}
with contributions $c_i^2$ of the different valence states, that depend on the strength of the hybridization and the energies of the different configurations.
When the hybridization is small compared to the energies the eigenstates will not differ significantly from the initial pure configurations, but an intermediate valence ground state will form when two configurations are close in energy.

\begin{SCfigure}
  \centering
  \includegraphics[width=0.5\textwidth]{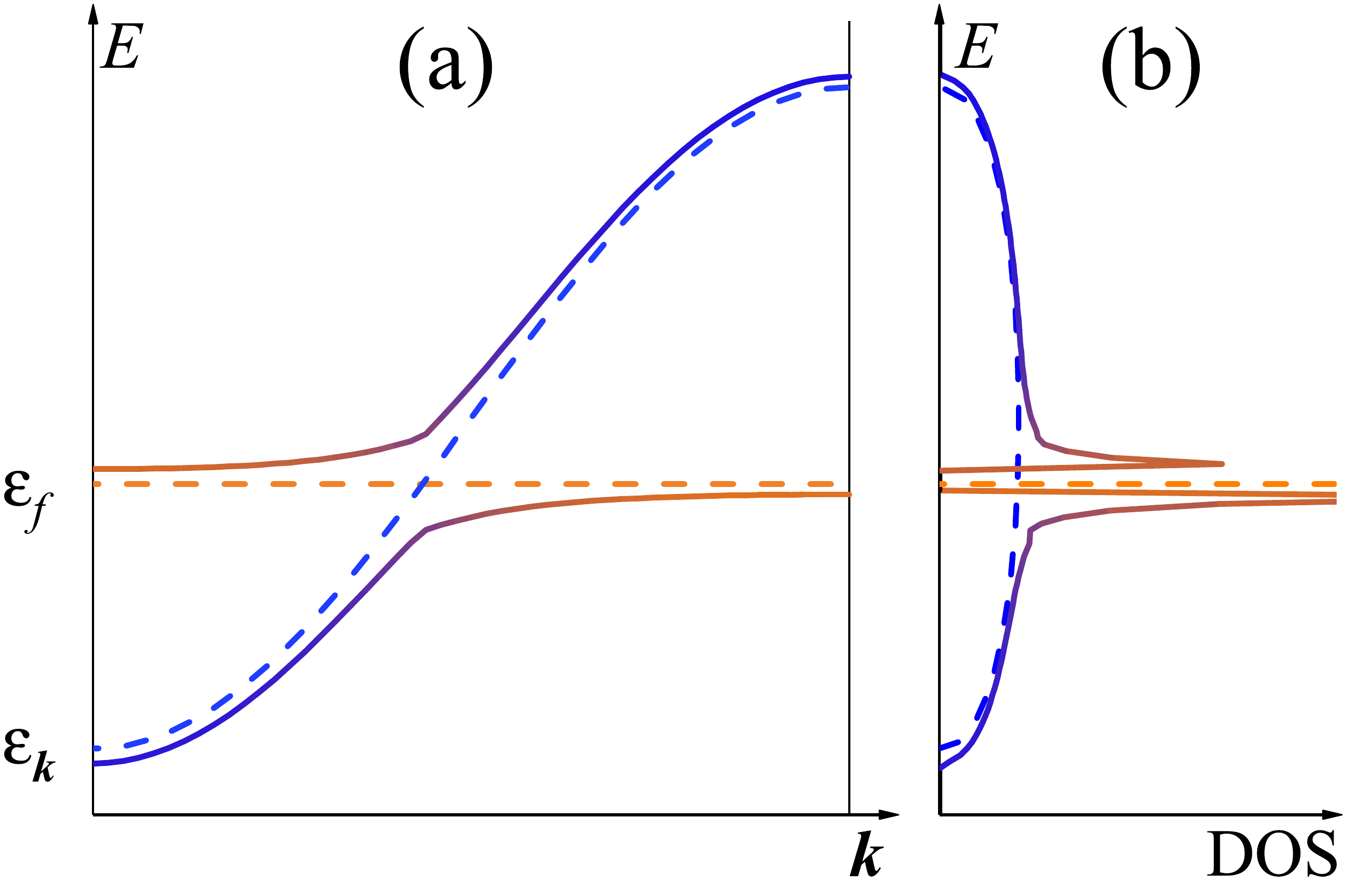}
  \caption{(a) Visualization of the eigenvalues $E(\vec{k})$ of the \AIM{} in \equ{eq:Hexc} for a Kondo lattice with equal amount of conduction and $f$ electrons, \onsiteU{}$\to$$\infty$, and a conduction band with finite dispersion. (b) The corresponding integrated density of states (\DOS{}). The color represents the character of the wavefunction, i.e.\ orange for $f$ and blue for $d$.}
  \label{fig:couplingscheme}
\end{SCfigure}
The \AIM{} hybridizes the $f$ electrons with the conduction electrons.
The hybridization modulates the energy levels of the states and a crossing of the hybridizing states will be avoided.
For example, the eigenenergies of \equ{eq:Hexc} are shifted within \Veff{}\textsup{2}/\Deltaf{} (\Veff{}\,$\ll$\,\Deltaf{}) and \Veff{} (\Veff{}\,$\gg$\,\Deltaf{}) for the $f^0$ and $f^1$ configuration in the limit \onsiteU{}~$\to$\,$\infty$.
\Fig[a]{fig:couplingscheme} shows the result of the \AIM{} for a Kondo lattice system with a dispersing conduction band and equal amount of $f$ and conduction electrons.
The dashed lines show the initial energy levels and the solid lines those of the hybridized states.
Here the crossing is lifted and a direct gap with $\Delta E$\,=\,2\Veff{} opens.
In addition an indirect gap $\Delta E$\,$\approx$\,2\Veff{}\textsup{2}/\bandwidth{} forms, which depends on the width of the conduction band \bandwidth{}.

\paragraph{Expanded Doniach phase diagram}

\begin{figure}
  \centering
  \includegraphics[width=0.9\textwidth]{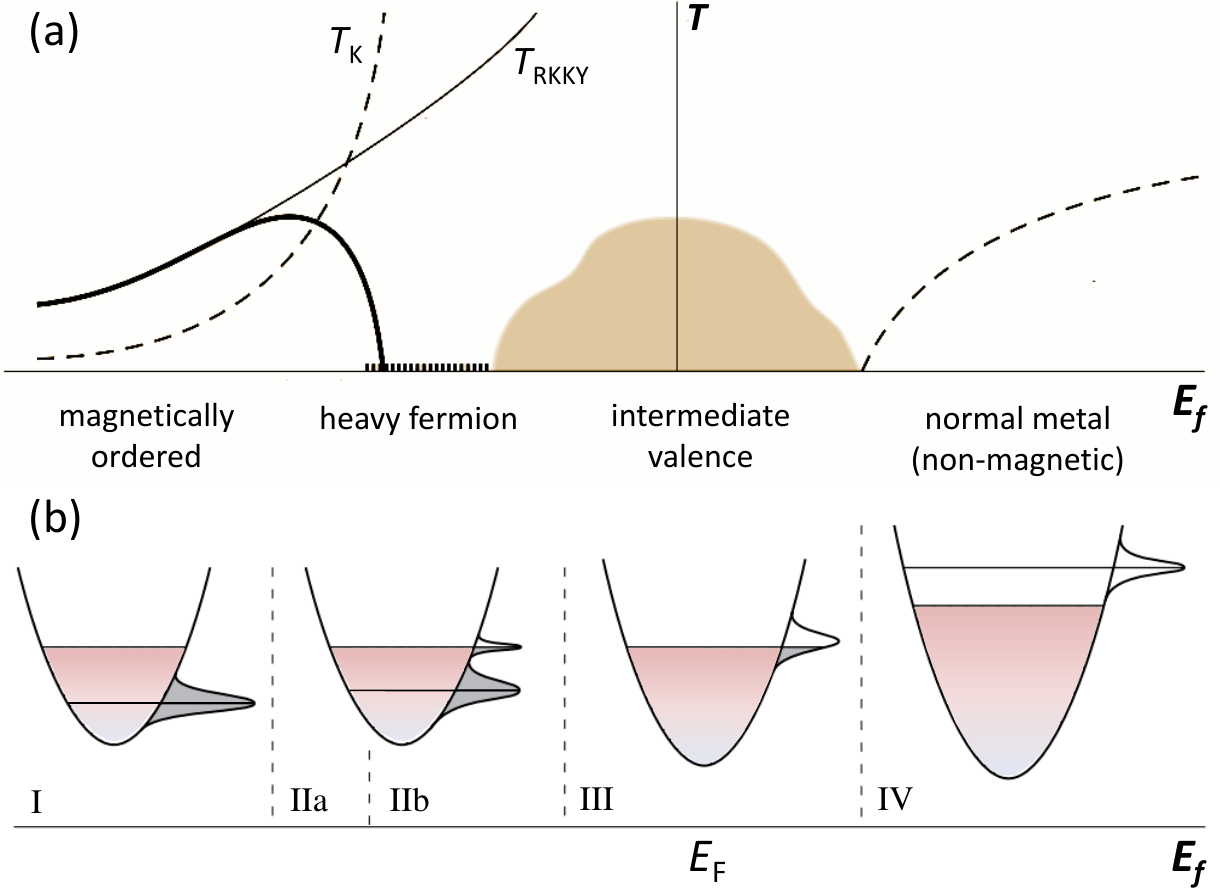}
  \caption{(a) Modified Doniach phase diagram expanded to the regime \Ef{}>\EFermi{}. (b) Sketch of the density of states with typical energy levels (black line) of \Ef{} for the different regimes, taken from Ref.\,\cite{StrigariPhD}.}
  \label{fig:exDoniachPD}
\end{figure}
\Fig[a]{fig:exDoniachPD} shows an expanded version of the Doniach phase diagram (\fig{fig:DoniachPD}).
Here the coupling strength \csJ{} is replaced by the $f$-electron binding energy (\Ef{}), which still describes the same phase diagram, as $J$ increases when the $f$ state at energy \Ef{} approaches \EFermi{}.
Below the phase diagram (\fig[b]{fig:exDoniachPD}) the \DOS{} is shown as a function of the same energy scale as in \fig[a]{fig:exDoniachPD}.
The regions I-IIb are identical to the previous diagram, starting with the magnetic ground state and the crossover to the heavy Fermi liquid state at the quantum critical point in (a) and the formation of the Kondo resonance peak in (b).
Region III expands the diagram by including the intermediate valence regime, which occurs when the $f$ state is close to the Fermi energy.
The ions are no longer magnetic and the system turns into a conventional metal when the $f$ state is finally above the Fermi energy as shown in region IV.

This phase diagram gives already a good overview of what kind of physics is happening, how the presented models relate, and how the different phases connect.
But this image is not complete as it misses the fine details, which are in the focus of modern research, like the superconducting phase at the \QCP{} and its unconventional nature or the large variety of ordering parameters for various magnetically ordered phases.
Other interesting phases turn the metallic host into an so-called Kondo insulator or exhibit multipolar order, which is invisible to many dipole techniques and thus it is also referred to as hidden order (\HO{}) phase.
The latter two material classes are in the focus of this thesis and will be discussed in the following.

\subsection{(Topological) Kondo insulators}

Some Kondo lattice systems undergo a transition from the metallic state into an insulating state upon cooling.
This transition can be interpreted by two concepts\,\cite{Dzero2016}:
One arises from the Kondo effect and the other is based on the \AIM{}.
In both pictures a gap opens when the Kondo effect and hybridization set in at low temperatures and if \EFermi{} is inside this gap, the system becomes insulating.

In the picture of the Kondo lattice effect, a gap in the conduction electron sea can open when the coupling \csJ{} is strong and the Kondo singlet state forms (see \fig{fig:Kondo}).
The two bands can be understood as an electron or hole doped Kondo insulator.
The system becomes insulating, i.e.\ the Fermi energy falls inside the gap, when every conduction electron is bound in the Kondo singlet state and no excess electrons or holes remain, i.e\ the conduction electrons can no longer move without breaking the Kondo singlet state.
In this picture the width of the gap is governed by the cost of breaking the Kondo singlet state, which is necessary to recover mobility $\Delta E$\,$\propto$\,\csJ{}.

In the picture of the \AIM{}, a crossing of the bands is avoided by the hybridization as shown in \fig[a]{fig:couplingscheme} and the respective \DOS{} in \fig[b]{fig:couplingscheme} becomes gapped.
Note that the gap can open fully only when for each conduction electron there is one $f$ electron to hybridize with, i.e.\ the number of electrons in the unit cell must be even.
Otherwise some finite \DOS{} remains inside the indirect gap.
The system becomes insulating when this condition is fulfilled and \EFermi{} is inside the gap.

The macroscopic behavior of Kondo insulators is still that of a semiconductor or semimetal, as in both models the gap size is governed by the Kondo scale.
The semimetallic behavior can occur because a dispersion of the $f$ states may change the size of the indirect gap, which may even turn the gap negative.
A high Kondo temperature and a high symmetry, as the reduced dimensionality of the band structure has less space to form pockets, favors the opening of a full gap.

When dealing with a more complex band structure, it becomes important to know which conduction bands can hybridize with the involved $f$ states.
This again greatly depends on the underlying symmetries of the states.

\begin{SCfigure}
  \centering
  \includegraphics[width=0.5\textwidth]{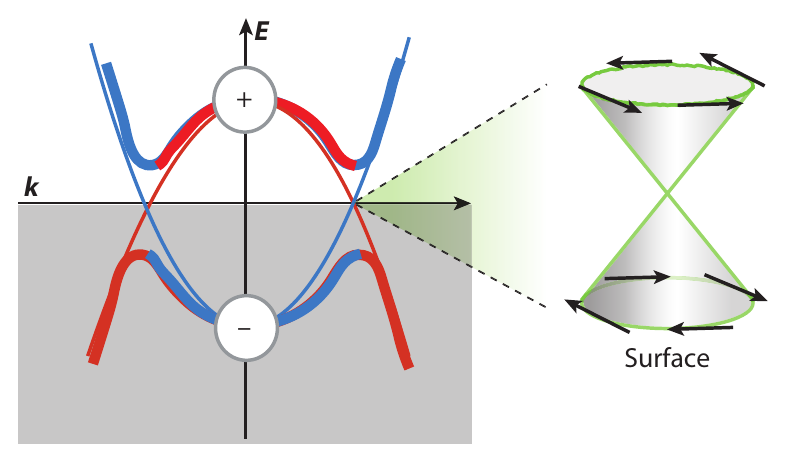}
  \caption{Scheme of a topological insulator taken from \cite[Fig.\,3b]{Dzero2016}. Two bands of opposite parity (red and blue) cross and open a gap at $E_\text{F}$ in the bulk (thick lines). The crossover creates topological protected, non dispersive surface states (thin lines) with a pinned spin texture (inset).}
  \label{fig:TKIsketch}
\end{SCfigure}

\paragraph{Topological aspects}

Even in systems with a fully open gap, as in SmB$_6$, a finite resistivity down to lowest temperatures has been observed.
Recent topological considerations present a promising explanation for the remaining metallicity. 
In the presence of strong \SOC{} and hybridizing states of opposite parity, topological protected states can occur, even if time reversal symmetry is not broken.
This is referred to as a \isingindex{} topological insulator.
Here Ref.\,\cite{Hasan2010} is summarized to give a general overview, but literature more specific to Kondo insulators is also suggested\,\cite{Dzero2016} to the interested reader.

In the presence of strong \SOC{} the states are at least 2-fold degenerate with half integer total momentum \qnJ{} (Kramers theorem).
The so-called Kramers doublets are protected as long as time reversal symmetry is present and can split between the time reversal invariant momenta in band structure.
The crossing of the Fermi level in between two time reversal invariant momenta is topologically protected for an odd number of spin polarized bands.
This is expressed by the Ising index \isingindex{}\,=\,-1 for topological insulators or \isingindex{}\,=\,+1 for conventional insulators.
This is just an easy example as in 3-dimensional systems with low symmetry (i.e\ without inversion symmetry) the calculation of the \isingindex{} becomes more difficult.

The second requirement (beside strong \SOC{}) is a different parity of the hybridizing states, such as the 4$f$ and 5$d$ of the rare earth or the 5$f$ and 6$d$ of the actinides.
This is needed to allow for \isingindex{}\,=\,-1 in the Kondo insulator without band crossings in the bulk.
The details of how the surface states cross the gap depend also on the character of the $f$ state which hybridizes with the conduction electrons and which is responsible for the formation of the Kondo insulating state.

\Fig{fig:TKIsketch} shows a sketch of a topological Kondo insulator.
The inversion of the states of opposite parity (red and blue) implies necessarily the existence of topological protected spin-polarized surface states.
The crossing in the bulk is prevented when the Kondo insulating state forms.
The topological surface states potentially have a linear dispersion (with a vanishing effective mass), forming a Dirac cone, and have a fixed spin texture (which prevents back scattering).

\subsection{Multipolar order}
Some heavy fermion materials show an ordering phase transition upon cooling, that goes along with the usual peak in specific heat and the loss of entropy, but without the observation of any usual symmetry breaking, e.g.\ due to crystal structure, charge order, or magnetic dipole ordering.
This is because ordering moments beyond dipole are hidden to dipole restricted techniques.
Such phases are often referred to as hidden order (\HO{}) because the identification of the ordering parameter is challenging.

\begin{SCfigure}
  \includegraphics[width=0.5\textwidth]{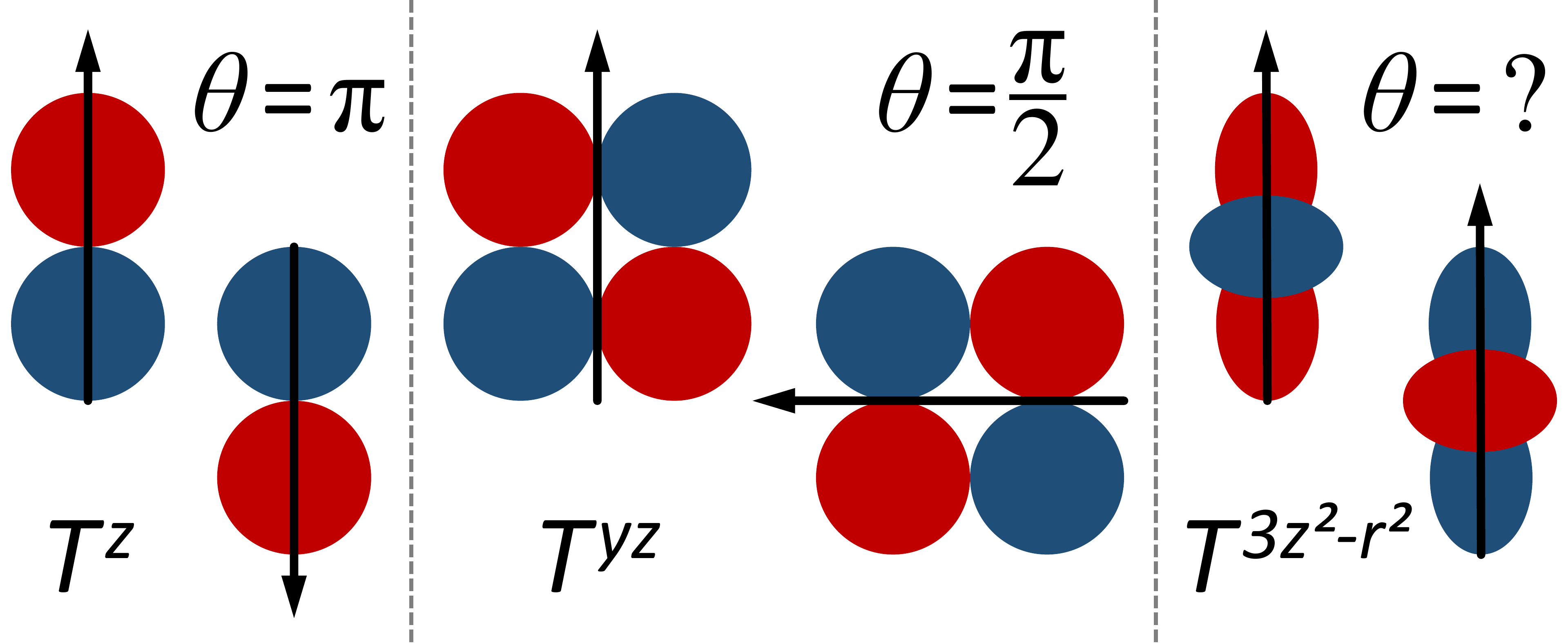}
  \caption{2D scheme of multipolar ordering reproduced from \cite[Fig.\,7]{Pi2014b} showing the usual axis flip of a dipole ($T^{z}$), rotation of a quadrupole ($T^{yz}$), and phase change of a quadrupole ($T^{3z^2-r^2}$) moment. The arrows are just guide to the eyes.}
  \label{fig:multipolarordering}
\end{SCfigure}

Multipolar order can occur in systems with both, orbital and spin, degrees of freedom in the presence of strong \SOC{}\,\cite{Santini2009}.
Multipolar order requires a high local point symmetry, which does not fully lift all orbital degrees of freedom.
It is not straightforward to see what actually orders.
Compared to an axis flip of a dipole moment, the multipolar moments can order similarly in terms of a rotation of the higher multipoles but also differently by a phase change, which no longer relates to a rotation (see \fig{fig:multipolarordering})\,\cite{Pi2014a, Pi2014b}.
Consequently, also the spin or charge densities may or may not be affected by the multipolar ordering.

Usually ingenious models have to be developed and must be tested by deduced, measurable parameters.
The characterization of \HO{} parameters is one of the big challenges that some heavy fermion compounds impose.
Insight to some of these \HO{} phases can be obtained by tuning the system, e.g.\ by breaking inversion symmetry with a magnetic field, but not all hidden orders could be identified this way (see e.g.\ URu$_2$Si$_2$).
The mathematical descriptions depend highly on the involved moments and their symmetries\,\cite{Santini2009}.

An alternative introduction to multipolar ordering is given for transition metals in Ref\,\cite[Sec.\,3]{WitczakKrempa2014}.
Their Ansatz is based on localized, atomic like states with strong \SOC{}; these criteria are typical for the rare earth and actinide $f$ electrons.

\clearpage
\section{Non-resonant Inelastic X-ray Scattering (NIXS)}\label{sec:expnixs}
X-ray spectroscopy is a handy tool to get insight into the microscopic nature of matter, whereby core-level spectroscopy on single crystals gives direct insight into the angular anisotropies of the charge densities\,\cite{Haverkort2007, Willers2012}.

In x-ray spectroscopy the interaction between electrons and photons causes a transfer of momentum and energy.
The process is called elastic if no energy is transfered and inelastic otherwise.
Elastic scattering yields insight into the crystal structure and other periodic textures such as ordering, whereas the inelastic response yields insight into the electronic structure.
A local probe of the electronic structure can be achieved by utilizing core-levels of the specific element.
Different methods, like different x-ray absorption spectroscopy (\XAS{}) techniques, resonant inelastic x-ray scattering (\RIXS{}), and photoelectron spectroscopy (\PES{}) are frequently used for their element specific local probe of the electronic structure of matter\,\cite{deGroot2008}.
However, they all involve resonant interactions.
For resonant interactions, the electrons are mostly driven by the electric photon field and the linear response of the electrons makes all these techniques subject to dipole selection rules.

With modern x-ray sources and optics, it became feasible to observe non-resonant inelastic interactions with x-rays, such as in optical Raman scattering.
Non-resonant inelastic x-ray scattering (\NIXS{}) complements the resonant techniques and provides valuable unique properties.
The non-resonant interactions are driven by the interaction of the electron with the potential, and not by the electrical field of the photon, such that hard x-rays allow one to overcome the dipole selection rules as introduced in \chap{sec:calcnixs}.
Being non-resonant requires photon energies well above the energies of the investigated transitions and the scattered photons can be observed directly.
In the following the experimental aspects of \NIXS{} will be introduced with the focus on core-level excitations.

\subsection{General aspects \label{subsec:nixs_setup}}

\begin{figure}
  \centering
  \includegraphics[width=0.85\textwidth]{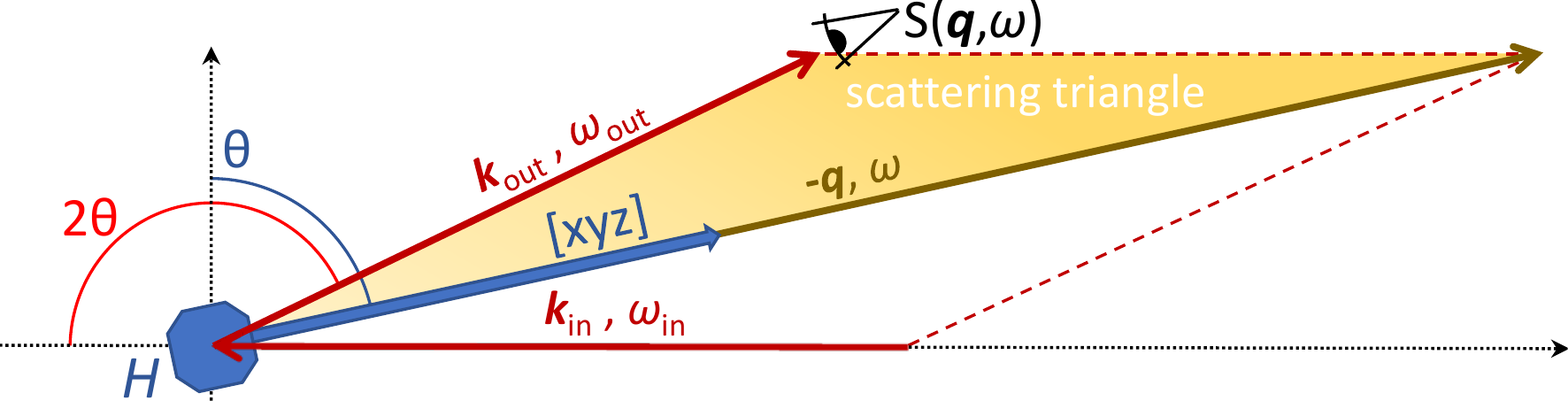}
  \caption{Sketch of the scattering geometry including important parameters of the setup (red) and the system (blue).}
  \label{fig:NIXSgeometry}
\end{figure}
\NIXS{} is a photon-in photon-out technique.
The incoming photon has a certain momentum \kin{} and energy $\hbar$\oin{}, whereas the observed photon has a different momentum \kout{} and energy $\hbar$\oout{}.
The difference is due to the momentum transfer \vecq{}\,=\,\kin{}\,-\,\kout{} and energy transfer $\hbar$\otr{}\,=\,$\hbar$\oin{}\,-\,$\hbar$\oout{} to the electron of the system $H$, fulfilling momentum and energy conservation.
This is graphically shown by the scattering triangle in \fig{fig:NIXSgeometry}, which defines the scattering plane.
The scattering angle 2\thS{} between the incoming and outgoing photon results in a momentum transfer along its bisector at an angle \thS{}, when \oin{}\,=\,\oout{} and one speaks of a so-called "\thS{}\,--\,2\thS{}" geometry.

The intensity of the photons within the energy window $\diff\hbar\omega$, that are observed within the solid angle $\diff\Omega$ is given by the double differential cross section
\begin{align}
\frac{\diff^2\sigma}{\diff\Omega \, \diff\hbar\omega} &= 
 \underbrace{ \vphantom{\sum_{\ket{f}}} \frac{\oout{}}{\oin{}} \left( \frac{\echarge{}^2}{2\emass{}} \right)^2 \abs{ \vec{\epsilon}_\text{out}^* \cdot \vec{\epsilon}_\text{in} }^2 }_{\text{\scalebox{1.4}{Thomson}}}
 ~ \underbrace{ \sum_{\ket{f}} \abs{ \bra{f} \opT{} \ket{i} }^2 ~ \del (E_\text{f} - E_\text{i} - \hbar\omega) }_{\text{\scalebox{1.4}{\sqw{}}}}.
\end{align}
It can be separated in the Thomson term and the dynamic structure factor (\sqw{}).
The Thomson term is structureless and almost constant in case of small variation of the photon energies.
It accounts for the photon polarization $\vec{\epsilon}$.
The physics of the investigated system is governed by \sqw{}.
In \sqw{} the first term represents the transition probability from the initial state \ket{i} to a final state \ket{f} and the $\del$-function takes care of the energy conservation.
The transition operator \opT{} of the scattering process allows for different final states \ket{f}, which give rise to specific spectroscopic features.\,\cite{Schuelke2007, Rueff2010, Kotani2001}

\subsection{Spectroscopic features of S(\textit{q},\,\(\omega\))\label{sec:nixsspectra}}
\begin{figure}
  \centering
  \includegraphics[width=0.8\textwidth]{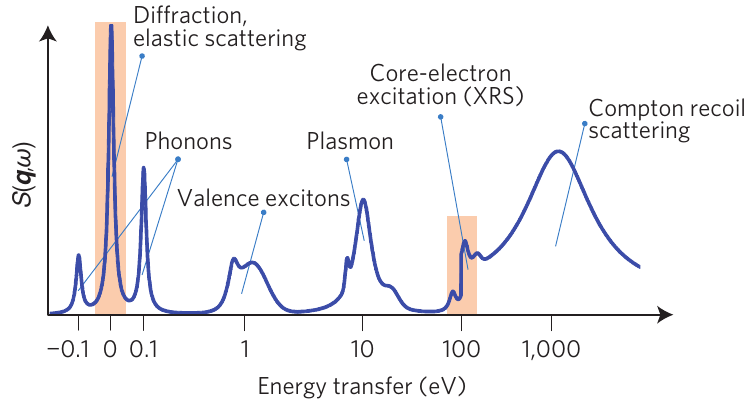}
  \caption{Illustration of the different features visible in a NIXS experiment from Ref.\,\cite{Huotari2011}. The scattering intensity is plotted versus the energy transferred to the system ($\hbar\omega_\text{in}$-$\hbar\omega_\text{out}$ in \Fig{fig:NIXSgeometry}).}
  \label{fig:XRSlogSpectrum}
\end{figure}
\Fig{fig:XRSlogSpectrum} gives an overview of the different phenomena and their typical energy scales that can be observed in a \NIXS{} experiment:

At zero energy transfer the scattering is elastic.
Only the states at the Fermi surface contributes in the diffraction process and the signal is greatly enhanced when the momentum transfer matches a reciprocal lattice vector, i.e.\ when the Bragg condition is fulfilled.

In valence \NIXS{} excitations which arise from the valence band are observed.
Here, low energy excitations are typically investigated using optical Raman scattering.
Typical features are vibrational excitations (phonons), magnetic excitations (magnons), local valence excitations, or collective charge oscillations (plasmons).

At higher energies core electrons are excited into valence states.
This gives rise to an element specific real space picture of the local charge densities.

Moreover, the emission of photo electrons can be observed by recoiled photons, via a quasi elastic scattering process between the electron and photon, where the transferred energy yields the kinetic energy of the electron in the vacuum state.
Note that the scattering still depends on the initial binding energy and momentum of the electron, thus giving rise to a very broad peak often overlapping with the core electron excitations.\footnote{For convenience, the position of the maximum and its width have been determined empirically from the performed experiments to 4\,eV$\cdot(\abs{\vec{q}}\,\AA)^2$ and \FWHM{}\,$\approx$\,10\,eV$\cdot(\abs{\vec{q}}\,\AA)^{1.5}$. The quadratic scaling of the position is supported by the linear scaling of the photon energy with the momentum, whereas it is quadratic for the kinetic energy of free electrons.}

\paragraph{Core level spectroscopy in NIXS}

The core-electron excitations in the \NIXS{} spectrum arise from the distinct atomic core levels of the individual elements and are identical to those observed in an \XAS{} experiment and can be well described by a  local many body approach.
The element specific probe creates a positively charged core hole, which allows for a local projection of the valence states, i.e.\ it creates excitonic states in the region of the onset of the unoccupied \DOS{} of the continuum states\,\cite{Soininen2005, Larson2007, Haverkort2007, Gordon2008, Gordon2009, Bradley2010, Caciuffo2010, Rueff2010, Hiraoka2011, SenGupta2011, Bradley2011, Laan2012}.

\begin{figure}
  \centering
  \begin{minipage}[b]{0.53\textwidth}
    \includegraphics[width=\textwidth]{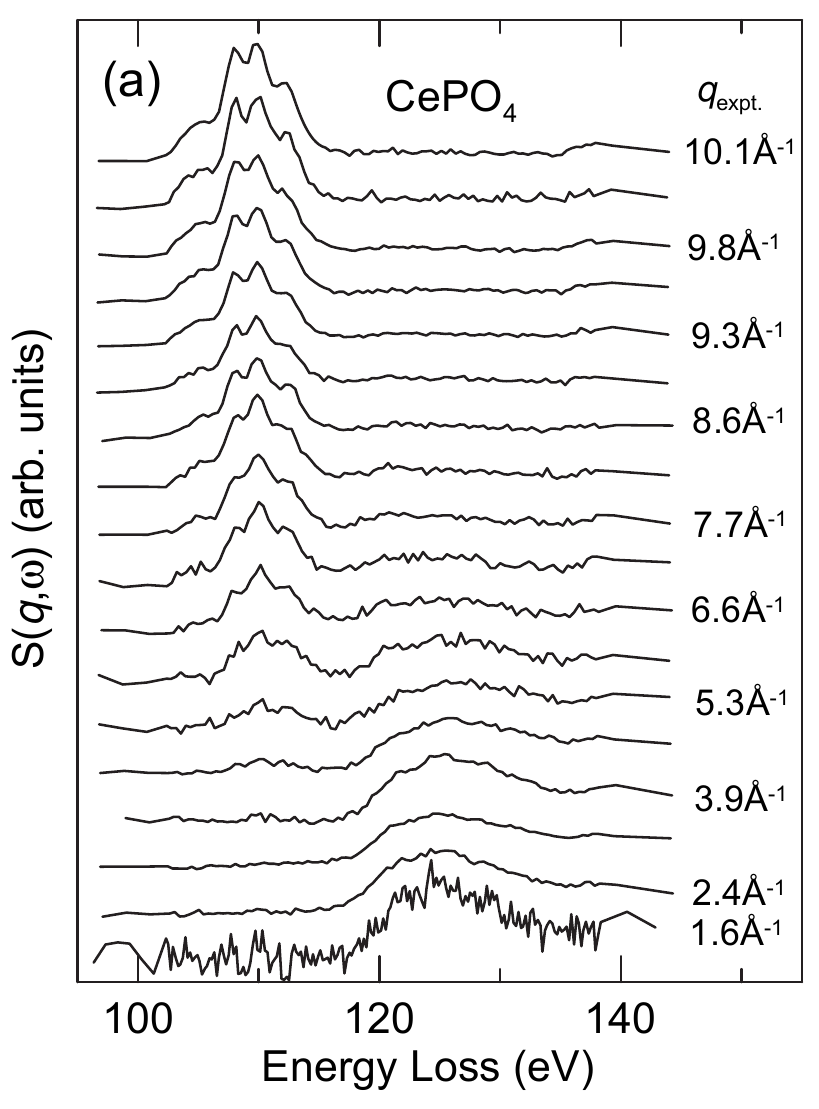}
	\vspace{3mm}\phantom{ }\\
    \includegraphics[width=\textwidth]{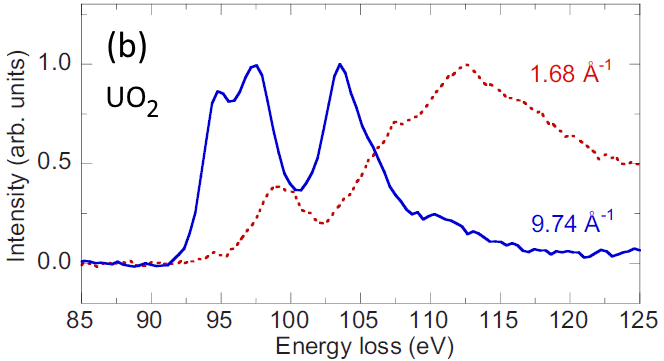}
	\vspace{-2mm}\phantom{ }
  \end{minipage}
  \begin{minipage}[b]{0.45\textwidth}
  \includegraphics[width=\textwidth]{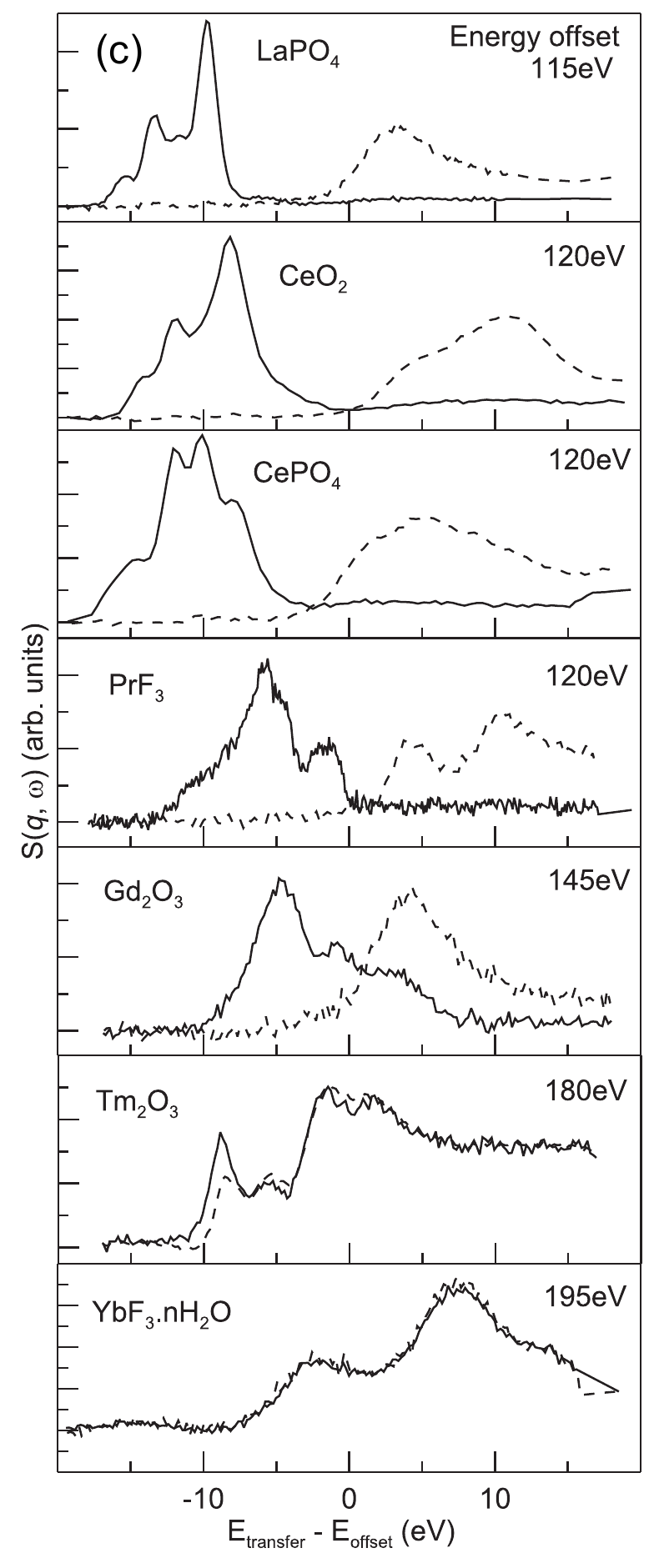}
  \end{minipage}
  \caption{Length of \vecq{} dependence of \sqw{} for some rare earth \edge{N}{4,5} and the \edge[U]{O}{4,5} edges: (a) \NIXS{} spectra of CePO$_4$ for various momentum transfers from Ref.\,\cite{Gordon2008}. (b) \NIXS{} spectra of UO$_2$ for small and large momentum transfers from Ref.\,\cite{Caciuffo2010}. (c) \NIXS{} spectra of different rare-earth compounds for low momentum transfers \absq{}\,<\,2\,$\AA^{-1}$ (dashed) an high \absq{}\,$\approx$\,8-10\,$\AA^{-1}$ (solid) from Ref.\,\cite{Gordon2011}.}
  \label{fig:NIXSmultipoles}
\end{figure}
\Fig[a]{fig:NIXSmultipoles} shows the \NIXS{} spectrum of the \edge[Ce]{N}{4,5} edges ($4d\rightarrow4f$) of polycristalline CePO$_4$ for different sizes of the momentum transfer (\absq{}).
Initially only a broad peak above 120\,eV is observed, but at \absq{}\,$\approx$\,5\,$\AA^{-1}$ this peak loses intensity and new features below 120\,eV occur which are strong at \absq{}\,$\approx$\,10\,$\AA^{-1}$.
The stronger bound lines at lower energy transfer appear to be narrower, i.e.\ they appear to have a longer lifetime.
The material is still the same, but the transition operator changes, as higher multipole transitions become accessible and other final states are reached.\,\cite{Soininen2005, Larson2007, Haverkort2007, Gordon2008, Gordon2009, Bradley2010, Caciuffo2010, Rueff2010, Hiraoka2011, SenGupta2011, Bradley2011, Laan2012}
\Fig[(b-c)]{fig:NIXSmultipoles} shows \NIXS{} spectra for low and high momentum transfer \absq{} of different \edge[RE]{N}{4,5} edges and of the \edge[U]{O}{4,5} edges.

The excitonic features can be understood by the lowest final state configuration $|\underline{c}f^{n+1}\rangle$ in the presence of the core hole $\underline{c}$ when mixed valence effects can be neglected.
The large \absq{} \NIXS{} spectrum (solid lines in \fig[c]{fig:NIXSmultipoles}) of Ce$^{4+}$ in CeO$_2$ looks alike the one of La$^{3+}$, as they have the same electronic configuration, but the CePO$_4$ spectrum with Ce$^{3+}$ configuration differs noticeably.
Hence \NIXS{} is sensitive to the configuration.
The \NIXS{} signal of the excitonic states gets weaker when the filling of the $f$ shell increases from La to Yb as the number of accessible final states with $|\underline{c}f^{n+1}\rangle$ configuration decreases.

Weakly bound ($\approx 100\,eV$) core levels appear to be suited best for \NIXS{}.
They show a rich multiplet structure due to on-site Coulomb interactions due to the core-electron excitonic state, which are exceeding the lifetime broadening.
The larger expansion of weaker bound core levels also gives rise to higher multipole transitions at momentum transfers as low as \absq{}\,$\approx$\,10\,$\AA^{-1}$, yet still before or at the onset of the Compton background.
Momentum transfers of \absq{}\,$\approx$\,10\,$\AA^{-1}$ require hard x-rays of the order of 10\,keV, such that this photon-in photon-out technique is truly bulk sensitive.
Probing depths are of the order of 10\,\(\mu\)m even for dense rare earth and actinide compounds (see \tab{tab:samppendepth}) and advanced sample environments become available as no vacuum is mandatory.
The interpretation of the data is as straight forward as for an \XAS{} experiment, i.e.\ no intermediate states as for resonant inelastic x-ray scattering are involved.

If the same experiment is performed on a single crystal, the directional dependence of \sqw{} provides information about the initial state symmetry\,\cite{Willers2012, Rueff2015} similar to a linear dichroism \LD{}-\XAS{} experiment\,\cite{Hansmann2008}.
Since with \NIXS{} higher than dipole signals can be reached even anisotropies of cubic materials can be detected and the observed features appear more excitonic.

In summary, \NIXS{} performed with a momentum transfer \absq{}\,$\approx$\,10\,$\AA^{-1}$ is bulk sensitive and enables higher multipole transitions.
These transitions excite into strongly bound states showing narrow features in the core level \NIXS{} spectra of $f$ electron ions and allow to measure anisotropies in systems with a higher than 2-fold symmetry.

\subsection{Experimental setup \label{subsec:nixs_setup}}
\begin{table}
  \centering
  \caption{Overview of the x-ray penetration depth for the investigated samples obtained from Ref.\,\cite{xrayinteractions} for 50\% transmission of 10\,keV photons.}
  \label{tab:samppendepth}
  \begin{tabular*}{\textwidth}{@{\extracolsep{\fill}}ccccccc}
    & CeB\textsub{6} & CeRu\textsub{4}Sn\textsub{6} & SmB\textsub{6} & UO\textsub{2} & URu\textsub{2}Si\textsub{2} & \\
    \hline
    & 13\,$\mu$m & 10\,$\mu$m & 16\,$\mu$m & 5\,$\mu$m & 14\,$\mu$m & \\
  \end{tabular*}
\end{table}

To make transitions beyond the dipole limit prominent, momentum transfers of the order of 10\,\(\text{\AA}\)\textsup{-1} are required.
On the other hand, spectroscopic features have to be observed and an energy resolution of $\lesssim$1\,eV is desired in addition.
In addition, the flux should be as high as possible, which for a given resolution is towards lower photon energies for an undulator source with given brilliance\,\cite{AlsNielsen2011}.
This implies a backscattering geometry, which maximizes the momentum transfer for a fixed photon energy and \absq{}\,$\approx$\,10\,$\AA^{-1}$ can be realized at photon energies as low as 10\,keV.

\begin{figure}
  \includegraphics[width=\textwidth]{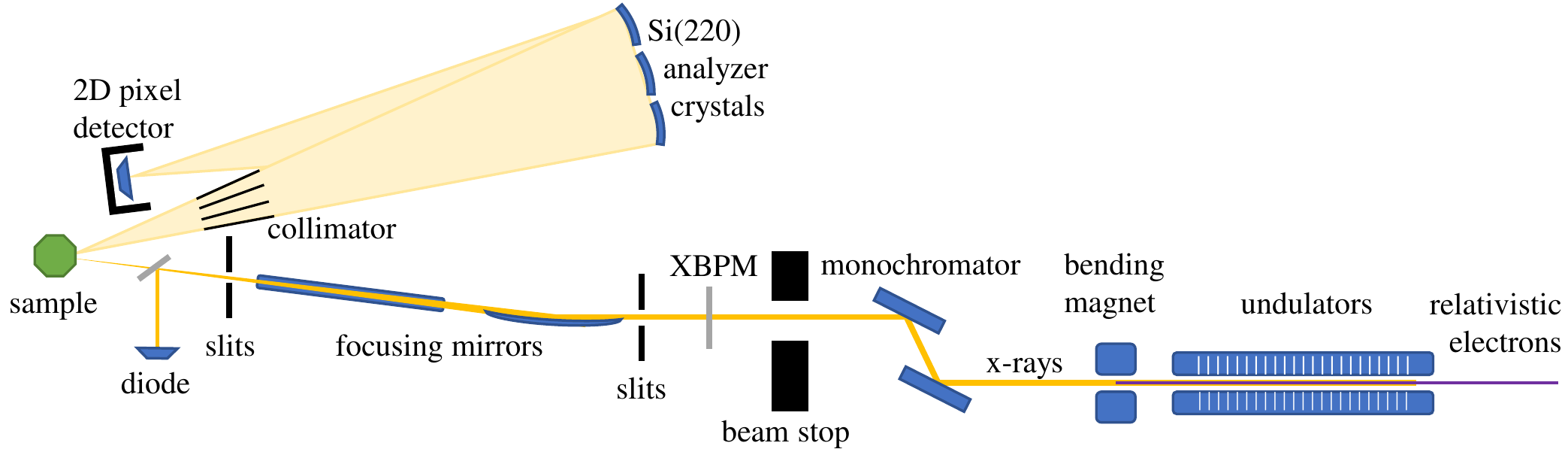}
  \caption{Sketch of a typical beamline design for NIXS without the vacuum parts.}
  \label{fig:beamline}
\end{figure}

The experiments were performed at the P01 beamline at the PETRA~III in Hamburg and the ID20 beamline at the ESRF in Grenoble.
\Fig{fig:beamline} shows the design of a typical \NIXS{} beamline.
The following list gives a summary of the utilized parts:
\begin{description}
\item[Relativistic electrons:] A beam of electrons at relativistic velocities is generated by a synchrotron accelerator and provided within a storage ring.
\item[Undulator:] A brilliant light source, providing a focused and polarized x-ray beam with high intensities. A periodic array of magnets forces the electron beam on an oscillating orbit. Hereby, Bremsstrahlung is generated and the high energy photons are mainly emitted along the electrons propagation direction. Interference of the periodically emitted light further enhances and focuses the generated beam intensities.
\item[Bending magnet:] It separates the electron and photon beam and keeps the relativistic electrons within the closed storage ring.
\item[Monochromator:] High quality single crystals. Bragg reflections are used to let pass only specific energies from the generated \textit{white} x-ray beam. At least two crystals are used. The second crystal keeps the beam parallel to the incoming beam and has a better resolution, as it experiences much less heat load. For the performed experiments the Si(311) reflection with about 300\,meV resolution at 10\,keV is used in most cases. Note that for soft x-ray beamlines below about 2\,keV gratings are used instead, as the wavelength becomes larger than crystalline lattice spacings.
\item[Beam stop:] Beside offering radiation protection, it blocks the \textit{white} x-ray beam and reduces stray radiation coming to the setup.
\item[XBPM:] The X-ray Beam Position Monitor feeds a feedback to the second monochromator crystal to keep the beam position fixed during energy scans.
\item[Slits:] As the beam stop they block divergent or scattered parts of the light.
\item[Focusing mirrors:] The divergence of x-rays generated by undulators yields typical beam sizes of few mm\textsup{2} at the sample. Total reflection condition on ellipsoidal bent, polished surfaces focuses the beam size to about 10x10\,\(\mu\)m\textsup{2}. Also they filter higher order Bragg reflections of the monochromator, which require lower angles for total reflection.
\item[Diode:] A 300\,\(\mu\)m silicon diode measures the scattered light just before the sample, monitoring the incoming beam for normalization.
\item[Collimator:] Beam stop for the scattered beam, which reduces the stray light hitting the analyzers by only giving line of sight to the sample.
\item[Analyzers:] Similar to the monochromator they are high quality single crystals using Bragg conditions to reflect only specific energies from the scattered light into the detector.
\item[Detector:] A 300\,\(\mu\)m silicon Medipix3 chip\cite{Ballabriga2010} based 2D pixel detector counts the number of scattered photons reflected from the analyzers.
The black box around indicates some shielding against x-rays, that otherwise scatter directly into the back of the detector.
\item[Vacuum:] Most optics are kept in vacuum of 10$^{-7}$\,mbar or below, to reduce the in-burning of adsorbents in the presence of the incident high intensity hard x-rays and the flightpath is kept in vacuum of about 10$^{-1}$\,mbar to reduce absorption and air scattering, e.g.\ in the analyzer box to reduce the background on the detector.
This can alternatively be achieved with a He atmosphere.
\item[Windows:] Windows holding the vacuum and facing the scattered beam can easily be made from 25-50\,\(\mu\)m polyimide (Kapton$^\text{\textregistered}$) for high transmission.
Windows passed by the incident beam should be more resistant, especially for a focused beam;
common workarounds are made of beryllium and aluminum.
Beryllium is hard to fabricate and needs special care to handle, whereas pure aluminum absorbs a lot.
A sandwich of 50\,\(\mu\)m polyimide and 12\,\(\mu\)m aluminum glued with \(\lesssim1\,\mu\)m EPO-TEK$^\text{\textregistered}$~301-2 has proven its ability to withstand the direct beam over weeks at decent transmission, whereas nitrogen or helium flow showed no effect to the polyimide lifetime.
\end{description}

The elements from source to sample show a rather general design for modern synchrotron beamlines.
The collection of the scattered beam is what is most characteristic for a \NIXS{} beamline, i.e.\ the geometry of analyzer and detector, which are going to be discussed in more detail.

\paragraph{NIXS spectrometer}
As motivated before, it is desired to operate in backscattering geometry.
Due to the horizontal polarization of the incoming photon the scattering plane is chosen vertical at the P01 beamline, to circumvent the geometric cos\textsup{2}\,\thS{} loss in intensity.
The experiments at the ID20 beamline were performed in the horizontal scattering plane, where the highest scattering angles could be realized.
Because of low count rates it helps to analyze and detect a large solid angle, but simultaneously the background should be kept small.
The resolution of a \NIXS{} instrument is designed to resolve spectroscopic features without unnecessary cutting the bandwidth, i.e.\ it should be smaller than the multiplet splitting of few eV, but not much smaller than the core hole life time broadening, which is a few hundred meV as empirically found for the investigated compounds.

\begin{figure}
  \centering
  \includegraphics[width=0.7\textwidth]{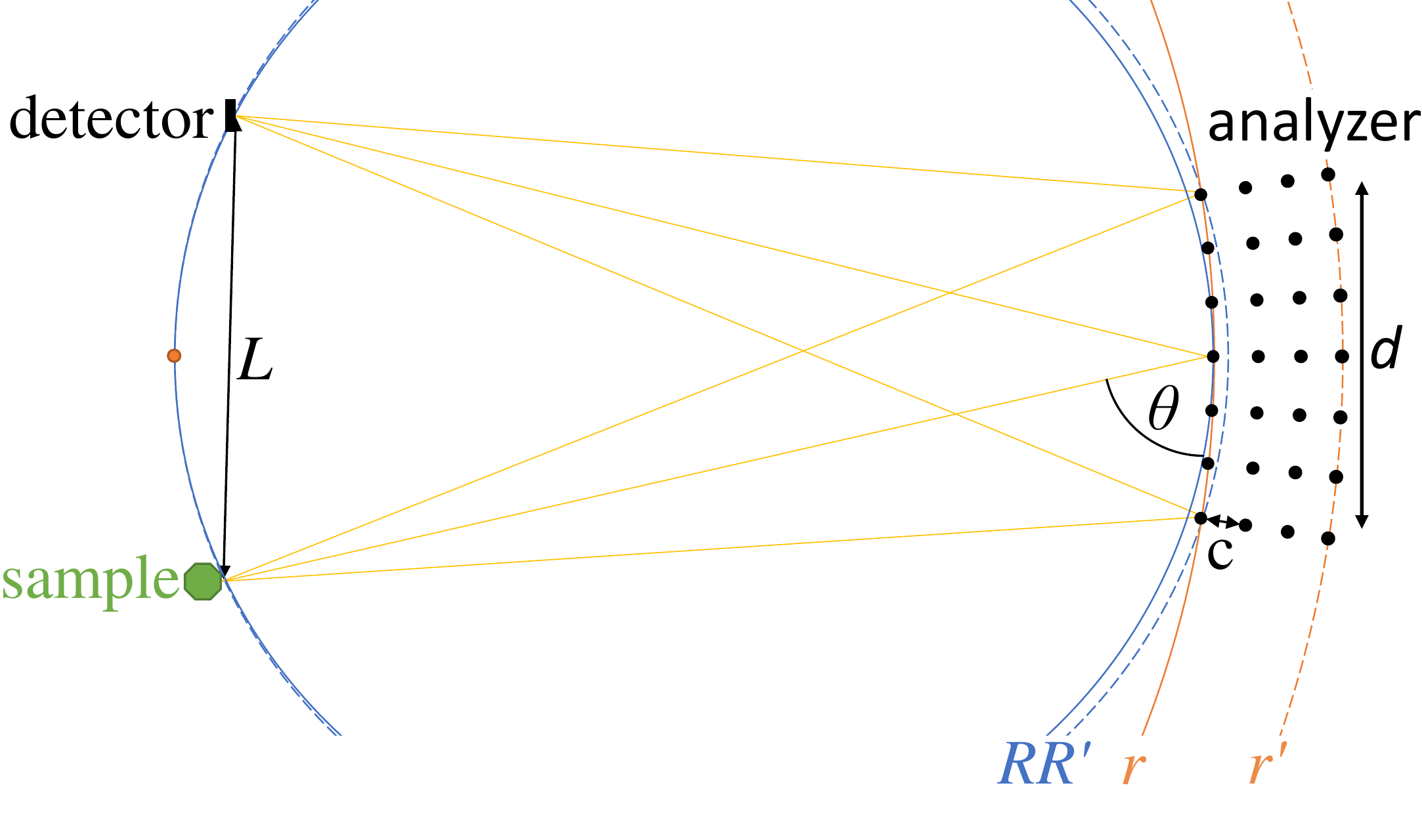}
  \caption{Sketch of the Rowland geometry.}
  \label{fig:Rowland}
\end{figure}

Spherically bent analyzer crystals with a bending radius $r$\,=\,1\,m optimize the focusing when placed on the Rowland circle with radius $R$=$\textfrac{r}{2}$ along with sample and detector as shown in \fig{fig:Rowland}\,\cite{Rueff2010}.
This maximizes the signal density on the detector allowing a more compact detector design and reducing the background collected within that area.
Both \NIXS{} beamlines use 12 Si(2\textit{n} 2\textit{n} 0) analyzer crystals close to backscattering (2\(\thB{}\approx155^\circ\)).
Following Rueff and Shukla\cite{Rueff2010} beside the intrinsic Darwin width of the Bragg reflection, three geometrical aspects affect the resolution:

\textbullet The variation of the Bragg angle $\thB{}$ due to finite source size $\Delta s\approx r\Delta\thB{}$
\begin{align}
\frac{\Delta\omega}{\omega} &= \Delta\thB{} \, \cot{\thB{}} &(< 3\cdot10^{-6}) \label{eq:specressource}
\intertext{\textbullet The Johann error caused by the deviation of the analyzer surface from the Rowland circle away from the center ($R'$-$R$ in \fig{fig:Rowland})} 
\frac{\Delta\omega}{\omega} &= \frac{d^2}{2 r^2} \, \cot^2{\thB{}}, &(\approx 1\cdot10^{-5}) \label{eq:specresjohann}
\intertext{\textbullet The change of $R$ and the lattice spacing $c$ due to the finite penetration $\Delta r$ into the bent analyzer crystal}
\frac{\Delta\omega}{\omega} &= 2 \frac{\Delta r}{r} \, \abs{\cot^2{\thB{}} - \nu}. &(\approx 1\cdot10^{-4}) \label{eq:specresbending}
\end{align}

Here $\nu$ is the material Poisson ratio ($\approx$0.22 for Si\cite{Dolbow1996}) describing the bending induced variation of $c$.
The values in the brackets are estimates for a silicon analyzer with bending radius \textit{r}=1\,m and diameter \textit{d}=0.1\,m, thickness \(\Delta\)\textit{r}=300\,\(\mu\)m, and a Bragg angle \thB{}\,$\approx$\,87$^\circ$.
The upper limit $\Delta s$<55\,$\mu$m is estimated by the detector pixel size.
The change of lattice spacing term dominates the overall resolution (here $\approx$0.7eV at the 9.7\,keV Si(660) reflection), which is desired in terms of maximizing intensity since this term collects the full energy window over the full solid angle.

The detector counts all events above a certain energy.
As the analyzer energy is kept fix, this detector threshold can be optimized accordingly: Between 1/2 and 2/3 of the selected analyzer energy have shown to be the ideal setting, reducing the background, without cutting much of the signal.

\subsection{Concept of the experiment}\label{sec:expnixsconcept}

The idea is to utilize the directional dependence of \vecq{} in the \NIXS{} core level spectra to obtain an element specific insight into the orbital symmetry of the valence states.
Single crystals are required, so that the directional dependence is not averaged out within the probed volume.
Ideally, the element of interest occupies only one single lattice site and the other elements in the compound should not have core levels within the energy range of the core level of interest.

For the experiment certain crystalline directions are of special interest -- typically the high symmetry points of the local point symmetry.
\NIXS{} spectra are acquired with these directions being aligned parallel to \vecq{}.
Note, for the higher multipole transitions in \NIXS{} more than 3 linear independent directions exist, such that e.g.\ in cubic symmetry no longer all linear independent directions must be the same.
Different spectra can be observed along different directions, e.g.\ for \qp{100}, \qp{110}, and \qp{111}.
Here \xyz{xyz} relates to the direct lattice vectors in real space, which is more closely related to the local point symmetry and the local basis (spherical harmonic functions).

Therefore, the isotropic spectrum in the limit of higher multipoles require more directions to be constructed of.
The knowledge of the isotropic spectrum helps to immediately identify the ratio of the different multipole contributions, but is not crucial for the analysis of the angular dependence.
In the following it will be referred to a pseudo-isotropic spectrum when the angular dependence is already small compared to the signal or can be made small by a certain linear combination of the measured directions.

The result will always relate to the initial state.
For investigations of the ground state it is important that no excited states are occupied, i.e.\ the sample should be investigated at low enough temperatures, so that the ground state is probed.

\paragraph{Sample preparation}
\begin{figure}
  \centering
  \begin{subfigure}[b]{0.48\textwidth}
    \centering
	(a)
	\includegraphics[height=34mm]{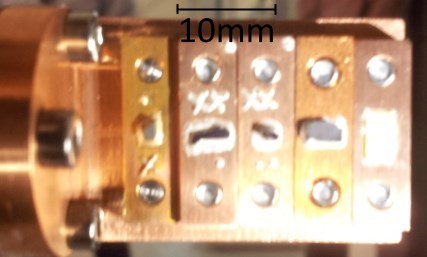}
	\phantom{00000000000000000}\\
    (b)
	\includegraphics[height=46mm]{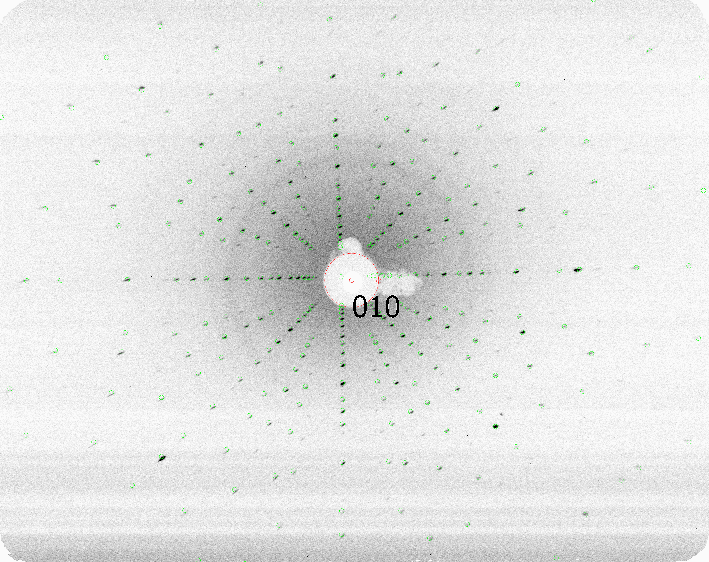}
  \end{subfigure}
  \hfill
  \begin{subfigure}[b]{0.48\textwidth}
    \centering
	(c)
	\includegraphics[height=40mm]{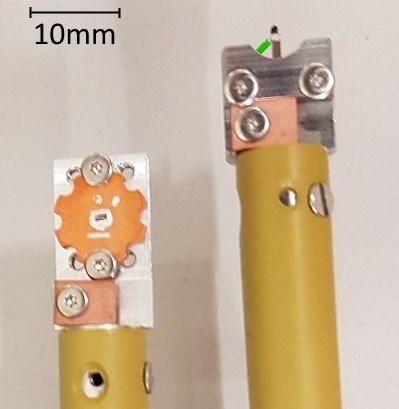}
	\phantom{00000000000000000}\\
	(d)
    \includegraphics[height=40mm]{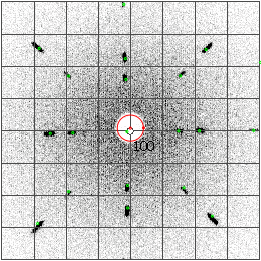}
  \end{subfigure}
  \caption{Samples mounted on the different sample holders. Left for the closed cycle cryostat: The closed cycle cryostat takes five sample holders; on the second holder from left CeB$_6$ is mounted with a \xyz{100} surface and on the third with a \xyz{110} surface aligned parallel to the surface of the holder. The rightmost holder carries a YAG for optical alignment. Below the Laue image of the mounted \xyz{100} sample is shown as obtained from the Laue diffractometer at the Institute of Physics II at the University of Cologne. The black dots represent the experimental Laue pattern and the green dots are calculated using CLIP and the structure parameters from ICSD\,\cite{clip, icsd, Blomberg1989}. Right for the He-flow cryostat: The image shows SmB$_6$ with \xyz{100} normal mounted on the original (left) and the modified (right) sample holder. The sample holders can be rotated around the surface normal in steps of 45$^\circ$ for the original and 15$^\circ$ for the modified case (as indicated by the green bar). Below is the Laue pattern of the mounted SmB$_6$ \xyz{100} normal sample as taken at the Laue diffractometer at MPI CPfS in Dresden and the CLIP calculation performed as for CeB$_6$\cite{clip, icsd, Funahashi2010}. For the usual sample distance of $\sim$10\,cm the central blind area amounts to $\sim$5$^\circ$ off the center for both diffractometers.}
  \label{fig:mount}
\end{figure}

The single crystal has to be aligned prior to the experiment.
The initial identification of these directions can be done by Laue diffraction.
Laue diffraction has been performed on a laboratory device that utilizes the broad energy range of an x-ray tube which yields sufficient intensity to measure the strong Bragg peaks in the elastic signal.
Note, that the elastic signal away from any Bragg peak, i.e.\ the dark areas in the Laue image, is still stronger than the \NIXS{} signal by orders of magnitude.
Here a broad energy range of the x-ray tube is desired to cover a wide range of momentum transfer, i.e.\ to observe a sufficiently large number of Bragg reflections in the solid angle observed by a 2D-pixel detector or photoactive plate.
\Fig[(b+d)]{fig:mount} show Laue images obtained for cubic CeB$_6$ and SmB$_6$ single crystals along one of the fourfold symmetry axis.
The blind area in the center is due to the collimated source.
The black dots represent the data and the green circles mark the simulated positions calculated with CLIP\,\cite{clip} based on the lattice parameters of the respective sample.
Intensities of the peaks can be calculated with e.g.\ QLaue\,\cite{qlaue}.
Note that higher multipole transitions, with order \qnQ{}, require higher accuracy of the alignment.
The alignment should be \qnQ{} times more accurate as for a dipole limited experiment (\qnQ{}\,=\,1), i.e.\ 2-3$^\circ$ accuracy for an experiment dominated by octupole (\qnQ{}\,=\,3) and dotriacontapole (\qnQ{}\,=\,5) transitions\footnote{The Legendre polynomial \Legendre{\qnQ{}}{x} shows \qnQ{} lobes within 180$^\circ$ (see also \chap{cap:nonresonanttransitionoperator}).}.

The samples were mounted inside a He cryostat to achieve sufficiently low temperatures.
Two types of cryostats have been utilized:
A closed cycle\,\cite{closedcycle} and a helium flow cryostat\,\cite{dynaflow}.
Both cryostats are designed such that the samples can be rotated inside the scattering plane.
The closed cycle cryostat consists of a single insulation vacuum chamber, where the sample is placed on a so-called cold finger.
The cooling is done by He decompression and the He is recovered.
The thermal exchange goes via the copper of the cold finger.
The periodic decompression here introduces vibrations of the order of 50\,\(\mu\)m at the sample position.
The helium flow cryostat has an inner chamber where the sample is surrounded by He gas, in addition to an outer insulation vacuum.
It requires an additional window plus liquid helium (about 1\,l/h at lowest temperatures and about 0.3\,l/h at or above 50\,K).
The advantages are a more direct cooling of the sample surface, especially helpful for low heat conducting samples, a faster cool-down, and less vibrations.
Two heaters, in form of electrical resistors, allow to stabilize temperatures up to room temperature.

The 10\,keV photon beam tends to burn holes into the cryostat entry windows.
Therfore, special care has to be taken.
For the incoming beam a polyimide-Al composite window is desired, while for the scattered beam pure polyimide can be used.
For the helium flow cryostat both windows should be made from the composite.
The outer shield avoids ice formation upon leakage.
The inner shield is less critical, but when leaking too much (typically after few days with pure polyimide) the insulation vacuum worsens and the temperature increases.

The design of the sample holder of the He-flow cryostat has been modified to adapt to the \NIXS{} experiment (see the right holder in \fig[c]{fig:mount}).
The sample can be mounted on top of a pin so that it is \textit{free standing}, which makes it easy to align the sample inside the cryostat\footnote{The sample can be aligned by only monitoring the projection of the x-ray beam of the analyzers on the 2D-detector without any optical windows, best when the energy is set to the peak of the Compton.} and avoids signals from the sample holder in case of thin samples.
Strong epoxy glue is not needed because no \textit{in-situ} cleaving is necessary (which is especially important for actinide compounds).
The surface can be prepared prior to mounting.
A solvable glue should be used which allows to retrieve the sample.
This makes is easier to prepare samples.

\paragraph{Acquisition of spectra}

The experiments were performed at the P01 beamline at the DESY synchrotron in Hamburg and ID20 beamline at the ESRF synchrotron in Grenoble.
At both beamlines the incident beam can be monochromatized with either a Si(111) or a Si(311) double crystal monochromator (\DCM{}), which yield an overall instrumental resolution with a Gaussian full width at half maximum \FWHMG{}\,$\approx$\,1.3 and 0.7\,eV, respectively, for the elastic signal of the Si(660) analyzer reflection at about 9.69\,eV.

\begin{figure}
  \centering
  \includegraphics[width=\textwidth]{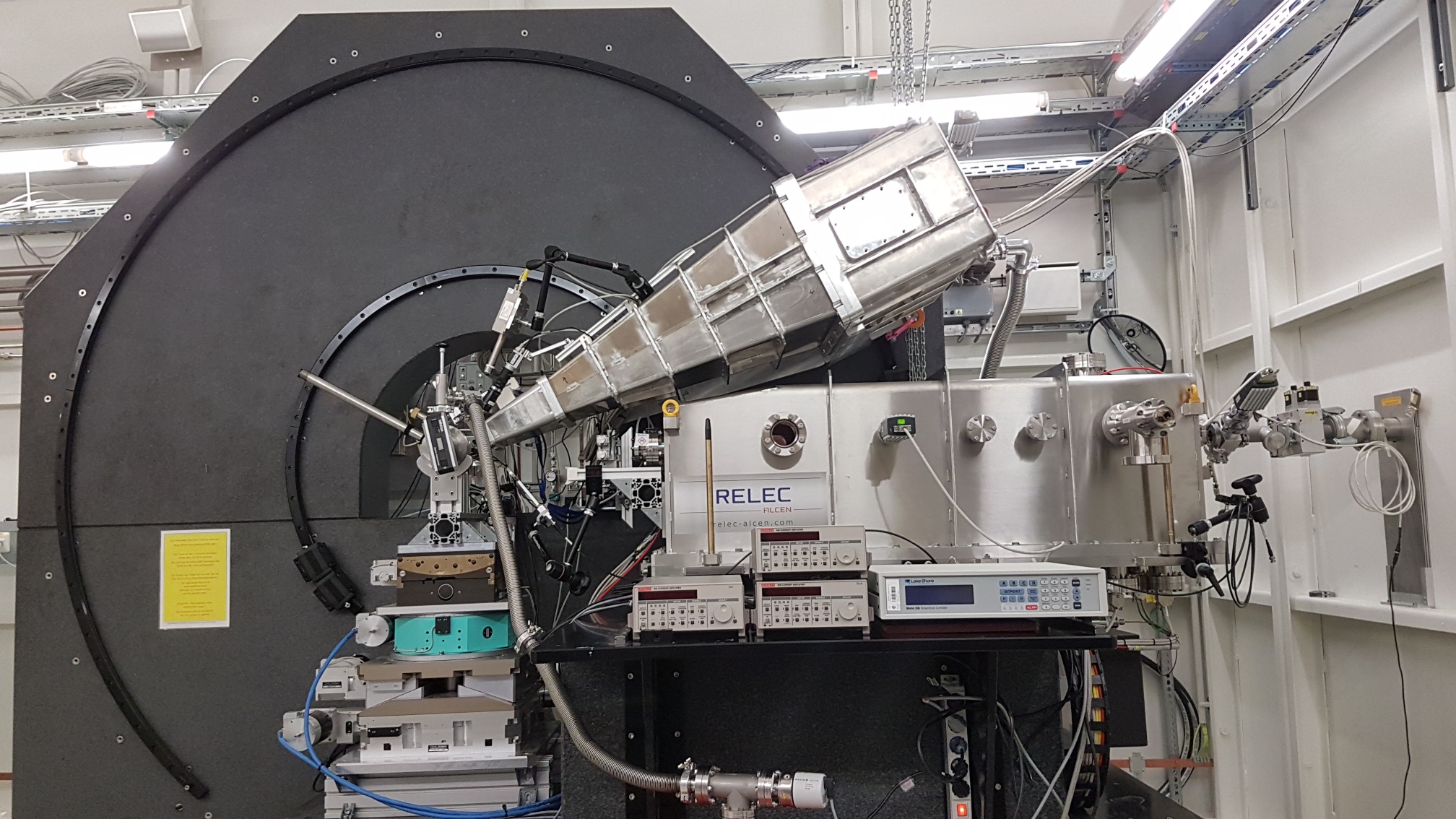}
  \caption{Image of the NIXS endstation at the P01 beamline at DESY in Hamburg. The monochromatized beam enter from the tube at the right hand side to the Kirkpatrick-Baez (KB) focusing mirror tank. Behind the tank it passes through a slit and is observed by a 300\,$\mu$m pin diode before it reaches the sample inside the He-flow cryostat on top of the translation and rotation stage. The scattered light enters the vacuum tank (here in backscattering geometry) through a collimator and the 12\,analyzers project the sample spot onto the 2D-Medipix3 Lambda pixel detector. (Further instruments for alignment or vacuum can be seen.)}
  \label{fig:P01setup}
\end{figure}

\begin{SCfigure}
  \includegraphics[width=0.5\textwidth]{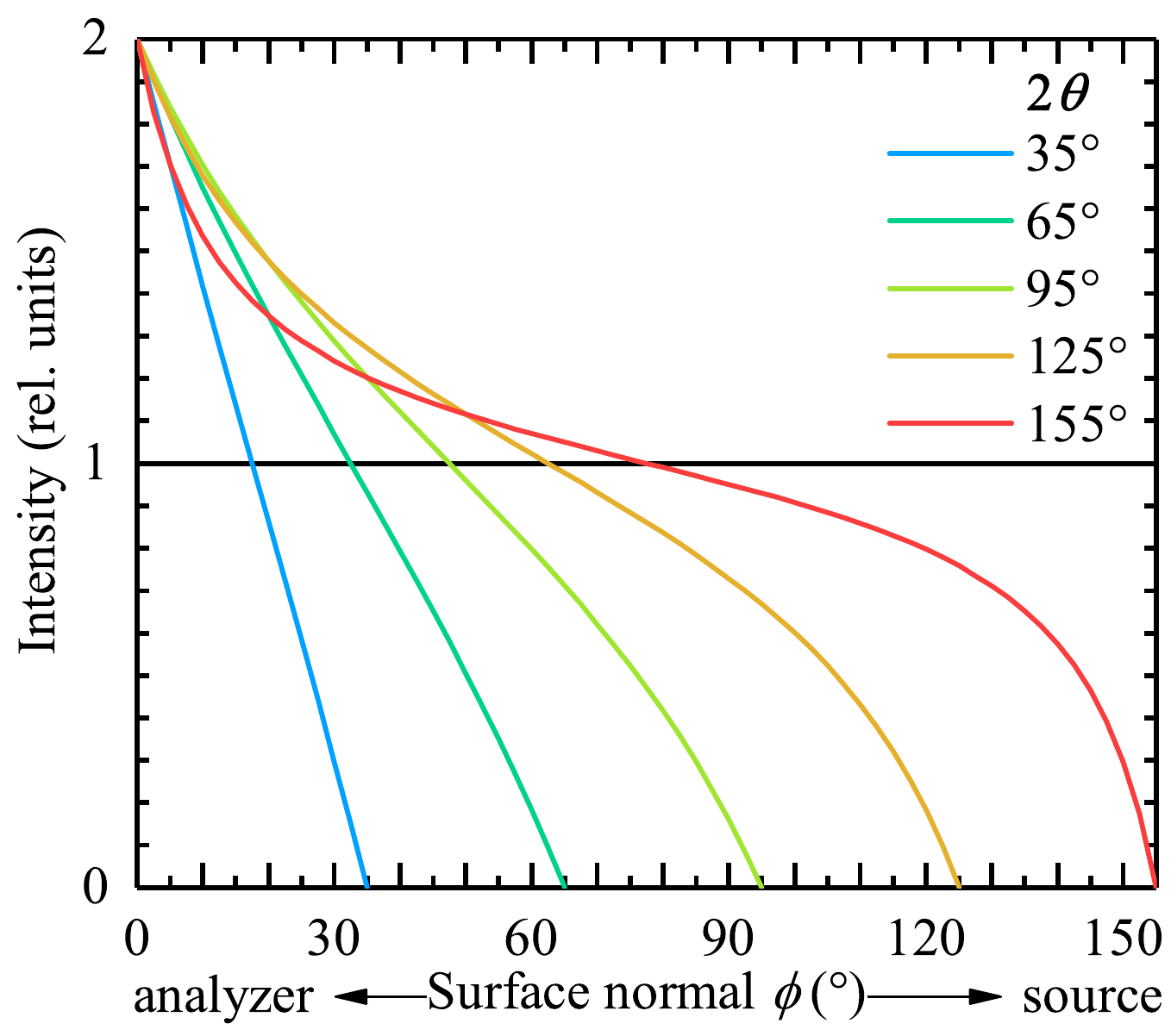}
  \caption{Scattering intensity relative to specular geometry as a function of the angle $\phi$ of the projection of the sample surface normal $\vec{n}$ onto the scattering plane for different scattering angels 2\thS{} due to the self absorption of the sample. In all cases the intensity is doubled for grazing incident beam ($\phi$\,=\,0) compared to specular geometry $\phi_0$\,=\,\thS{} (shown in \fig{fig:NIXSgeometry}, where \xyz{xyz}\,$||$\,$\vec{n}$) and goes to zero for grazing scattered beam ($\phi$\,=\,2\thS{}).}
  \label{fig:IntensitySurfacenormal}
\end{SCfigure}

For the experiment the cryostat is placed at the beamline such, that the contained single crystalline sample is in the focal point of the focusing mirrors, which is also on the Rowland circle of the analyzers (see \fig{fig:beamline}).
The cryostat is mounted on a motorized stage to adjust the sample position.
\Fig{fig:P01setup} shows an image of this setup.
The sample should be mounted with the desired direction (\xyz{xyz}) pointing along the momentum transfer as shown in \fig{fig:NIXSgeometry}, i.e.\ along the bisector of the line pointing to the center of the analyzers and the incident beam.
Specular geometry is realized, when the aligned direction reflects the surface normal.

A complete set of data usually contains three kinds of spectra:
A long scan with coarse energy steps (1--5\,eV) that covers the entire energy range in \fig{fig:XRSlogSpectrum} and scans with finer energy steps across the elastic line and across the core-level excitations of interest (see highlighted areas in \fig{fig:XRSlogSpectrum}).
The rough overview scan helps to normalize the data and to verify a clean sample by the absence of emission lines which do not belong to the sample.
The about 10\,eV wide scan across the elastic line with fine energy steps of typically 0.1\,eV is used to calibrate the elastic line of the different analyzers so that the energy transfer can be determined accurately.
And finally the edge scan in fine energy steps that covers the structure of the core-electron excitation.
Typical values for the integration time are 1\,s or even less for the elastic line and 10\,s otherwise.
The edge scan is taken several times until satisfactory statistics is achieved.

One of the major issues is to detect spurious scattering.
Since the scattering intensities are low, this is sometimes spotted only after several hours of acquisition time.
Spurious scattering may arise from Bragg scattering in the sample that enters the detector after secondary scattering events.
Usually it helps to turn the sample by 1--2$^\circ$ in order to avoid this signal.

Self absorption effects are energy independent because \NIXS{} is non-resonant.
This has been experimentally verified in this thesis.
Thus, going away from specular geometry (surface normal bisecting \kin{} and \kout{}) does not affect the signal, except for an overall change of the intensity \( I(\thS{},\phi) \propto 2\int_0^\infty \text{exp}(-x) \text{exp}(-x\sin(\phi)/\sin(2\thS{}-\phi))\,\diff x \) depending on the angle $\phi$ of the projection of the sample surface normal $\vec{n}$ onto the scattering plane.
This is shown in \fig{fig:IntensitySurfacenormal} for different scattering angles 2\thS{}.
Here, the backscattering geometry (large 2\thS{}) allows to measure a wider range of directions on the same surface.
The changing footprint of the beam does not affect the resolution (see \equ{eq:specressource}\,--\,\ref{eq:specresbending}) and it is beneficial in terms of intensity to rotate the surface normal away from the source when the alignment can be realized in multiple ways.

\clearpage
\section{Theoretical NIXS cross-section}\label{sec:calcnixs}
The interaction between electrons and photons is well understood.
With this knowledge the information of the valence states contained in the spectra can be extracted.
To make this connection more transparent a multipole expansion of the transition operator is performed.
Here will be shown how the different multipole transitions can be chosen.
In the end a formalism is presented which separates the experimental properties from those of the investigated system.

\subsection{Electron photon interaction}

Following Sch\"ulke\,\cite{Schuelke2007}, the interaction between electrons and photons is calculated using nonrelativistic perturbation theory upon the Hamiltonian of the electrons.
The photon is expressed by a vector field \vec{A} and potential $\phi$, with the respective electric and magnetic fields
\begin{align}
\vec{E}(\vec{r},t) &= -\frac{\partial\vec{A}(\vec{r},t)}{\partial t} - \vec{\nabla} \phi(\vec{r},t) \label{eq:Efield}\\
\vec{B}(\vec{r},t) &= \vec{\nabla} \times \vec{A}(\vec{r},t).
\end{align}
The important part of the perturbed Hamiltonian for $N$ electrons is then given by
\begin{align}
H &= \sum_{j=1}^{N} ~~ \frac{1}{2\emass{}} \left( \vec{p}_j + \echarge{} \vec{A} \right)^2
 + \dots \label{eq:Hkin}
\end{align}
with $\vec{p}_j$ being the momentum operator acting on one electron $j$.

Terms describing the interaction with the spin, as well as all terms for the unperturbed system are not shown for simplicity.
For \NIXS{} the interaction with the electron spin can be neglected as it scales compared to the interaction with the charge density with \( \textfrac{\hbar\omega}{\emass{} \sol{}^2}\) (e.g.\ \(\approx\)\,0.02 at \(\hbar\omega\)\,=\,10\,keV)\footnote{Usually much smaller considering also the form factors.}.

Using the Coulomb gauge without source ($\phi$\,=\,0 and $\nabla\vec{A}$\,=\,0) for describing the photon field, $H$ can be simplified further.
\begin{align}
H &= \sum_{j=1}^{N} ~~ \frac{\vec{p}_j^2}{2\emass{}} ~+~ \underbrace{ \frac{\echarge{}}{\emass{}} \vec{p}_j \cdot \vec{A}  }_{ \text{\scalebox{1.4}{resonant}} }
 ~ + \underbrace{ \frac{\echarge{}^2}{2\emass{}} \vec{A}^2 }_{ \text{\scalebox{1.7}{non}} \atop \text{\scalebox{1.7}{resonant}}  } +~ \dots \label{eq:HkinSplit}
\end{align}
In the resonant case only one photon is involved in a first order transition.
In the non-resonant case two photons are involved.
The resonant term of $H$ will describe the probability of absorbing the photon, as for \XAS{} or \PES{}.
The non-resonant term of $H$ will describe the probability of scattering the photons, elastically or inelastically.
Note, that the re-emission of a photon as in \XAS{} in the fluorescence yield or \RIXS{} is a two step process and involves an intermediate state, as introduced e.g.\ in Ref.\,\cite{Ament2011}.
Its cross-section, however, is small compared to non-resonant scattering as long as the energy of the incoming photon is not matching the energy of an intermediate state.

The vector field for the photon with polarization \vec{\epsilon} can be written as
\begin{align}
\vec{A}(\vec{r},t) &= \sum_{\vec{k},\vec{\epsilon}} \frac{1}{\sqrt{2}} \left( \vec{\epsilon} \, \exp{\imath ( \vec{k} \vec{r} - \omega t )} \, a_{\vec{k},\vec{\epsilon}} + \vec{\epsilon}^* \exp{-\imath ( \vec{k} \vec{r} - \omega t )} \, a_{\vec{k},\vec{\epsilon}}^\dagger \right).
\end{align}
The rate of a transition between the initial state \ket{i} and the final state \ket{f} is given by the overlap integral with the part of the Hamiltonian, that describes the interaction.
In an experiment the system is exposed to well defined incoming photons with energy $\hbar\omega_\text{in}$, wave vector $\vec{k}_\text{in}$, and polarization $\vec{\epsilon}_\text{in}$.
The outgoing photons with energy $\hbar\omega_\text{out}$, wave vector $\vec{k}_\text{out}$, and polarization $\vec{\epsilon}_\text{out}$ are observed within a certain direction \(\vec{k}_\text{out} \in \diff\Omega\) and energy range \(\vec{w}_\text{out} \in \diff\omega\).
Therefore the experiment can be well described in the form of a double differential scattering cross-section.
For the non resonant term of $H$, it is given by
\begin{align}
\frac{\diff^2\sigma}{\diff\Omega \, \diff\hbar\omega} &= 
 \underbrace{ \vphantom{\sum_{\ket{i},\ket{f}} \frac{\exp{\frac{-E_\text{i}}{\kB{}\T{}}}}{\pfZ{}}} \frac{\oout{}}{\oin{}} \left( \frac{\echarge{}^2}{2\emass{}} \right)^2 \abs{ \vec{\epsilon}_\text{out}^* \cdot \vec{\epsilon}_\text{in} }^2 }_{\text{\scalebox{1.4}{Thomson}}}
 ~ \underbrace{ \sum_{\ket{i},\ket{f}} \frac{\exp{\frac{-E_\text{i}}{\kB{}\T{}}}}{\pfZ{}} ~ \abs{ \bra{f} \sum_{j=1}^N \exp{\imath \vecq{} \vec{r}_j} \ket{i} }^2 ~ \del (E_\text{f} - E_\text{i} - \hbar\omega) }_{\text{\scalebox{1.4}{\sqw{}}}}.
\end{align}
A sum over the initial states \ket{i} is introduced to account for their partial occupation, i.e.\ due to Boltzmann population with the partition function \pfZ{}.
The $\del$-function takes care of the energy conservation with the energy of the initial states $E_\text{i}$ and the energies of all final eigenstates $E_\text{f}$.
The energy transfer $\hbar\omega$ and the momentum transfer \vecq{} are defined as (see also \fig{fig:NIXSgeometry})
\begin{align}
  \omega &\equiv \oin{}-\oout{}
  \hspace{1cm}\text{and}\hspace{1cm}
  \vecq{} \equiv \kin{}-\kout{}. \label{eq:energymomentumtransfer}
\end{align}

For small relative variation of the photon energies, \oin{}\,$\approx$\,\oout{}, which is valid across a single edge, the Thomson scattering cross-section can be considered as a constant prefactor and the signal is proportional to the dynamic structure factor
\begin{align}
 \sqw{} &= \sum_{\ket{i},\ket{f}} \frac{\exp{\frac{-E_\text{i}}{\kB{}\T{}}}}{\pfZ{}} ~ \abs{ \bra{f} \sum_{j=1}^N \exp{\imath \vecq{} \vec{r}_j} \ket{i} }^2 ~ \del (E_\text{f} - E_\text{i} - \hbar\omega). \label{eq:sqw}
\end{align}
For the interpretation of the spectra, the Hamiltonian of the electron system and the transition rates given by the operator $\exp{\imath \vecq{} \vec{r}}$ need to be calculated.

\subsection{Expansion of the transition operator}\label{cap:nonresonanttransitionoperator}

The core electrons involved are well described as atomic orbitals, e.g.\ on the basis of spherical harmonics.
For the observation of core-electron excitations the $f$ states can still be approximated by atomic orbitals.
Therefore, it is convenient to expand the transition operator on the basis of spherical harmonics, as well.

The multipole expansion of the plane wave \(\exp{\imath \vecq{}\vec{r}}\) is given by the spherical Bessel function \(j_{\qnQ{}}(\abs{{\vecq{}}}\abs{\vec{r}})\), which accounts for the size of the momentum transfer, and the Legendre polynomial \Legendre{\qnQ{}}{\cos{\theta_{\vecq{}\vec{r}}}}  with the multipolar order \qnQ{}, which accounts for the direction of the plane wave with the angle \(\theta_{\vecq{}\vec{r}}\) between \(\vecq{}\) and \(\vec{r}\) and can be expressed in terms of renormalized spherical harmonics \Clm{\qnQ{}}{$m$}\,\cite{Haverkort2007, Gordon2008, Mehrem2011}.
\begin{align}
\exp{\imath \vecq{}\vec{r}} &= \sum\limits_{\qnQ{}=0}^{\infty} \imath^{\qnQ{}} \, (2\qnQ{}+1) \, \Legendre{\qnQ{}}{ \cos\,\theta_{\vecq{}\vec{r}} } \, j_{\qnQ{}}(\absq{}\abs{\vec{r}}) \label{eq:planewaveexpansion} \\
\Legendre{\qnQ{}}{ \cos{\theta_{\vecq{}\vec{r}}} } &= \sum\limits_{m=-\qnQ{}}^{\qnQ{}} {\Clm{\qnQ{}}{$m$}}^\ast(\theta_{\vecq{}},\phi_{\vecq{}}) \, \Clm{\qnQ{}}{$m$}(\theta_{\vec{r}},\phi_{\vec{r}}) \label{eq:legendrepolinominal}
\end{align}
This returns the form as presented in Ref.\,\cite{Haverkort2007} which represents the angular momentum transfer of the transition.
Inside \sqw{} the overlap integral
\begin{align}
\bra{f} \exp{\imath \vecq{}\vec{r}} \ket{i} &=
\sum\limits_{\qnQ{},m}^{} \imath^{\qnQ{}} \, (2\qnQ{}+1) ~ \Clm{\qnQ{}}{$m$}^{\ast}(\theta_{\vecq{}},\phi_{\vecq{}}) ~ \bra{f} ~ j_{\qnQ{}}(\absq{}\,\abs{\vec{r}}) ~ \Clm{\qnQ{}}{$m$}(\theta_{\vec{r}},\phi_{\vec{r}}) ~ \ket{i} \label{eq:transitionopgeneral}
\intertext{can now be further simplified when dealing with states that can be expressed on the spherical harmonics basis, like atomic orbitals}
\bra{f} \exp{\imath \vecq{}\vec{r}} \ket{i} &= \sum\limits_{\qnQ{}}^{} \imath^{\qnQ{}} \, (2\qnQ{}+1) ~ \bra{$\qnn{}_\text{f},\qnl{}_\text{f}$} ~ j_{\qnQ{}}(\absq{}\,r) ~ \ket{$\qnn{}_\text{i},\qnl{}_\text{i}$} \nonumber \\
 &\times  \Clm{\qnQ{}}{$m_\text{f}-m_\text{i}$}^{\ast}(\theta_{\vecq{}},\phi_{\vecq{}})
 ~ \bra{$\qnl{}_\text{f},m_\text{f}$} ~ \Clm{\qnQ{}}{$m_\text{f}-m_\text{i}$}(\theta_{\vec{r}},\phi_{\vec{r}}) ~ \ket{$\qnl{}_\text{i},m_\text{i}$}. \label{eq:transitionopfinal}
\end{align}

The second integral in \equ{eq:transitionopfinal} contains the angular momentum conservation and restricts \qnQ{} such, that
\begin{align}
\bra{$\qnl{}_\text{f},m_\text{f}$} ~ \Clm{\qnQ{}}{$m_\text{f}-m_\text{i}$}(\theta_{\vec{r}},\phi_{\vec{r}}) ~ \ket{$\qnl{}_\text{i},m_\text{i}$} = 0, ~ \text{if} ~ \qnQ{} \notin \{\abs{\qnl{}_\text{f}-\qnl{}_\text{i}},~\abs{\qnl{}_\text{f}-\qnl{}_\text{i}}+2,~...~,~\qnl{}_\text{f}+\qnl{}_\text{i}\} \label{eq:NIXSselectionrules}
\end{align}
and there is no need for expansions beyond \qnQ{}\,=\,6 considering all elements known so far.

\begin{SCfigure}
  \includegraphics[width=0.55\textwidth]{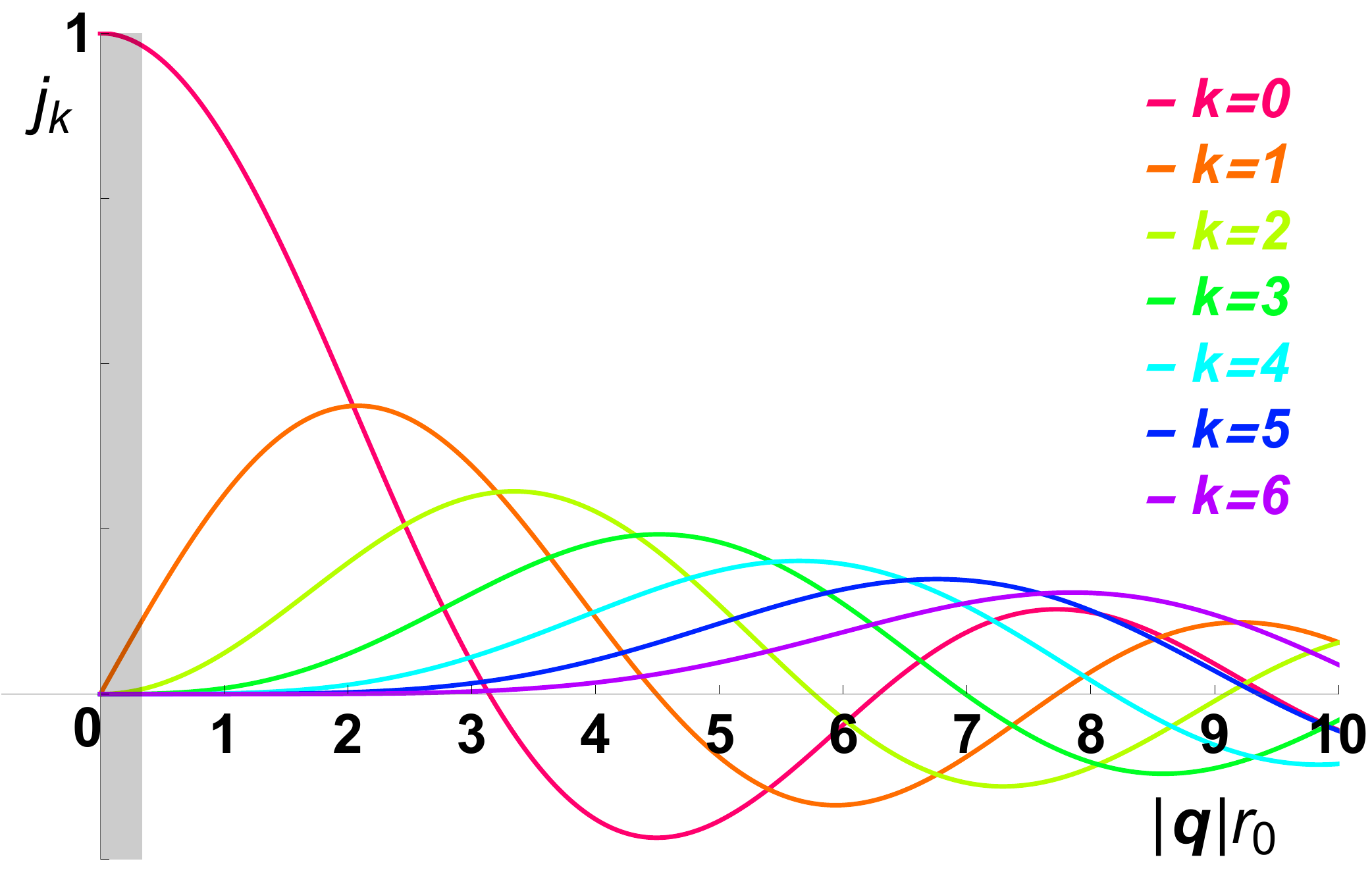}
  \caption{Visualization of the spherical Bessel functions (\equ{eq:planewaveexpansion}) for the in \NIXS{} allowed multipole transitions \qnQ{}. The momentum range available to resonant transitions is marked gray.}
  \label{fig:sphericalbesselj}
\end{SCfigure}

\subsubsection{Length dependence of \(\vec{\textit{q}}\)}
\begin{figure}
  \centering
  \includegraphics[width=\textwidth]{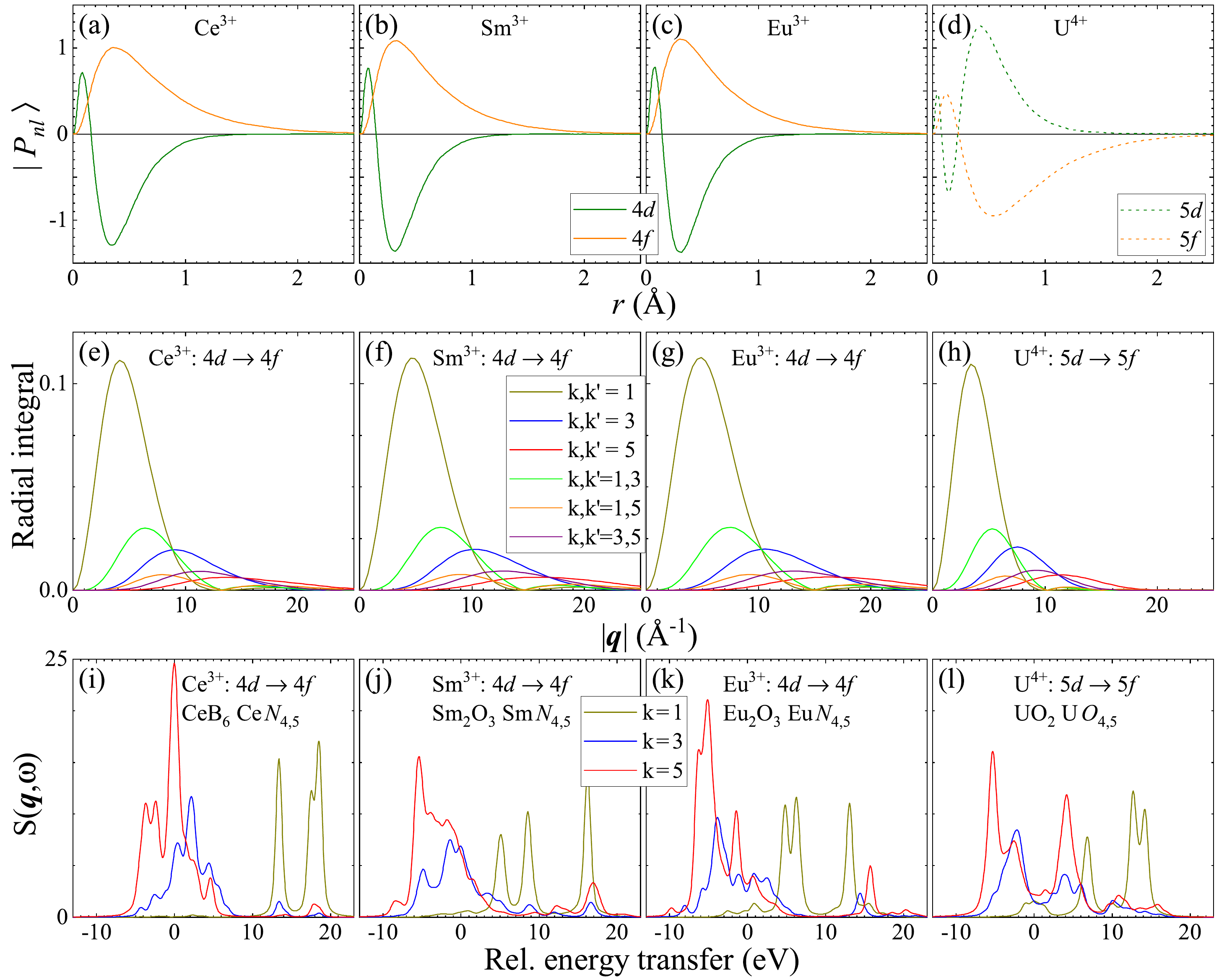}
  \caption{(a-d) Radial expansion \ket{\Pnl{\qnn{}}{\qnl{}}} of the $3d$ core shell and $4f$ valence shell of Ce\textsup{3+}, Sm\textsup{3+}, Eu\textsup{3+}, and of the $4d$ and $5f$ of U\textsup{4+}. (e-h) Radial integrals $\prod_{l\in \qnQ{},\qnQ{}'} \abs{\bra{\Rnl{$\qnn{}_c$}{$\qnl{}_c$}}\,j_l(\absq{} r)\,\ket{\Rnl{$\qnn{}_v$}{$\qnl{}_v$}}}$ of the respective core $c$ and valence $v$ shell as a function of \absq{}. (i-l) Isotropic dynamic structure factor \( \iint_{4\pi} \sqw{} \sin(\theta) \, \diff \theta \, \diff \phi \) calculated for the single multipole contributions \qnQ{}, each with defining $\abs{\bra{\Rnl{$\qnn{}_c$}{$\qnl{}_c$}}\,j_{\qnQ{}}(\absq{} r)\,\ket{\Rnl{$\qnn{}_v$}{$\qnl{}_v$}}}$\,=\,1. For better visibility the spectra have been normalized to the initial number of holes in the valence shell. All calculations are based on the atomic parameters calculated with Cowans Atomic Structure Code\,\cite{Cowan1981} (see also \app{app_atomicvalues}). For the spectra the parameters in the tables \ref{tab:CeB6parameters}, \ref{tab:SmB6parameters}, and \ref{tab:UO2parameters} of the respective compounds have been used.}
  \label{fig:rjrofall}
\end{figure}

After the multipole expansion, \absq{} remains only inside the spherical Bessel function.
\Fig{fig:sphericalbesselj} shows the spherical Bessel function for all possibly allowed values of \qnQ{} as a function of \absq{}.
Considering the mean radial expansion $r_0$ to be about the Bohr radius \rBohr{}, the higher multipoles peak around \absq{}\,=\,(3\,--\,8)\,\rBohr{}$^{-1}$, i.e.\ at about 10\,$\AA^{-1}$.
The different multipole orders \qnQ{} peak at different momentum transfers \absq{} and the higher \qnQ{} the larger is \absq{} for the maximum.
From the relation between the momentum transfer and the photon energies
\begin{align}
\sol{}^2\,\absq{}^2 &= \otr{}^2 + 2 \, \oin{} \oout{} \left(1-\cos{2\thS{}}\right) \label{eq:cabsq}
\end{align}
\absq{} can be approximated to be independent of \otr{} for \otr{}\,$\ll$\,\oin{}:
\begin{align}
\absq{} &\approx \frac{\oin{}+\oout{}}{\sol{}} \sin{\thS{}} \\
&\approx 0.5 ~ (\Ein{}+\Eout{}) \sin{\thS{}} \, \frac{\AA^{-1}}{\text{keV}}. \nonumber
\end{align}
For backscattering, as a rule of thumb, every 1\,keV incident energy gives 1\,$\AA^{-1}$ extra momentum transfer.
\absq{} can be increased while keeping \otr{} fixed.

For an $s$ type core hole excitation with $\qnl{}_\text{i}$\,=\,0 the multipole order \qnQ{} is uniquely defined by the final state angular momentum (see \equ{eq:NIXSselectionrules}) and \absq{} can be used to maximize the cross-section.
The initial core state and the final valence state of the scattered electron have to be considered in the integral in \equ{eq:transitionopgeneral} or, if applicable, in \equ{eq:transitionopfinal}.
Here, the radial expansions \Pnl{\qnn{}}{\qnl{}}\,=\,$r$\Rnl{\qnn{}}{\qnl{}}\,$\equiv$\,$r$\,\ket{\qnn{},\qnl{}} of these states have been calculated using Cowans atomic structure code\,\cite{Cowan1981} in the Hatree-Fock approximation.
They are equal for all states inside one shell.
\Fig[(a-d)]{fig:rjrofall} show the calculated radial expansions of the wave functions of the investigated ions.

\Fig[(e-h)]{fig:rjrofall} show the \absq{}-dependent integrals \( \prod_{k\in \qnQ{},\qnQ{}'} \abs{\langle \qnn{}_\text{f},\qnl{}_\text{f} | ~ j_k(\absq{}\,r) ~ | \qnn{}_\text{i},\qnl{}_\text{i} \rangle} \) of the investigated ions.
All these plots show indeed a clear separation of the different multipole contributions.
So with the length of \vecq{}, it is even possible to select the multipole moment \qnQ{} of the transition between the same shells.
It also becomes visible that the different \qnQ{} peak at slightly different values of \absq{} for the different elements, due to different radial expansions of the wave functions.
\Fig[(i-l)]{fig:rjrofall} show the isotropic responses of the \edge[RE]{N}{4,5} and \edge[U]{O}{4,5} edges in \sqw{} for the allowed transition with momentum transfer \qnQ{}.
Note that responses with \qnQ{}\,$\neq$\,$\qnQ{}'$ vanish in the isotropic case.

The choice of the best \qnQ{} to focus on in the experiment will be discussed based on the directional \vecq{} dependence.

\subsubsection{Directional dependence of \(\vec{\textit{q}}\)}\label{ch:vecqdependence}

\begin{figure}
\newcommand*\sioplot[1]{%
\raisebox{-9mm}[10mm][0mm]{%
\hspace{-1.5mm}%
\includegraphics[height=20mm]{cqvi_#1.png}%
\hspace{-3.5mm} } }
\centering
\begin{tabular}{|cc|cccccccc|}
\hline
    & && $s$ & $p_z$ & $d_{xy}$ & $d_{3z^2-r^2}$ & $f_{xyz}$ & $f_{5z^3-3zr^2}$ &\\
\hline
 & \qnQ{}\,=\,0 && \sioplot{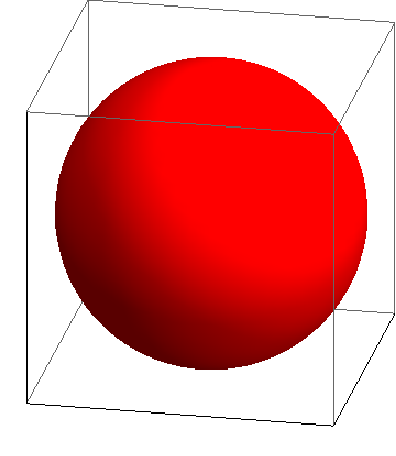} & \--- & \--- & \--- & \--- & \--- & \vphantom{\raisebox{12mm}{}}\\
\raisebox{-6mm}[0mm][0mm]{{\Large $s$}} & \qnQ{}\,=\,1 && \--- & \sioplot{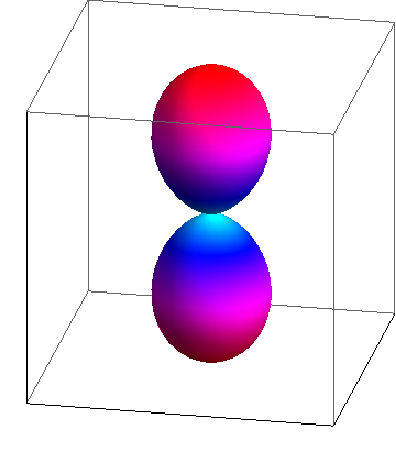} & \--- & \--- & \--- & \--- &\\
 & \qnQ{}\,=\,2 && \--- & \--- & \sioplot{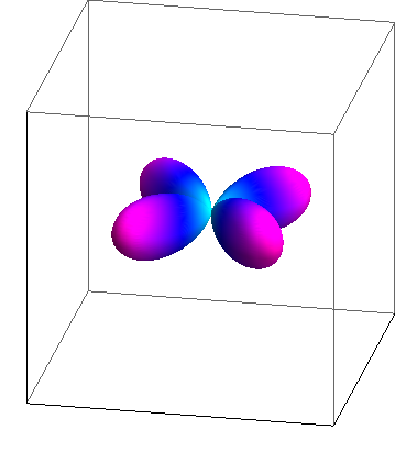} & \sioplot{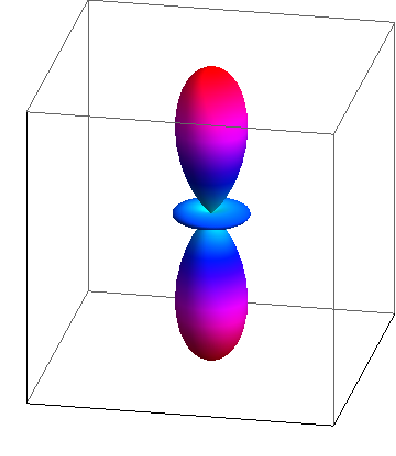} & \--- & \--- &\\
 & \qnQ{}\,=\,3 && \--- & \--- & \--- & \--- & \sioplot{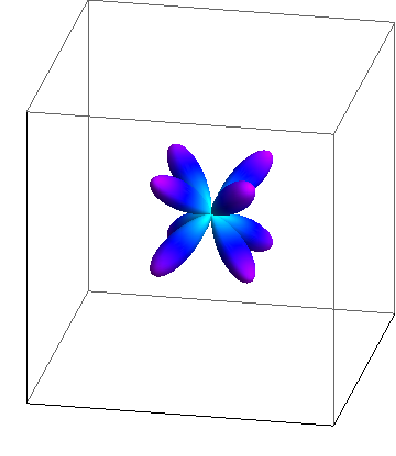} & \sioplot{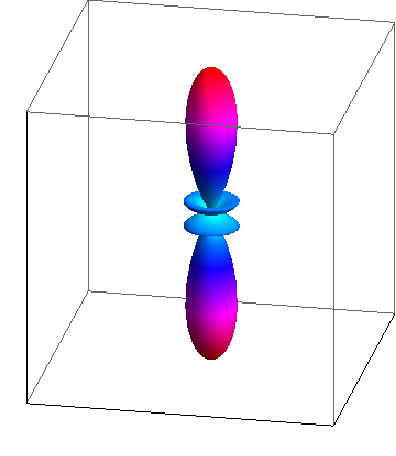} &\vphantom{\raisebox{-9mm}{}}\\
\hline
 & \qnQ{}\,=\,0 && \--- & \sioplot{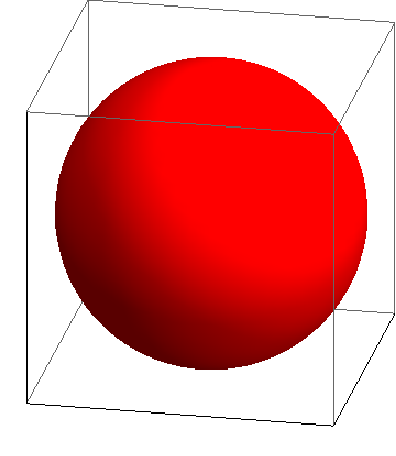} & \--- & \--- & \--- & \--- &\vphantom{\raisebox{12mm}{}}\\
 & \qnQ{}\,=\,1 && \sioplot{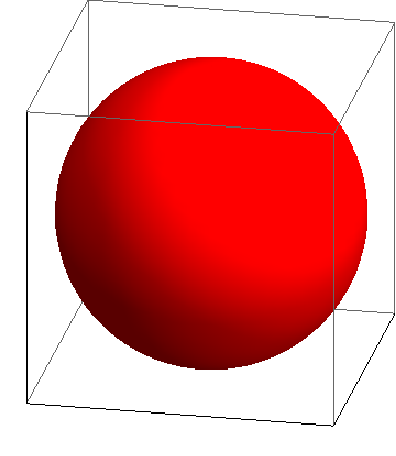} & \--- & \sioplot{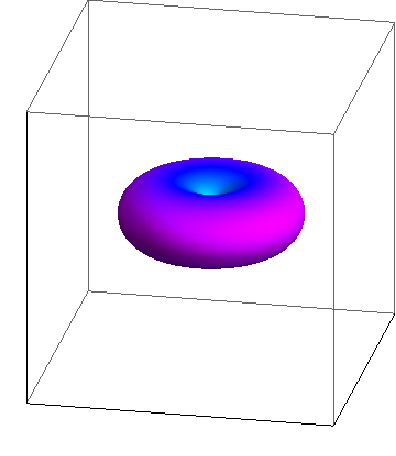} & \sioplot{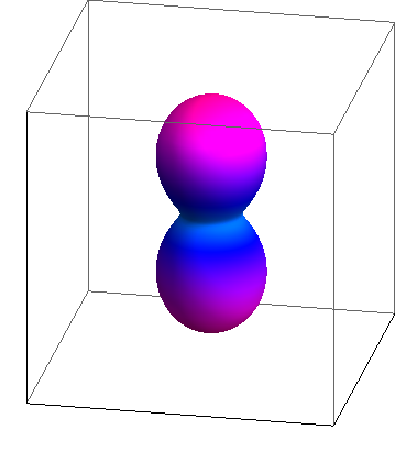} & \--- & \--- &\\
{\Large $p$} & \qnQ{}\,=\,2 && \--- & \sioplot{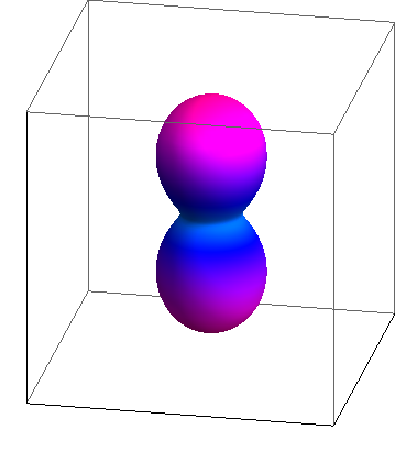} & \--- & \--- & \sioplot{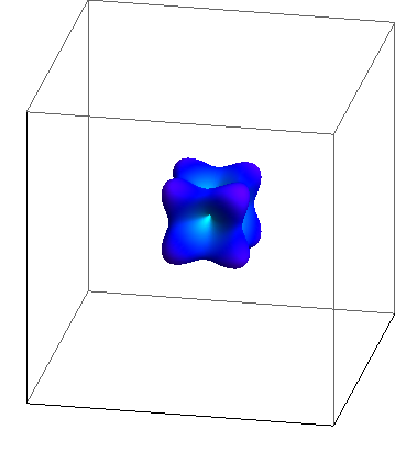} & \sioplot{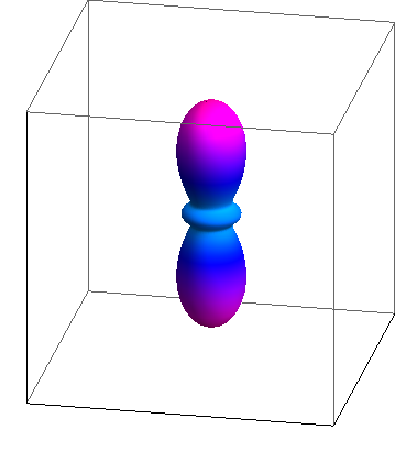} &\\
 & \qnQ{}\,=\,3 && \--- & \--- & \sioplot{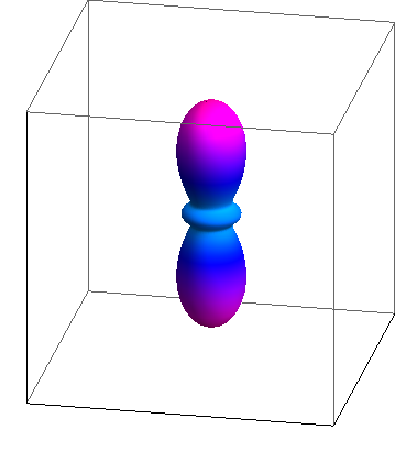} & \sioplot{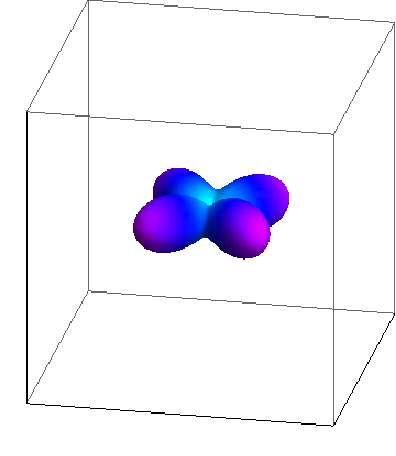} & \--- & \--- &\\
 & \qnQ{}\,=\,4 && \--- & \--- & \--- & \--- & \sioplot{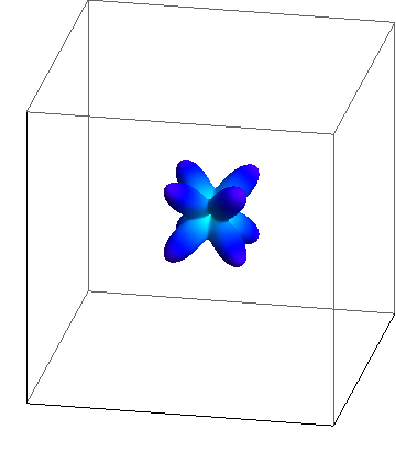} & \sioplot{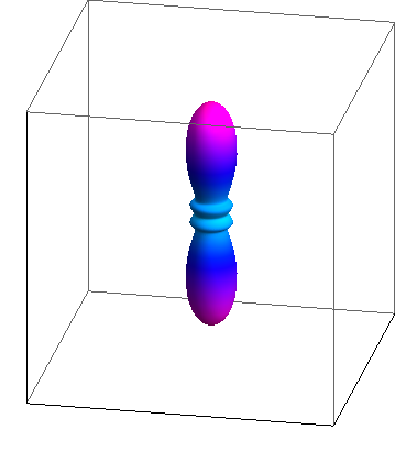} &\vphantom{\raisebox{-9mm}{}}\\
\hline
 & \qnQ{}\,=\,0 && \--- & \--- & \sioplot{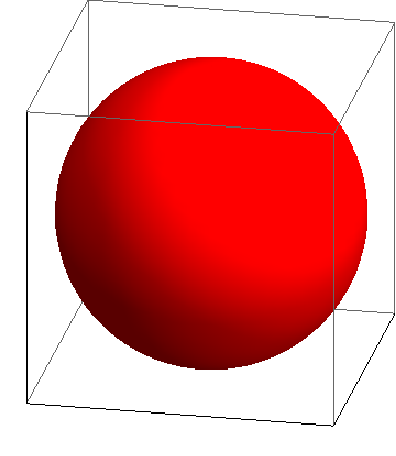} & \sioplot{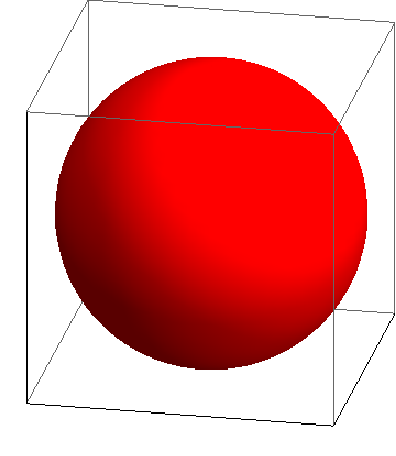} & \--- & \--- &\vphantom{\raisebox{12mm}{}}\\
 & \qnQ{}\,=\,1 && \--- & \sioplot{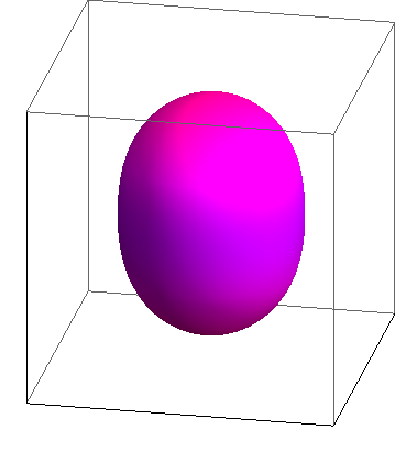} & \--- & \--- & \sioplot{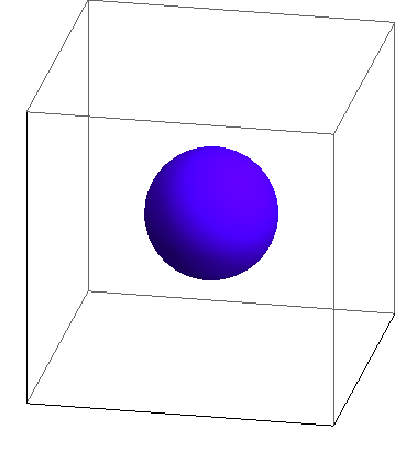} & \sioplot{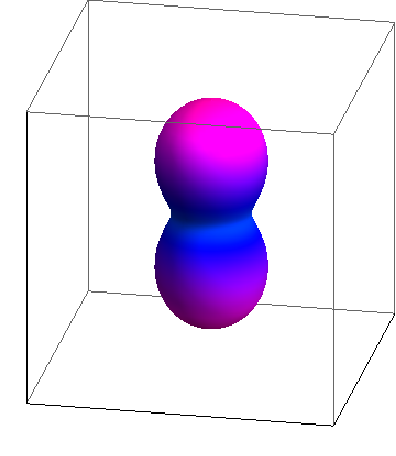} &\\
\raisebox{-6mm}[0mm][0mm]{{\Large $d$}} & \qnQ{}\,=\,2 && \sioplot{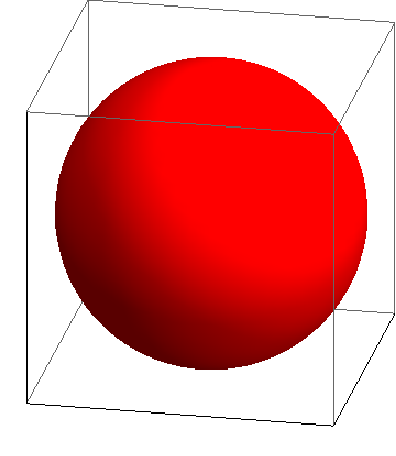} & \--- & \sioplot{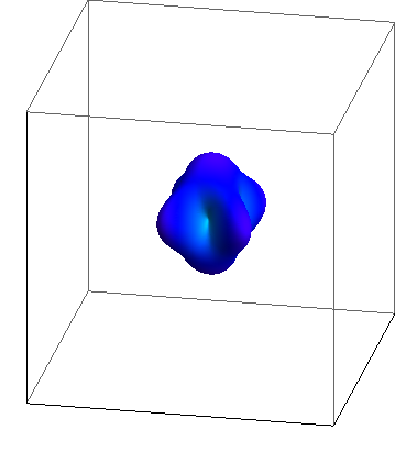} & \sioplot{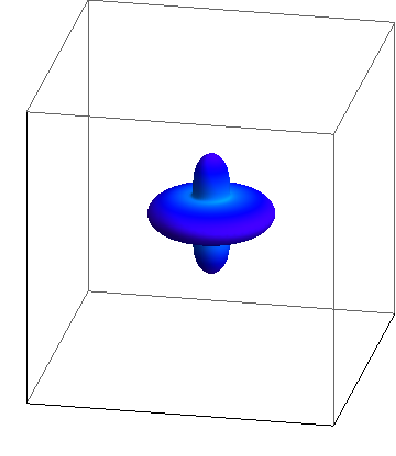} & \--- & \--- &\\
 & \qnQ{}\,=\,3 && \--- & \sioplot{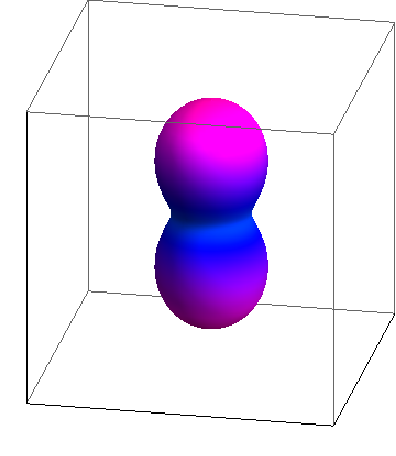} & \--- & \--- & \sioplot{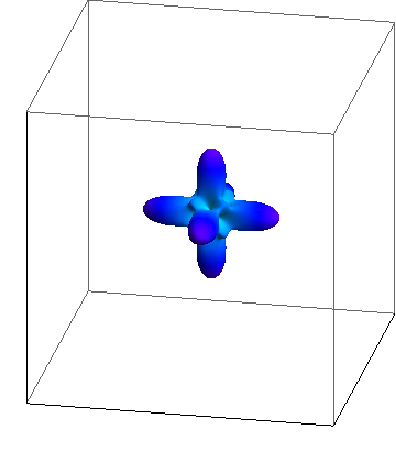} & \sioplot{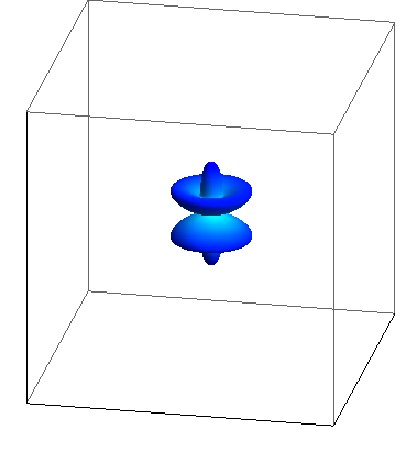} &\\
 & \qnQ{}\,=\,4 && \--- & \--- & \sioplot{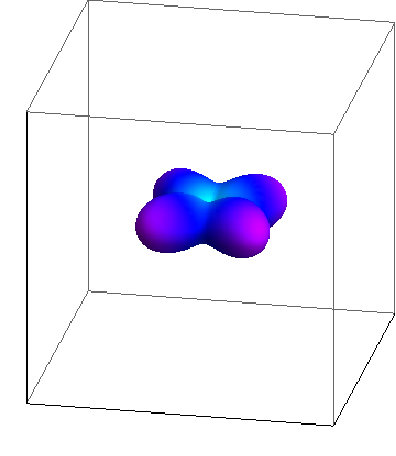} & \sioplot{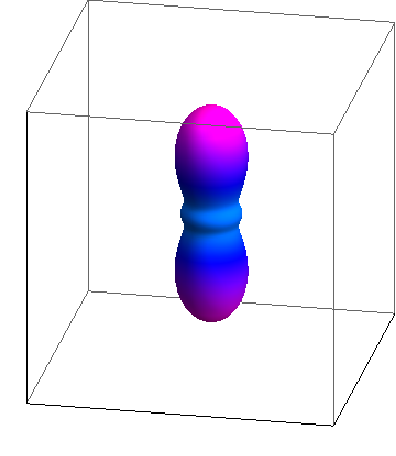} & \--- & \--- &\\
 & \qnQ{}\,=\,5 && \--- & \--- & \--- & \--- & \sioplot{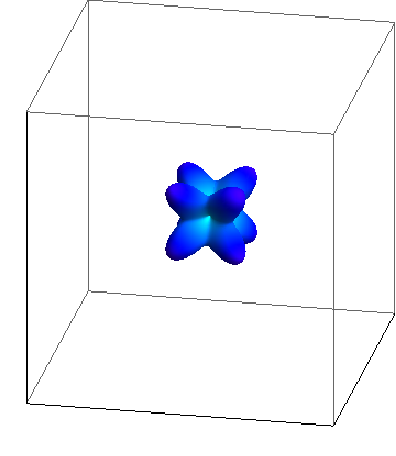} & \sioplot{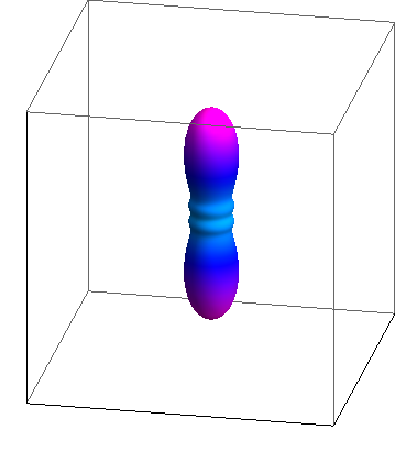} &\vphantom{\raisebox{-9mm}{}}\\
\hline
\end{tabular}
\caption{Plot of the direction of $\vecq{}$ dependence of transitions into specific valence states \ket{$\qnl{}_\text{f},\qnml{}_\text{f}$} as defined in \tab{tab:cubicbasis} (top ruler). It shows the result of \mbox{\( (2\qnQ{}\text{+}1) \, \abs{\sum_{m=-\qnQ{}}^{\qnQ{}} \bra{$\qnl{}_\text{f},\qnml{}_\text{f}$}\,\Clm{\qnQ{}}{$m$}\,\ket{$\qnl{}_\text{i},\qnml{}_\text{f}$-$m$} \, \Clm{\qnQ{}}{$m$}^{\ast} }^2 \)} for each multipole order \qnQ{} and the core level $\qnl{}_\text{i}$ up to $d$ (left).}
\label{fig:angdepplots}
\end{figure}

\begin{figure}
\newcommand*\sioplot[1]{%
\raisebox{-9mm}[10mm][0mm]{%
\hspace{-1.5mm}%
\includegraphics[height=20mm]{cqqvi_#1.png}%
\hspace{-3.5mm} } }
\centering
\begin{tabular}{|cc|ccccccc|}
\hline
    & && $p_z$ & $d_{xy}$ & $d_{3z^2-r^2}$ & $f_{xyz}$ & $f_{5z^3-3zr^2}$ &\\
\hline
 & \qnQ{}\,=\,0, $\qnQ{}'$\,=\,2 && \sioplot{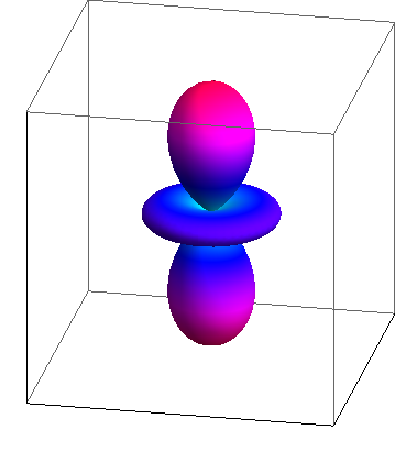} & \--- & \--- & \--- & \--- &\vphantom{\raisebox{12mm}{}}\\
{\Large $p$} & \qnQ{}\,=\,1, $\qnQ{}'$\,=\,3 && \--- & \sioplot{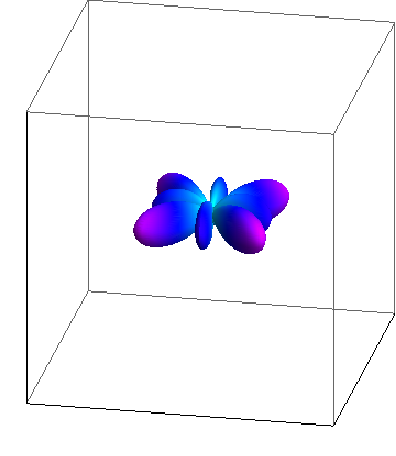} & \sioplot{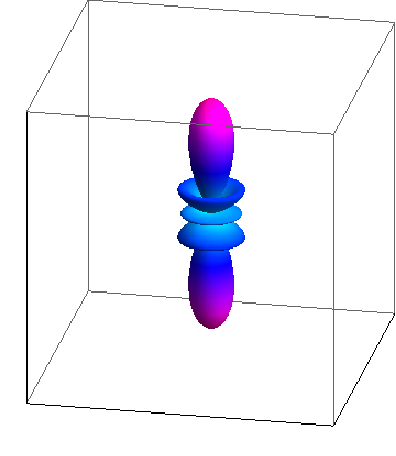} & \--- & \--- &\\
 & \qnQ{}\,=\,2, $\qnQ{}'$\,=\,4 && \--- & \--- & \--- & \sioplot{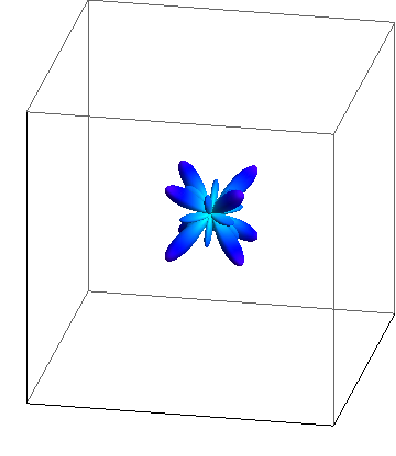} & \sioplot{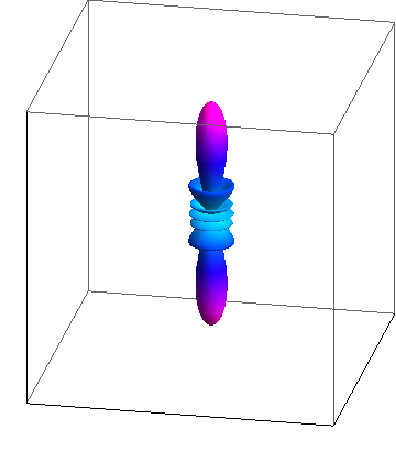} &\vphantom{\raisebox{-9mm}{}}\\
\hline
 & \qnQ{}\,=\,0, $\qnQ{}'$\,=\,2 && \--- & \sioplot{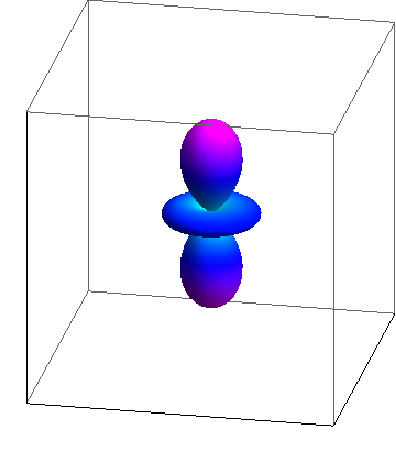} & \sioplot{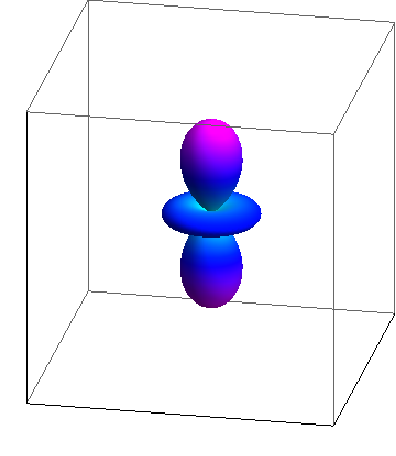} & \--- & \--- &\vphantom{\raisebox{12mm}{}}\\
 & \qnQ{}\,=\,1, $\qnQ{}'$\,=\,3 && \sioplot{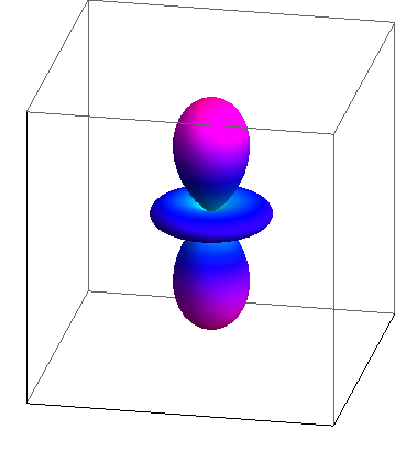} & \--- & \--- & \sioplot{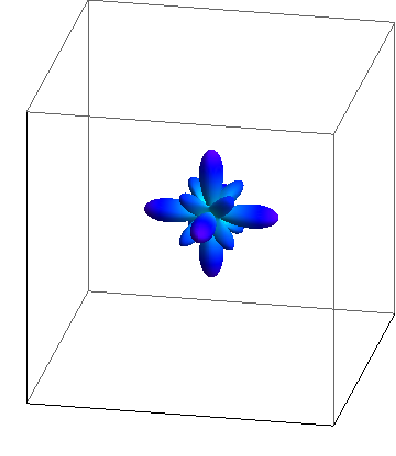} & \sioplot{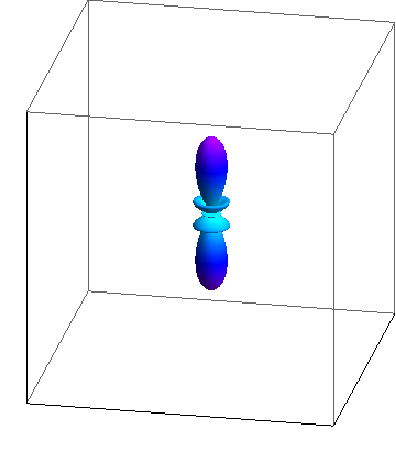} &\\
\raisebox{-6mm}[0mm][0mm]{{\Large $d$}} & \qnQ{}\,=\,2, $\qnQ{}'$\,=\,4 && \--- & \sioplot{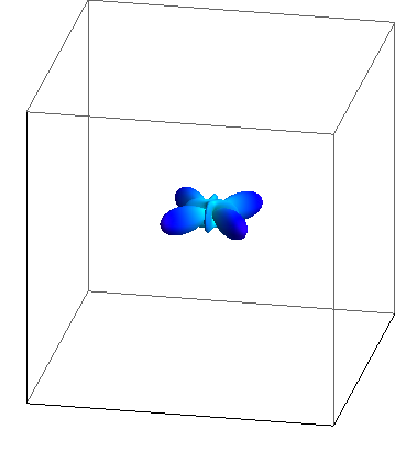} & \sioplot{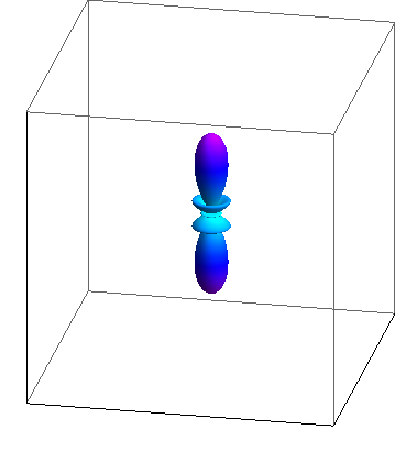} & \--- & \--- &\\
 & \qnQ{}\,=\,3, $\qnQ{}'$\,=\,5 && \--- & \--- & \--- & \sioplot{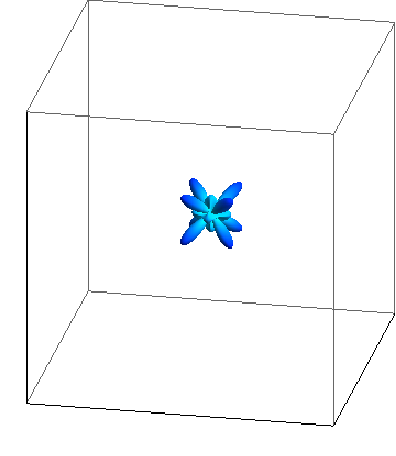} & \sioplot{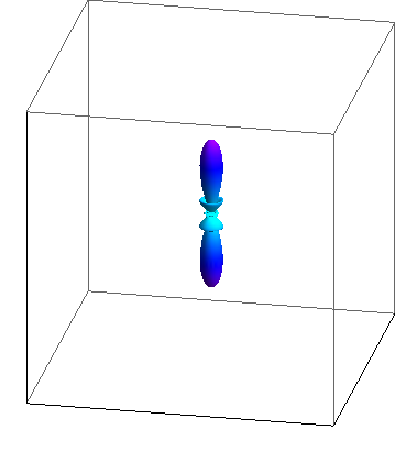} &\\
 & \qnQ{}\,=\,0, $\qnQ{}'$\,=\,4 && \--- & \sioplot{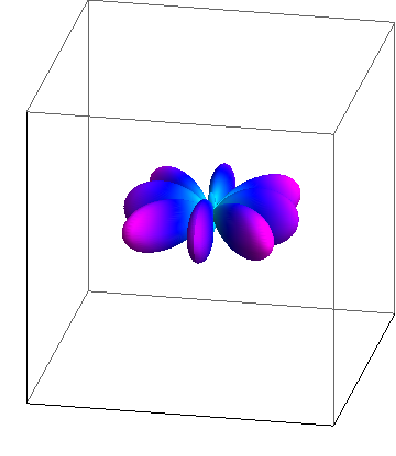} & \sioplot{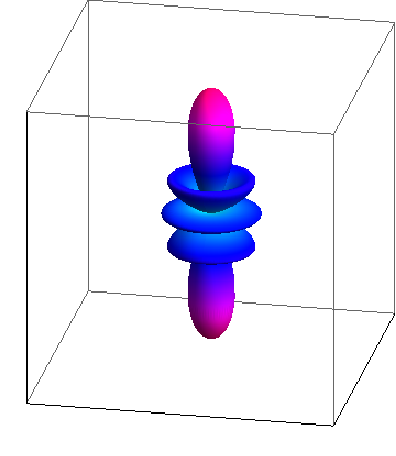} & \--- & \--- &\\
 & \qnQ{}\,=\,1, $\qnQ{}'$\,=\,5 && \--- & \--- & \--- & \sioplot{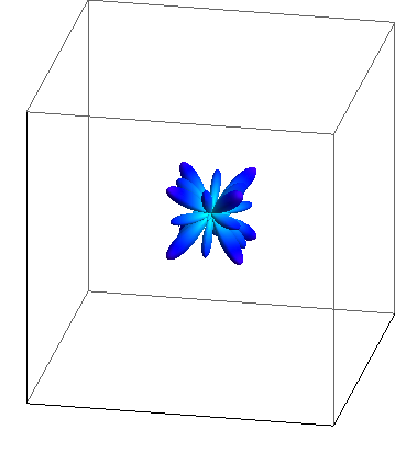} & \sioplot{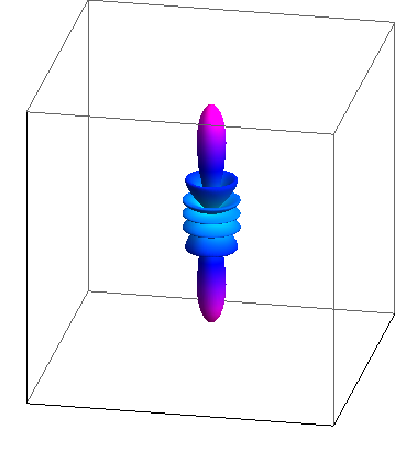} &\vphantom{\raisebox{-9mm}{}}\\
\hline
\end{tabular}
\caption{Plot of the direction of $\vecq{}$ dependence of transitions into specific valence states \ket{$\qnl{}_\text{f},\qnml{}_\text{f}$} as defined in \tab{tab:cubicbasis} (top ruler). It shows the result of \mbox{\(\sqrt{(2\qnQ{}\text{+}1)(2\qnQ{}'\text{+}1)} \sum_{\qnml{}_\text{i}=-\qnl{}_\text{i}}^{\qnl{}_\text{i}} \bra{$\qnl{}_\text{i},\qnml{}_\text{i}$}\,\Clm{\qnQ{}}{$\qnml{}_\text{i}-\qnml{}_\text{f}$}^{\ast}\,\ket{$\qnl{}_\text{f},\qnml{}_\text{f}$} \bra{$\qnl{}_\text{f},\qnml{}_\text{f}$}\,\Clm{\qnQ{}'}{$\qnml{}_\text{f}-\qnml{}_\text{i}$}\,\ket{$\qnl{}_\text{i},\qnml{}_\text{i}$} \)} \mbox{\( \cdot \Clm{\qnQ{}}{$\qnml{}_\text{i}-\qnml{}_\text{f}$} \, \Clm{\qnQ{}'}{$\qnml{}_\text{f}-\qnml{}_\text{i}$}^{\ast} \)} for the mixing term between different multipole orders \qnQ{}, $\qnQ{}'$ and the core level $\qnl{}_\text{i}$ up to $d$ (left).}
\label{fig:angdepplots2}
\end{figure}

The directional \vecq{} dependence is contained in the second line of \equ{eq:transitionopfinal}.
The transition rate depends on the direction of \vecq{}.
The directional dependent prefactor $\Clm{\qnQ{}}{$\qnml{}_\text{f}-\qnml{}_\text{i}$}^{\ast}(\theta_{\vecq{}},\phi_{\vecq{}})$ is sensitive to the magnetic quantum number of the states involved.
For different directions of \vecq{} different valence states are probed, which yields the information of the initial \CF{} state as one can only excite the core electron into unoccupied states.
For example: for a dipole transition with \vecq{}\,||\,\vec{z} (\qnQ{}\,=\,1 and $m_{\qnQ{}}$\,=\,0) $\qnml{}_\text{i}$\,=\,$\qnml{}_\text{f}$ must be fulfilled.
When exciting from a $s$ level \ket{$\qnl{}_\text{i}$\,=\,0, $\qnml{}_\text{i}$\,=\,0} the electron must end in the $p_z$ state \ket{$\qnl{}_\text{f}$\,=\,1, $\qnml{}_\text{f}$\,=\,0} and the signal is proportional to the number of $p_z$ holes.

The variation of the cross-section for transitions from fully occupied $s$, $p$, and $d$ core levels into single selected cubic harmonic states are shown in \fig{fig:angdepplots} for single multipole order and in \fig{fig:angdepplots2} for interfering multipole orders.
Note that the other cubic harmonic states, which are not shown, can be obtained by rotation of the states presented.

The directional \vecq{} dependence strongly depends on \qnQ{}.
The monopole never shows a directional dependence, but the higher \qnQ{} the more structured it becomes.
The dipole fails to differentiate between all $f$ states in a $d$ to $f$ transition, but the octupole (\qnQ{}\,=\,3) and dotriacontapole (\qnQ{}\,=\,5) do so.
The higher the multipoles the more information can be obtained.

Coming back to the question which \qnQ{} to choose for the experiment, \qnQ{} should be at least so high, that the states of interest can be resolved.
If multiple \qnQ{} are applicable, one may look for the one with the best contrast.
One drawback of higher \qnQ{} is the higher energy needed for realizing larger \absq{}.
A good choice for the investigation of $f$ states with $d$ core levels is to optimize for \qnQ{}\,=\,3.
For \qnQ{}\,=\,3 the seven linear independent directions of \Clm{3}{$m$}$(\theta_{\vecq{}},\phi_{\vecq{}})$ can uniquely represent the seven $f$ orbitals.
This is one of the reasons that the measurements in this thesis are performed around \absq{}\,$\approx$\,10\,$\AA^{-1}$ (see \fig{fig:rjrofall}).

\subsection{Resonant vs.\ non-resonant interactions}\label{sec:resvsnres}
\begin{figure}
  \begin{subfigure}[t]{0.48\textwidth}
    \includegraphics[width=\textwidth]{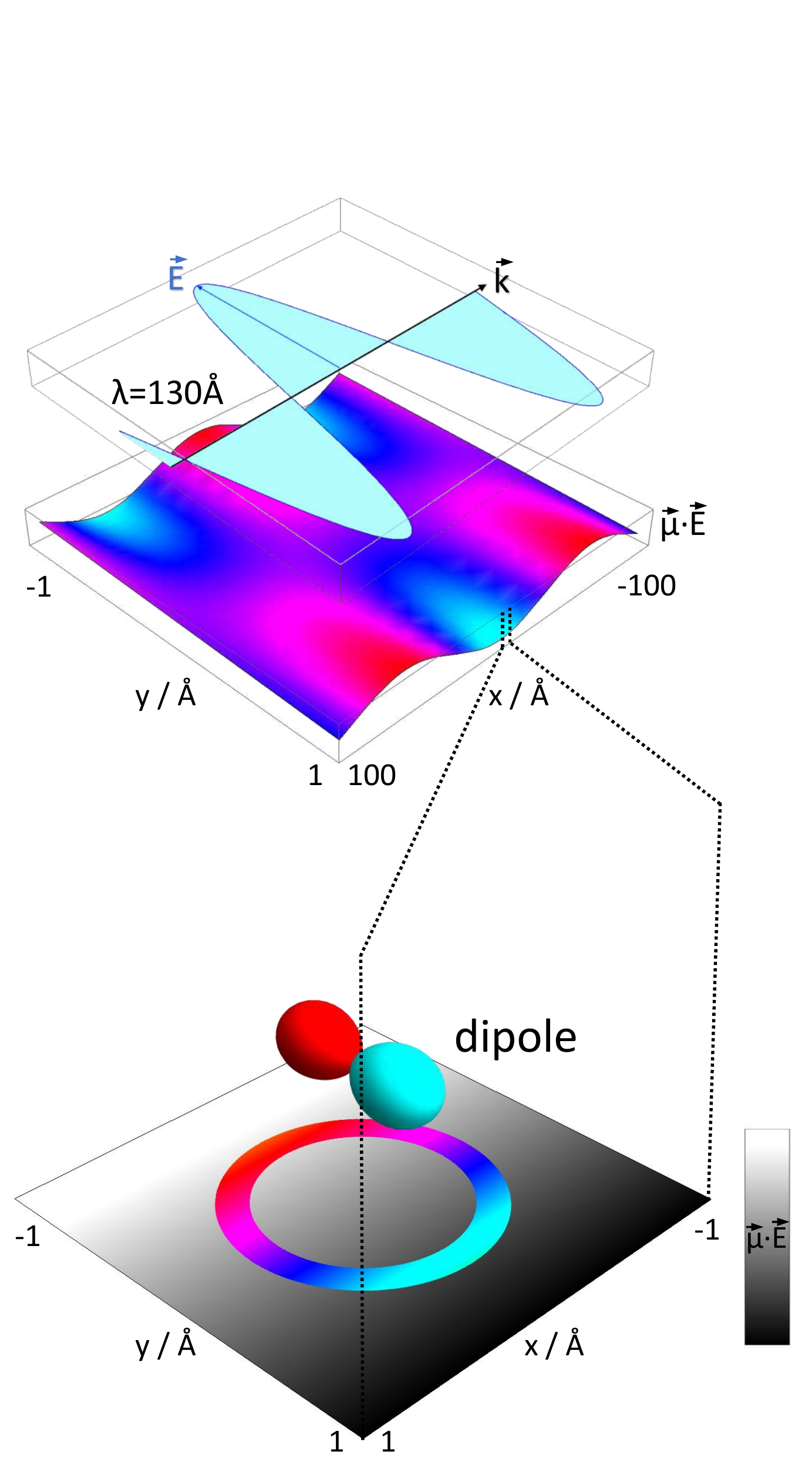}
    \caption{\XAS{} scheme for y-polarized light with $\lambda$\,=\,130\,$\text{\AA}$ (h$\nu$$\approx$10$^2$\,eV) and \vec{k}\,||\,\vec{x}. The top panel shows the photon and the panel below shows the product of the electrical dipole moment \vec{\mu} of an ion and the electrical field \vec{\epsilon} of the photon. The bottom indicates the profile of the \vec{\mu}$\cdot$\vec{\epsilon} part on an atomic scale. An annulus with \rBohr{} radius is highlighted on a color scale and compared to a dipolar $p_y$ tesseral harmonics function.}
  \end{subfigure}
  \hfill
  \begin{subfigure}[t]{0.48\textwidth}
    \includegraphics[width=\textwidth]{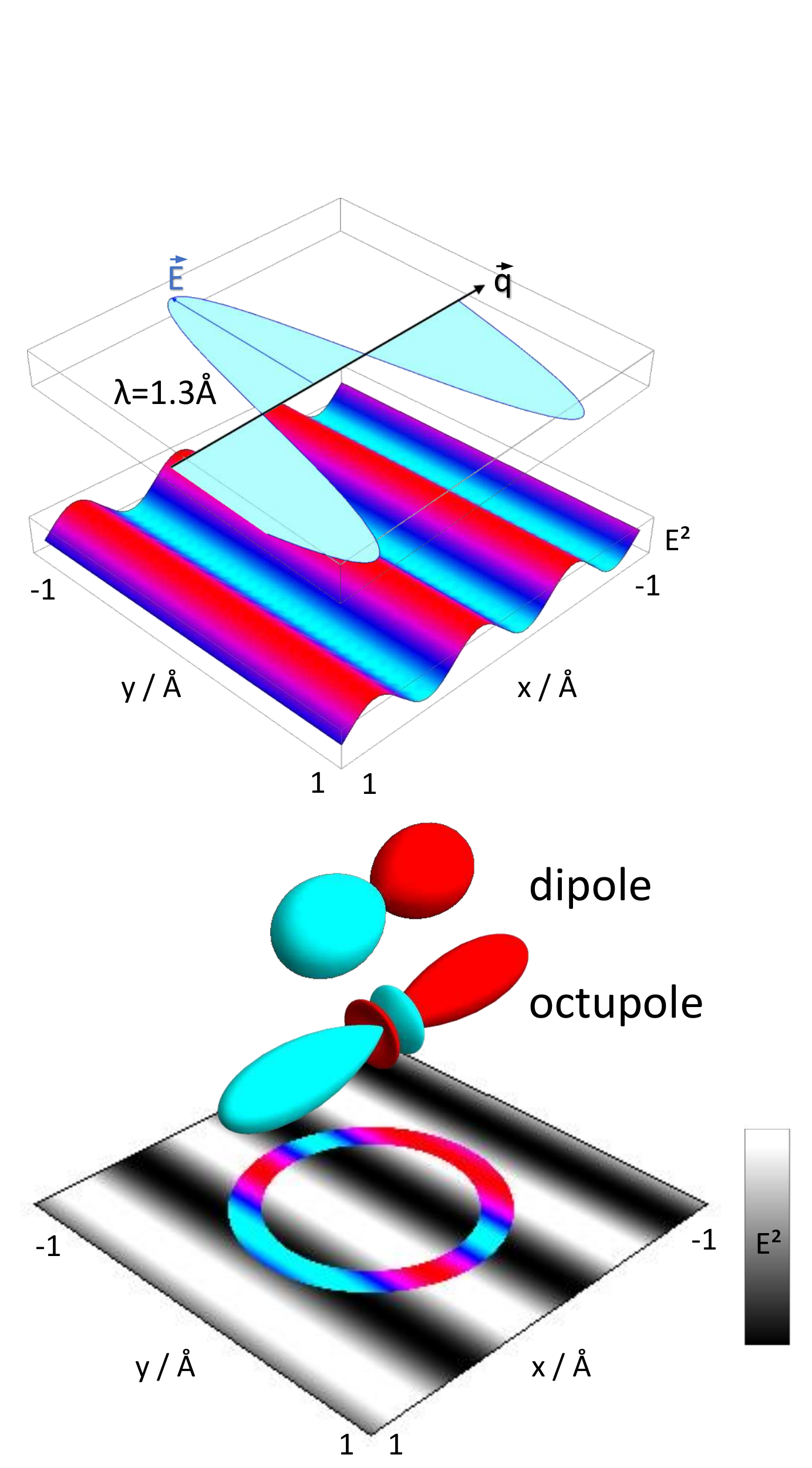}
    \caption{\NIXS{} scheme for y-polarized light with $\lambda$\,=\,1.3\,$\text{\AA}$ (h$\nu$$\approx$10$^4$\,eV) and $\vecq{}\,||\,\vec{x}$. The top panel shows the photon and the panel below shows the spatial distribution of the energy density $\vec{\epsilon}^2$ of the electric field. The bottom indicates the profile of the $\vec{\epsilon}^2$ distribution. An annulus with \rBohr{} radius is highlighted on a color scale and compared to the dipolar $p_x$ and the octupolar $f_{5x^3-3x}$ tesseral harmonic functions.}
  \end{subfigure}
  \caption{Sketch of the single step transition processes for the resonant (\XAS{}) and non-resonant (\NIXS{}) scenario. Note that the plotted wave functions result from the Legendre polynomials in \equ{eq:legendrepolinominal}.}
  \label{fig:xasvsnixs}
\end{figure}

\XAS{} can be sensitive to the occupation of specific valence states in a similar manner as \NIXS{}.
But there are some conceptional differences.
Most obvious, \XAS{} is a absorption, whereas \NIXS{} is a scattering process.
In the following, it is pointed out why higher multipoles are easily realized in \NIXS{} but not in \XAS{} and why the polarization in \XAS{} plays the role of momentum transfer \vecq{} of \NIXS{}.

For the \emph{non-resonant} term was shown that two fields, of the incoming and of the scattered photon, are involved.
The product of the two related polarizations ($\vec{\epsilon}_\text{out} \cdot \vec{\epsilon}_\text{in}$), enter the scattering function as a geometric prefactor.
The transitions between the states are quasi independent of the polarization and are simplified to \( \exp{\imath \vecq{}\vec{r}} \).
Due to the free choice of \absq{} in a scattering experiment, the exponent may become large and the full expansion of the plane wave has to be considered.
Here the higher multipole orders come into play.

For the \emph{resonant} term the electric polarization \vec{E} interacts with the orbital momentum \(\vec{\mu} \propto \vec{p}\,|i\rangle\) (see \equ{eq:HkinSplit}).
Here $\abs{\vec{k}}$ of the absorbed photon is fixed by its energy $\hbar\omega$.
This energy must fit the electron binding energy $\Delta E$.
For an electron bound to an atom the exponent of \(\exp{\imath \vec{k} \vec{r}}\) can therefore be estimated to:
\begin{align}
  \abs{\vec{k}} r_0 &\approx \frac{\qnZ{} \echarge{}^2}{2\hbar \sol{}} = \frac{\qnZ{}}{274} \ll 1 &\text{with } \abs{\vec{k}} = \frac{\Delta E}{\hbar \sol{}}, \label{eq:resonantexponent}
\end{align}
where \qnZ{} is the atomic number, \( \vec{r}_0 \) is the radial extension, and \(\Delta E\) is the binding energy of the electron equal to the photon energy\,\cite[Eq.\,15.100b]{Fayyazuddin1990}.
The gray region in \fig{fig:sphericalbesselj} indicates the \absq{}-range available to resonant processes.
Thus, \equ{eq:resonantexponent} justifies the dipole approximation
\begin{align}
\exp{\imath \vec{k}\vec{r}} &\approx 1. \label{eq:dipoleapproximation}
\end{align}
The transition operator can now be written as
\begin{align}
\vec{\mu} \, \vec{\epsilon} \, \exp{\imath \vec{k}\vec{r}} \approx \vec{\mu} \, \vec{\epsilon} = \abs{\vec{\mu}} \, \abs{\vec{\epsilon}} \, \cos{}\,\theta_{\vec{\epsilon}\vec{\mu}}, \label{eq:XAStransitionoperator}
\end{align}
where \(\theta_{\vec{\epsilon}\vec{\mu}}\) is the angle between the electric field vector \(\vec{\epsilon}\) and the induced orbital momentum \(\vec{\mu}\).
Realizing that \(\Legendre{k=1}{\cos{\theta}} \equiv \cos{\theta}\) explains why it is called dipole approximation.
\Fig{fig:xasvsnixs} shows a graphical interpretation for both cases, resonant (\XAS{}) and non-resonant (\NIXS{}).

\subsection{Scattering tensor}\label{sec:scatteringtensor}

Although \vecq{} can be tuned continuously in an infinite number of directions, the provided information remains finite.
Each multipole transition has 2\qnQ{}+1 linear independent directions.
This means that it is possible to derive \sqw{} for all directions from a finite set of calculations with different \vecq{}.

For resonant dipole transitions (\XAS{}) the polarization \vec{\epsilon} can be usually expressed as a linear combination of ($\epsilon_x$, $\epsilon_y$, $\epsilon_z$), e.g.\ ($\epsilon_{m=-1}$, $\epsilon_{m=0}$, $\epsilon_{m=1}$).
This forms a complete basis for the angular momentum of 1.
Here all experiments can be described by 
\begin{align}
S(\vec{\epsilon},\omega) &= \vec{\epsilon}^{\ast}\cdot\sigma(\omega)\cdot\vec{\epsilon} \label{eq:tensornotation} \\
 &= (\epsilon_x,\epsilon_y,\epsilon_z) \left(
{\setlength\arraycolsep{2pt}\setstretch{1}\begin{array}{ccc}
 \sigma_{xx}(\omega) & \sigma_{xy}(\omega) & \sigma_{xz}(\omega) \\
 \sigma_{yx}(\omega) & \sigma_{yy}(\omega) & \sigma_{yz}(\omega) \\
 \sigma_{zx}(\omega) & \sigma_{zy}(\omega) & \sigma_{zz}(\omega) \\
\end{array}}
\right)\left(
{\setlength\arraycolsep{2pt}\setstretch{1}\begin{array}{c}
 \epsilon_x \\
 \epsilon_y \\
 \epsilon_z \\
\end{array}}
\right) \nonumber
\end{align}
with the rank 2 conductivity tensor $\sigma(\omega)$.
It has 3$\times$3 dimensions, related to the 3 independent directions of $\epsilon$.
With these nine elements every resonant process is described.

When working on a local basis, the directional \vecq{} dependence in \NIXS{} can be simplified using the same trick.
Here comes the multipole order back into play.
As described in the second line of \equ{eq:transitionopfinal}, the scattering vector \vecq{} is already projected onto \vec{\varepsilon}(\vecq{})\,=\,\Clm{\qnQ{}}{$m$}$(\theta_{\vecq{}},\phi_{\vecq{}})$.
For the dipole case (\qnQ{}\,=\,1) \vecq{} can consequently be expressed by ($\varepsilon_x$, $\varepsilon_y$, $\varepsilon_z$), just as before.
For quadrupole transitions (\qnQ{}\,=\,2) the projection of the momentum transfer \vec{\varepsilon}(\vecq{})\,=\,\Clm{\qnQ{}=2}{$m$}$(\theta_{\vecq{}},\phi_{\vecq{}})$ has 5 independent elements in its irreducible representation, e.g.\ ($\varepsilon_{x^2-y^2}$, $\varepsilon_{3z^2-r^2}$, $\varepsilon_{xz}$, $\varepsilon_{yz}$, $\varepsilon_{xy}$).
For \qnQ{}\,=\,3 it has 7, for \qnQ{}\,=\,4 it has 9, and so on.

Note, the cubic harmonic functions \Klm{\qnQ{}}{$m$} are used instead of the renormalized spherical harmonics.
Any complete basis can be used that fulfills \equ{eq:legendrepolinominal}, like
\begin{align}
\Legendre{k}{ \cos{\theta} } &= \frac{4\pi}{2\qnQ{}+1} \sum\limits_{m=-\qnQ{}}^{\qnQ{}} \Klm{\qnQ{}}{$m$}^\ast(\theta_{\vecq{}},\phi_{\vecq{}}) \, \Klm{\qnQ{}}{$m$}(\theta_{\vec{r}},\phi_{\vec{r}}). \label{eq:cubiclegendrepolinominal}
\end{align}
A scattering tensor or extended conductivity tensor can be defined in analogy to the conductivity tensor, so that the observed response of the system can be calculated by
\begin{align}
\sqw{} &= \vec{\varepsilon}^{\ast}(\vecq{})\cdot\sigma(\omega)\cdot\vec{\varepsilon}(\vecq{}) \label{eq:tensornotation}
\end{align}
for a given \ket{$\qnn{}_c,\qnl{}_c$} and valence shell \ket{$\qnn{}_v,\qnl{}_v$} as well.
For this all allowed multipole orders have to be considered.
The dimension of $\varepsilon(\vecq{})$ is given by \(\sum_{\qnQ{}}\)\,2\qnQ{}+1.
$\qnQ{}_j$ and $m_j$ is used to index all allowed elements in this basis with multiple multipole orders.
The single elements of the tensor in \equ{eq:tensornotation}, when defined as
\begin{align}
\sigma_{j,k}(\omega) &=  \sum_{\ket{i},\ket{f}} \frac{\exp{\frac{-E_\text{i}}{\kB{}\T{}}}}{\pfZ{}} ~ (4\pi)^2 \imath^{\qnQ{}_j+\qnQ{}_k} ~ \langle \qnl{}_c,m_\text{i} | ~ \Klm{$\qnQ{}_j$}{$m_j$}^{\ast}(\theta_{\vec{r}},\phi_{\vec{r}}) ~ | \qnl{}_v,m_\text{f} \rangle \nonumber\\
 &\times \langle \qnl{}_v,m_\text{f} | ~ \Klm{$\qnQ{}_k$}{$m_k$}(\theta_{\vec{r}},\phi_{\vec{r}}) ~ | \qnl{}_c,m_\text{i} \rangle ~ \del (E_\text{f} - E_\text{i} - \hbar\omega) \label{eq:sigmaofw} \\
\varepsilon_j(\vecq{}) &= \langle \qnn{}_v,\qnl{}_v | ~ j_{\qnQ{}}(\absq{}\,r) ~ | \qnn{}_c,\qnl{}_c \rangle ~ \Klm{$\qnQ{}_j$}{$m_j$}(\theta_{\vecq{}},\phi_{\vecq{}}) \label{eq:epsilonofq}
\end{align}
will return the correct dynamical structure factor
\begin{align}
\sqw{} &=
 \sum_{\ket{i},\ket{f}} \frac{\exp{\frac{-E_\text{i}}{\kB{}\T{}}}}{\pfZ{}}
 ~ |\sum\limits_{\qnQ{}}^{} \sum\limits_{m=-\qnQ{}}^{\qnQ{}} 4\pi \, \imath^{\qnQ{}} ~ \langle \qnn{}_v,\qnl{}_v | ~ j_{\qnQ{}}(\absq{}\,r) ~ | \qnn{}_c,\qnl{}_c \rangle ~ \Klm{\qnQ{}}{$m$}^{\ast}(\theta_{\vecq{}},\phi_{\vecq{}}) \nonumber \\
 &\times \langle \qnl{}_v,m_\text{f} | ~ \Klm{\qnQ{}}{$m$}(\theta_{\vec{r}},\phi_{\vec{r}}) ~ | \qnl{}_c,m_\text{i} \rangle|^2 ~ \del (E_\text{f} - E_\text{i} - \hbar\omega).
\end{align}

The above formula considers only transitions into a single atomic shell.
If more shells are involved or states cannot be projected on atomic orbitals, one may not be able to separate the integration over $j_{\qnQ{}}(\absq{}r)$ anymore.
Here it is still possible to use
\begin{align}
\sigma'_{j,k}(\omega) &=  \sum_{\ket{i},\ket{f}} \frac{\exp{\frac{-E_\text{i}}{\kB{}\T{}}}}{\pfZ{}} ~ (4\pi)^2 \imath^{\qnQ{}_j+\qnQ{}_k} ~ \bra{i} ~ j_{\qnQ{}_j}(\absq{}\,r) ~ \Klm{$\qnQ{}_j$}{$m_j$}^{\ast}(\theta_{\vec{r}},\phi_{\vec{r}}) ~ \ket{f} \nonumber\\
 &\times \bra{f} ~ j_{\qnQ{}_k}(\absq{}\,r) ~ \Klm{$\qnQ{}_k$}{$m_k$}(\theta_{\vec{r}},\phi_{\vec{r}}) ~ \ket{i} ~ \del (E_\text{f} - E_\text{i} - \hbar\omega) \\
\varepsilon'_j(\vecq{}) &= \Klm{$\qnQ{}_j$}{$m_j$}(\theta_{\vecq{}},\phi_{\vecq{}}).
\end{align}

But now the scattering tensor is restricted to a single value of \absq{} and an infinite number of different tensors exists.
Furthermore, the restrictions for \qnQ{} have to be reconsidered.
This makes it harder to use this scattering tensor to account for certain experimental settings, i.e.\ the variation of \vecq{} within the area of the analyzers cannot be performed exactly.
The integration over solid angles for fixed \absq{} is still possible, which, for example, allows to extract the isotropic spectrum from a single initial state independent of its manifold.

\begin{figure}
  \centering
  \begin{subfigure}[t]{0.26\textwidth}
    \includegraphics[width=\textwidth]{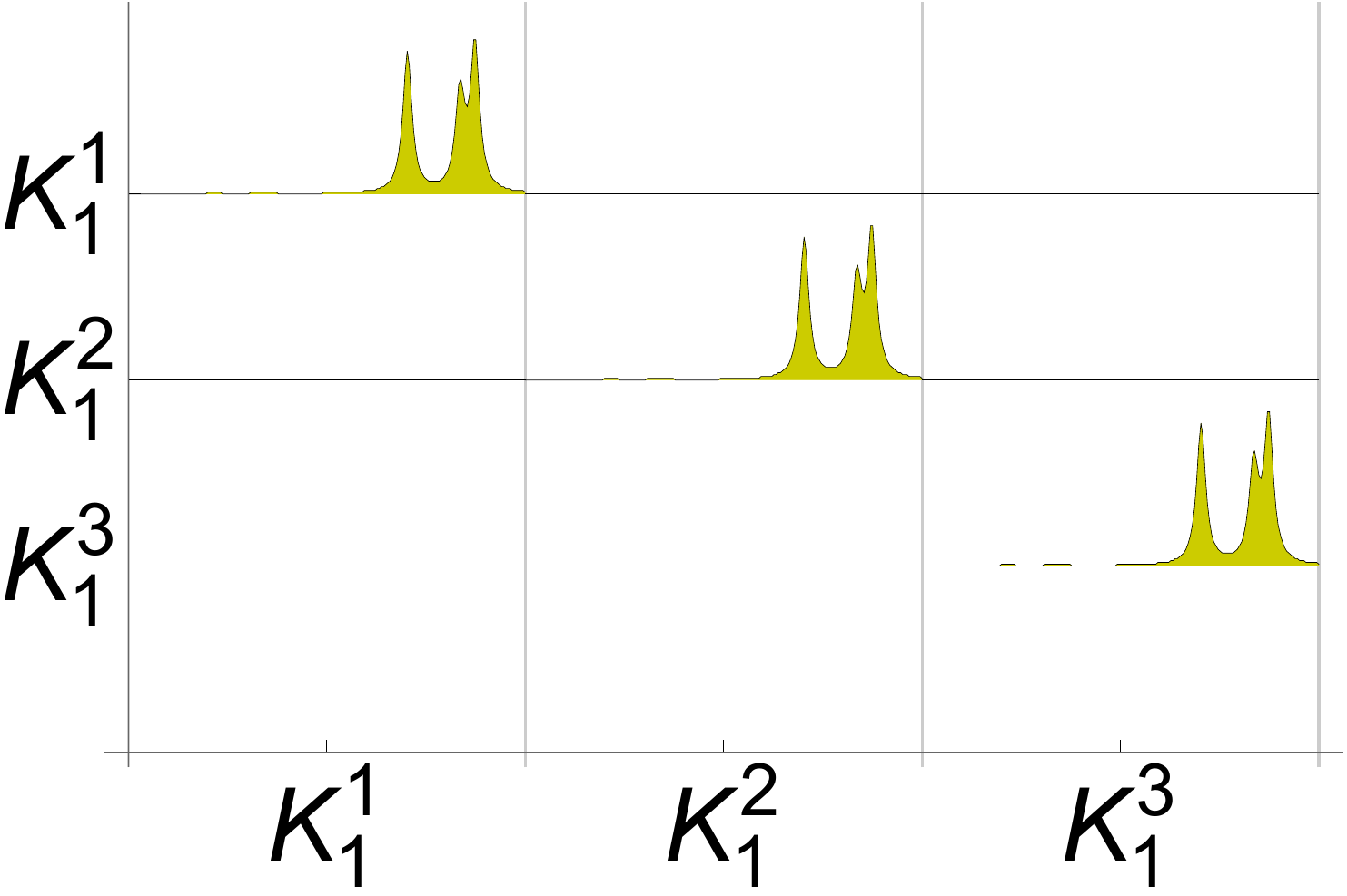}
	\caption{}
  \end{subfigure}\\
  \begin{subfigure}[t]{0.58\textwidth}
    \includegraphics[width=\textwidth]{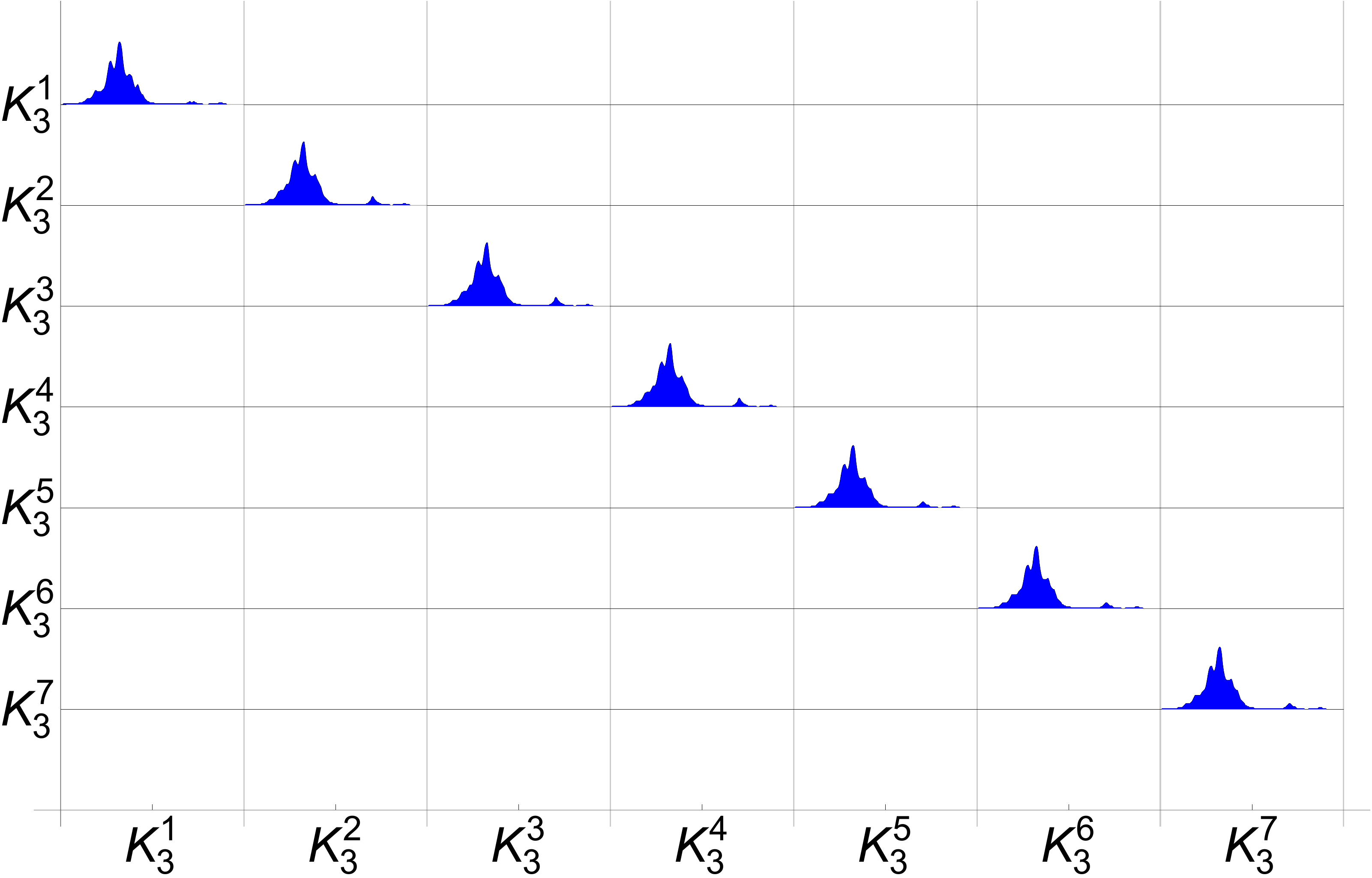}
	\caption{}
  \end{subfigure}\\
  \begin{subfigure}[t]{0.9\textwidth}
    \includegraphics[width=\textwidth]{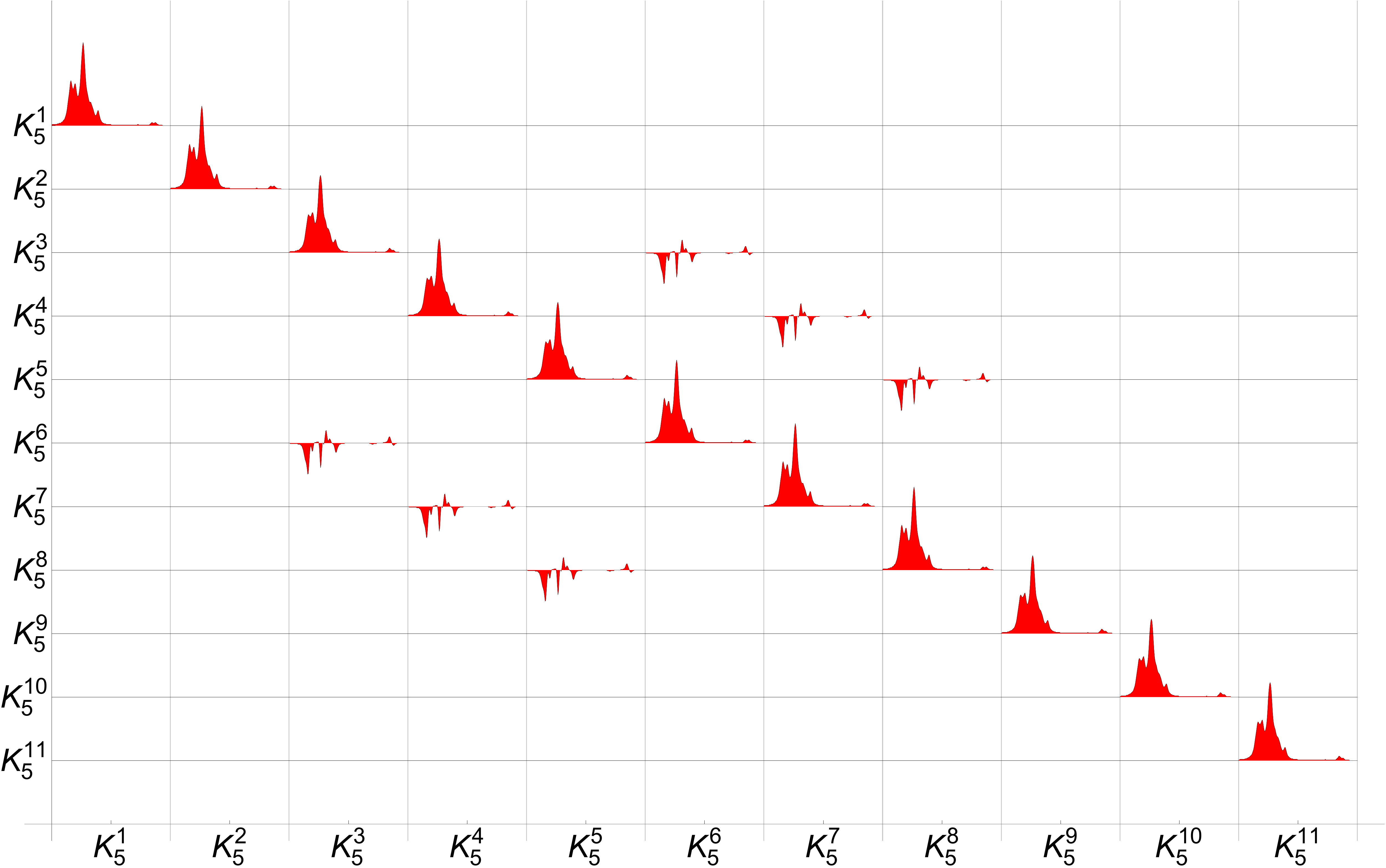}
	\caption{}
  \end{subfigure}
  \caption{Scattering tensor of cubic CeB$_6$ ($f^1$) \edge[Ce]{N}{4,5} edges ($3d\to4f$) in the basis of cubic harmonics. The different plots show the tensor for the single multipole transitions (a) \qnQ{}\,=\,1, (b) \qnQ{}\,=\,3, and (c) \qnQ{}\,=\,5 transitions. The off-diagonal elements in (c) have been multiplied by 10 for better visibility.}
  \label{fig:CeB6tensorSingleQ}
\end{figure}
\Fig[(a-c)]{fig:CeB6tensorSingleQ} show conductivity tensors for pure \qnQ{}\,=\,1, 3, and 5 transitions at the example of the CeB$_6$ \edge[Ce]{N}{4,5} edges (\chap{sec:CeB6}).
All values of $\varepsilon_j(\vecq{})$ are purely real for static \vecq{}, due to the cubic harmonics basis.
Because of this, it is sufficient to consider only the negative imaginary part of $\sigma(\omega)$\,=\,$\frac{1}{\pi}$\bra{i}\,$T \frac{1}{\omega-H-\imath\Gamma} T$\,\ket{f}, as shown in \fig{fig:CeB6tensorSingleQ}.
The dimension increases from 3$\times$3 for the dipole transition (a) to 7$\times$7 for the octupole (b) and 11$\times$11 for the dotriacontapole (c).
For \qnQ{}\,=\,5, in addition to the diagonal elements, some off-diagonal elements are present.
This is because the cubic harmonic basis functions contain two t\textsub{2g} eigenstates in cubic O\textsub{h} symmetry and they can mix.

\begin{figure}
  \centering
  \ifodd\value{page} \includegraphics[angle=-90,width=0.9\textwidth]{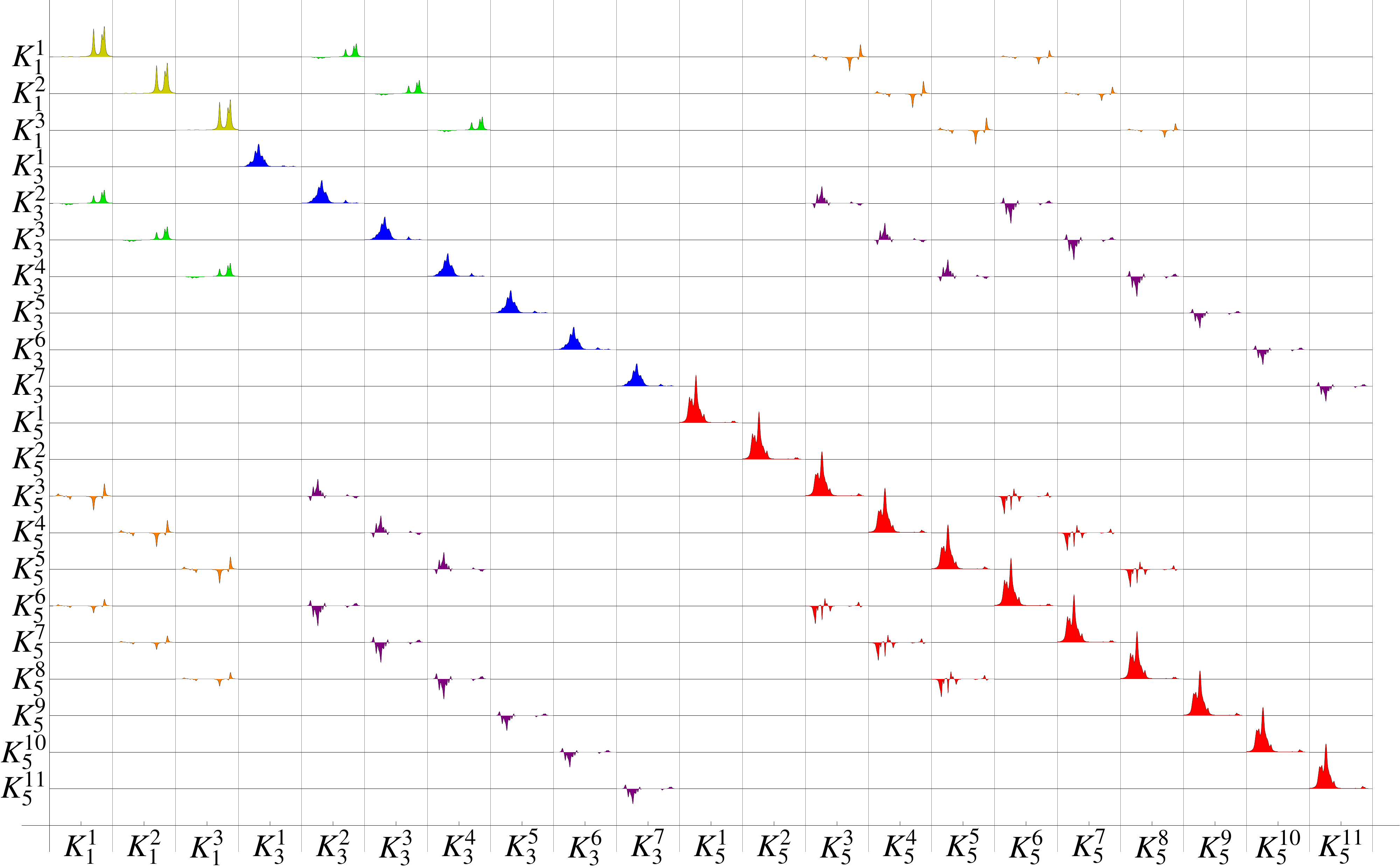} 
  \else \includegraphics[angle=90,width=0.9\textwidth]{CeB6_tensor_full.pdf} \fi
  \caption{Full scattering tensor of cubic CeB$_6$ ($f^1$) \edge[Ce]{N}{4,5} edges ($3d\to4f$) in the basis of cubic harmonics. The different elements are filled according to the different multipole moments: Dipole (\qnQ{}\,=\,1) in yellow, octupole (\qnQ{}\,=\,3) in blue, and dotriacontapole (\qnQ{}\,=\,5) in red. Note also the interference terms: \qnQ{},$\qnQ{}'$\,=\,1,3 in green, \qnQ{},$\qnQ{}'$\,=\,1,5 in orange, and \qnQ{},$\qnQ{}'$\,=\,3,5 in purple. The off-diagonal elements have been multiplied by 10 for better visibility.}
  \label{fig:CeB6tensorfull}
\end{figure}
\Fig{fig:CeB6tensorfull} shows the full conductivity tensor for the CeB$_6$ \edge[Ce]{N}{4,5} edges.
It contains all information of the \NIXS{} experiment.
The real part of the tensor is not required here due to the real basis and the static \vecq{}.
The dimension of the full tensor is 21$\times$21 (3+7+11 for \qnQ{}\,=\,1,3, and 5, respectively).
In addition to the elements already seen in \fig{fig:CeB6tensorSingleQ}, additional off-diagonal elements between different \qnQ{} occur (green, orange, and purple).
As before, they appear only between basis functions which have the same symmetry.
They play an important role for the angular dependence in \sqw{}, such that summing up spectra of independent multipole transitions will yield wrong results.

The symmetry is also present in a way that elements are identical when the basis functions have the same eigenvalues of the symmetry operator.
Here the combination of the cubic harmonic basis and cubic O\textsub{h} symmetry makes the tensor readable without any further math.
Only the symmetries of the cubic harmonic functions (\tab{tab:cubicbasis} in \app{app_atomicvalues}) in O\textsub{h} point symmetry have to be taken into account:
t\textsub{1u} for \Klm{1}{1-3},
a\textsub{2u} for \Klm{3}{1},
t\textsub{1u} for \Klm{3}{2-4},
t\textsub{2u} for \Klm{3}{5-7},
e\textsub{u}  for \Klm{5}{1-2},
t\textsub{1u} for \Klm{5}{3-5},
t\textsub{1u} for \Klm{5}{6-8}, and
t\textsub{2u} for \Klm{5}{9-11}.
Off-diagonal elements only occur between elements containing equal symmetries.
In this representation, also the spectra of elements of same symmetry and same \qnQ{} are equal.

\begin{figure}
  \centering
  \includegraphics[width=\textwidth]{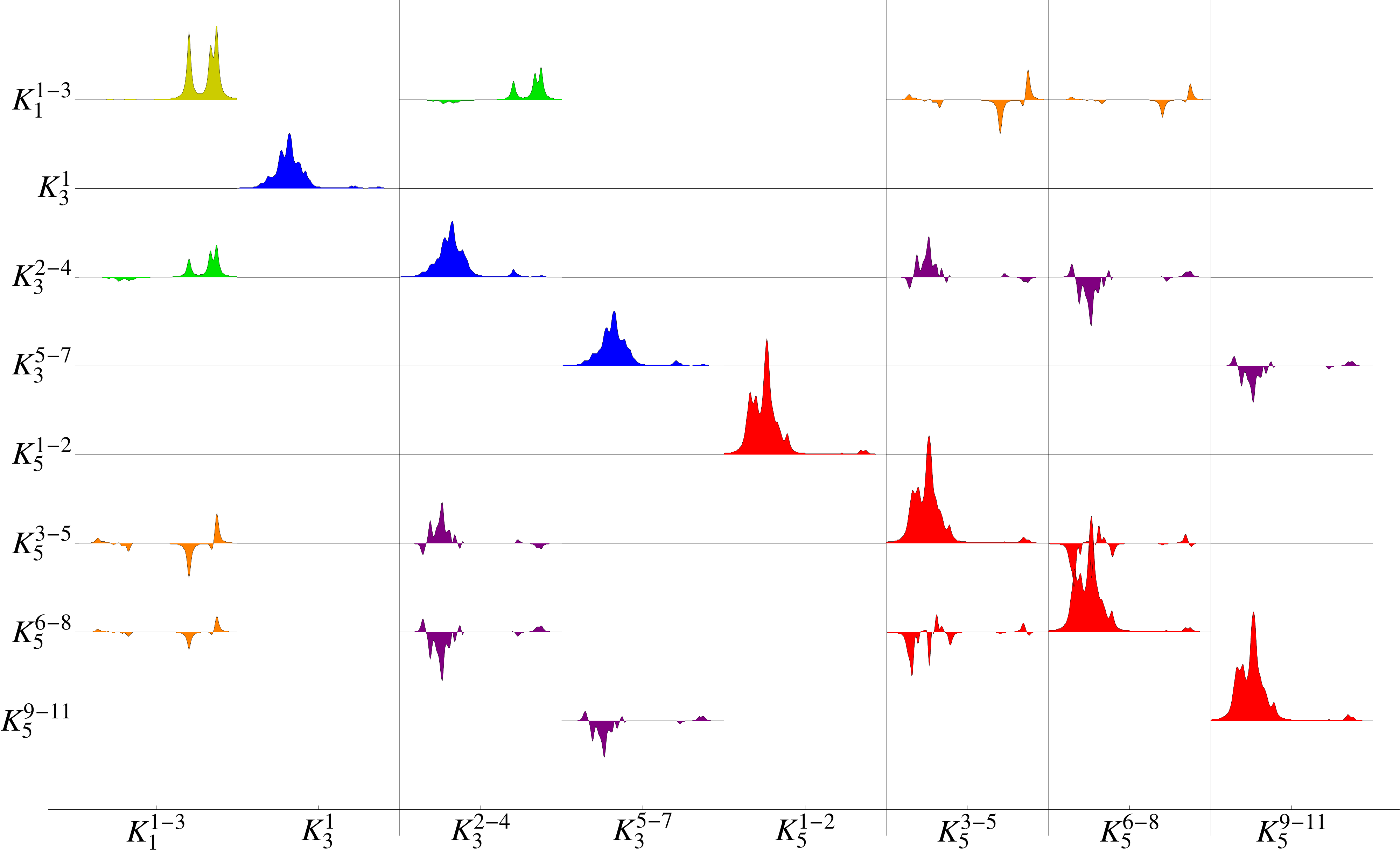}
  \caption{Independent elements of the full scattering tensor (\fig{fig:CeB6tensorfull}) with the same color coding.}
  \label{fig:CeB6tensor}
\end{figure}
\Fig{fig:CeB6tensor} shows the tensor after uniting the identical elements.
The tensor has 15 independent non-zero elements, since it is symmetric.
The variation of the spectra on the diagonal with same \qnQ{} is rather small due to the large number of 13\,holes in the initial state.

\clearpage
\section{Atomic multiplet calculations}\label{sec:calcH}
\label{ch:multiplet}

In this chapter, it is introduced how the multiplet structures of the absorption edges can be calculated, following Carl J.\ Ballhausen\,\cite{Ballhausen1962}.
At first the origin of the different absorption edges will briefly be discussed.
Then spin-orbit coupling (\SOC{}) and Coulomb repulsion (\CI{}) will be introduced.
These two interactions can directly be observed in spectra with a line width of several hundred meV.
Crystal-field (\CF{}) potentials lift the remaining degeneracy of the states further and play an important role for the macroscopic behavior.
In the case of 4$f$ electrons, the \CF{} splittings are to small to be observed in core-level absorption spectra and can be neglected in the final state.
In the initial state, however the symmetry of the \CF{} ground state is crucial.
\SOC{} and \CI{} are rotational invariant and the directional dependence of \vecq{} in the core-level \NIXS{} spectra is due to the anisotropy of the initial \CF{} ground state.
At last, the effect of mixed valence is discussed, which can also affect the spectral shape, which is utilized in \chap{chap:CeRu4Sn6_HAXPES} and \chap{chap:CeRu4Sn6_PFYXAS}.
Interactions, that are much smaller then several hundred meV and do not alter the \CF{} ground state, can neither be resolved in the spectroscopic features nor do they affect the directional dependence.
Therefore, these interactions can be neglected in the interpretation of a \NIXS{} experiment.

In the following, first only the interaction between the electrons with the nucleus will be introduced and later \SOC{}, \CI{}, and \CF{} will be added in order of decreasing energy scale for rare earth and actinides.
The operators are presented in second quantization formalism, which is also the language of the quantum many body script language \Quanty{}.
With the step by step introduction of the operators, the related quantum numbers are introduced.
The calculations of the \SOC{} strength $\zeta_{\qnn{},\qnl{}}$ and the Slater integrals $F^k(i,j)$, $G^k(i,j)$, and $R^k(i,j)$ were performed in the Hartree-Fock approximation with the atomic structure code of Cowan\,\cite{Cowan1981} and are listed in \app{app_atomicvalues}.

\subsection{Atomic shells}

Electrons are bound to the atoms by the strong Coulomb attraction of the charge of the nucleus \qnZ{}\echarge{}.
The energy of an electron which is solely affected by this potential can be expressed by the Hamiltonian
\begin{align}
  H_\text{n} &= - \frac{\hbar^2 \nabla^2}{2\emass{}} - \frac{\qnZ{}\echarge{}^2}{r} \nonumber\\
  &= \sum\limits_{\qnn{},\qnl{},\qnml{},\qnms{}} E_{\qnn{},\qnl{}} ~ \hat{c}^\dag_{\qnn{},\qnl{},\qnml{},\qnms{}}\hat{c}_{\qnn{},\qnl{},\qnml{},\qnms{}}. \label{eq:H_Schroedinger}
\end{align}
Here $\hat{c}^\dag_{\qnn{},\qnl{},\qnml{},\qnms{}}$ and $\hat{c}_{\qnn{},\qnl{},\qnml{},\qnms{}}$ denote the creation and annihilation of the state with the corresponding quantum numbers: the principal quantum number \qnn{}, the angular momentum quantum number \qnl{}, the projection of the angular momentum \qnml{}, and the projection of the spin \qnms{}.
The solution of the Schr\"odinger equation can be expressed on a local basis centered at the nucleus.
This projection yields the famous eigenfunctions, which are expressed by a product of the radial function \Rnl{\qnn{}}{\qnl{}} by the orthogonal spherical harmonic functions \Ylm{\qnl{}}{\qnml{}}, characterized by respective quantum numbers\footnote{The quantum numbers define the character of the wavefunction in terms of the number of nodes in the three dimensions, when transformed to the real tesseral harmonic functions (see \app{app_basesdefinitions}): \qnn{}-\qnl{}-1 nodes in $0<r<\infty$, \qnl{}-\qnml{} nodes in $0\leq\theta<\pi$, and \qnml{} nodes in $0\leq\phi<\pi$}.
The energy of $H_\text{n}$ only depends on \Rnl{\qnn{}}{\qnl{}} and gives rise to the splitting into the atomic shells quantized by \qnn{} and \qnl{}.
The number of fermionic states of these shells is determined by the number of states of \Ylm{\qnl{}}{\qnml{}} and the spin degree of freedom, i.e.\ 2(2\qnl{}+1).

The different shells of strongly bound electrons are well separated such that different core levels excitations can be observed at well separated energies.
They give rise to the different absorption edges: \edge{K}{} for \qnn{}\,=\,1, \edge{L}{1} for \qnn{}\,=\,2 and \qnl{}\,=\,0, \edge{L}{2,3} for \qnn{}\,=\,2 and \qnl{}\,=\,1, and so forth.
These edges are element specific, due to the different charges of different nuclei.
The indexes 2 and 3 of the \edge{L}{2,3} edges refer to the two shells that form because for \qnl{}\,>\,0 the strong \SOC{} of core levels lifts the degeneracy further, with \qnj{}\,=\,1/2 and 3/2 for 2 and 3, respectively (see below).

\subsection{Spin-orbit coupling}
For a single electron in a shell with \qnl{}>0, the electron spin \ops{} will couple with the angular momentum \opl{} with a coupling strength $\zeta_{\qnn{},\qnl{}}$.
In the case of pure \SOC{} \ops{} aligns either parallel or anti-parallel to \opl{}, resulting in 2 possible total momenta \qnj{}\,=\,\qnl{}+\qnms{}, with the spin quantum number \qns{}\,=\,$\frac{1}{2}$ and the projection of the spin \qnms{}\,=\,$\pm\frac{1}{2}$.
These two shells contain 2\qnj{}+1 states for which the total momentum \qnj{} and its projection \qnmj{} turn into good quantum numbers and \(\opl{} \cdot \ops{} = \frac{1}{2} [ \opj{}^2 - \opl{}^2 - \ops{}^2 ]\).
For given \qnl{} the energy increases with increasing \qnj{}.
\begin{align}
 H &= H_\text{n}+H_{\SOC{}}\label{eq:H_SO_only} \\
 H_{\SOC{}} &= \zeta_{\qnn{},\qnl{}} \, \opl{} \cdot \ops{} \nonumber \\
 &= \sum\limits_{\qnml{},\qnms{}} \frac{\zeta_{\qnn{},\qnl{}}}{2} \, \left( \qnj{}(\qnj{}+1) - \qnl{}(\qnl{}+1) - \qns{}(\qns{}+1) \right) \hat{c}^\dag_{\qnn{},\qnl{},\qnml{},\qnms{}}\hat{c}_{\qnl{},\qnml{},\qnms{}} \label{eq:H_SO}
\end{align}
Here $\hat{c}^\dag_{\qnn{},\qnl{},\qnml{},\qnms{}}$ and $\hat{c}_{\qnl{},\qnml{},\qnms{}}$ denote the creation and annihilation of the state with the corresponding quantum numbers \qnl{}, \qnml{}, and \qnms{}.

\subsection{Coulomb interaction}

For shells with more than one electron or hole or two partially filled shells, the \CI{} between the electrons has to be considered.
For simplicity the \SOC{} shall be neglected at first (this will be discussed in the next section).

A shell with orbital momentum quantum number \qnl{} and \textit{N} electrons contains \(\binom{4\qnl{}+2}{N}\) states and can be described by the total spin \opS{}, total orbital momentum \opL{}, and the total momentum \opJ{}, with
\begin{align}
 \opS{} = \sum_{i=1}^{N} \ops{}_i
 \hspace{1cm}\text{and}\hspace{1cm}
 \opL{} = \sum_{i=1}^{N} \opl{}_i
 \hspace{1cm}\text{and}\hspace{1cm}
 \opJ{} = \opL{} + \opS{}.
\end{align}
Similar to the electric potential of the Schr\"odinger equation, the eigenvalues of the Coulomb operator \(H_{\CI{}}\) depend on the total angular momentum quantum number \qnL{} with (2\qnS{}+1)(2\qnL{}+1) fold degenerate states, like the atomic shells before.
Note the total spin quantum number \qnS{} can take any half integer value and the spin degeneracy 2\qnS{}+1 has to be given explicitly.
The Hamiltonian can be written as
\begin{align}
 H &= H_\text{n}+H_{\CI{}} \label{eq:H_CR_only} \\
 H_{\CI{}} &= \sum_{i>j} \frac{\echarge{}^2}{r_{ij}} \nonumber \\
  &= \sum_{k} R^k(\qnl{}_1\qnl{}_2\qnl{}_3\qnl{}_4) \sum\limits_{\qnml{}_1=-\qnl{}_1}^{\qnl{}_1}\sum\limits_{\qnml{}_2=-\qnl{}_2}^{\qnl{}_2}\sum\limits_{\qnml{}_3=-\qnl{}_3}^{\qnl{}_3} \langle \qnl{}_1\qnml{}_1 | \Clm{$k$}{$\qnml{}_1$-$\qnml{}_3$} | \qnl{}_3\qnml{}_3 \rangle \langle \qnl{}_4\qnml{}_4 | \Clm{$k$}{$\qnml{}_4$-$\qnml{}_2$} | \qnl{}_2\qnml{}_2 \rangle \nonumber \\
  &\times \, \hat{c}^\dag_{\qnn{}_1,\qnl{}_1,\qnml{}_1,\qnms{}}\hat{c}^\dag_{\qnn{}_2,\qnl{}_2,\qnml{}_2,\qnms{}'}\hat{c}_{\qnn{}_3,\qnl{}_3,\qnml{}_3,\qnms{}}\hat{c}_{\qnn{}_4,\qnl{}_4,\qnml{}_4,\qnms{}'}, \label{eq:H_CR}
\end{align}
where \Clm{\qnl{}}{\qnml{}} are the renormalized spherical harmonics, and \(\qnml{}_4 = \qnml{}_1 + \qnml{}_2 - \qnml{}_3\) and the factors \bra{\qnl{}\qnml{}}\,\Clm{$k$}{\qnml{}-$\qnml{}'$}\,\ket{$\qnl{}'\qnml{}'$} can be evaluated using the Wigner 3j-symbols (see \app{app_ThreeJSymbols}).
The $R^k$ denote the corresponding Slater integrals
\begin{align}
R^k(\qnl{}_1\qnl{}_2\qnl{}_3\qnl{}_4) &= \echarge{}^2 \hspace{-4pt}\iint\limits_{0}^{\infty} \hspace{-2pt} \frac{\text{Min}(r_1,r_2)^k}{\text{Max}(r_1,r_2)^{k+1}} \Rnl{$\qnn{}_1$}{$\qnl{}_1$}(r_1) \Rnl{$\qnn{}_2$}{$\qnl{}_2$}(r_2) \Rnl{$\qnn{}_3$}{$\qnl{}_3$}(r_1) \Rnl{$\qnn{}_4$}{$\qnl{}_4$}(r_2) r_1^2 r_2^2 \, \diff r_1 \diff r_2.
\end{align}

\begin{figure}
  \centering
  \begin{subfigure}{0.3\textwidth}
    \centering
    \begin{subfigure}{\textwidth}
      \centering
      \includegraphics[height=0.1\textheight]{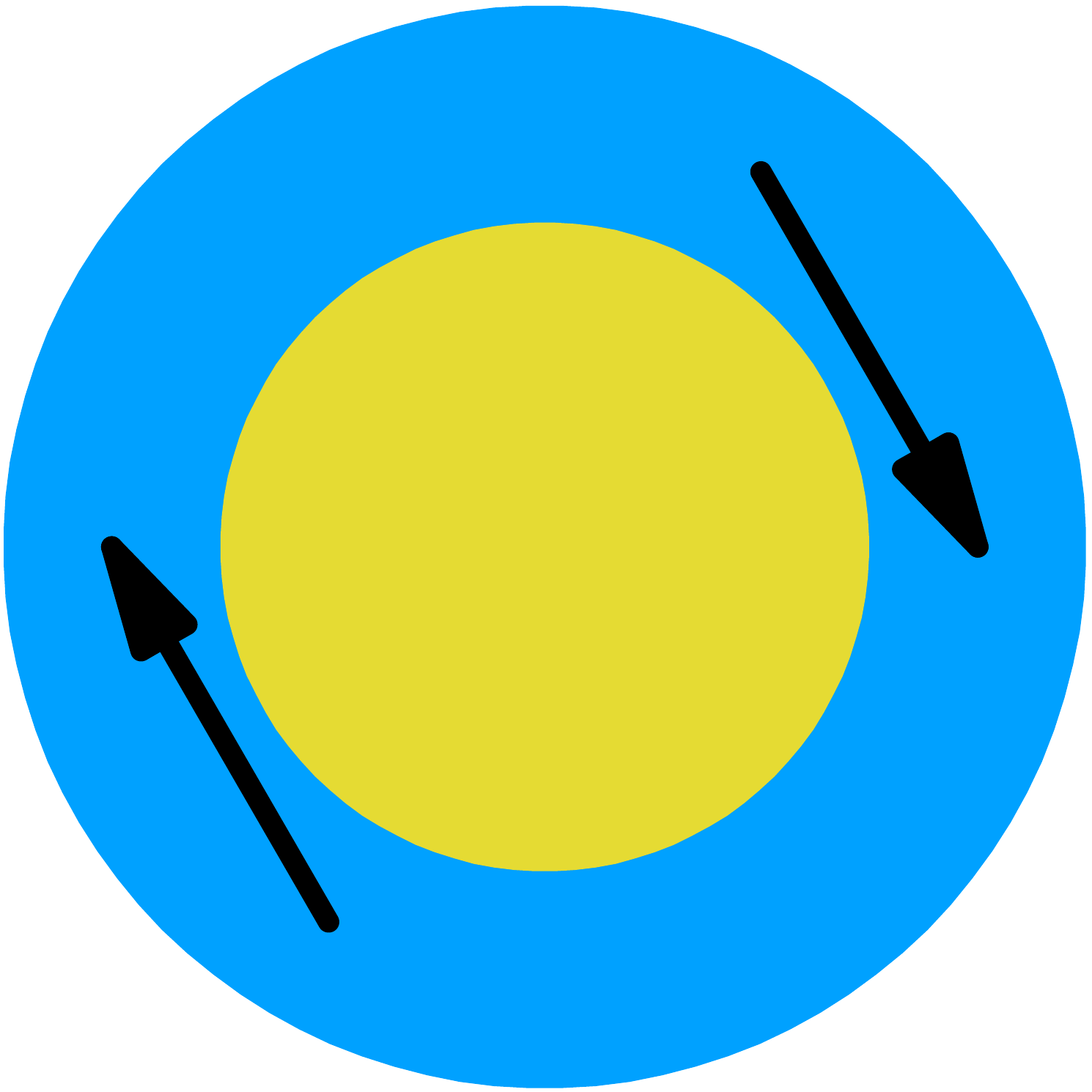}
	  \caption{$F^k(\qnl{}_1,\qnl{}_1)$}
      \label{fig_Rksketch_a}
    \end{subfigure}
  \end{subfigure}
  \begin{subfigure}{0.3\textwidth}
    \centering
    \begin{subfigure}{\textwidth}
      \centering
      \includegraphics[height=0.1\textheight]{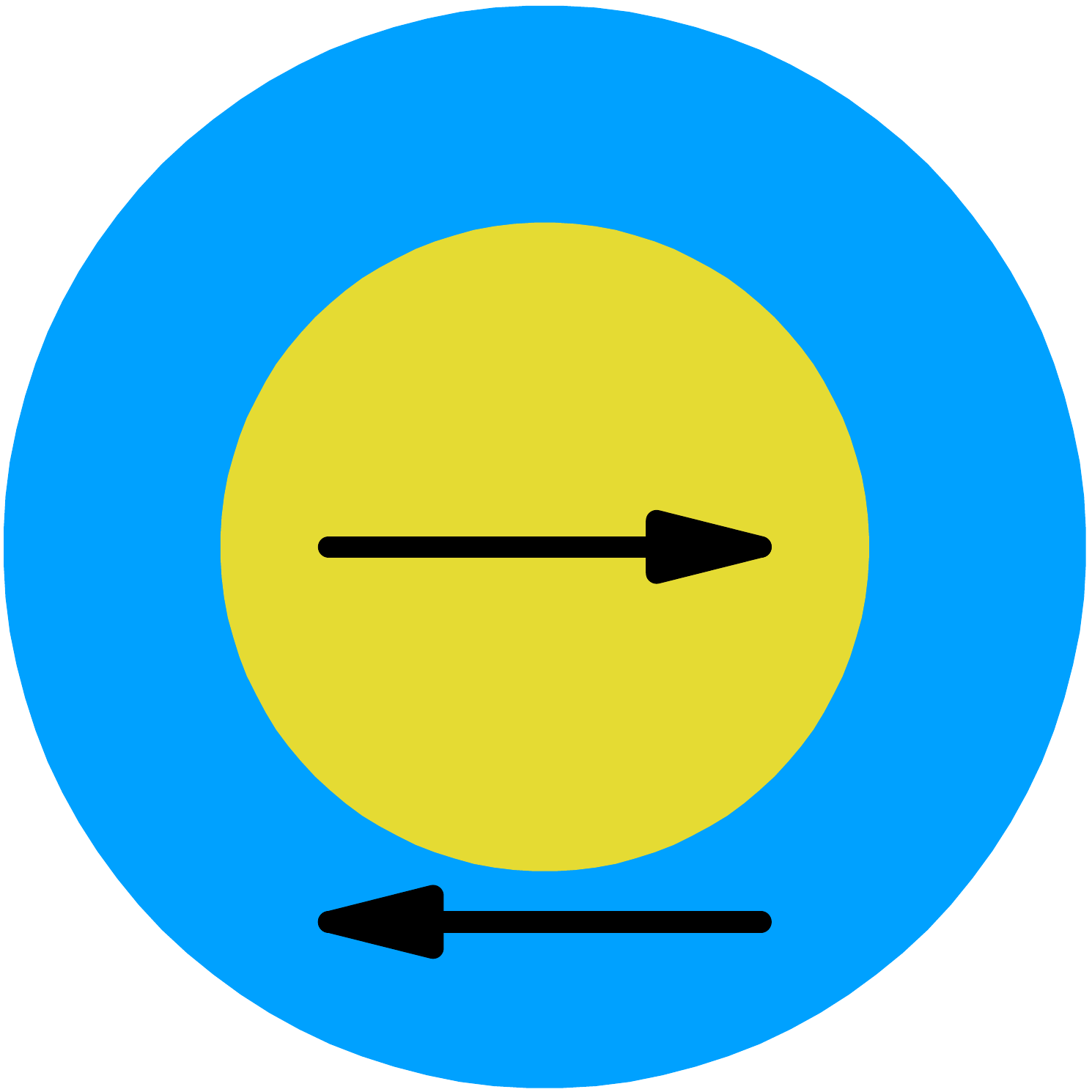}
      \caption{$F^k(\qnl{}_1,\qnl{}_2)$}
      \label{fig_Rksketch_b}
    \end{subfigure}
	    \begin{subfigure}{\textwidth}
      \centering
      \includegraphics[height=0.1\textheight]{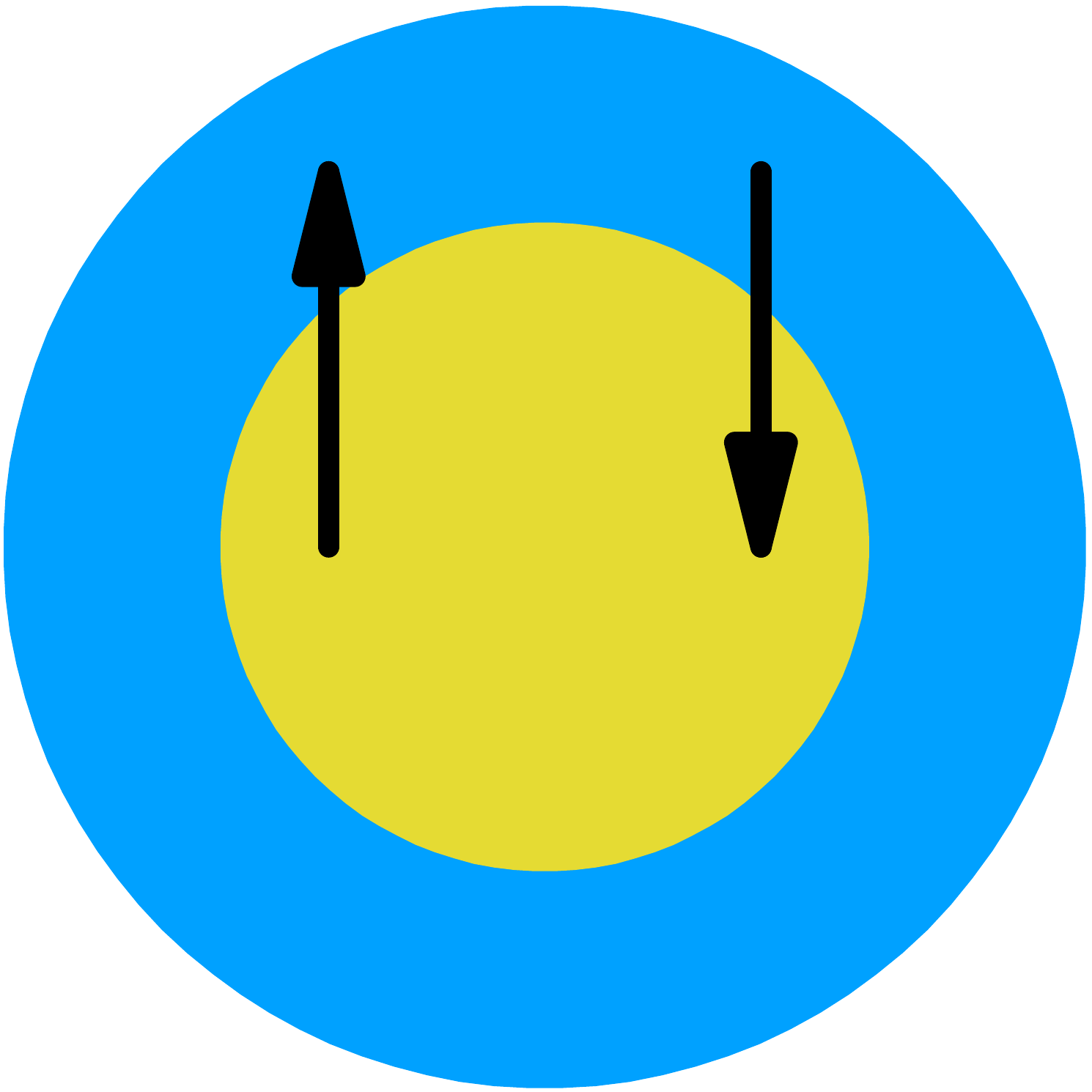}
	  \caption{$G^k(\qnl{}_1,\qnl{}_2)$}
      \label{fig_Rksketch_c}
    \end{subfigure}
  \end{subfigure}
  \begin{subfigure}{0.3\textwidth}
    \centering
    \begin{subfigure}{\textwidth}
      \centering
      \includegraphics[height=0.1\textheight]{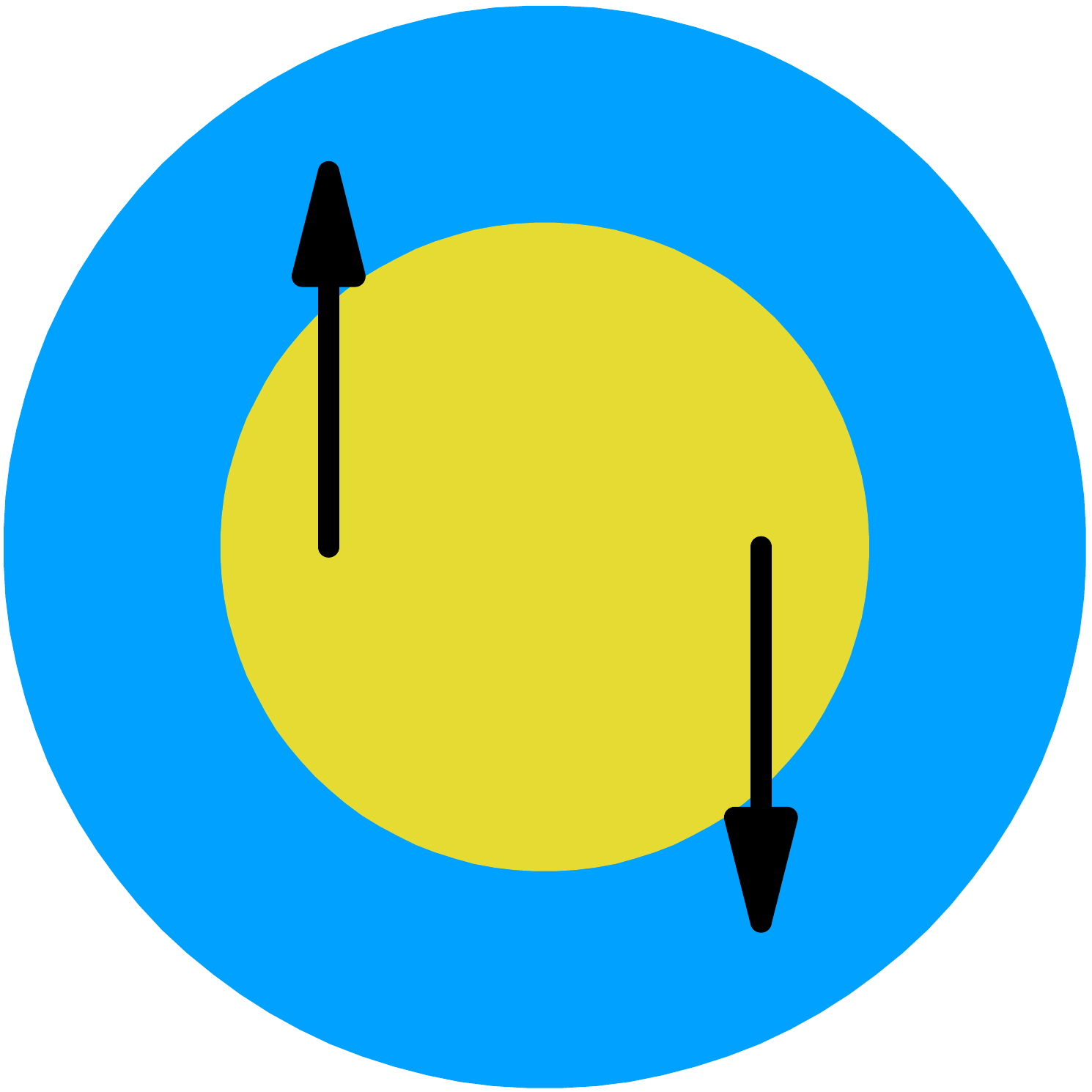}
	  \caption{$R^k(\qnl{}_1,\qnl{}_2)$}
      \label{fig_Rksketch_d}
    \end{subfigure}
    \begin{subfigure}{\textwidth}
      \centering
      \includegraphics[height=0.1\textheight]{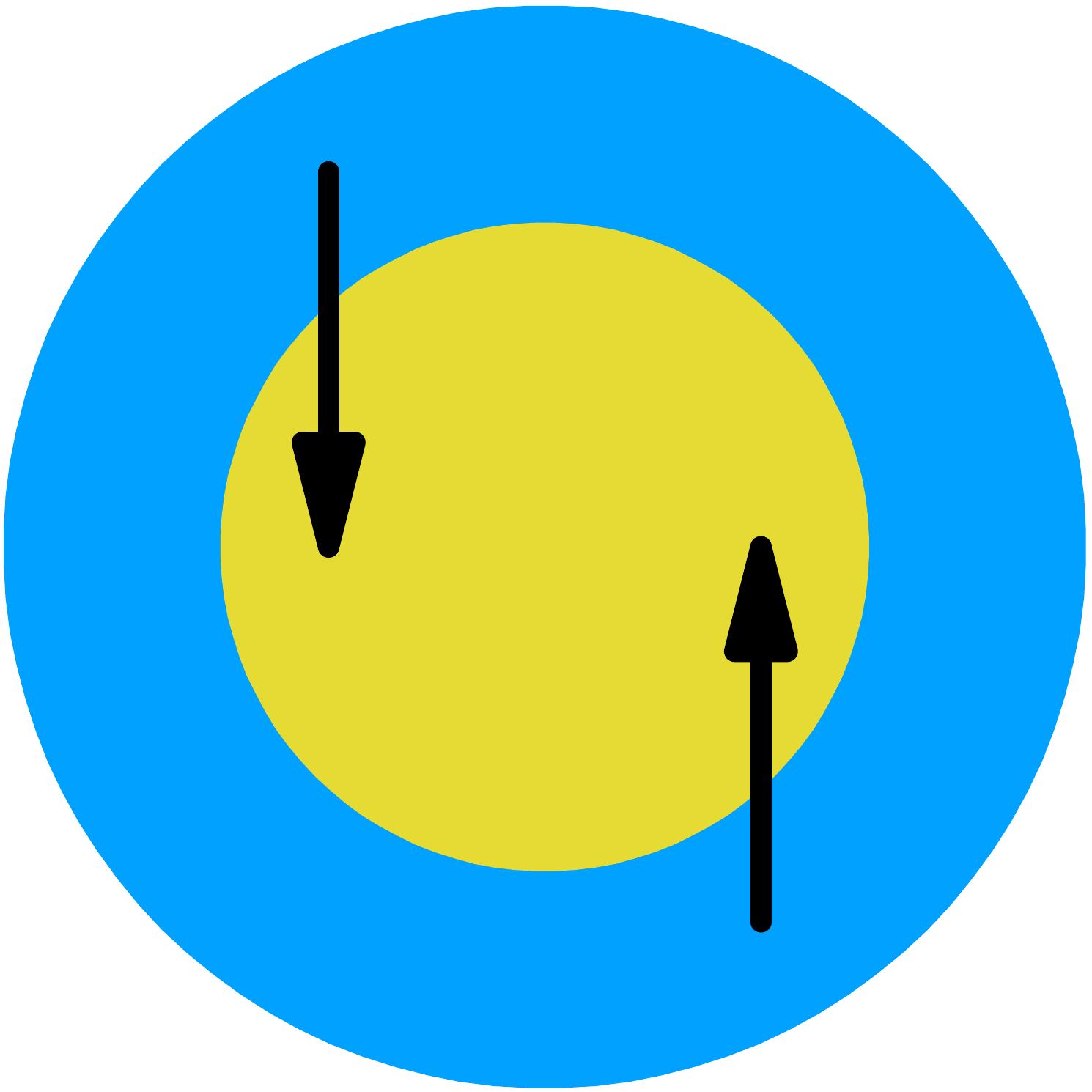}
	  \caption{$R^k(\qnl{}_2,\qnl{}_1)$}
      \label{fig_Rksketch_e}
    \end{subfigure}
  \end{subfigure}
  \caption{Sketch of the different possible Coulomb scattering processes within one (a) and two (b-e) shells. The anti-parallelism of the vectors indicates the momentum conservation. Note, that \(R^k\) changes the occupation of the shells.}
  \label{fig_Rksketch}
\end{figure}

For two partially filled shells \textit{i} and \textit{j} three types of \CI{} exist:
\begin{align}
F^k(i,j) &= R^k(ijij) &\text{with }~ 0 \leq k \leq 2\,\text{Min}(\qnl{}_i,\qnl{}_j) \nonumber\\
G^k(i,j) &= R^k(ijji) &\text{with }~ \abs{\qnl{}_i-\qnl{}_j} \leq k \leq \abs{\qnl{}_i+\qnl{}_j} \label{eq:FkGkRkSI}\\
R^k(i,j) &= R^k(iijj) &\text{with }~ \abs{\qnl{}_i-\qnl{}_j} \leq k \leq \abs{\qnl{}_i+\qnl{}_j}\nonumber
\end{align}
The momentum $k$ of the scattering electrons is restricted by the angular momenta of the shells involved, just as it is for the restrictions of the multipolar order \qnQ{}.
The restrictions are given in \equ{eq:FkGkRkSI} and $k$ must increase stepwise by 2 because of parity.
The processes described by the different terms are sketched in \fig{fig_Rksketch}.
$F^k$ describes the scattering were the electrons remain in the same shell, $G^k$ were they exchange the shells, and $R^k$ only appears in models allowing configuration interactions, were both electrons can scatter from one shell into the other.
Consequently, \CI{} within a single shell are solely described by $F^k(i,i)$.

\subsection{Spin-orbit coupling plus Coulomb interaction}\label{sec:SOplusCoulmb}
\begin{align}
 H &= H_\text{n}+H_{\CI{}}+H_{\SOC{}} \label{eq:H_ER_SO}
\end{align}
\CI{} yields states quantized by \qnL{}, \qnLz{}, \qnS{}, and \qnSz{}.
Weak \SOC{} appears to act on \opS{} and \opL{} as before on \ops{} and \opl{}.
This means every energy level with total orbital momentum quantum number \qnL{} created through the \CI{} will split into 2\qnS{}+1 levels with \qnJ{}\,=\,\qnL{}+\qnSz{} and a degeneracy of 2\qnJ{}+1.
This is known as LS coupling, where \qnS{} and \qnL{} are still considered to be good quantum numbers and the term symbols \termsymbol{2\qnS{}+1}{\qnL{}}{\qnJ{}} are used.

In fact the \SOC{} still acts on each pair (\opl{}\(_i\), \ops{}\(_i\)).
The stronger the \SOC{} relative to the \CI{} becomes, the more \qnL{} and \qnS{} may mix and deviate from their original integer value.
In this intermediate coupling regime \qnJ{} is still, but \qnL{} and \qnS{} are no longer good quantum numbers.

Finally for strong \SOC{} \opl{} and \ops{} become good quantum numbers, as for the single particle operator, and \(\opJ{} = \sum_{i=1}^{N} \opj{}_i = \sum_{i=1}^{N} (\opl{}_i+\ops{}_i) = (\opL{} + \opS{})\), known as jj-coupling.

\begin{figure}
  \centering
  \includegraphics[width=0.75\textwidth]{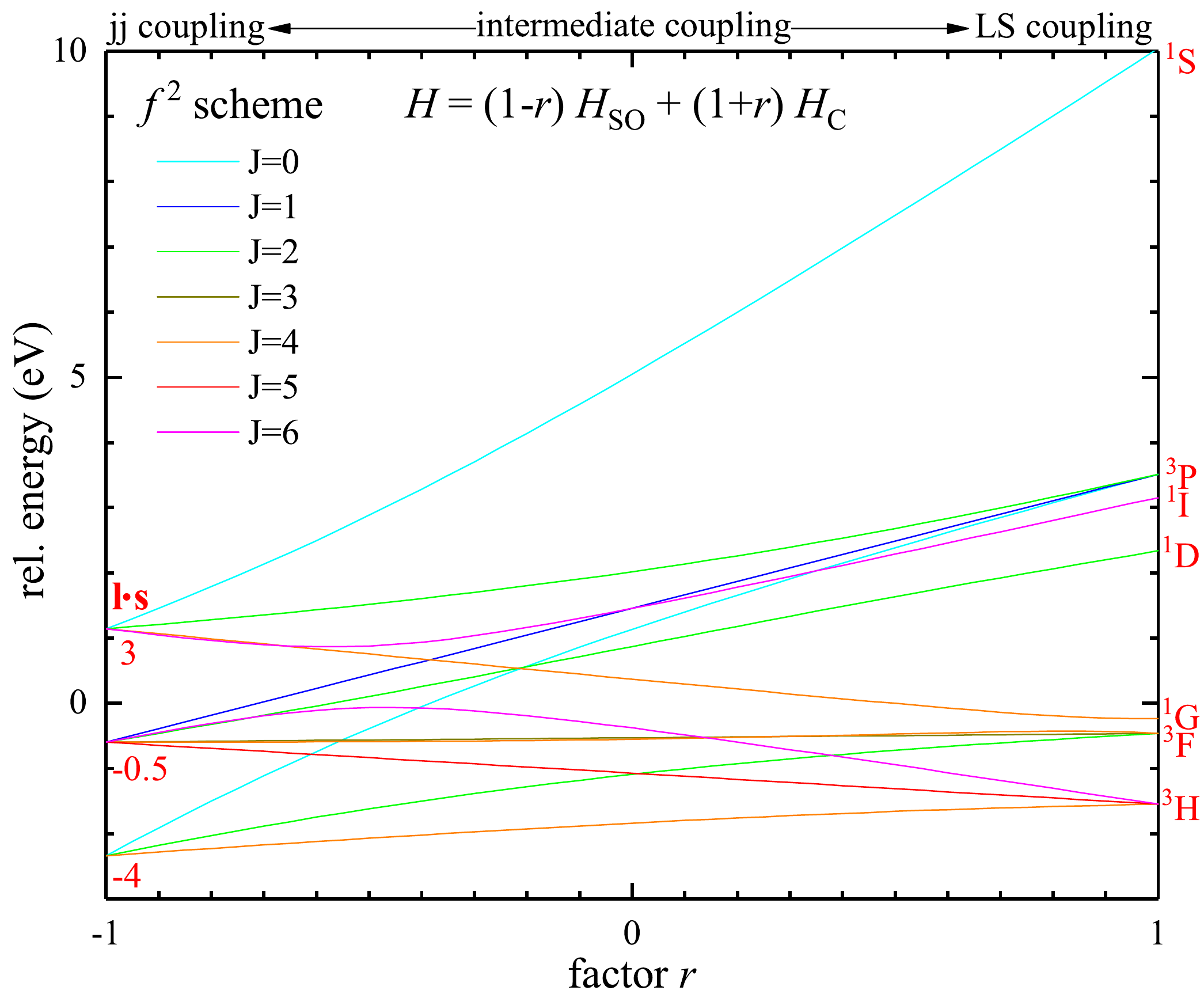}
  \caption{$f^2$ energy level diagram: Eigenvalues calculated for the atomic U\,5\textit{f}\textsup{2} configuration. The x-axis varies linearly a prefactor for \SOC{} and \CI{}, such that left is pure \SOC{}, zero is atomic values, and right is pure \CI{}. Red insets label the good quantum numbers (\opl{}$\cdot$\ops{} for pure \SOC{} and \textsup{2\qnS{}+1}\qnL{} for pure \CI{}).}
  \label{fig_f2ELD}
\end{figure}

\Fig{fig_f2ELD} shows the energy levels of an $f$\textsup{2} configuration from pure \SOC{} ($r$\,=\,-1) to pure \CI{} ($r$\,=\,1).
The bending of lines is related to the mixing of different \opl{}$\cdot$\ops{} when starting from the left hand side and also of different \qnL{} and \qnS{} when starting from the right hand side.

In the case of rare earth and actinides, \SOC{} and \CI{} are about equally strong and \qnL{}, \qnS{}, and \opl{}$\cdot$\ops{} should no longer be treated as good quantum numbers.
One of the consequences is that the electron charge densities deviate from the ones belonging to the corresponding term symbol \(^{2\qnS{}+1}\qnL{}_{\qnJ{}}\).
This has a significant effect on the angular dependence of \NIXS{} spectra and must not be neglected when determining the initial state.

\subsection{Crystal field}\label{sec:calccf}
The angular anisotropy of the $f$ electrons is due to the crystalline environment, which breaks rotational invariance, and the local point symmetry of the ion determines the possible symmetry (character) of the $f$ electrons.
States with different or even same character will split in energy by this \CF{} and exhibit a different angular dependence, which can be observed even if the \CF{} splittings are too small to be observed.

Within a given symmetry a finite number of symmetry operations will project the wave functions back to itself, with a possible change in phase.
In cubic symmetry this is the identity E, the 2$\pi$ rotation R, the inversion I, the rotations C\textsub{4} around \xyz{100}, C\textsub{2} around \xyz{100}, C\textsub{2}\textquotesingle\ around \xyz{110}, and  C\textsub{3} around \xyz{111}.
From these operations the character tables are build up (see \tab{tab:chartabOh} as cubic example).
On the left hand side the irreducible representations are given using the Mulliken symbols and on the right hand side in the Bethe notation.
Here, Ref.\,\cite{chartabonline} gives a nice overview of the character tables of the different point groups.
\begin{table}
  \centering
  \caption{Character table for O\textsub{h} point group appended by the double group ($\Gamma_6$ to $\Gamma_8$), reproduced from Ref.\,\cite{Ballhausen1962}. The top row shows the irreducible symmetry operators and the left column shows the allowed symmetries for this point group. The characters represent the expectation value when applying a symmetry operation to a symmetry.}
  \label{tab:chartabOh}
  \begin{tabular}{  l | r r r r r r r | l}
             &  E & R &  I  & C$_4^{[100]}$ & C$_2^{[100]}$ & C$_2^{[110]}$ & C$_3^{[111]}$ & \\
	\hline
	A$_{1g}$ &  1 & 1 &  1  &     1 &     1 &      1 &     1 & $\Gamma_1$ \\
	A$_{2g}$ &  1 & 1 &  1  &    -1 &     1 &     -1 &     1 & $\Gamma_2$ \\
	E$_{g}$  &  2 & 2 &  2  &     0 &     2 &      0 &    -1 & $\Gamma_3$ \\
	T$_{1g}$ &  3 & 3 &  3  &     1 &    -1 &     -1 &     0 & $\Gamma_4$ \\
	T$_{2g}$ &  3 & 3 &  3  &    -1 &    -1 &      1 &     0 & $\Gamma_5$ \\
	\hline                
	A$_{1u}$ &  1 & 1 & -1  &     1 &     1 &      1 &     1 & $\Gamma_1$ \\
	A$_{2u}$ &  1 & 1 & -1  &    -1 &     1 &     -1 &     1 & $\Gamma_2$ \\
	E$_{u}$  &  2 & 2 & -2  &     0 &     2 &      0 &    -1 & $\Gamma_3$ \\
	T$_{1u}$ &  3 & 3 & -3  &     1 &    -1 &     -1 &     0 & $\Gamma_4$ \\
	T$_{2u}$ &  3 & 3 & -3  &    -1 &    -1 &      1 &     0 & $\Gamma_5$ \\
	\hline
	E$_2$    &  2 & -2 & 0  & $\pm\sqrt{2}$ & 0 & 0 & $\pm$1 & $\Gamma_6$ \\
	E$_3$    &  2 & -2 & 0  & $\mp\sqrt{2}$ & 0 & 0 & $\pm$1 & $\Gamma_7$ \\
	G        &  4 & -4 & 0  &     0 &     0 &     0 & $\mp$1 & $\Gamma_8$ \\
  \end{tabular}
\end{table}
The symmetries for a set of states can then be found calculating the sum of their expectation values for the various rotation operators, i.e.\ the trace of the transition operator \(\exp{2 \imath \pi \qnml{}/\rotn{}}\)
\begin{align}
  \text{Tr}(\text{C}_{\rotn{}}) = \sum\limits_{\qnml{}=-\qnl{}}^{\qnl{}} \exp{2 \imath \pi \qnml{}/\rotn{}} = \lim_{\Pi\to\pi}\sin\left(\frac{(2\qnl{}+1){\scriptstyle\Pi}}{\rotn{}}\right)\sin^{-1}\left(\frac{{\scriptstyle\Pi}}{\rotn{}}\right). \label{eq:traceT}
\end{align}
Note that E, R, and I can be expressed by C$_{1/2}$, C$_1$, and S$_2$, respectively.
\Tab{tab:f2cubicrepresentations} shows these expectation values for all integer and half-integer momenta up to 23/2.
There will be one unique set of representations for each momentum.
The representations of momenta higher than 23/2 can be easily determined due to the periodicity of \equ{eq:traceT}.
\Equ{eq:traceT} as function of \qnl{} is linear or constant with a periodicity of \rotn{} and the least common multiple of \rotn{} in O\textsub{h} is 12.
E.g.\ for $M$\,=\,12 E, R, and I each yield 25 and each C yields 1, which gives the irreducible representation 2A$_{1g}$+A$_{2g}$+2E$_{g}$+3T$_{1g}$+3T$_{2g}$.

\begin{table}
  \centering
  \caption{Irreducible representations in O\textsub{h} symmetry. Depending on the relative strength of interactions $M$ may be replaced by \qnl{}, \qnL{} or \qnJ{} (see text).\label{tab:f2cubicrepresentations}}
  \begin{tabular}{  r | r r r r r r r | l  }
 $M$ &  E &  R &  I  & C$_4^{\xyz{100}}$ & C$_2^{\xyz{100}}$ & C$_2^{\xyz{110}}$ & C$_3^{\xyz{111}}$ & irreducible representation \\
	\hline
	   0  &  1 &  1 &   1 &     1 &     1 &      1 &     1 & A$_{1g}$ \\
	   2  &  5 &  5 &   5 &    -1 &     1 &      1 &    -1 & E$_{g}$+T$_{2g}$ \\
	   4  &  9 &  9 &   9 &     1 &     1 &      1 &     0 & A$_{1g}$+E$_{g}$+T$_{1g}$+T$_{2g}$ \\
	   6  & 13 & 13 &  13 &    -1 &     1 &      1 &     1 & A$_{1g}$+A$_{2g}$+E$_{g}$+T$_{1g}$+2T$_{2g}$ \\
	   8  & 17 & 17 &  17 &     1 &     1 &      1 &    -1 & A$_{1g}$+2E$_{g}$+2T$_{1g}$+2T$_{2g}$ \\
	  10  & 21 & 21 &  21 &    -1 &     1 &      1 &     0 & A$_{1g}$+A$_{2g}$+2E$_{g}$+2T$_{1g}$+3T$_{2g}$ \\
	 $\vdots$ &  &  &     &       &       &        &       & \\
	 +12  & +24& +24& +24 &    +0 &    +0 &     +0 &    +0 & +A$_{1g}$+A$_{2g}$+2E$_{g}$+3T$_{1g}$+3T$_{2g}$ \\
	\hline            
	   1  &  3 &  3 &  -3 &     1 &    -1 &     -1 &     0 & T$_{1u}$ \\
	   3  &  7 &  7 &  -7 &    -1 &    -1 &     -1 &     1 & A$_{2u}$+T$_{1u}$+T$_{2u}$ \\
	   5  & 11 & 11 & -11 &     1 &    -1 &     -1 &    -1 & E$_{u}$+2T$_{1u}$+T$_{2u}$ \\
	   7  & 15 & 15 & -15 &    -1 &    -1 &     -1 &     0 & A$_{2u}$+E$_{u}$+2T$_{1u}$+2T$_{2u}$ \\
	   9  & 19 & 19 & -19 &     1 &    -1 &     -1 &     1 & A$_{1u}$+A$_{2u}$+E$_{u}$+3T$_{1u}$+2T$_{2u}$ \\
	  11  & 23 & 23 & -23 &    -1 &    -1 &     -1 &    -1 & A$_{2u}$+2E$_{u}$+3T$_{1u}$+3T$_{2u}$ \\
	 $\vdots$ &  &  &     &       &       &        &       & \\
	 +12  & +24& +24& -24 &    +0 &    +0 &     +0 &    +0 & +A$_{1u}$+A$_{2u}$+2E$_{u}$+3T$_{1u}$+3T$_{2u}$ \\
	\hline
	  1/2 &  2 & -2 &   0 &$\pm\sqrt{2}$&0&      0 & $\pm$1& E$_2$ \\
	  3/2 &  4 & -4 &   0 &     0 &     0 &      0 & $\mp$1& G \\
	  5/2 &  6 & -6 &   0 &$\mp\sqrt{2}$&0&      0 &     0 & E$_3$+G \\
	  7/2 &  8 & -8 &   0 &     0 &     0 &      0 & $\pm$1& E$_2$+E$_3$+G \\
	  9/2 & 10 &-10 &   0 &$\pm\sqrt{2}$&0&      0 & $\mp$1& E$_2$+2G \\
	 11/2 & 12 &-12 &   0 &     0 &     0 &      0 &     0 & E$_2$+E$_3$+2G \\
	 13/2 & 14 &-14 &   0 &$\mp\sqrt{2}$&0&      0 & $\pm$1& E$_2$+2E$_3$+2G \\
	 15/2 & 16 &-16 &   0 &     0 &     0 &      0 & $\mp$1& E$_2$+E$_3$+3G \\
	 17/2 & 18 &-18 &   0 &$\pm\sqrt{2}$&0&      0 &     0 & 2E$_2$+E$_3$+3G \\
	 19/2 & 20 &-20 &   0 &     0 &     0 &      0 & $\pm$1& 2E$_2$+2E$_3$+3G \\
	 21/2 & 22 &-22 &   0 &$\mp\sqrt{2}$&0&      0 & $\mp$1& E$_2$+2E$_3$+4G \\
	 23/2 & 24 &-24 &   0 &     0 &     0 &      0 &     0 & 2E$_2$+2E$_3$+4G \\
	 $\vdots$ &  &  &     &       &       &        &       & \\
	 +12  & +24& -24&  +0 &    +0 &    +0 &     +0 &    +0 & +2E$_2$+2E$_3$+4G \\
  \end{tabular}
\end{table}

What happens upon adding the \CF{} to the previous interactions?
For a \CF{} stronger than the \CI{} and \SOC{} the moment $M$ is given by the angular momentum \qnl{}, as for light covalent compounds.
Otherwise, for \CF{} stronger than \SOC{}, i.e.\ for \CI{}\,>\,\CF{}\,>\,\SOC{}, $M$ is given by the total angular momentum \qnL{}, as for many transition metals.
Finally, for \CF{} weaker than \SOC{} $M$ is given by the total momentum \qnJ{}, as typical for rare earth.
The splitting of the \CF{} states lifts the 2\qnJ{}+1 degenerate states shown in \fig{fig_f2ELD}.
Otherwise the diagram remains unchanged to a great extent, because the \CF{} is weaker than \SOC{} and \CI{} for the rare earth and the actinides still.
A full multiplet calculation takes care of how the various quantum numbers mix when going from one to the other case, given that all parameters are known.

In the case of half integer \qnJ{} the character table can be extended by the double group.
These symmetries are characteristic by their change of sign upon a 2$\pi$ rotation.
Inversion immediately tells whether the moment is even (when positive), odd (negative), or belongs to the double group (zero real part).

Knowing all the symmetries, an operator lifting the degeneracy of these states should be defined on the local basis.
For this a static potential due to ligand ions can be assumed, which must reproduce under all symmetry operations, i.e\ it must have A\textsub{1g} symmetry.
Expanding the potential on the basis of spherical harmonics leads to the \CF{} Hamiltonian in the Wybourne formalism\,\cite{Wybourne1965}
\begin{align}
H_{\CF{}} &= \sum\limits_{k}^{} \sum\limits_{m=-k}^{k} A_{k,m} \, \opClm{$k$}{$m$} \nonumber \\
&= \sum\limits_{k=0}^{2\qnl{}} \sum\limits_{\qnml{},\qnml{}'=-\qnl{}}^{\qnl{}} A_{k,\qnml{}-\qnml{}'} ~ \bra{\qnl{}\qnml{}}\,\Clm{$k$}{\qnml{}-$\qnml{}'$}\,\ket{\qnl{}$\qnml{}'$} ~ \hat{c}^\dag_{\qnn{},\qnl{},\qnml{},\qnms{}}\hat{c}_{\qnn{},\qnl{},\qnml{}',\qnms{}}. \label{eq:CEFpotential}
\end{align}
\App{app_stevensformalism} gives the relation to the Stevens notation.
A\textsub{1g} symmetry only appears for even $k$.
So, \(A_{k,m}\) must be zero if $k$ is odd (and also if $k$\,=\,2 in the present case).
\(A_{0,0}\) is one trivial solution shifting all states of a given \qnl{} and \qnml{} by the same amount.
For $k\geq4$, the symmetry operators can be applied to the corresponding spherical harmonic functions to identify the invariant, non-vanishing linear combinations.
Examples for the identification of such linear combinations are presented e.g.\ in Ref.\,\cite[Chap.\,4]{Ballhausen1962}.

\subsection{Hybridization}
So far only interactions between local states were accounted for and hybridization effects as introduced in \chap{subsec:AIM} were neglected.
In the following it will be motivated why for the calculation of \NIXS{} spectra hybridization is expected to play a subordinate role and why special care should be taken when comparing the valence of the same system determined at different absorption edges or by different techniques.

Here the effect is summarized based on the effective \AIM{} Hamiltonian (\equ{eq:Hexc}) in the limit \onsiteU{}\,$\to$\,$\infty$.
The ground state is then given by \ket{0}\,=\,$c_0$\ket{$f^0$}\,+\,$c_1$\ket{$f^1 \underline{L}$} with $c_0^2$\,=\,$\sqrt{1-c_1^2}$.
In case that hybridization is suppressed in the final state, the observation of an absorption experiment would reflect the response of the two finial state multiplet structures of the \ket{f$_0$}\,=\,\ket{$\underline{c}f^1$} and \ket{f$_1$}\,=\,\ket{$\underline{c}f^2\underline{L}$} configuration and the ratio of these responses is given by the overlap of the initial state with the final states |\bra{f$_0$}\(\hat{c}^\dag_{f}\hat{c}_{c}\)\ket{0}|$^2$\,=\,|$c_0$|$^2$ and |\bra{f$_1$}\(\hat{c}^\dag_{f}\hat{c}_{c}\)\ket{0}|$^2$\,=\,|$c_1$|$^2$.
But the effect of hybridization in the final state must not be neglected.

Hybridization in the final state can be treated similar to the initial as introduced by Gunnarsson and Sch{\"o}nhammer\,\cite{Gunnarsson1983}.
Consequently also the final states are given by \ket{f$_0$}\,=\,$c'_0$\ket{$\underline{c}f^1$}\,+\,$c'_1$\ket{$\underline{c}f^2\underline{L}$} and \ket{f$_1$}\,=\,$c'_1$\ket{$\underline{c}f^1$}\,$-$\,$c'_0$\ket{$\underline{c}f^2\underline{L}$}, with the contributions $c'_0{}^2$ and $c'_1{}^2$ of the integer valence states.
The observed responses scale depending on the overlap of the initial an final states |\bra{f$_0$}\(\hat{c}^\dag_{f}\hat{c}_{c}\)\ket{0}|$^2$\,=\,|$c_0c'_0+c_1c'_1$|$^2$ and |\bra{f$_1$}\(\hat{c}^\dag_{f}\hat{c}_{c}\)\ket{0}|$^2$\,=\,|$c_0c'_1-c_1c'_0$|$^2$.
The size of $c'_0$ and $c'_1$\,=\,$\sqrt{1-c'_0{}^2}$ depends on the energy difference $\Delta'_f$\,=\,$\Deltaf{}+\onsiteU{}-U_{fc}$ between the two final states in the presence of the core hole, when assuming that the hybridization remains equal to that of the initial state ($V'_\text{eff}$\,=\,\Veff{}).

Here one can consider two special cases:
$U_{fc}\to0$, which yields $c'_0\to1$ so that the scenario without final state hybridization is recovered.
And $U_{fc}\to\onsiteU{}$, resulting in $c'_0\to c_0$.
In the latter case the overlaps are given by |\bra{f$_0$}\(\hat{c}^\dag_{f}\hat{c}_{c}\)\ket{0}|$^2$\,=\,1 and |\bra{f$_1$}\(\hat{c}^\dag_{f}\hat{c}_{c}\)\ket{0}|$^2$\,=\,0, i.e.\ only the \ket{f$_0$} final state multiplet is observed independently on how strongly intermediate valence the system is.
This also gives an explanation for the different results of different absorption experiments presented for the valence of U in URu$_2$Si$_2$ (see \chap{sec:URu2Si2}).

\begin{figure}
  \centering
  \includegraphics[width=\textwidth]{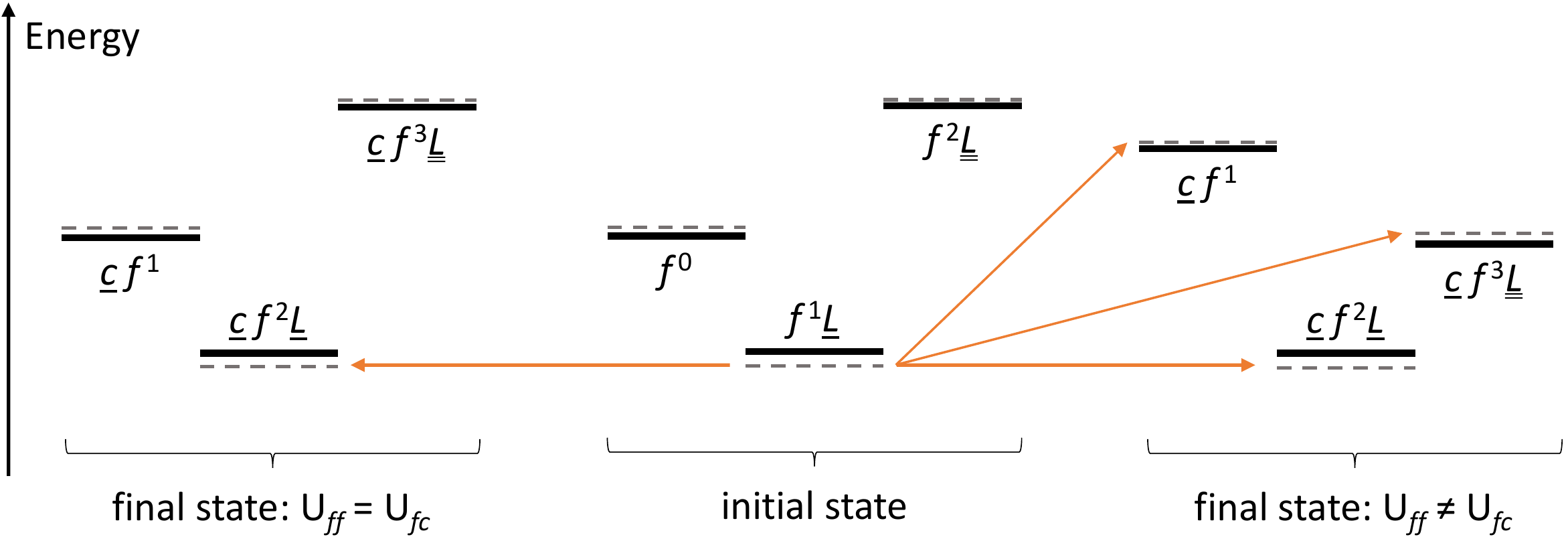}
  \caption{Sketch of the hybridization in the initial state (middle) and after an absorption process with $\onsiteU{}=U_{cf}$ (left) and for the general case (right) following the Gunnarsson-Sch{\"o}nhammer scheme of the Anderson impurity model\,\cite{Gunnarsson1983}. Dashed lines indicate the hybridized states and the orange arrows indicate the transitions that are observed.}
  \label{fig:AIMGS}
\end{figure}

Based on these considerations the analysis of the \HAXPES{} and \edge{L}{3}-edge \XAS{} data of CeRu$_4$Sn$_6$ (\chap{sec:CeRu4Sn6}) are performed considering the observation of three configurations \ket{$\underline{c}f^1$}, \ket{$\underline{c}f^2\underline{L}$}, and \ket{$\underline{c}f^3\underline{\underline{L}}$} ($U_{fc}\not\approx\onsiteU{}$).
Also the \edge[Sm]{N}{4,5} of the SmB$_6$ \NIXS{} experiment shows spectral features of two configurations (Sm$^{2+}$ and Sm$^{3+}$).
The \edge[Ce]{N}{4,5} edges and \edge[U]{O}{4,5} observed by \NIXS{}, on the other hand, are reproduced in great detail by the single \ket{$\underline{c}f^2\underline{L}$} final-state configuration ($U_{fc}$\,$\approx$\,\onsiteU{}).
This concept is sketched in \fig{fig:AIMGS}.
The prediction of which scenario applies is somewhat difficult due to strong screening of the on-site Coulomb repulsions $U$.

\paragraph{Configuration interaction model}\label{subsec:CIM}

For \HAXPES{} a configuration interaction model (\CIM{}) has been used to correct for the modulation of the spectral weights due to final state effects.

In the \CIM{} the \AIM{} Hamiltonian (\equ{eq:Hexc}) is used in both the initial state and final state to identify the initial and final state contributions of the configurations $c_i^2$ and $c'_i{}^2$.
In the final state each $f$ electron experiences the potential of the core hole $U_{fc}$, such that the effective \textit{f}-electron binding energy changes in the final state (\(\Delta_f'\)\,=\,\Deltaf{}\,+\,$U_{fc}$).
The other parameters are assumed constant in this model so that the \CIM{} Hamiltonian for \HAXPES{} is
\begin{align}
H_{\CIM{}} &= \left( \begin{array}{ccc}
  -\Deltaf{}-U_{fc} & \Veff{} & \Veff{} \\
   \Veff{}            & 0       & 0 \\
   \Veff{}            & 0       & -2\Deltaf{}-2U_{fc}+\onsiteU{} \\
\end{array} \right). \label{eq:Hcim}
\end{align}

For the three configurations, $f^1$, $f^0$, and $f^2$, the two intensity ratios and two energy distances of the three spectral features uniquely reflect the four model parameters.
In this way the initial state contributions $c_i^2$ can be determined.

\clearpage
\section{CeB\textsub{6}} \label{sec:CeB6}
\begin{center}
\textbf{
\scalebox{1.2}{The quartet ground state in CeB\textsubscript{6}:} \scalebox{1.2}{An inelastic x-ray scattering study}}\\
\hyperref[CeB6_2017]{\scalebox{1.2}{Europhysics Letters \textbf{117} 1, 17003 (2017)}}\end{center}
\FloatBarrier
\begin{SCfigure}[][h]
  \centering
  \includegraphics[width=0.48\textwidth]{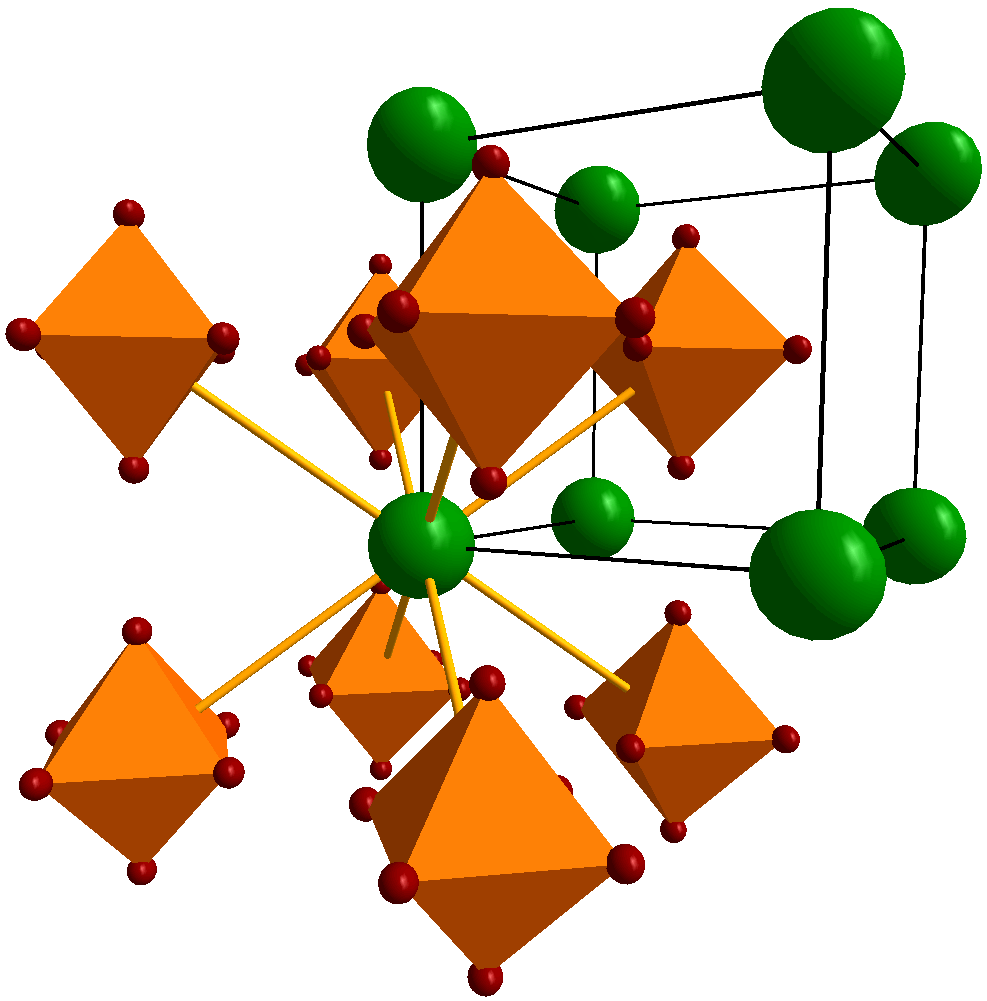}
  \caption{Unit cell (black box) and chemical environment of the Ce\textsup{3+} ion in cubic CeB$_6$. Ce (green spheres), B octahedra (red spheres with polygons), and the Ce nearest neighbor connections (yellow bars) are shown. Structure parameters from ICSD\cite{icsd,Blomberg1989}.}
  \label{fig:CeB6structure}
\end{SCfigure}
\FloatBarrier

\subsection{Introduction}

\begin{SCfigure}
	\centering
\includegraphics[width=0.6\columnwidth]{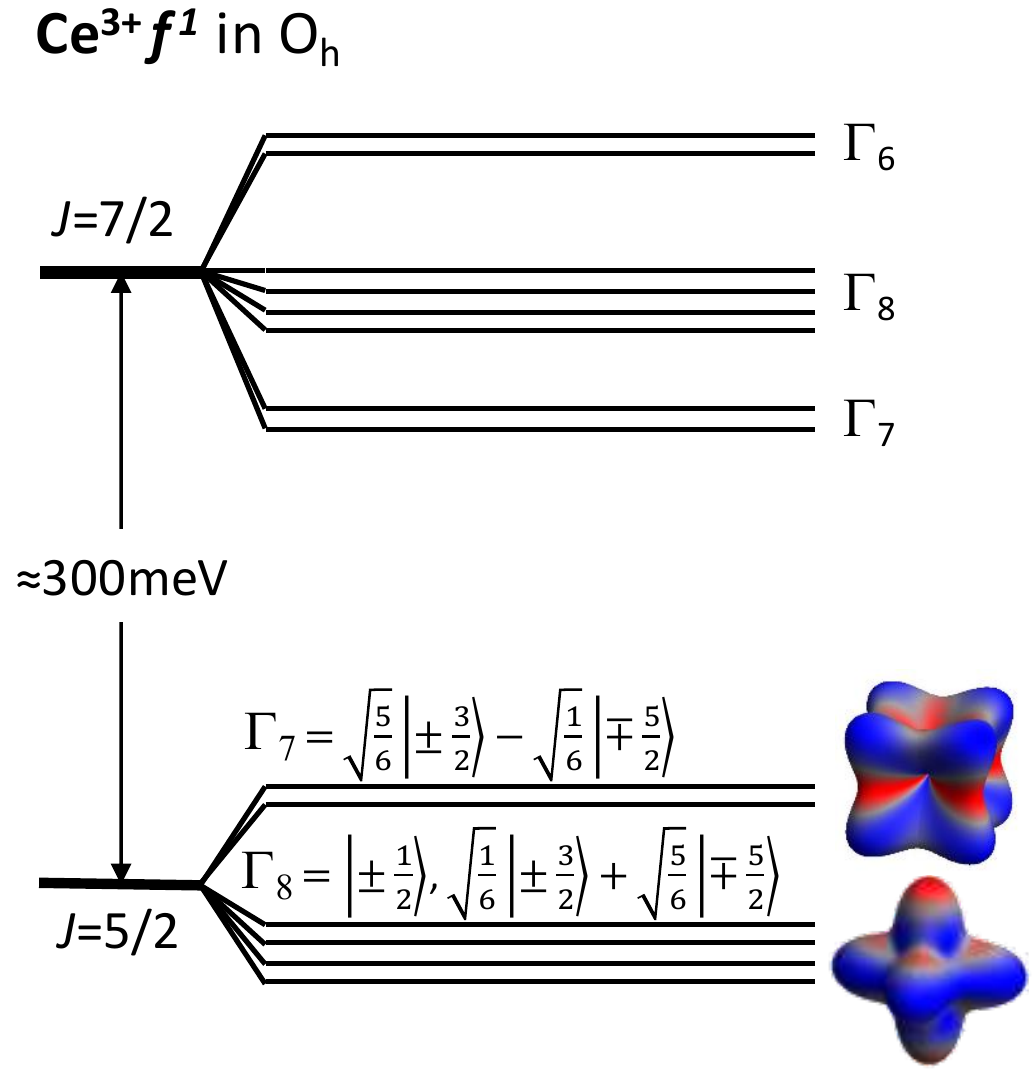}
\caption{Scheme of the low lying Ce\,4$f$ energy levels for Ce$^{3+}$ in cubic O\textsub{h} point symmetry  (see \tab{tab:f2cubicrepresentations}). For the Hund's rule ground state with \qnJ{}\,=\,5/2 the electron charge densities and the \ket{\qnJz{}} representation of the CF eigenstates are shown.}
\label{fig:CeB6scheme}
\end{SCfigure}

CeB$_6$ is a well studied rare earth hexaboride.
CeB$_6$ orders magnetically, i.e.\ it is located in the localized regime of the Doniach phase diagram.
Ce is trivalent (Ce$^{3+}$) and one 4$f$ state is occupied.
In the high cubic (Pm$\bar{3}$m) space group (see \fig{fig:CeB6structure}) with the O\textsub{h} point symmetry at the Ce site the sixfold degenerate \qnJ{}\,=\,5/2 Hund's rule ground state is split into one doublet $\Gamma_7$ and one quartet $\Gamma_8$ as shown in \fig{fig:CeB6scheme}.
The $\Gamma_8$ ground state in cubic CeB$_6$ is well established.
It can therefore serve as benchmark system for demonstrating the validity of the quantitative analysis of \NIXS{} spectra.

\begin{SCfigure}
  \centering
  \includegraphics[width=0.4\columnwidth]{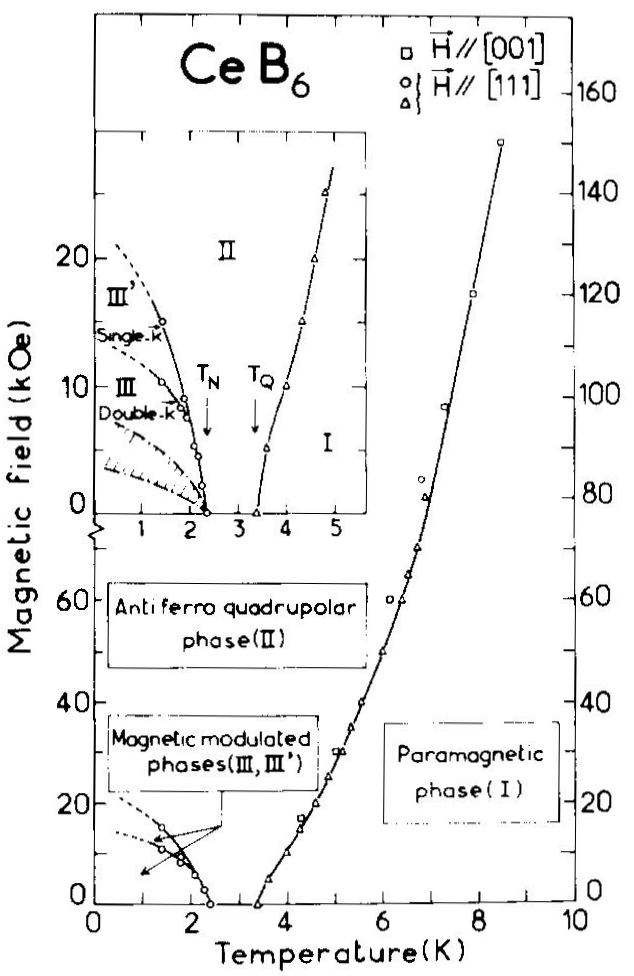}
  \caption{Magnetic phase diagram of CeB$_6$ taken from Ref.\,\cite{Effantin_1985}. In zero field CeB$_6$ enters a \AFM{} quadrupolar order below $T_\text{Q}$\,=\,3.2\,K before reaching its \AFM{} groundstate $T_\text{N}$\,=\,2.4\,K. With external parameters (here magnetic field) even more phases arise: In phase III' the magnetic unit cell reduces from 4 to 2 Ce sites, due to the loss of some ordering vector $\vec{k}$.}
  \label{fig_CeB6phasediagram}
\end{SCfigure}

CeB$_6$ has a rich magnetic phase diagram despite its simple cubic CsCl structure\cite{Effantin_1985} (see \fig{fig_CeB6phasediagram}).
It enters a \HO{} phase below 3.2\,K and an antiferromagnetic (\AFM{}) phase at 2.4\,K.
In the \HO{} phase no symmetry breaking was observed with e.g.\ neutron or standard x-ray diffraction at zero field.
The induced dipole moment of an applied magnetic field points towards a quadrupolar order\,\cite{Erekelens1987}.
Theory suggests that the multipolar moments of the localized 4$f$ electrons interact with each other via the itinerant 5$d$ conduction electrons.
This interaction breaks the fourfold ground state degeneracy of the Ce\,4$f$ wave function in the cubic crystal field, thus stabilizing an antiferro-quadrupolar (AFQ) order\,\cite{Thalmeier1997}.
This concept for the hidden magnetic order has received credibility from resonant x-ray diffraction\,\cite{Matsumura2009, Lovesey2002}.

The observation of a spin resonance in the inelastic neutron data of CeB$_6$\,\cite{Friemel2012, Portnichenko2016, Cameron2016} shows the importance of strong coupling between $f$ electrons and conduction electrons for the formation of the magnetic and multipolar order\,\cite{Thalmeier2012}, the former being supported by electronic structure investigations of CeB$_6$\,\cite{Neupane2015, Koitzsch2016}.
Inelastic neutron scattering finds, in agreement with Raman scattering, a \CF{} excitation at 46\,meV and it is generally accepted that the intriguing magnetic properties of CeB$_6$ evolve out of the fourfold degenerate $\Gamma_8$ ground state.
The quartet ground state had been originally deduced from an unusual low temperature shift of the \CF{} excitation in Raman and inelastic neutron scattering data\,\cite{Zirn1984,Loew1985}.
The energy shift was interpreted as a splitting of the quartet ground state in the low temperature phase in accordance with electron paramagnetic resonance measurements\,\cite{Terzioglu2001}.
A quartet ground state is also consistent with findings of the magnetic anisotropy\,\cite{Sato1984}, magnetic neutron form factor measurements\,\cite{Givord2003} as well as x-ray diffraction measurements of the electron density distribution at low temperatures and 300\,K\,\cite{Tanaka2002} with an unconfirmed claim that a level inversion may occur at higher temperatures\,\cite{Makita2007}.

A \NIXS{} experiment has been performed on CeB$_6$ single crystals at the \edge[Ce]{N}{4,5} edges ($4d\rightarrow4f$), which directly probes the 4$f$ states.
In order to distinguish between the $\Gamma_7$ and $\Gamma_8$ states in cubic symmetry it is sufficient to measure \sqw{} for two independent directions.
In this work, three independent directions have been measured (\qp{100}, \qp{110}, and \qp{111}), so that the results could be confirmed unambiguously.

\subsection{Experimental}

\begin{figure}
	\centering
	\includegraphics[width=0.9\columnwidth]{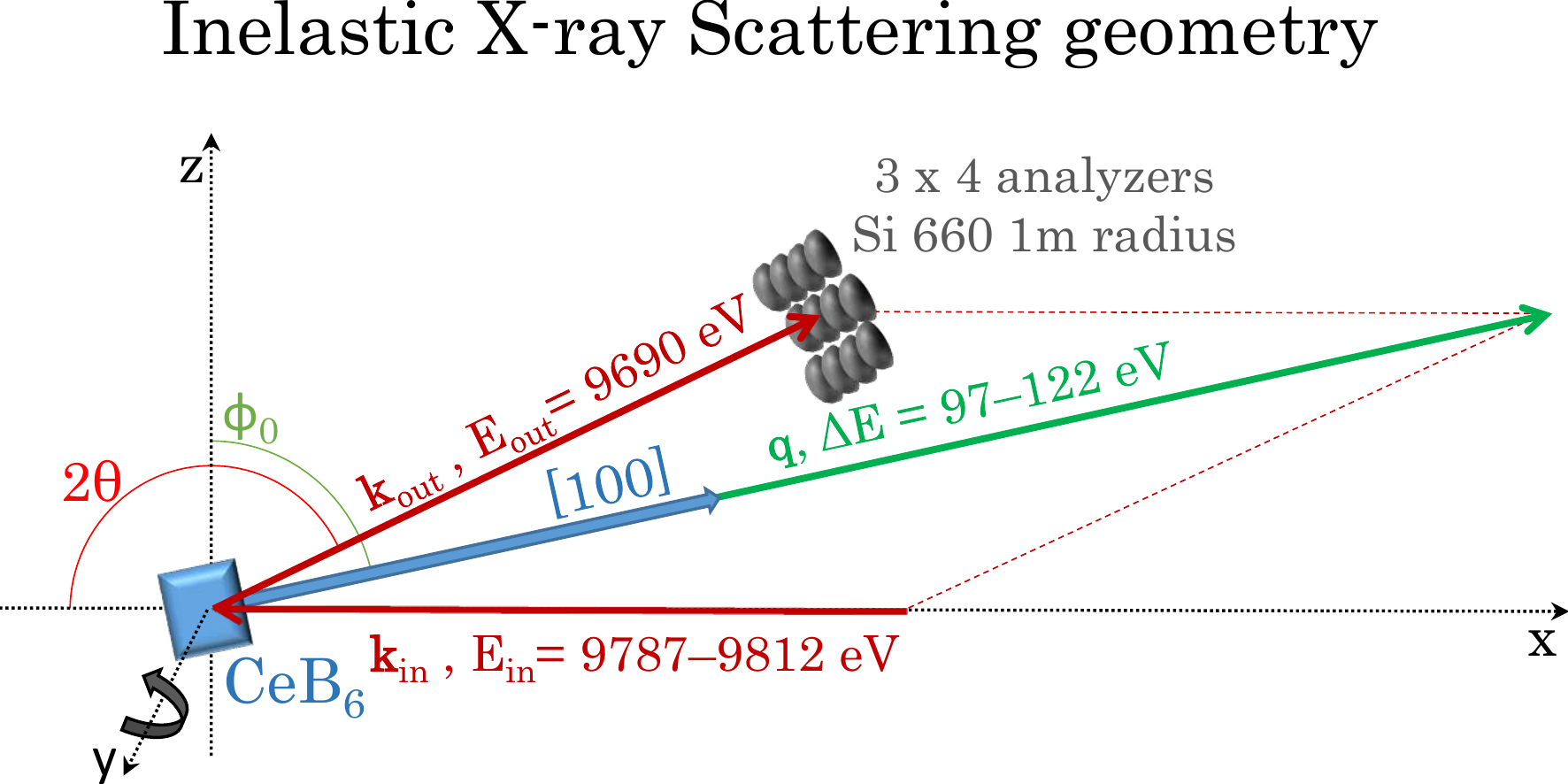}
	\caption{Scattering geometry of the \NIXS{} experiment, here for a scan with \qp{100} showing the energies scanned over the \edge[Ce]{N}{4,5} edges.}
\label{fig:CeB6geometry}
\end{figure}

CeB$_6$ single-crystals were provided by H.~Lee and Z.~Fisk from University of California, Irvine.
They were grown with the Al-flux method.
Typically 0.7\,g of CeB$_6$ (as the elements) are heated with 60\,g of high purity Al\,(59) to 1\,450\,$^\circ$C, held there for 8\,h and then cooled to 1\,000\,$^\circ$C at 2\,K/h, when the furnace is shut off.
The crystals are leached from the Al in NaOH solution.

Two samples have been prepared for the \NIXS{} experiment: One with a polished \xyz{100} and a second one with a polished \xyz{110} surface.
The alignment was controlled using Laue diffraction, both for the surface normal and the in-plane rotation.
Both crystals were mounted in the cryostat so that for \qp{100} and \qp{110} a so-called specular scattering geometry was realized, i.e.\ with the surface normal parallel to the momentum transfer.
This is realized when the sample angle $\phi_0$ is equal to the scattering angle $\theta$, $\phi_0$\,=\,$\theta$ (see \fig{fig:CeB6geometry}).

The \NIXS{} measurements on the CeB$_6$ \edge[Ce]{N}{4,5} edges ($4d\to4f$) were performed at the beamline P01 using the closed cycle cryostat (see \chap{subsec:nixs_setup} for detailed information) together with Kai Chen, technical support from Hasan Yava\c{s}, and under supervision of Andrea Severing and Liu Hao Tjeng.
For the experiment a backscattering geometry with 2\thS{}\,=\,150$^\circ$, 155$^\circ$, and 160$^\circ$ of the three analyzer columns was chosen.
In this geometry the momentum transfer amounts to \absq{}\,=\,$(9.6\pm0.1)\AA^{-1}$ around the elastic energy of 9.69\,keV of the Si(660) analyzer reflection.

A sketch of the scattering geometry, showing the incoming and outgoing photons (\kin{} and \kout{}) as well as the transferred momentum \vecq{}, is given in \fig{fig:CeB6geometry} for \qp{100} in specular geometry (angle in = angle out).
Here, the surface normal is also parallel to \vecq{} since $E_\text{in}\approx E_\text{out}$.
The \qp{110} configuration was realized in two ways: First by using the crystal with polished \xyz{110} surface in specular geometry and second by rotating the \xyz{100} crystal 45$^\circ$ towards \xyz{010} around the axis that is normal to the scattering plane ($y$, as in \fig{fig:CeB6geometry}).
In the second case the x-ray paths in and out of the sample are different.
Yet, both data sets match.
This is an important finding.
It confirms bulk sensitivity and has important implications for future experiments.
Finally, the \qp{111} configuration was realized by rotating the \xyz{110} crystal 35$^\circ$ ($\approx$ arctan($\sqrt{1/2}$) around $y$ towards \xyz{001}.

During this experiment the cryostat window was changed from bare polyimide to a polyimide-Al composite, due to the short (few hours) lifetime of polyimide in the 10\,keV beam behind the focusing mirrors.
With the Al-composite the lifetime of the windows increased considerably and measurements over many days could be performed with a stable insulation vacuum of about 10$^{-6}$\,mbar.

\subsection{Data processing}\label{sec:CeB6Dataprocessing}

\begin{figure}
  \centering
  \begin{subfigure}[t]{0.44\textwidth}
    \centering
    \includegraphics[height=60mm]{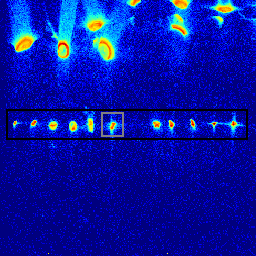}
    \caption{Typical detector image (\textit{starmap}): Color map of the single pixel intensities summed over the whole energy range.} \label{fig:starmap}
  \end{subfigure}
  \hfill
  \begin{subfigure}[t]{0.52\textwidth}
    \centering
    \includegraphics[height=67mm]{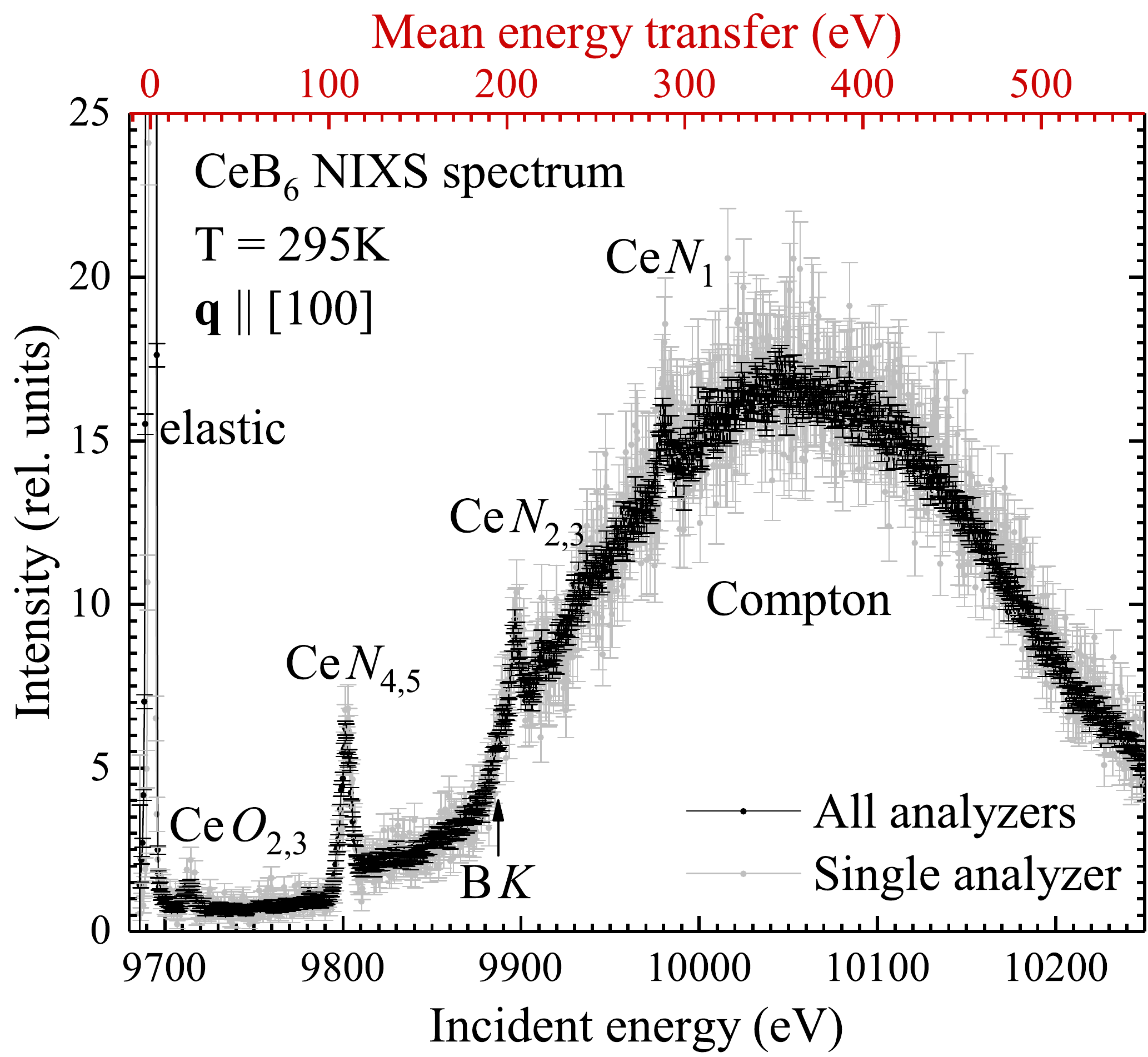}
    \caption{Experimental NIXS spectra of CeB$_6$. Counts of 10\,s/point normalized to the incident intensity vs.\ the incident energy for a single (gray, rescaled) and all analyzers (black).} \label{fig:CeB6widescan}
  \end{subfigure}
  \caption{Detector signal of NIXS spectra for CeB$_6$ at \T{}=295\,K with \qp{100}.}
  \label{fig:CeB6wide}
\end{figure}

Experiments are performed with fixed final energy and the incident energy was scanned.
The signal of each of the 12 analyzers is focused to the Medipix3-based 300$\mu$m\,Si coated 2D pixel-detector\cite{lambda2012} by aligning the analyzers accordingly.
In the following is introduced in detail how the energy dependent spectra are obtained.

\paragraph{Detector image and NIXS spectra}

\Fig{fig:starmap} shows such a detector signal as false-color image (\textit{starmap}).
Colors are logarithmic from dark blue (zero intensity) to red (maximum intensity).
The 11 spots in the black box are the sample signals, which are focused by the 11 analyzers into the 2D-detector.\footnote{One analyzer could not be focused due to motor control problems and its diffuse signal is moved out of the detector.}
The analyzers are tilted so that the sample signal appears on a horizontal line in the 2D-detector.
The larger spots on the top of the image are reflections from the cryostat window.
They are out of focal condition and appear larger.
The horizontal arrangement of the focal spots avoids the overlap of the two signals.
The blurry stripes above the signals of the cryostat window are generated by the air outside the cryostat.

Only the signals inside the black framed region are of interest.
The frames can be chosen freely.
For example, the large black frame sums up the signals of all 11 analyzers and yields the black spectrum in \fig{fig:CeB6widescan}.
Alternatively, the signal of only one analyzer can be extracted (see gray frame in \fig{fig:starmap}) yielding the gray spectrum in \fig{fig:CeB6widescan}.
The advantage of being able to freely choose the so-called region of interest (\ROI{}) is that spurious scattering, that does not correspond to the inelastic scattering process in \fig{fig:CeB6geometry}, can be eliminated.
This way the background can be reduced by adding up 11 single \ROI{}s, each set tightly around the signal. 

The counts are normalized to the incoming intensity, that is measured with a pin diode just before the cryostat.
This ensures that the signal is not falsified due to variation of the incoming intensity, e.g.\ due to so-called monochromator \textit{glitches} (another than the desired Bragg reflection of the monochromator single crystal scatters into a different direction at a specific Bragg angle and thus reduces the intensity of the desired reflection).

\Fig{fig:CeB6widescan} shows the typical features of non-resonant inelastic scattering (as introduced in \fig{fig:XRSlogSpectrum}):
Phonon and valence excitations are hidden inside the tail of the elastic line that has a \FWHMG{} of about 1\,eV\footnote{before the pixelwise energy calibration of the analyzer signals, see below.}.
The minimum around 40\,eV above the elastic line is likely given by fading of the plasmon excitations and the onset of Compton scattering (see \chap{sec:expnixs}).
The Compton scattering is the huge background that peaks at around 10.05\,keV (i.e.\ 350\,eV above the elastic line) in the present set-up.
On top of the Compton scattering the different Ce core electron excitations are visible:
The \edge[Ce]{N}{4,5} ($4d$$\rightarrow$$4f$), \edge{N}{2,3} ($4p$$\rightarrow$$4f$), and \edge{N}{1} ($4s$$\rightarrow$$4f$).
The \edge[B]{K}{} edge is only poorly resolved.

The information about the 4$f$ charge density is given by the small angular \vecq{} dependence of a Ce core excitation.
Comparing the error bars of the single and summed analyzer signals shows the demand of multiple analyzers.
With 10\,s/point and 0.5\,eV step size, this 600\,eV wide energy scan takes more than 3:20\,hours.
The most promising Ce core level in the wide scan in \fig{fig:CeB6widescan} is the $4d$ (\edge[Ce]{N}{4,5} edges), since it has the best signal to background ratio.

\begin{figure}
  \centering
    \centering
    \includegraphics[width=\textwidth]{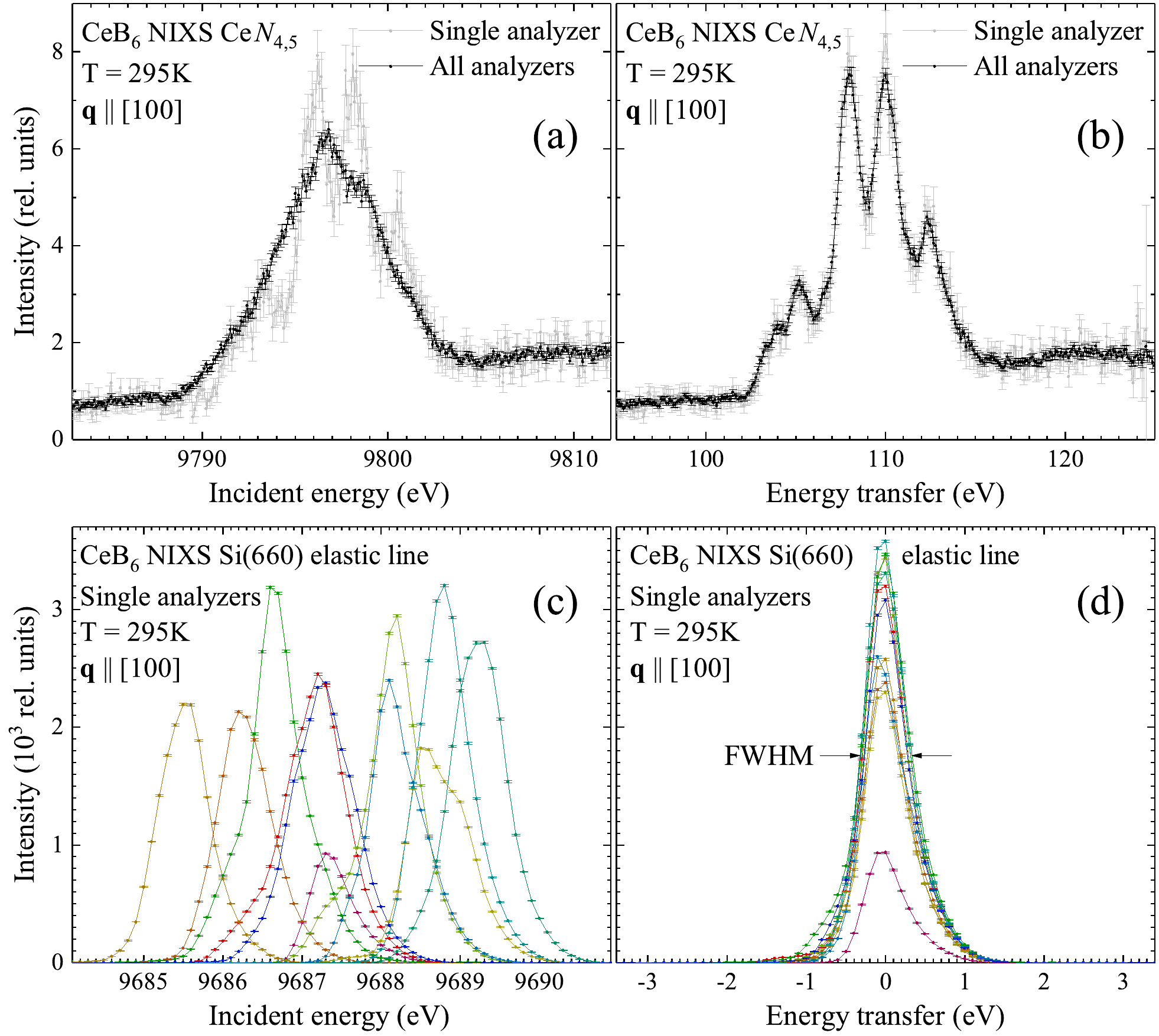}
  \caption{Energy calibration of the \NIXS{} data. (a) Single normalized \edge[Ce]{N}{4,5} edges (10\,s/point) vs incident energy \--- as measured \--- before the energy calibration for a single and all analyzers. (b) Signal of one elastic scan (1\,s/point) vs the incident energy for the different analyzers, which is used for the energy calibration. (c) Same \edge[Ce]{N}{4,5} edges, but after the energy calibration and vs energy transfer. (d) Same elastic scan after the energy calibration and vs energy transfer.}
  \label{fig:CeB6calibration}
\end{figure}

\paragraph{Incident energy and energy transfer}

\Fig[a]{fig:CeB6calibration} shows the spectrum of the \edge[Ce]{N}{4,5} edges of a single (gray) and the summed up analyzers (black) for the 11 \ROI{}s in \fig{fig:CeB6wide} as function of the measured incident energy.
\Fig[b]{fig:CeB6calibration} shows the same data as a function of the energy transfer.
The differences arise from the correction of the position of the elastic energy.

The single analyzer exhibits two peaks and two shoulders but the summed signal only shows one broad peak.
The reason becomes obvious when looking at the elastic energies of all analyzers in \fig[c]{fig:CeB6calibration}.
Their energies are shifted by few eV, so that in the summed up edge the signal is washed out.
The energy position of the elastic lines should be corrected for before summing the signals of different analyzers.
For every analyzer the energy scale is shifted to bring the center of gravity of the elastic line to zero.
After the correction, it is convenient to change to the energy transfer scale, relative to the elastic signal, which is now the identical for all signals.
The result is shown in \fig[d]{fig:CeB6calibration} for the elastic and in \fig[b]{fig:CeB6calibration} for the summed up \edge[Ce]{N}{4,5} edges.

Note that the single lines in \fig[d]{fig:CeB6calibration} also appear more narrow than in \fig[c]{fig:CeB6calibration}.
This is because the energy calibration is done for every detector pixel.
This improves the total instrumental \FWHMG{} of each Si(660) analyzer signal from about 1\,eV to almost 0.6\,eV.
Only pixels with sufficient number of counts can be taken into account.
For the energy calibration the elastic line should be measured regularly, especially after every intervention.
Here, the elastic signal was measured before each scan with a new setting, as well as every few hours in between collecting data of a particular edge.

\paragraph{Normalization}

The measured and summed up spectra of CeB$_6$ are shown for different \vecq{} directions in \fig[(a+b)]{fig:CeB6normalize} for \T{}\,=\,17\,K and \T{}\,=\,295\,K, respectively.
Three sample directions have been measured with \qp{100}, \qp{110}, and \qp{111}, whereby the \qp{110} was realized in two ways: with the crystal with the polished \xyz{110} surface and by turning the \xyz{100} crystal.
The latter is labeled \qp{100}*.

\begin{figure}
  \centering
  \includegraphics[width=\textwidth]{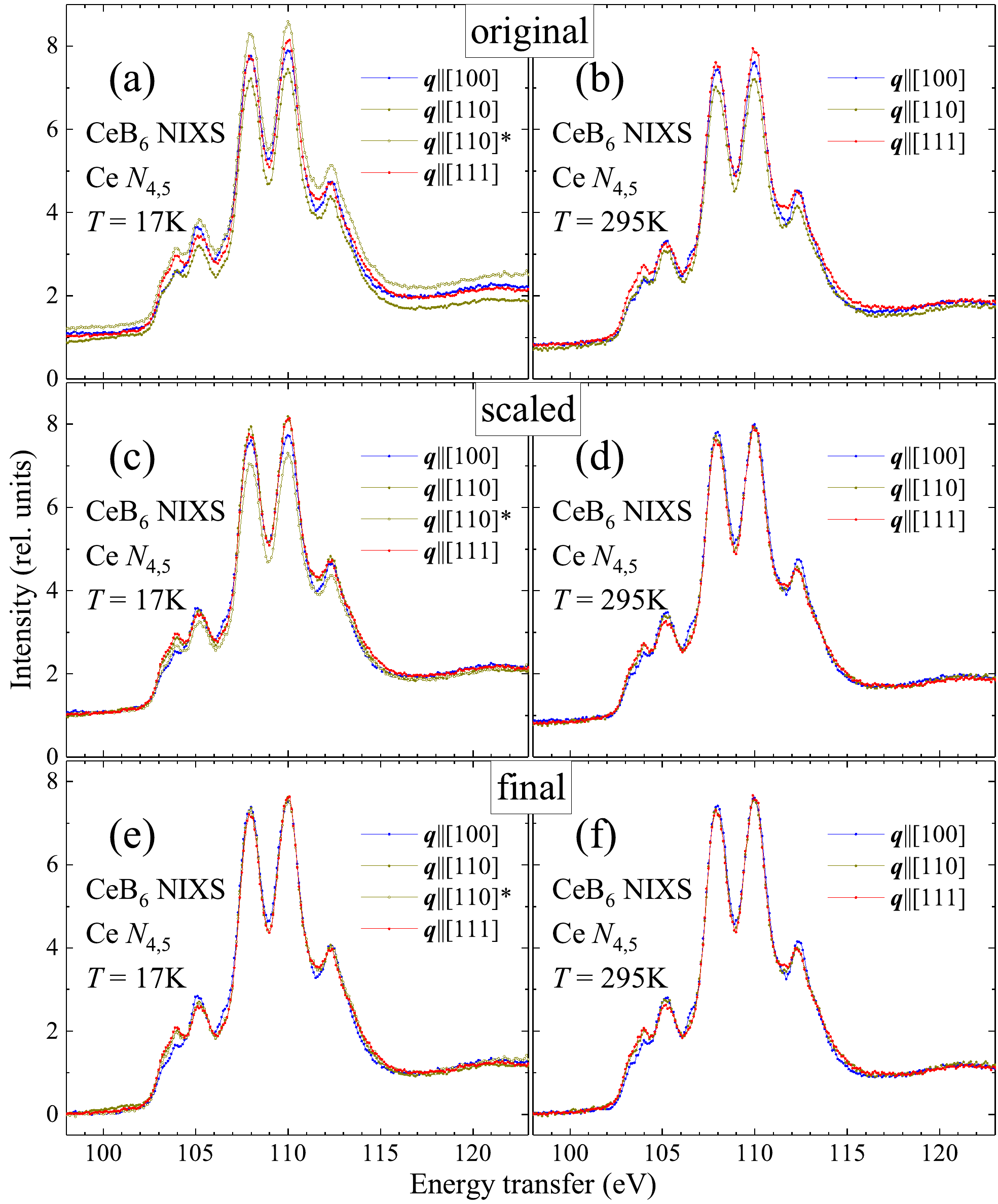}
  \caption{Data of the \edge[Ce]{N}{4,5} edges taken at \T{}\,=\,17\,K (left) and \T{}\,=\,295\,K (right) after the energy calibration. The colors indicate different measured \vecq{} directions. (a+b) as measured, (c+d) scaled to the preedge region around 100\,eV energy transfer, and (e+f) after subtraction of the preedge and scaling to the integrated edge intensity up to 116\,eV. The \qp{111} and \qp{110}* direction have been measured with \vecq{} away from the surface normal.}
  \label{fig:CeB6normalize}
\end{figure}

In \fig[a]{fig:CeB6normalize} the preedge scattering as well as the background beyond the edge do not fall on top of each other.
The same is true for the room temperature data in \fig[b]{fig:CeB6normalize}, but the differences are smaller.
This can happen due to different surface quality, e.g.\ roughness, or if the geometry changes (i.e.\ not all directions were measured specular).
\Fig[(c+d)]{fig:CeB6normalize} show the same data scaled to the \qp{111} direction, in the energy range of 100\,eV energy transfer.
This scaling works very well for the data at 295\,K in (d) but not for the data at 17\,K in (c).
It turned out, that condensates on the top of the sample contribute to the background, especially at low temperatures.
\Fig[(e+f)]{fig:CeB6normalize} show the same data after also subtracting a constant background.
This seems justified since the condensates on the sample (N, O, H) will give rise to a flat background in the region of the \edge[Ce]{N}{4,5} edges.
The background subtracted and scaled spectra in \fig[(e+f)]{fig:CeB6normalize} show that the two measurements of \qp{110} match.
They show further, that the \qp{110} and \qp{111} direction are very similar, while the \qp{100} direction deviates in the energy range 103-107\,eV.
In the next section will be shown that this difference is due to the $\Gamma_8$ symmetry of the \CF{} ground state.

\subsection{NIXS results}

Here the directional \vecq{} dependence of S(\vecq{},$\omega$) of the \edge[Ce]{N}{4,5} \NIXS{} spectra of CeB$_6$ in \fig[(e+f)]{fig:CeB6normalize} is used to find the systems initial state charge density.

Calculations are performed to understand the angular dependence of the cross-sections of the $4d^{10}4f^1$\,$\to$\,$4d^{9}4f^2$ transition.
The cross-sections are derived as introduced in \chap{sec:calcnixs} and the energy positions of the lines are derived using the Hamiltonian introduced in \chap{sec:calcH}.
This \edge[Ce]{N}{4,5} \NIXS{} full multiplet calculation is performed using the code \textsl{Quanty}\cite{Haverkort2016} which includes the local Coulomb interactions as well as the \SOC{}.
The atomic values are derived by \textit{Cowans code}\cite{Cowan1981} for Ce$^{3+}$ as given in \tab{tab:atomicvalues} in \app{app_atomicvalues}.

\paragraph{Multiplet structure}

Before modeling the \vecq{} dependence, the so called isotropic spectrum is calculated and compared to the data.
It is used for adjusting the atomic parameters and the Lorentzian lifetime broadening.
The isotropic spectrum is calculated from the trace of the scattering tensor from \chap{sec:scatteringtensor}, which represents an integration over all \vecq{} directions.
A Gaussian line broadening of 0.7\,eV has been used to account for the instrumental broadening, as obtained from the elastic line (\fig[d]{fig:CeB6calibration}).

\begin{figure}
  \centering
  \includegraphics[width=0.75\textwidth]{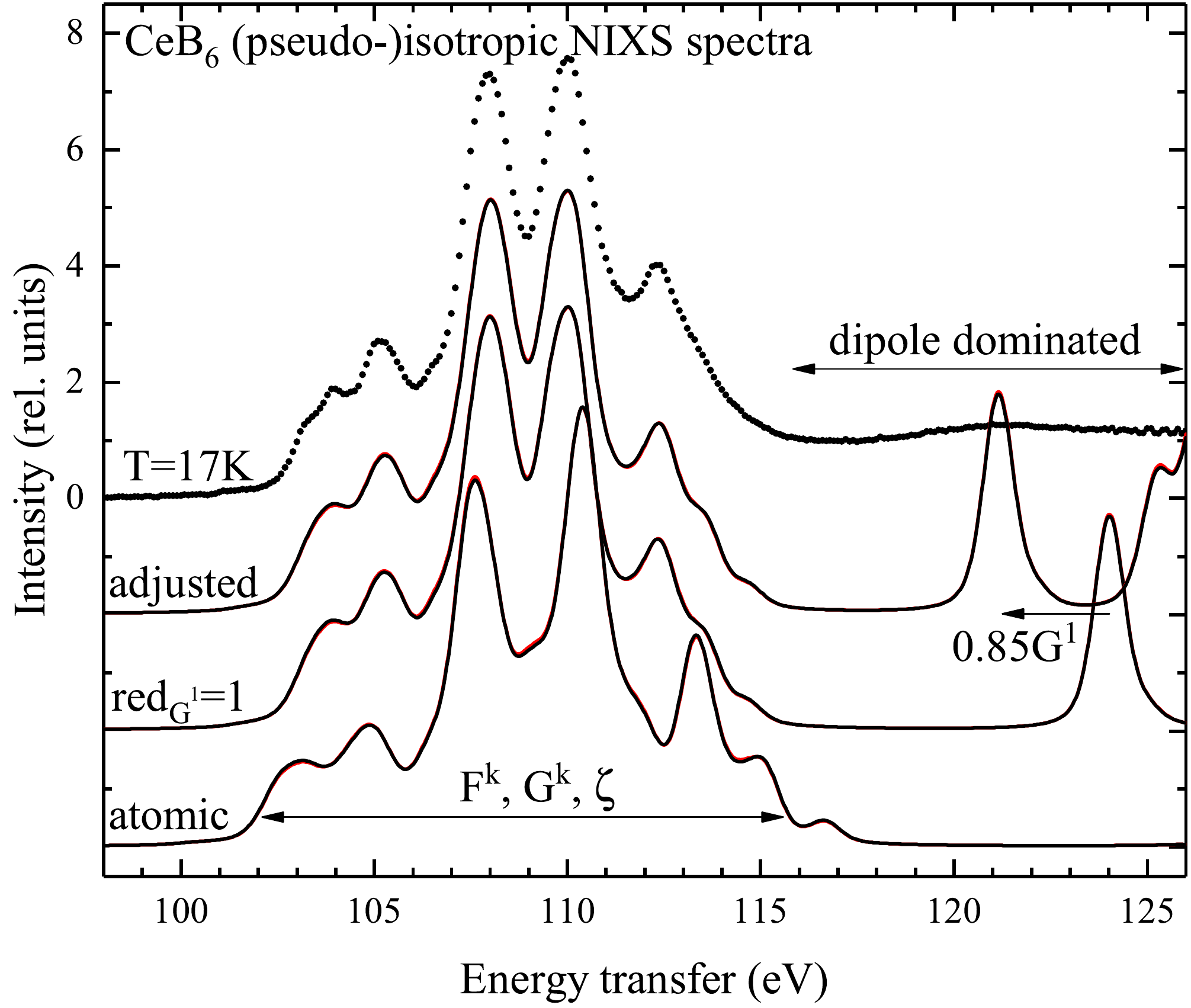}
  \caption{Experimental (pseudo-)isotropic spectrum obtained from linear combination of different directions (dots) and calculated isotropic spectra (lines) for atomic values and for the adjusted parameters (see \tab{tab:CeB6parameters}) with and without additional reduction of \redG{}.}
  \label{fig:CeB6isotropic}
\end{figure}

The bottom black line in \fig{fig:CeB6isotropic} shows the calculation of the isotropic spectrum for the atomic values of Ce$^{3+}$.
For comparison, a pseudo-isotropic spectrum has been calculated from the experimental data (dots).
This pseudo-isotropic spectrum is a linear combination of the three measured directions that yields an isotropic spectrum for the multipole scattering with \qnQ{}\,=\,1 and \qnQ{}\,=\,3.
The prefactors depend on the point group of the measured ion in the crystal, the measured directions, and the values of \qnQ{} present in the transition.
In the present case for \qnQ{}\,=\,3:
\begin{align}
I_\text{iso} &= \left(40\times I(\qp{100})+32\times I(\qp{110})+27\times I(\qp{111}) \right)/99
\label{eq:OhQ3isotropic}
\end{align}
For the 4 independent diagonal elements of the \qnQ{}\,=\,5 transition in the conductivity tensor shown in \fig{fig:CeB6tensor} the measured directions are not sufficient to provide the true isotropic spectrum.
The error of this approximations has been estimated by also calculating the pseudo-isotropic spectrum of the $\Gamma_8$ initial state (\fig{fig:CeB6isotropic} underlying red lines).
The differences are so small, that they are barely noticeable on the plot.

A comparison of the calculation in \fig{fig:CeB6isotropic} with the experiment shows that the calculated spectrum is spread on a wider energy range.
This is due to the Coulomb interactions (F$^k$, G$^k$) and \SOC{} ($\zeta$).
The atomic calculation does not include boundary conditions due to the crystalline environment.
Also Coulomb interactions with the surrounding ions are not included in the atomic values.
As a consequence the values for the solid will deviate from the atomic model and the atomic parameters need to be adjusted.
These corrections tend to screen the interactions (at most Coulomb) so that the adjusted values are smaller than the atomic ones.
Also \absq{} may be adjusted, since the radial extension of the wave function may differ from the atomic one as well.

\begin{table}
  \centering
  \caption{Parameters used for calculations. The values represent the best set found to describe the pseudo-isotropic spectrum in \fig{fig:CeB6isotropic}. The values in the brackets are the variation step size of the parameters.}
  \label{tab:CeB6parameters}
  \begin{tabular*}{\columnwidth}{@{\extracolsep{\fill}}ccccc||ccc}
    \hline
	\redff{} & \redfc{} & \redG{} & \redZc{} & \redZv{} & \absq{}/$\AA^{-1}$ & \FWHML{} & \FWHMG{} \\
	 0.7\,(1) & 0.78\,(5) & 0.85\,(5) & 1.00\,(5) & 0.95\,(5) & 9.2\,(2) & 0.4\,(1)\,eV & 0.7\,eV \\
	\hline
  \end{tabular*}
\end{table}

The middle line in \fig{fig:CeB6isotropic} shows the calculated isotropic spectrum using reduced atomic values.
The prefactors scaling the atomic values are given in \tab{tab:CeB6parameters}.
The Gaussian full width at half maximum \FWHMG{} is fixed to the width of the elastic line.
The Lorentzian full width at half maximum \FWHML{} has been chosen to fit best the mean line width of the two main peaks at 108 and 110\,eV energy transfer.
The reductions of the atomic parameters \redff{}, \redfc{}, \redZc{}, and \redZv{} were chosen to fit the various peak positions in the range 102-115\,eV.
The adjustment of the spin-orbit coupling with the given resolution of the spectra gives only a minor improvement of the line shape.
The reduction factors \redff{}, \redfc{} have the biggest impact.
Finally the momentum transfer \absq{} is adjusted.
Instead of the instrumental value \absq{}\,=\,(9.6$\pm$0.1)\,\AA$^{-1}$ the experimental branching ratios of the different lines in the isotropic spectrum is reproduced best by \absq{}\,=\,9.2\,\AA$^{-1}$.
All these adjustments account in a phenomenological way for the changes of the atomic wave functions inside the crystalline environment.

In the region of higher multipoles (below $\approx$118\,eV) the simulation reproduces the experiment in great detail.
Remaining discrepancies are due to the lifetime broadening, which is kept constant in the calculation over the entire energy range, but it appears to increase with increasing energy transfer in the experiment.
As a result, fine structures in the data at $\approx$103\,eV are not reproduced in the calculation, but the shoulder at 113\,eV is less structured in the data than the calculation suggests.
The line broadening towards higher energy transfer is due to the proximity of the continuum states (see e.g.\ \chap{sec:nixsspectra}).

All orders of the Coulomb interaction within the same shells are reduced with the same prefactor so far.
The top line in \fig{fig:CeB6isotropic} shows the same isotropic spectrum with reduced atomic values, but with the extra reduction of the Coulomb dipole interaction Slater integral G\textsup{1} (see \tab{tab:CeB6parameters}).
This indicates, that ionic dipole moments can be screened more effectively by the surrounding charges than the higher order moments.
There may likely be a relation between the stronger lifetime broadening and the stronger Coulomb reduction.

The final calculations are performed with the parameters as in \tab{tab:CeB6parameters}, incl.\ \redG{}.
The brackets behind the values give the final step width of the variation of the last digits in the process of adjusting the parameters.

\paragraph{Directional dependence of \vecq{}}

Now \vecq{} dependent (opposed to isotropic) calculations are compared to the directional dependent data.

For the analysis of the model calculation with the adjusted parameters for CeB$_6$ from \fig{fig:CeB6isotropic} in \tab{tab:CeB6parameters} is compared with the experiment for the different directions.

\begin{figure}
  \centering
  \includegraphics[width=0.48\textwidth]{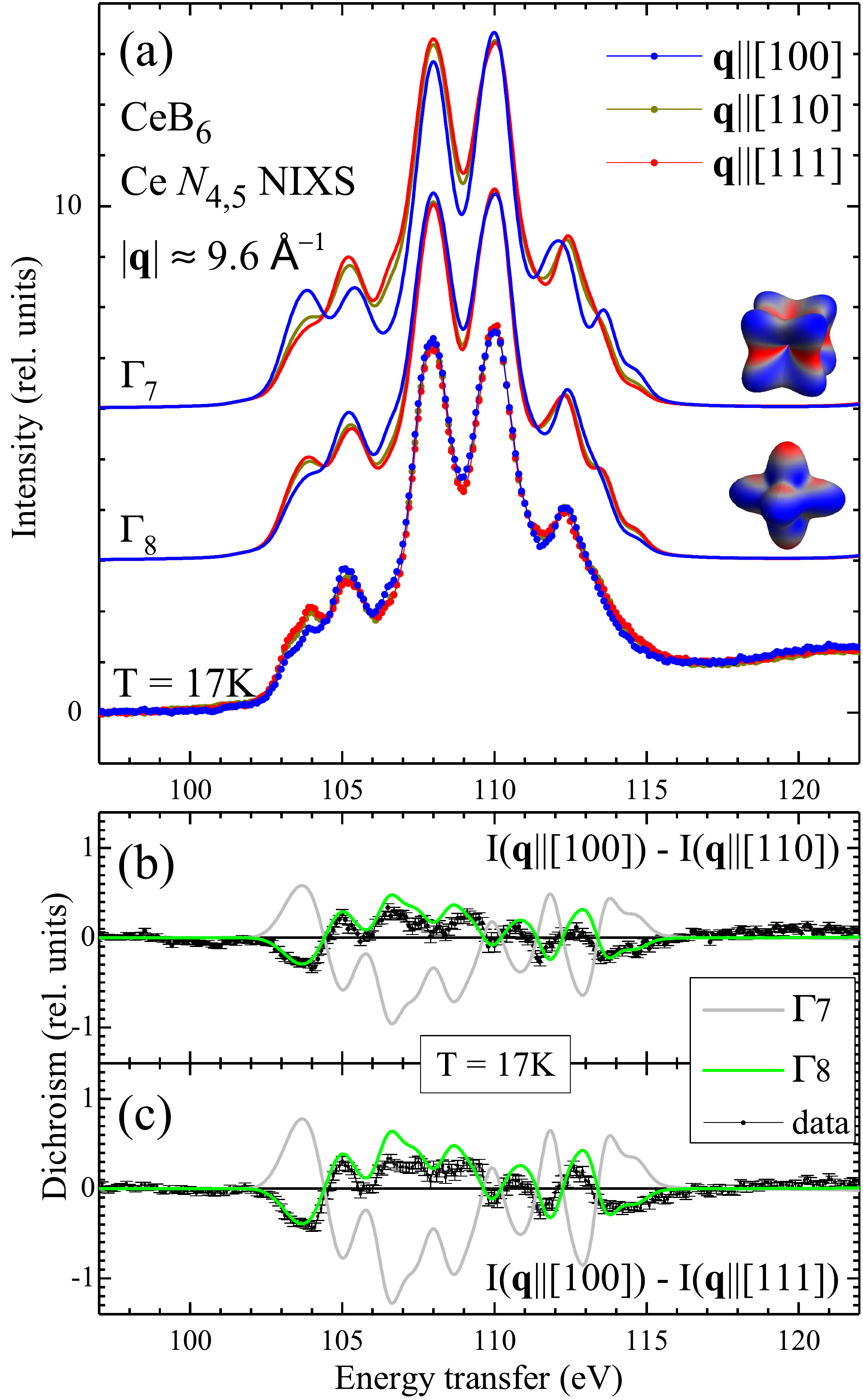}
  \hfill
  \includegraphics[width=0.48\textwidth]{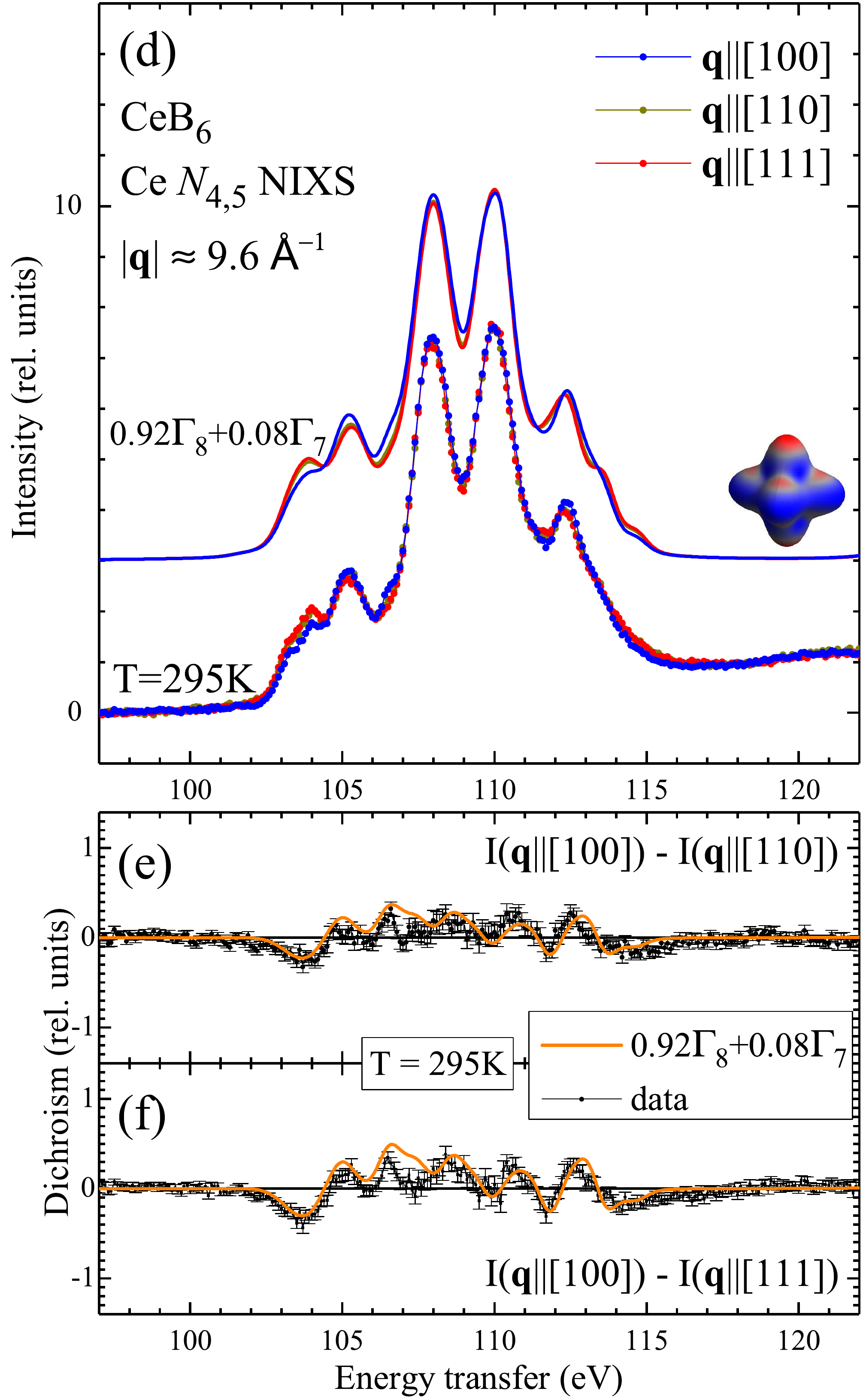}
  \caption{Angular dependent \edge[Ce]{N}{4,5} \NIXS{} spectra. Data (dots) same as in \fig[(c+f)]{fig:CeB6normalize} and calculation (lines) obtained from different $\epsilon(\vec{q})$.}
  \label{fig:CeB6N45}
\end{figure}

\Fig[a]{fig:CeB6N45} shows the calculated spectra of the $\Gamma_7$ doublet and the $\Gamma_8$ quartet for CeB$_6$.
Note that the reduction factors change the positions of the emission lines, but do not effect their directional dependence.
The \vecq{}-directional dependence in CeB$_6$ is clearly reproduced with the $\Gamma_8$ quartet initial state charge density, whereas the $\Gamma_7$ doublet should give rise to an opposite and stronger difference spectrum.
To be more specific, the spectra show red over blue on the peak at lowest energy transfer around 104\,eV, and blue over red on the second.
This directional dependence is reproduced only by the $\Gamma_8$ quartet initial state.
For a better visualization \fig[(b+c)]{fig:CeB6N45} show the difference spectra for \qp{100} and \qp{110} (top) and \qp{100} and \qp{111} (bottom).
Here the data (black dots) are plotted on top of both calculations, $\Gamma_7$ (gray) and  $\Gamma_8$ (orange).
For the $\Gamma_8$ the difference spectra show an overlap over the entire energy transfer range, but a clear opposite behavior for the $\Gamma_7$.
Also the maxima and minima in the spectra are reproduced in the calculation.
Minor deviations are likely due to the approximation with effective reduction factors and the Stevens approximation.
This model, however, is sufficient for determining the \CF{} symmetry especially since the question of the $\Gamma_7$ or $\Gamma_8$ ground state is reduced to a boolean condition, i.e.\ it is a red over blue or blue over red experiment.

\Fig[(d-f)]{fig:CeB6N45} shows a similar comparison for the data at \T{}\,=\,295\,K.
The difference in the data is visibly smaller than at 17\,K, yet still present.
Instead of comparing with the pure $\Gamma_7$ or $\Gamma_8$ spectra the calculation is performed including both states with the Boltzmann occupation of the $\Gamma_7$ state for a the \CF{} splitting of 46\,meV as reported in Ref.\,\cite{Zirn1984,Loew1985}.
The almost 8\% occupation of the $\Gamma_7$ reduces the difference in the signal by about 23\%.
Thus, the experimental reduction in the difference spectra is in good agreement with the \CF{} splitting of 46\,meV.

\paragraph{Summary}

The $\Gamma_8$ quartet \CF{} ground-state symmetry measured at 20\,K of the cubic \HO{} compound CeB$_6$ has been unambiguously confirmed using \edge[Ce]{N}{4,5} core level \NIXS{}.

The high signal to background ratio of the \edge[Ce]{N}{4,5} \NIXS{} signal indicates that this bulk sensitive and element specific spectroscopic technique is a powerful method to study the local electronic structure of the rare-earth ions in rare-earth borides.
The result agrees with the well-established description of the ground state of CeB$_6$, so that this particular \NIXS{} experiment proves the validity of the presented \NIXS{} analysis.
It is therefore an important \textit{benchmark} result.

\clearpage
\section{SmB\textsub{6}} \label{sec:SmB6}
\begin{center}\textbf{
\scalebox{1.2}{4\textit{f} crystal-field ground state of the strongly correlated}\\
\scalebox{1.2}{topological insulator SmB\textsub{6}}}\\
\hyperref[SmB6_2017]{\scalebox{1.2}{Physical Review Letters \textbf{120} 016402 (2018)}}
\end{center}

\FloatBarrier
\begin{SCfigure}[][h]
  \centering
  \includegraphics[width=0.48\textwidth]{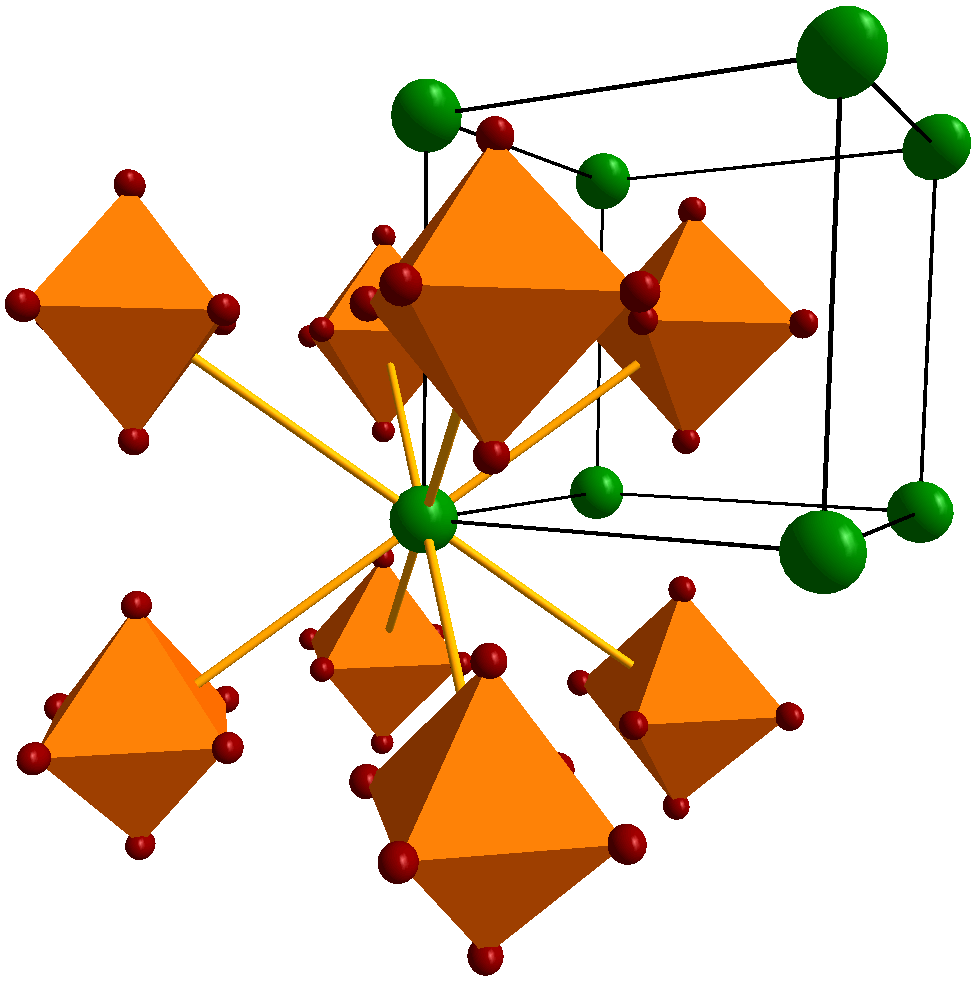}
  \caption{Unit cell (black box) and chemical environment of the Sm ion in cubic SmB$_6$. Sm (green spheres), B octahedra (red spheres with polygons), and the Sm nearest neighbor connections (yellow bars) are shown. Structure parameters from ICSD\,\cite{icsd,Funahashi2010}.}
  \label{fig:SmB6structure}
\end{SCfigure}
\FloatBarrier

\subsection{Introduction}

\begin{figure}
    \centering
    \includegraphics[width=0.9\columnwidth]{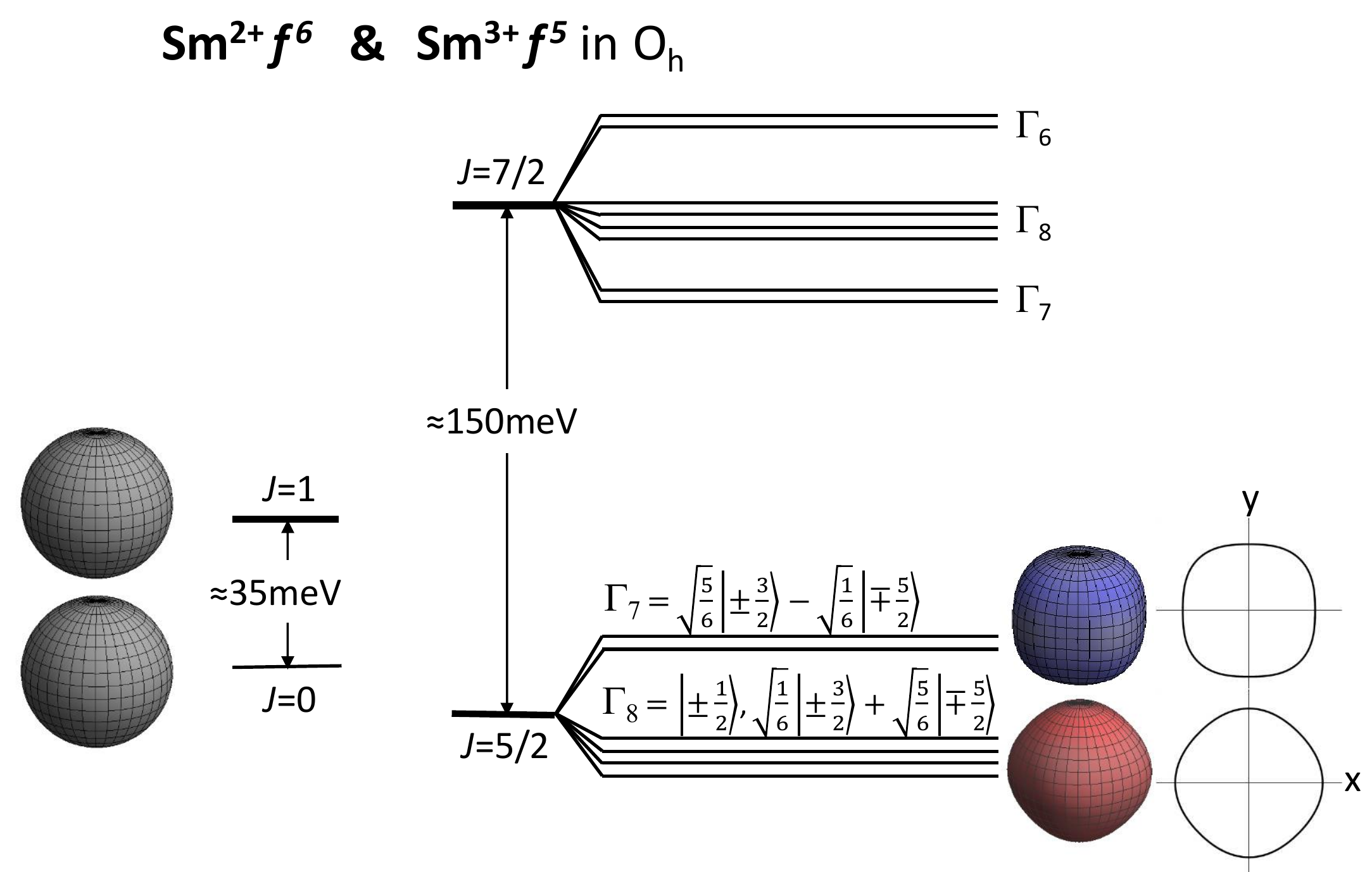}
    \caption{Sm$^{2+}$ and Sm$^{3+}$ energy level diagram. The Sm$^{2+}$ configuration is split into a \qnJ{}\,=\,0 and \qnJ{}\,=\,1, and the Sm$^{3+}$ into a \qnJ{}\,=\,5/2 and \qnJ{}\,=\,7/2 multiplet. The Sm$^{3+}$ 2\qnJ{}+1-fold degenerate multiplets are further split by the cubic O\textsub{h} \CF{} into different $\Gamma_i$ states (see \tab{tab:f2cubicrepresentations}). The insets show the corresponding charge densities for six and five electrons and their 2D projections, respectively. For Sm$^{3+}$ \qnJ{}\,=\,5/2 the \ket{\qnJz{}} representation of the CF states is given.}
    \label{fig:SmB6scheme}
\end{figure}

\begin{SCfigure}
    \centering
    \includegraphics[width=0.5\columnwidth]{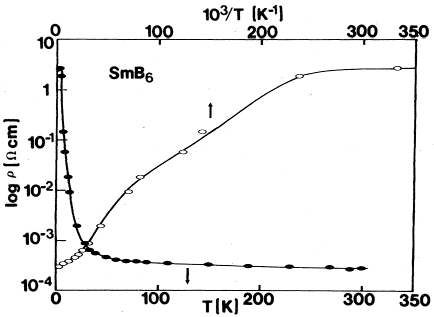}
    \caption{Temperature dependent resistivity curve of SmB$_6$ adapted from Ref.\,\cite{Allen1979}. Below $\approx$5\,K the resistivity flattens.}
    \label{fig:SmB6_resisitivity}
\end{SCfigure}

SmB$_6$ is an intermediate valent Kondo insulator\,\cite{Menth1969, Cohen1970, Allen1979, Gorshunov1999, Riseborough2000} with a cubic (Pm$\bar{3}$m) crystal structure (see \fig{fig:SmB6structure}).
The strong hybridization of the low lying 4$f$ states with conduction band $d$ states gives rise to a hybridization gap of the order of 20\,meV\,\cite{Xu2013, Zhu2013, Neupane2013, Jiang2013, Denlinger2013, Denlinger2014}.
It also causes a strong mixing of the Sm$^{2+}$ and Sm$^{3+}$ configurations, leading to a valence of 2.5 to 2.7\,\cite{Allen1980, Tarascon1980, Mizumaki2009, Hayashi2013, Lutz2016, Butch2016, Utsumi2017}.
\Fig{fig:SmB6scheme} shows the total energy level diagram of the two Sm configurations, that contribute to the ground state of SmB$_6$.
At low temperatures, when the hybridization is active and the hybridization gap opens, SmB$_6$ becomes insulating, since the unit cell has an even number of electrons and the Fermi energy lies in the gap.
The resistivity $\rho$(\T{}), however, does not diverge for \T{}$\rightarrow$0.
It remains finite forming a plateau below \T{}\,=\,10\,K, even in the purest samples.
\Fig{fig:SmB6_resisitivity} shows the resistivity of SmB$_6$.
This low temperature finite resistivity kept scientists puzzled for decades.
In 2010, an appealing explanation was introduced:
SmB$_6$ could be a \textit{strongly correlated topological insulator}, where the surface states give rise for the finite conductivity at low \T{}\,\cite{Dzero2010, Takimoto2011, Dzero2012, Lu2013, Dzero2013, Alexandrov2013}.
SmB$_6$ has all ingredients for the non-trivial topology:
Strong spin-orbit coupling of the 4$f$ electrons and bands of opposite parity, the 4$f$ and 5$d$.
This caused a flurry of experimental and theoretical efforts in the recent years using angle-resolved photoelectron spectroscopy\,\cite{Xu2013, Zhu2013, Neupane2013, Jiang2013, Denlinger2013, Denlinger2014, Xu2014natcomm, Xu2014, Hlawenka2015}, scanning tunneling spectroscopy\,\cite{Yee2013, Roessler2014, Ruan2014, Roessler2016, Jiao2016}, resistivity and surface conductance measurements\,\cite{Hatnean2013, Zhang2013, Kim2013, Wolgast2013, Kim2014, Wolgast2015, Thomas2016, Nakajima2016} and the prediction of novel phenomena in this new material class\,\cite{Vishwanath2013, Bonderson2013, Wang2013, Fidkowski2013}.

Information of the surface topology can be unambiguously inferred from the symmetries and parities of the bulk states involved.
Knowledge about the \CF{} ground state symmetry of SmB$_6$ therefore plays an essential role.
For example, theoretical predictions for the spin texture of the sought-after topological surface states depend very much whether the ground state of the $f^5$ \qnJ{}\,=\,$5/2$ configuration is the $\Gamma_8$ quartet or the $\Gamma_7$ doublet \CF{} state.
The resulting type or direction of the winding of the spin with wave vector is opposite\,\cite{Baruselli2015, Legner2015, Baruselli2016}. 

Surprisingly, after forty years of research, the \CF{} scheme of SmB$_6$ has still to be determined.
The classical tool inelastic neutron scattering has not been able to identify the \CF{} states, possibly due to the superposition of both Sm $f^5$ and $f^6$ configurations in this mixed valence compound and the strong neutron absorption despite double isotope samples\,\cite{Alekseev1993, Alekseev1995, Fuhrman2015}.
From inelastic neutron scattering, a sharp excitation at 14\,meV close to the hybridization gap was reported.
It was assigned to a spin exciton and not to a \CF{} excitation since its intensity does not follow the 4$f$ magnetic form factor. Further magnetic intensities at about 35\,meV, 115\,meV, and 85\,meV have been assigned to the inter-multiplet transitions of the Sm$^{2+}$ configuration and of the \CF{} split Sm$^{3+}$ configuration (see \fig{fig:SmB6scheme}), and to some magnetoelastic coupling, respectively.
In-gap transitions at about 15\,meV in Raman spectra could be interpreted as \CF{} excitations but Raman does not yield the information about which state forms the ground state\,\cite{Nyhus1995, Nyhus1997}.
A semi-empirical extrapolation method can predict \CF{} parameters across the rare earth series for highly diluted systems\,\cite{Frick1986}.
Applying such an extrapolation to the measured \CF{} schemes of $RE$B$_6$ with $RE$\,=\,Ce, Pr and Nd\,\cite{Zirn1984, Loewenhaupt1986} yields for SmB$_6$ a \CF{} splitting of the order of 15\,meV with the $\Gamma_8$ quartet as the ground state.
The Kondo insulator SmB$_6$, however, is not a highly diluted system and it is definitely not an ionic system but has a highly intermediate valence instead, questioning the validity of such an extrapolation.

The local ground state of SmB$_6$ is almost equally governed by the two configurations $4f^6$ for Sm$^{2+}$ and $4f^5$ for Sm$^{3+}$ due to its strong hybridization.
The Hund's rule ground states of these configurations are given by the total momenta of \qnJ{}\,=\,0 and 5/2, respectively.
In cubic \CF{} symmetry the \qnJ{}\,=\,5/2 is split into a $\Gamma_7$ doublet and $\Gamma_8$ quartet.
The next higher \qnJ{}\,=\,7/2 multiplet of the Sm$^{3+}$ is about 150\,meV higher in energy.
The next higher \qnJ{}\,=\,1 multiplet of Sm$^{2+}$ is 35\,meV above the \qnJ{}\,=\,0 singlet state.
Neither the \qnJ{}\,=\,0 singlet nor the \qnJ{}\,=\,1 triplet split in cubic O\textsub{h} symmetry.
Only the degeneracy of states with \qnJ{}\,$\geq$\,2 gets lifted by the O\textsub{h} \CF{} (see \tab{tab:f2cubicrepresentations} or Ref.\,\cite{Lea1962}).

\Fig{fig:SmB6scheme} shows the low lying energy levels of both Sm configurations in SmB$_6$, which are important for the analysis.
The full energy level diagram can be found in \app{app_Sm3pELD}.
The insets show the corresponding electron charge densities, calculated for \redff{}\,=\,0.8.
For \redff{}\,=\,0.8 the admixture of the \qnJ{}\,=\,5/2 ground state multiplet is given by 95\% of $^6$H$_{5/2}$ and 4\% of the energetically lowest $^4$G$_{5/2}$ multiplet, with only minor contributions of the other 26 term symbols with \qnJ{}\,=\,5/2.
Since the \CF{} does not split the Sm$^{2+}$ states, they remain spherical, whereas the Sm$^{3+}$ multiplets charge densities are anisotropic.
These anisotropies are much smaller than for CeB$_6$.
Note, that the lobes of the $\Gamma_8$ point towards the \xyz{100} direction like for CeB$_6$, which is in contrast to the single particle picture (\qnJ{}\,=\,5/2 with 1 electron or 1 hole).
Hence it is not directly obvious in the case of Sm$^{3+}$ whether the \CF{} ground state is $\Gamma_7$ or $\Gamma_8$, even if the \CF{} potential was known.
See \app{app_Sm3pELD} for more detailed information.

\subsection{Experimental}

The SmB$_6$ single crystals were grown by the aluminum flux method by the group of Z.~Fisk at University of California, Irvine\,\cite{Kim2014}.
Commercial polycrystalline Sm$_2$O$_3$ and Eu$_2$O$_3$ powder with 99.9\,\% and 99.99\,\% purity, respectively, has been pressed into pellets.
These have been used as references for the $4f^5$ and $4f^6$ configuration.

Two SmB$_6$ single crystals with \xyz{100} and \xyz{110} surfaces were mounted in a closed cycle cryostat with Kapton windows.
The two samples were oriented such that for \qp{100} and \qp{110} a specular scattering geometry was realized.
For the \qp{111} direction one of the crystals was turned accordingly with respect to the scattering triangle.
The two reference compounds were measured under the same conditions in the closed cycle cryostat at 18\,K.

The \NIXS{} measurements with focus on the Sm and Eu\,$N_{4,5}$ edges ($4d^{10}4f^5$\,$\rightarrow$\,$4d^{9}4f^6$ and $4d^{10}4f^6$\,$\rightarrow$\,$4d^{9}4f^7$, respectively) were performed at the P01 beamline at PETRA-III.
The incident energy was selected with a Si(311) \DCM{}.
The fixed final energy was about 9690\,eV.
The analyzers were positioned at scattering angles of 2\thS{}\,$\approx$\,150$^\circ$, 155$^\circ$, and 160$^\circ$ which provide an averaged momentum transfer of \absq{}\,=\,(9.6\,$\pm$\,0.1)\,\AA$^{-1}$.
The scattered beam was detected by a position sensitive custom-made Lambda detector, based on a Medipix3 chip.
The elastic line was consistently measured and a pixel wise calibration yields an instrumental energy resolution of $\approx$0.7\,eV \FWHM{}.

\subsection{NIXS results}

\begin{SCfigure}
    \centering
    \includegraphics[width=0.5\columnwidth]{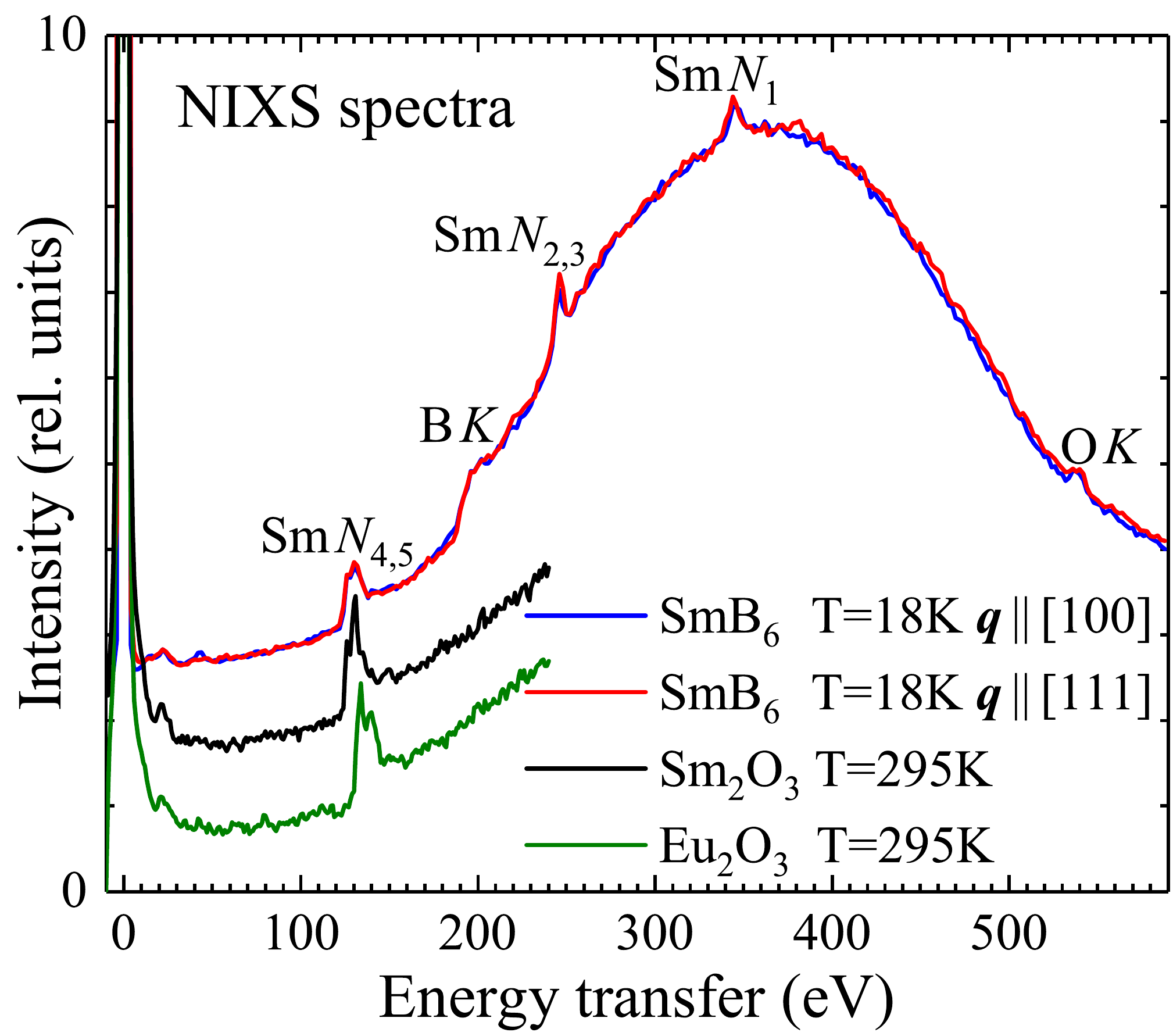}
    \caption{Energy scans of SmB$_6$ for \qp{100} and \qp{111}, Sm$_2$O$_3$ and Eu$_2$O$_3$ measured at \absq{}\,=\,(9.6\,$\pm$\,0.1)\,\AA$^{-1}$ and with a constant final energy of 9690\,eV over a wide energy ranges, for SmB$_6$ with 0.5\,eV energy steps, for Sm$_2$O$_3$ and Eu$_2$O$_3$ with 0.2\,eV; shifted for clarity.}
    \label{fig:SmB6_wide}
\end{SCfigure}

\Fig{fig:SmB6_wide} shows \NIXS{} spectra measured over a large energy range of SmB$_6$ (blue and red) and of the two reference compounds Sm$_2$O$_3$ (black) and Eu$_2$O$_3$ (olive) measured up to 250\,eV energy transfer.
The strongest signal is the elastic line, followed by shallow Sm/Eu core resonances (O-edges) and the \edge[Sm/Eu]{N}{4,5} edges at about 130 and 135\,eV sitting on top of the rising Compton background.
The spectra are offset on the $y$-axis by 1 unit after scaling to the Compton background.
For SmB$_6$ the signal of the \edge[Sm]{N}{4,5} edges yields the best signal to background ratio of the observed Sm edges.
For the reference compounds the signal to background ratio is even better, due to the lower number of electrons per Sm/Eu in these compounds.

\paragraph{Comparison with Sm\textsub{2}O\textsub{3} and Eu\textsub{2}O\textsub{3}}

\begin{figure}
    \centering
    \includegraphics[width=\textwidth]{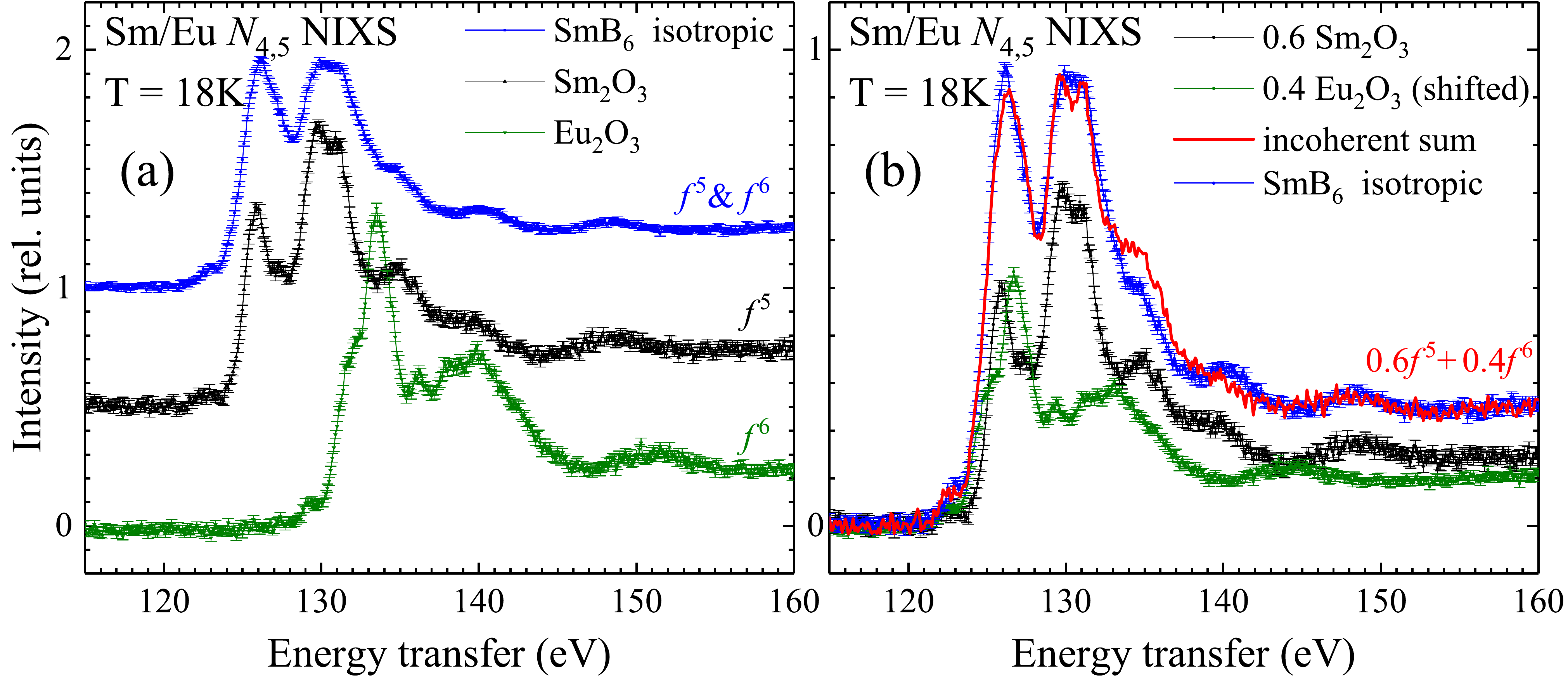}
    \caption{(a) Energy scans at the \edge[Sm]{N}{4,5} edges of SmB$_6$, Sm$_2$O$_3$ and Eu$_2$O$_3$ measured with a constant final energy  of 9690\,eV and 0.2 eV energy steps at \absq{}\,=\,(9.6\,$\pm$\,0.1)\,\AA$^{-1}$. A linear background has been subtracted. For SmB$_6$ the pseudo-isotropic spectrum (see \equ{eq:OhQ3isotropic}) is shown instead of the different directions of \vecq{}. (b) Without offset in y, the reference data of Sm$_2$O$_3$ and Eu$_2$O$_3$ have been scaled by 0.6 and 0.4, respectively, and the Eu$_2$O$_3$ were shifted by 6.8\,eV. The red line shows the incoherent sum of the Sm$_2$O$_3$ and Eu$_2$O$_3$ data.}
    \label{fig:SmB6references}
\end{figure}

\Fig[a]{fig:SmB6references} shows the \NIXS{} spectra across the \edge[Sm]{N}{4,5} edges of SmB$_6$ (blue) and of the two reference compounds Sm$_2$O$_3$ ($f^5$) and Eu$_2$O$_3$ ($f^6$) (black and olive) after scaling to the integrated \edge{N}{4,5} edges intensity and subtraction of a linear background.
For SmB$_6$ the \qnQ{}\,=\,3 pseudo-isotropic spectrum has been calculated as for CeB$_6$ (\equ{eq:OhQ3isotropic}).
The Eu edge appears at a higher energy transfer than in the case of Sm because Eu has a higher atomic number. 

At first it will be clarified whether the SmB$_6$ spectrum can be reconstructed by the two reference compounds Sm$_2$O$_3$ and Eu$_2$O$_3$.
For this purpose a spectrum is constructed from the weighted sum of Sm$_2$O$_3$ $f^5$ and Eu$_2$O$_2$ $f^6$.
The Eu$_2$O$_3$ spectrum is shifted by 6.8\,eV to lower energies in order to compensate for its higher atomic number.

\Fig[b]{fig:SmB6references} shows the scaled reference spectra and the incoherent sum on top of the experimental data.
The Sm$_2$O$_3$ spectrum weighted with a factor 0.6 plus the Eu$_2$O$_2$ spectrum weighted with a factor of 0.4 reproduce the SmB$_6$ spectrum very satisfactorily.
This reflects a valence of $\approx$2.6, when neglecting final state effects, and is in good agreement with literature.
This provides confidence to carry out further analysis using full multiplet calculations based on the 4$f^5$ and 4$f^6$ configurations of Sm.

\FloatBarrier
\paragraph{Multiplet structure}

\begin{table}[t]
  \centering
  \caption{Parameters used for all calculations. The values represent the best set found to describe the pseudo-isotropic spectra in \fig{fig:SmB6reffit}. The values in the brackets are the variation step size of the parameters.}
  \label{tab:SmB6parameters}
  \begin{tabular*}{\columnwidth}{@{\extracolsep{\fill}}ccccc||ccc}
    \hline
	\redff{} & \redfc{} & \redG{} & \redZc{} & \redZv{} & \absq{}/$\AA^{-1}$ & \FWHML{} & \FWHMG{} \\
	 0.8\,(1) & 0.80\,(5) & 0.85\,(5) & 1.00\,(5) & 1.00\,(5) & 9.8\,(2) & 0.4\,(1)\,eV & 0.7\,eV \\
	\hline
  \end{tabular*}
\end{table}

\begin{figure}
    \centering
    \includegraphics[width=\columnwidth]{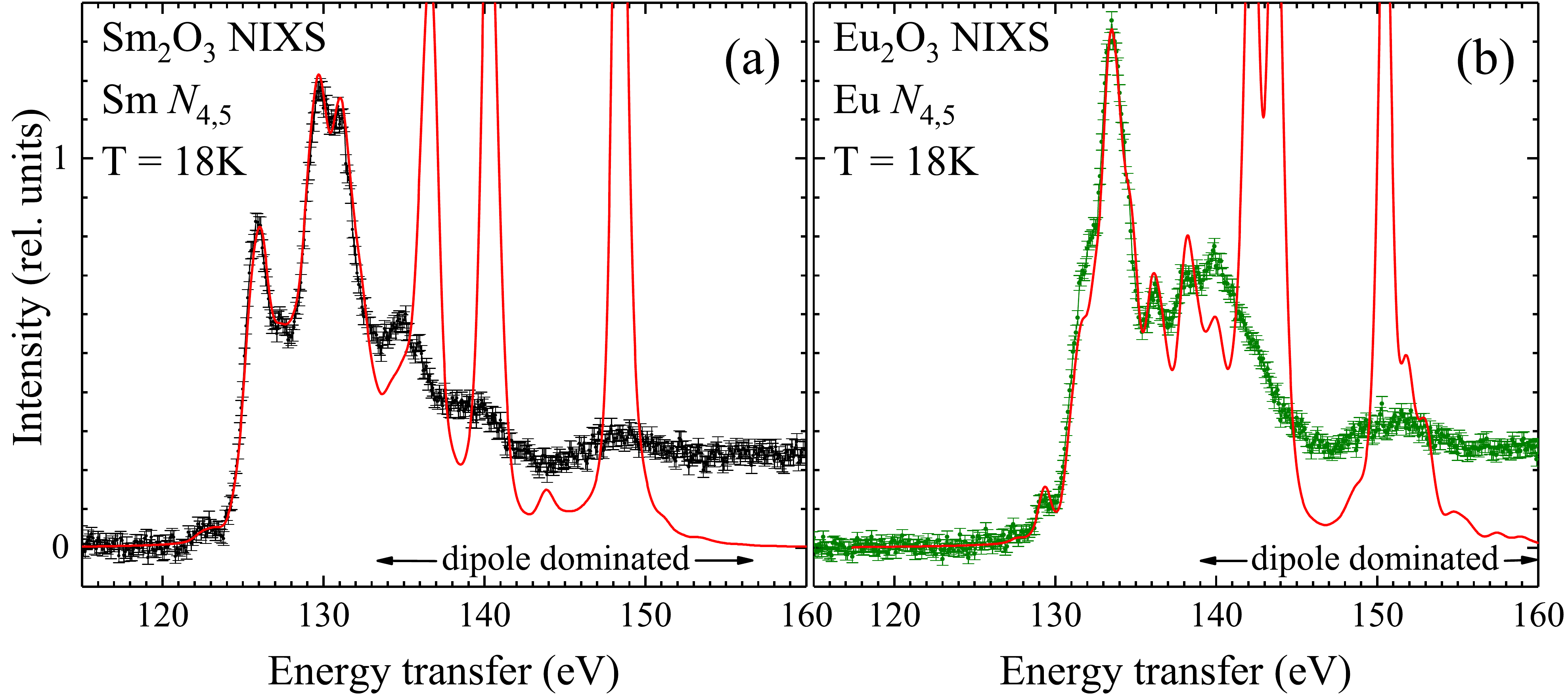}
    \caption{Comparison of the \edge{N}{4,5} edges data (dots) of Sm$_2$O$_3$ (a) and Eu$_2$O$_3$ (b) with the simulation (red lines).}
    \label{fig:SmB6reffit}
\end{figure}

In the following, it is checked to what degree the full multiplet calculation captures the multiplet structures of Sm$_2$O$_3$ ($4f^5$) and Eu$_2$O$_3$ ($4f^6$).

\Fig[a]{fig:SmB6reffit} shows the calculation of the \edge[Sm]{N}{4,5} edges for Sm$^{3+}$ compared to the Sm$_2$O$_3$ \NIXS{} data and \fig[b]{fig:SmB6reffit} shows the \edge[Eu]{N}{4,5}-edges for Eu$^{3+}$ compared to the Eu$_2$O$_3$ \NIXS{} data.
Both calculations use the same reduction and lineshape parameters, shown in \tab{tab:SmB6parameters}.
The individual atomic parameter sets (before reduction) of Sm$^{3+}$ ($f^5$) and Eu$^{3+}$ ($f^6$) have been used, for the respective calculations.
In the multipole region (120-135\,eV) the agreement between calculation and data is excellent.
In the dipole dominated region above $\approx$135\,eV the simulation yields unrealistically sharp features since, as for CeB$_6$, interference with continuum states are not included in the calculation.

For the adjustment of the parameters in \tab{tab:SmB6parameters} first the \FWHMG{} was fixed to the instrumental resolution, obtained from the elastic line.
The Lorentzian lifetime broadening \FWHML{} is chosen to fit the observed line width of the excitonic \edge{N}{4,5} multipole region.
The reduction factors \redff{}, \redfc{}, \redZc{}, and \redZv{} are given by the peak positions in the multipole region.
\redG{} is adjusted according to the dipole dominated peaks above 135\,eV to the broad humps.
Finally, the momentum transfer of \absq{}\,=\,9.8\,\AA$^{-1}$ has been used for the simulations rather than the experimental value of 9.6\,$\pm$\,0.1\,\AA$^{-1}$ in order to reproduce best the experimental peak intensity ratio of the two main features around 126 and 130\,eV.

\Fig[a]{fig:SmB6directions} shows the full multiplet calculation of the \edge[Sm]{N}{4,5} edges for Sm$^{3+}$ contribution of SmB$_6$ (black).
It results from the fit of the Sm$_2$O$_3$ data, i.e.\ it is identical to the calculation in \fig[a]{fig:SmB6reffit}, but scaled with the previous determined factor 0.6.
The \edge[Sm]{N}{4,5} edges of the SmB$_6$ Sm$^{2+}$ contribution (olive) is calculated using the same parameters as in \tab{tab:SmB6parameters}, but the atomic parameters of Sm$^{2+}$, and is scaled by a factor of 0.4.
The shape differs slightly from the Eu$^{3+}$ calculation.
The weighted sum (60\%\,Sm$^{3+}$ and 40\%\,Sm$^{2+}$) of the calculated curves (red) describes the SmB$_6$ spectrum (blue dots) very well in the energy range of the higher multipole scattering dominated region, i.e.\ between 120 and 135\,eV.
Now it is possible to interpret the \vecq{}-directional dependence of the \edge[Sm]{N}{4,5} edges of SmB$_6$.

\paragraph{Directional dependence of \(\vec{\textit{q}\,}\)\!}

\begin{figure}
    \centering
    \includegraphics[width=\textwidth]{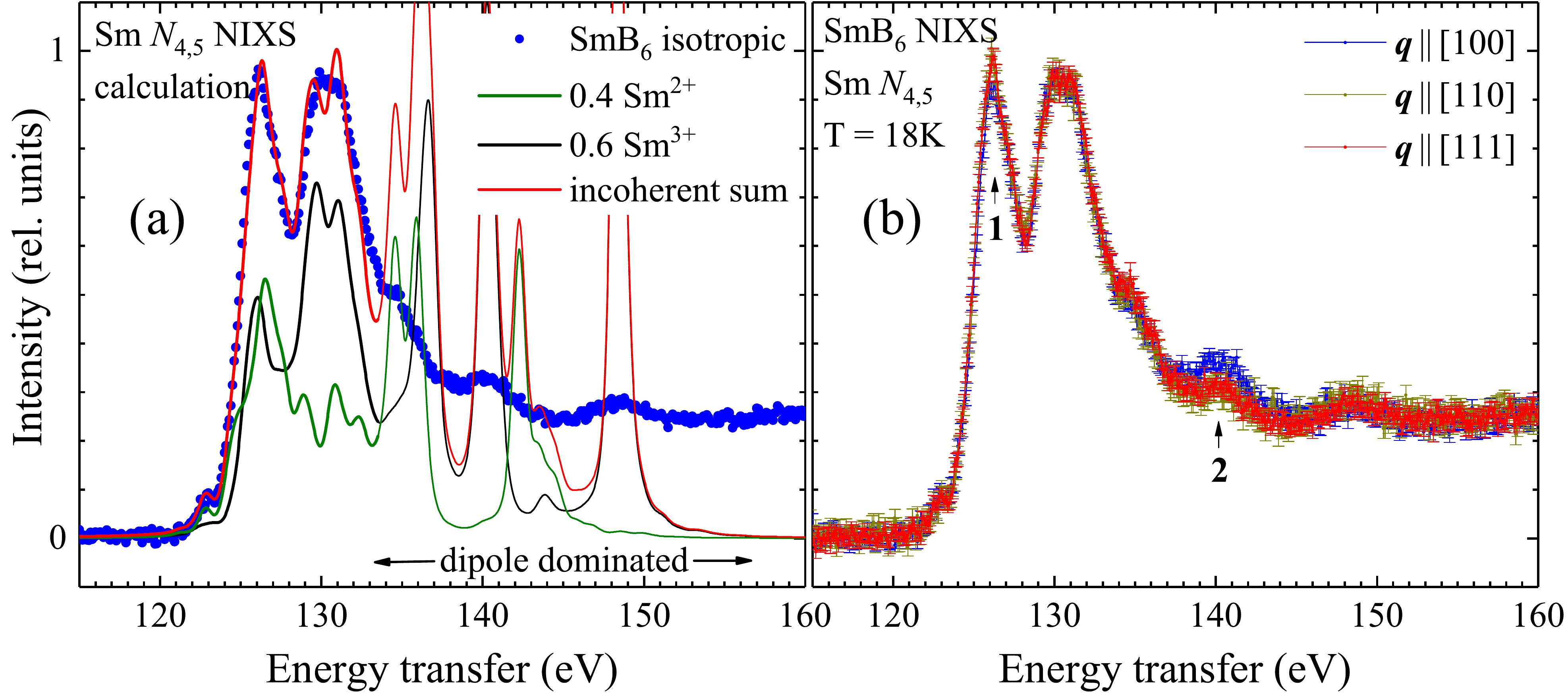}
    \caption{(a) Full multiplet simulation of Sm$^{3+}$ (black line) and Sm$^{2+}$ spectra (olive line) and their weighted sum (red line). For comparison the pseudo-isotropic data from \fig{fig:SmB6references} are shown below. (b) SmB$_6$ \NIXS{} data at 16\,K for \qp{100} (blue), \qp{110} (dark yellow), and \qp{111} (red).}
    \label{fig:SmB6directions}
\end{figure}

\Fig[b]{fig:SmB6directions} shows the directional dependence of the \edge[Sm]{N}{4,5} edges of SmB$_6$.
Although the effect is small, there are clear differences between the spectra in the energy regions marked with the arrows.
At about 126\,eV energy transfer the scattering of the \qp{110} and \qp{111} (dark yellow and red) directions are both stronger than for the \qp{100} (blue), and at about 140\,eV it is opposite.
To show these directional differences in a more transparent manner, \fig[b]{fig:SmB6dichroism} shows the difference spectrum between the \qp{100} and \qp{111} (black dots): this so-called dichroic spectrum has unambiguously a negative peak at 126\,eV whereas it displays positive intensity in a broader region around 140\,eV \-- again indicated by the arrows.

\begin{figure}
    \centering
    \includegraphics[width=\textwidth]{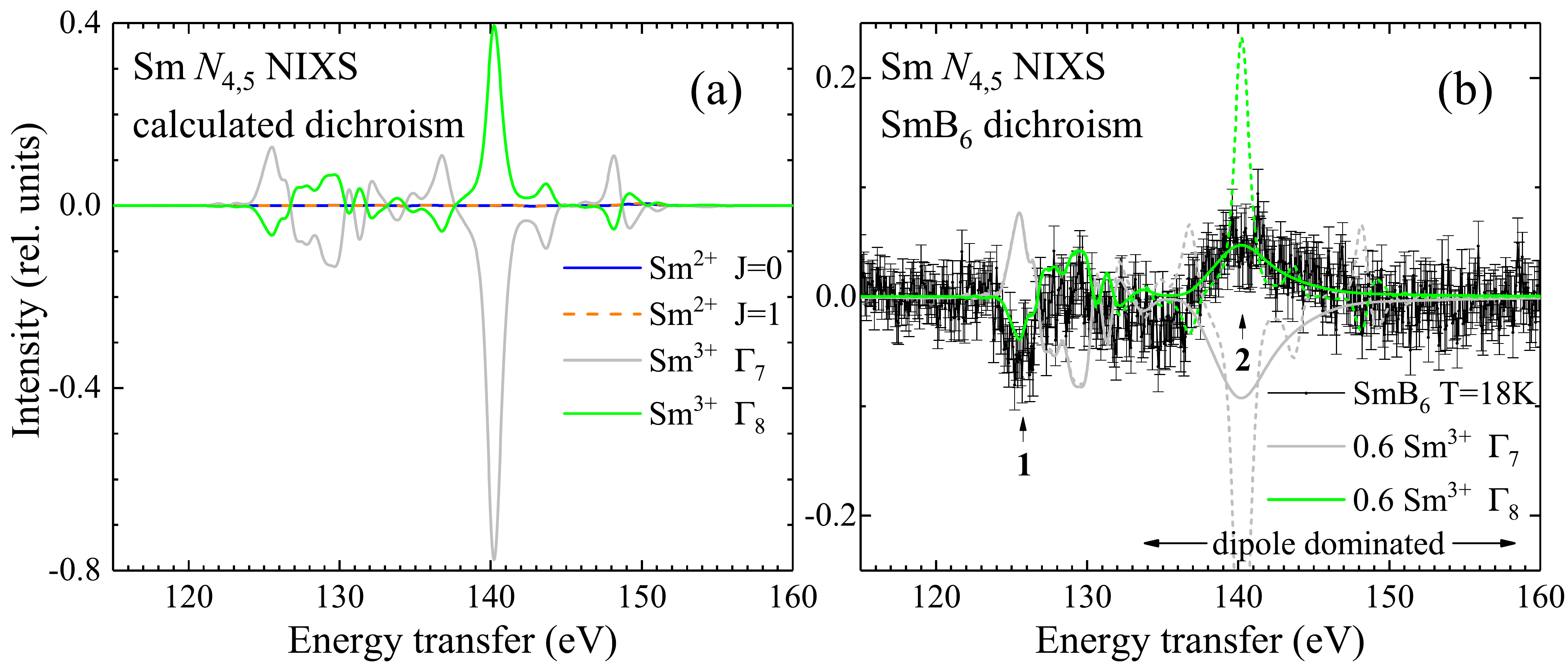}
    \caption{(a) Simulation of the dichroic spectrum (I(\qp{100})\,-\,I(\qp{111})) for the \qnJ{}\,=\,0 (blue) and \qnJ{}\,=\,1 (orange) states of the Sm$^{2+}$ configuration as well as for the $\Gamma_7$ doublet (gray) and $\Gamma_8$ quartet (green) of the \qnJ{}\,=\,5/2 Sm$^{3+}$ configuration; (b) Experimental dichroic spectrum (black) and simulated dichroic spectra for the $\Gamma_8$ and $\Gamma_7$ scaled with the factor of 0.6 to account for the Sm$^{3+}$ component of the ground state; dashed lines with energy independent broadening, solid lines with extra broadening in the dipole-dominated region.}
    \label{fig:SmB6dichroism}
\end{figure}

To interpret the observed direction dependence, it is important to know how each \CF{} state or multiplet component contributes to the dichroic signal.
\Fig[a]{fig:SmB6dichroism} shows the difference spectrum (I(\qp{100})\,-\,I(\qp{111})) of the four lowest multiplets: the \qnJ{}\,=\,0 singlet and the \qnJ\,=\,1 triplet of the Sm$^{2+}$ $f^6$ configuration and the $\Gamma_7$ doublet and the $\Gamma_8$ quartet of the Sm$^{3+}$ $f^5$ configuration\footnote{The excited \qnJ{}\,=\,7/2 multiplet is not considered because it will not contribute since it is too high in energy.}.
Both multiplets of the Sm$^{2+}$ configuration do not split in a cubic \CF{}, are still spherical, and thus have no dichroism (see blue and orange lines).
Hence, the observed direction dependence of the signal is solely due to the Sm$^{3+}$ Hund's rule ground state.
Another important finding is that the $\Gamma_8$ and $\Gamma_7$ \CF{} states exhibit different and opposite dichroism (see green and gray lines), consistent with their opposite anisotropy in the charge densities (see \fig{fig:SmB6scheme}).
These aspects facilitate the straight forward determination of the \CF{} ground state of Sm$^{3+}$ in SmB$_6$.

The dashed lines in \fig[b]{fig:SmB6dichroism} show the calculated difference spectrum of the $\Gamma_7$ and $\Gamma_8$ scaled by the factor of 0.6 to account for the Sm$^{3+}$ component in intermediate valence compound SmB$_6$.
One can clearly observe that in the regions of pronounced dichroism, at 125 and 140\,eV, the sign of the experimental dichroic signal is correctly explained by the $\Gamma_8$ (green line) but not at all by the $\Gamma_7$ (gray line).
In addition, the $\Gamma_8$ reproduces the experimental dichroism quantitatively at 125\,eV, i.e.\ in the region dominated by higher multipole scattering.
The dichroism also fits quantitatively in the dipole-dominated region at 140\,eV when an extra broadening of \FWHML{}\,=\,4\,eV is applied beyond $\approx$135\,eV energy transfer to mimic the interference with continuum states (solid lines in \fig[b]{fig:SmB6dichroism}).
Note, that in the dipole-dominated energy range the dichroism still arises from the higher multipole scattering.
Here the dipole and the higher multipole signals are equally broad due to the vicinity of the continuum states.
The dashed lines correspond to the calculation without the extra broadening.
These results unambiguously establish that the \CF{} ground state of the Sm $f^5$ component in SmB$_6$ is the $\Gamma_8$ quartet. 

It should be pointed out that the down scaling to 60\% of the Sm\,$f^5$ component gives a good quantitative agreement in the magnitude of the dichroic signal.
This provides confidence that the \NIXS{} method is indeed reliable since this 60\% number is also fully consistent with literature.
The findings are very robust especially because the Sm$^{2+}$ contribution does not provide any dichroism.

\paragraph{Trials with the Sm\,\textit{N}\textsub{1}-edge}

\begin{SCfigure}
    \centering
    \includegraphics[width=0.49\columnwidth]{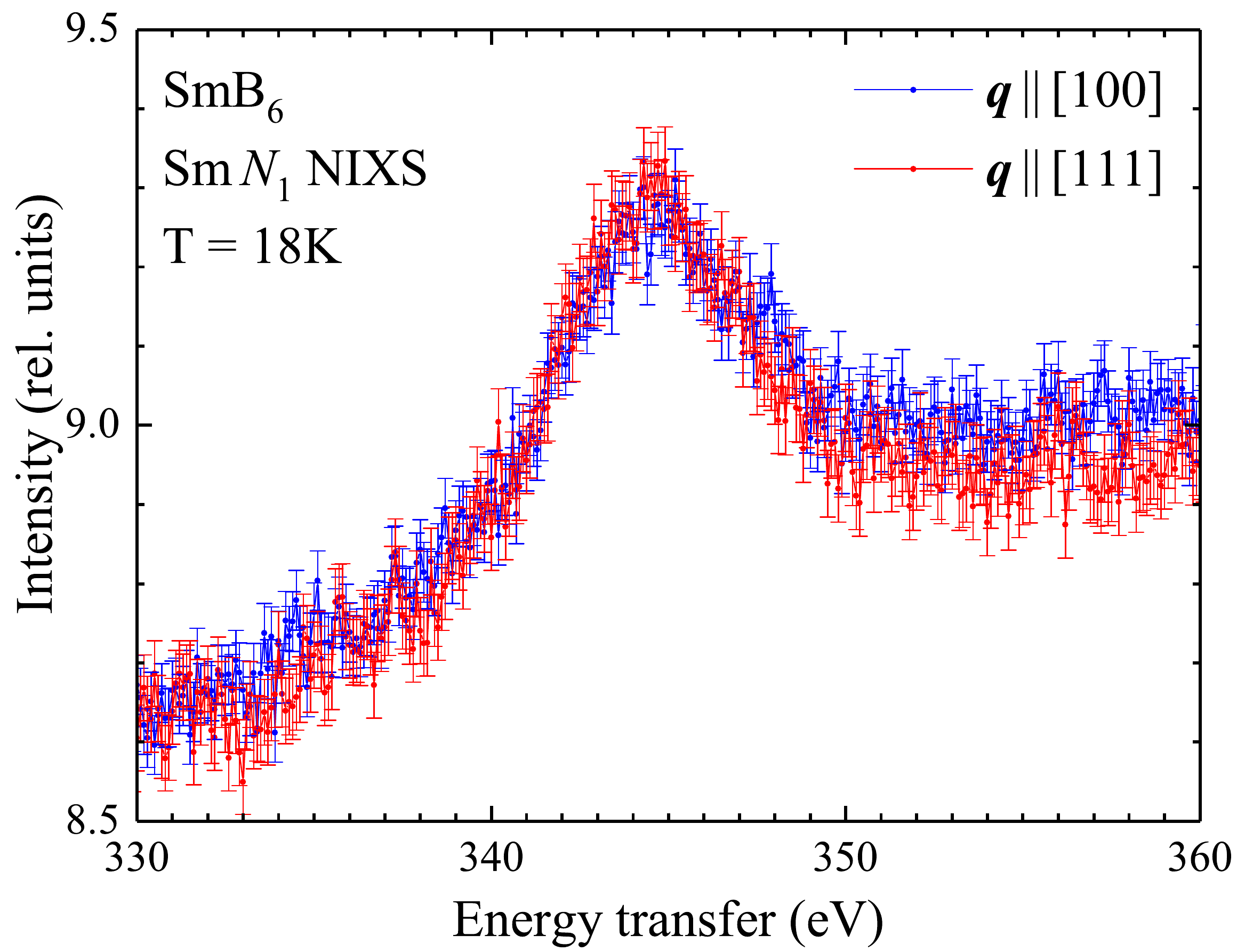}
    \caption{Experimental spectrum of the SmB$_6$ \edge[Sm]{N}{1} edge for \qp{100} (blue dots) and \qp{111} (green dots) scaled according to the corresponding long scans in \fig{fig:SmB6_wide}.}
    \label{fig:SmB6N1}
\end{SCfigure}

The \edge[Sm]{N}{1} edge ($4s\rightarrow4f$) of SmB$_6$ has also been measured for the two directions \qp{100} and \qp{111} at \T{}\,=\,18\,K.
The spectra are shown in \fig{fig:SmB6N1} after normalization to the Compton background.
The \edge[Sm]{N}{1} edge is a pure octupole transition (\qnQ{}\,=\,3) and according to \fig{fig:SmB6N1} it does not show any fine structures.
It is just a broad peak.

The integrated intensity still contains information about the initial charge density.
From the results of the \edge[Sm]{N}{4,5} edges, that proved the $\Gamma_8$ ground state, the expected ratio of the integrated intensities of I(\qp{100})/I(\qp{111}) amounts to 0.96-0.97.
This is due to the small angular variation of the $f^5$ charge density\footnote{The Sm$^{3+}$ charge density with the experimentally found \redff{} has an $\approx$4 times weaker angular dependence than expected from the single particle picture (see \app{app_Sm3pELD}).}, and the 40\% contribution of the isotropic $f^6$ configuration (see \fig{fig:SmB6scheme}).
The energy of the \edge[Sm]{N}{1} edge coincides with the peak of the Compton background (\fig{fig:SmB6_wide}) so that the signal to background ratio is very unfavorable in the present setup.
This hampered the analysis of the directional dependence of the \edge[Sm]{N}{1} edge.

\paragraph{Summary}

The SmB$_6$ \edge[Sm]{N}{4,5} \NIXS{} signal is well described by an incoherent sum of the \NIXS{} signals of 40\% of the Sm$^{2+}$ reference Eu$_2$O$_3$ and 60\% of the Sm$^{3+}$ reference Sm$_2$O$_3$.
The resulting valence of 2.6 is in good agreement with other literature.

The \CF{} ground state symmetry of the Sm$^{3+}$ \qnJ{}\,=\,5/2 multiplet was unambiguously determined to be the $\Gamma_8$.
The problem was an either-or question, as it was in CeB$_6$, because the Sm$^{2+}$ \qnJ{}\,=\,0 and \qnJ{}\,=\,1 configurations do not contribute to any dichroism.

An analysis of the \edge[Sm]{N}{1} edge is hampered by the small signal to background ratio.

\paragraph{Discussion}

Follow-up many body calculations should use this experimental result as input for explaining the spin texture of the surface states as well as the excitations of SmB$_6$.
The $\Gamma_8$ quartet ground state supports very much the results of spin and angle resolved \PES{}\,\cite{Xu2014natcomm}.
They find spin polarized surface states, fulfilling time reversal as well as crystal symmetry, that have spins locked to the crystal momenta $k$ such that at opposite momenta the surface states have opposite spins.
The anticlockwise spin texture is in agreement with spin expectation values that are calculated by Baruselli and Vojta for a $\Gamma_8$ ground state\,\cite{Baruselli2015, Baruselli2016}.
For the $\Gamma_7$ the spin directions should be reversed. 

The result of a $\Gamma_8$ local ground-state symmetry contradicts the outcome of several density functional band structure calculations\,\cite{Yanase1992, Antonov2002, Lu2013, Kang2015}.
For example, Kang \textsl{et al.} reported for the $X$-point an unoccupied $4f$ state of $\Gamma_7$ origin\,\cite{Kang2015}.
Also their $k$-integrated $4f$ \qnJ{}\,=\,5/2 partial density of states (pDOS) shows the hole residing in the $\Gamma_7$ band, in line with the fact that the center of gravity of the $\Gamma_7$ pDOS is higher in energy than that of the $\Gamma_8$, and despite the fact that the $\Gamma_7$ band is lower than the $\Gamma_8$ at the $\Gamma$ point.

\clearpage
\section{CeRu\textsub{4}Sn\textsub{6}} \label{sec:CeRu4Sn6}
\begin{center}\textbf{
\scalebox{1.2}{CeRu\textsub{4}Sn\textsub{6}: a strongly correlated material}
\scalebox{1.2}{with nontrivial topology}}\\
\hyperref[CeRu4Sn6_2015]{\scalebox{1.2}{Scientific Reports \textbf{5}, 17937 (2015)}}\end{center}
\FloatBarrier
\begin{SCfigure}[][h]
  \centering
  \includegraphics[width=0.48\textwidth]{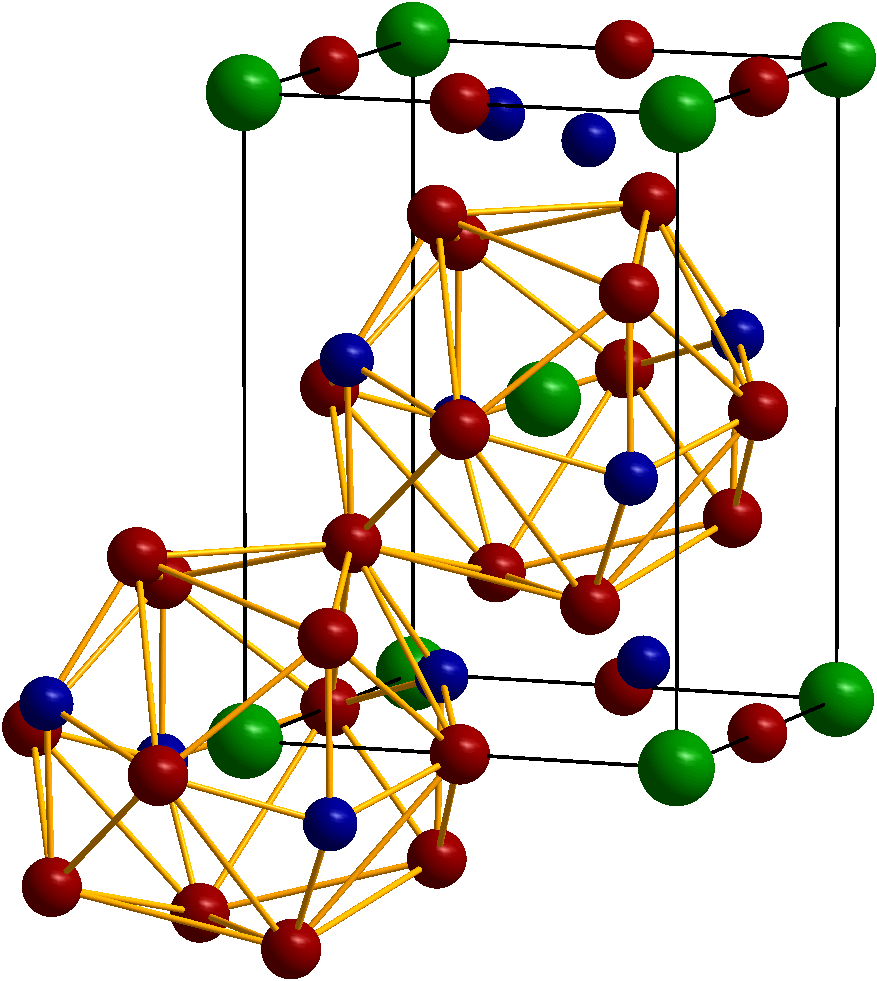}
  \caption{Unit cell (black box) and chemical environment of the Ce$^{3+}$ ion in tetragonal CeRu$_4$Sn$_6$. Ce (green spheres), Ru (blue spheres), and Sn (red spheres). The Ce is embedded inside a cage of Ru and Sn (yellow bars) with local D\textsub{2d} point symmetry. Structure parameters from ICSD\,\cite{icsd, Pottgen_1997}.}
  \label{fig:CeRu4Sn6structure}
\end{SCfigure}
\FloatBarrier

\subsection{Introduction}

CeRu$_4$Sn$_6$ is a Kondo semimetal with a tetragonal, non-centrosymmetric ($I\bar{4}2m$) crystal structure (see \fig{fig:CeRu4Sn6structure}) with a Ce$^{3+}$ Hund's rule ground state of \qnJ{}\,=\,5/2, as in CeB$_6$.
A hybridization-induced gap was reported first in 1992\,\cite{Das_1992}, supported further by several experimental studies\,\cite{Strydom_2005, Paschen_2010, Bruning_2010}, and later estimated to be of the order of 100\,K from single crystal resistivity data\,\cite{Paschen_2010, Winkler_2012}.
There is finite metallicity at low temperatures according to optical conductivity measurements\,\cite{Guritanu_2013} but nevertheless, the low temperature resistivity is in the m$\Omega$\,cm range, indicating clearly the depletion of charge carriers\,\cite{Das_1992, Paschen_2010, Winkler_2012, Guritanu_2013}.
Magnetization and transport properties are highly anisotropic (see \fig[(a+b)]{fig:CeRu4Sn6transport}) thus strongly suggesting the \qnJ{}\,=\,5/2 Hund's rule ground state is split by the tetragonal crystal field\,\cite{Paschen_2010, Winkler_2012}.
$\mu$SR measurements did not find magnetic order down to 50\,mK, thus pointing out the importance of Kondo-type interactions in this material\,\cite{Strydom_2007}.

\begin{figure}
  \centering
  \includegraphics[height=40mm]{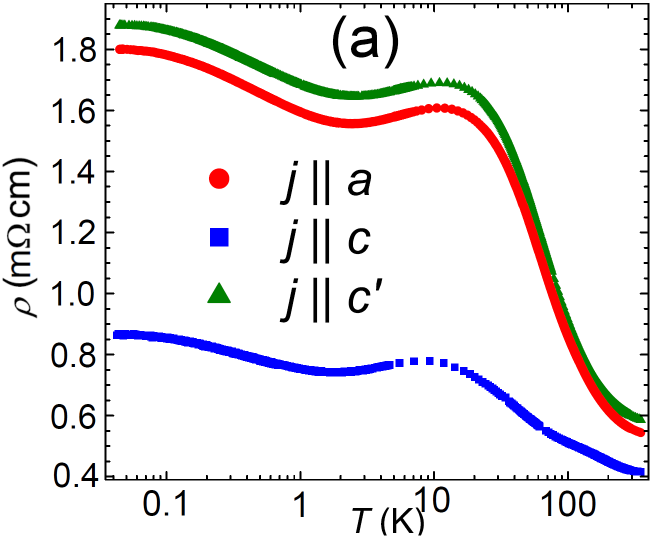}
  \includegraphics[height=40mm]{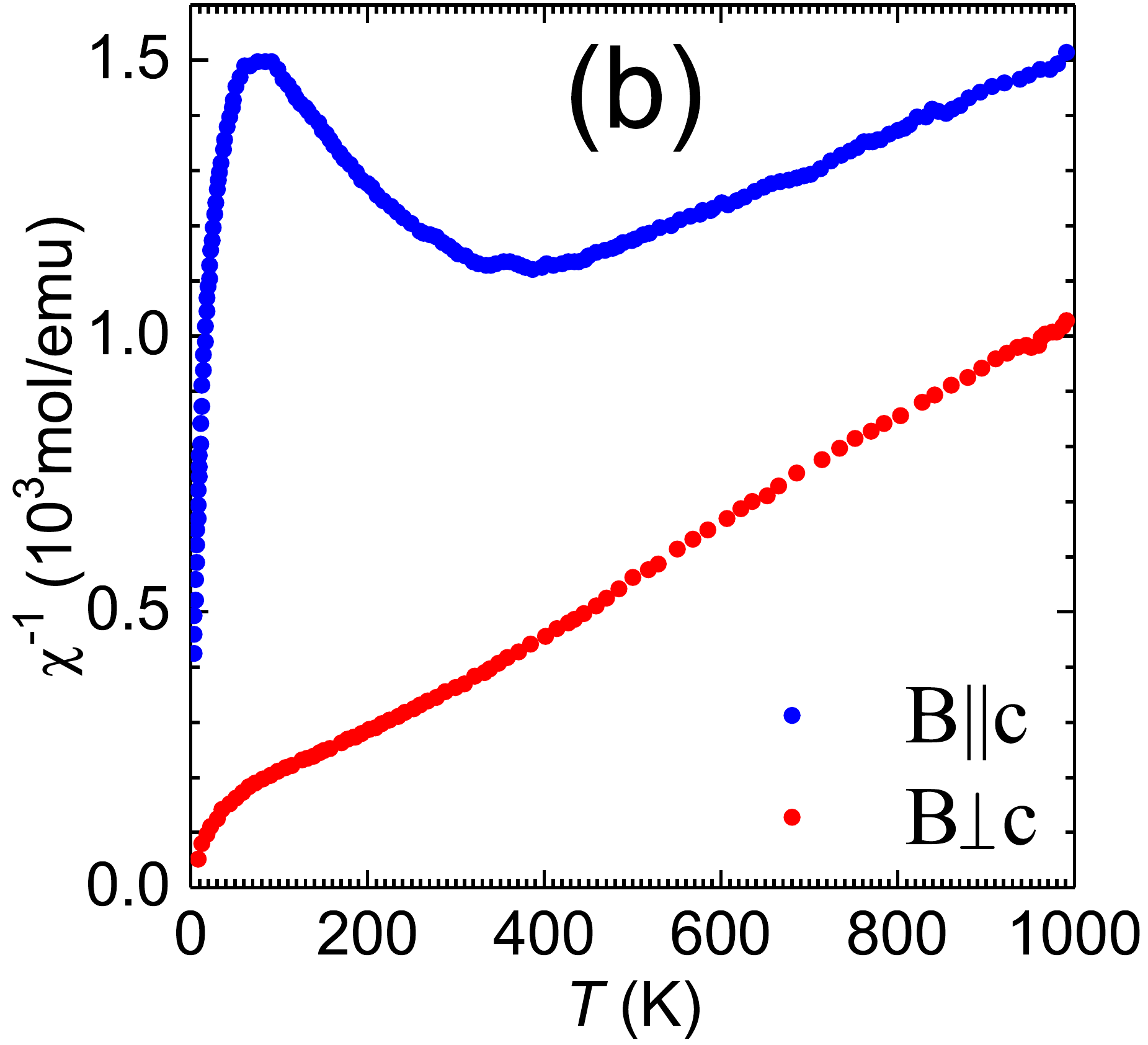}
  \includegraphics[height=40mm]{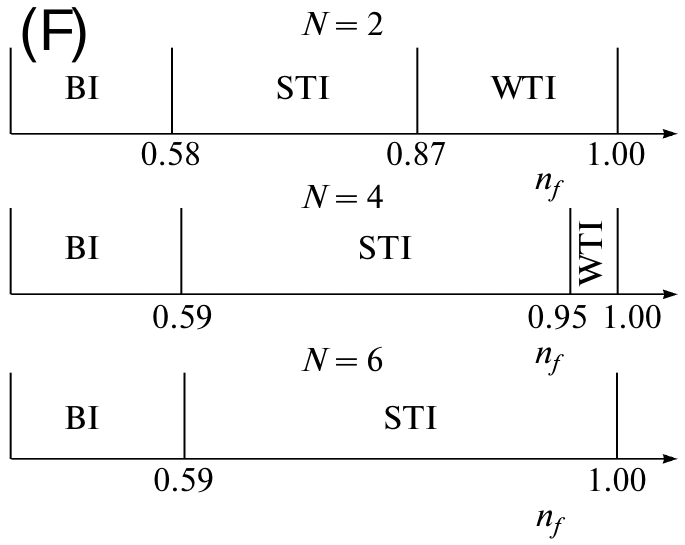}
  \caption{(a) Temperature dependent resistivity of CeRu$_4$Sn$_6$ single crystals along $c$ and in perpendicular directions $a$ and $c$' adapted from Ref.\,\cite{Winkler_2012}. (b) Temperature dependent inverse susceptibility of CeRu$_4$Sn$_6$ single crystals along $c$ and perpendicular to $c$ from Ref.\,\cite{Amorese2018}. (c) Topological phase diagram for different degeneracy $N$ of the ground state multiplet from Ref.\,\cite{Dzero2013}. Depending on the $f$ occupation $N$, mean field calculations predict a weak topological insulator (WTI), a strong topological insulator (STI), or a bulk insulator (BI).}
  \label{fig:CeRu4Sn6transport}
\end{figure}

\begin{SCfigure}
  \centering
  \includegraphics[width=0.6\columnwidth]{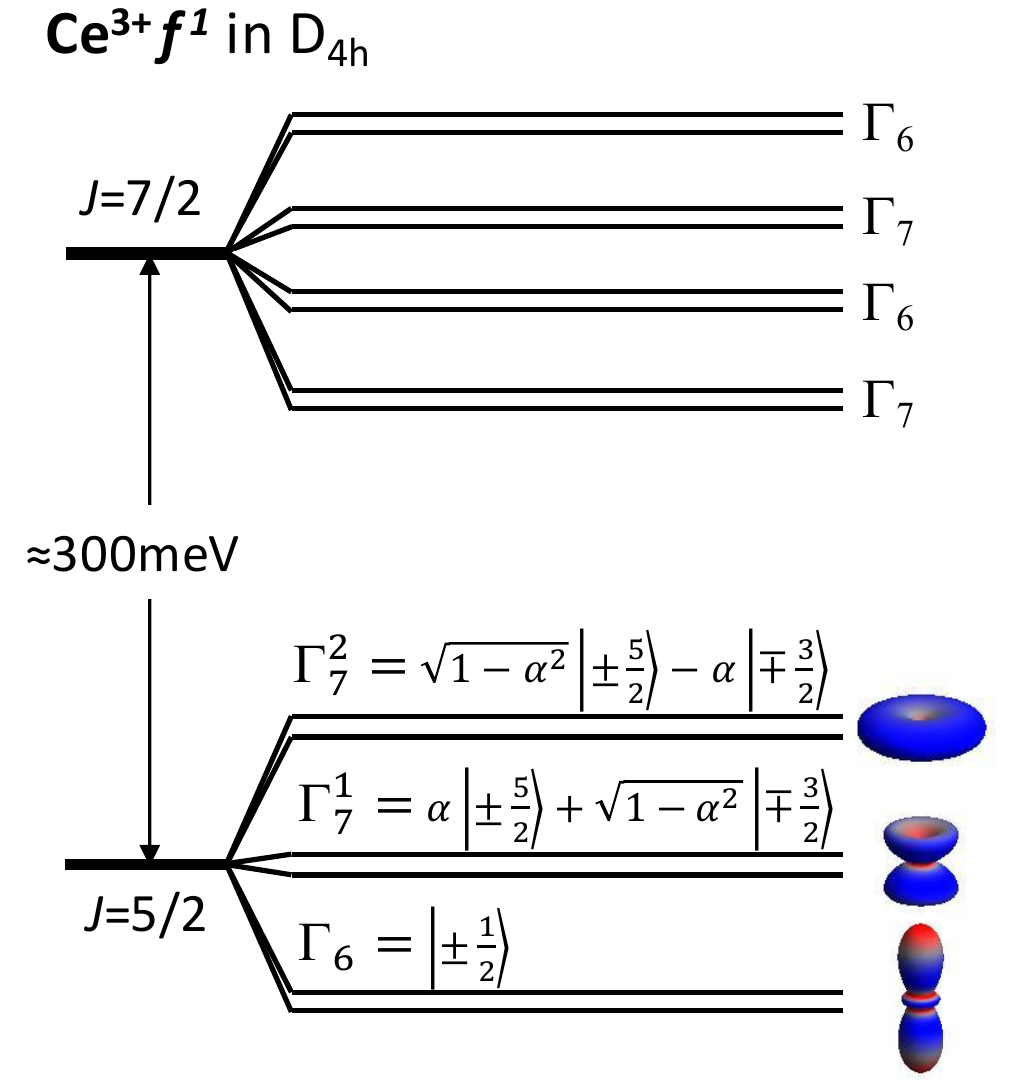}
  \caption{Scheme of the Ce\,4$f$ energy levels for Ce$^{3+}$ in D\textsub{2d} point symmetry  (see \tab{tab:irD2d}). For the low laying \qnJ{}\,=\,5/2 multiplet the \ket{\qnJz{}} representation and the electron density plots are shown as well. $\alpha$ in the \ket{\qnJz{}} representation is set to zero for the plots, resulting in pure |\qnJz{}| states.}
  \label{fig:CeRu4Sn6scheme}
\end{SCfigure}

Here shall be investigated to what extend the conditions for non-trivial topology are fulfilled so that the hybridization induced gap may give rise to non-trivial topological properties as in SmB$_6$.
CeRu$_4$Sn$_6$ has a hybridization induced gap, bands of opposite parity (4$f$ and 5$d$ bands), and also the strong spin-orbit coupling of rare earths.
Dzero and Galitzky investigated tetragonal Kondo insulators and predict the likelihood of finding a weak or strong topological insulator (STI or WTI) or a trivial insulator (BI) as function of the 4$f$ occupancy for different ground state degeneracies (see \fig[c]{fig:CeRu4Sn6transport})\,\cite{Dzero2013}.
CeRu$_4$Sn$_6$ is tetragonal, hence the \qnJ{}\,=\,5/2 multiplet is split into three Kramers doublets, i.e.\ the ground state degeneracy is $N$\,=\,2.
According to \fig{fig:CeRu4Sn6transport}, CeRu$_4$Sn$_6$ could qualify as a STI if the 4$f$ occupation is less than 87\%.
In Ref.\,\cite{Dzero2013} it is further shown that a \CF{} ground state with \qnJz{}\,=\,1/2 symmetry is most favorable for the opening of a hybridization gap.
\Fig{fig:CeRu4Sn6scheme} shows the total energy level diagram.

The effective model case of Dzero and Galitzky has been further investigated specifically for CeRu$_4$Sn$_6$ by D.~Kasinathan, who performed the band structure calculations with the full-potential non-orthogonal local orbital code (FPLO)\,\cite{Koepernik_1999}\,\ref{CeRu4Sn6_2015}.
These local density approximation (LDA)+\SOC{} calculations also include a constant potential to mimic the Ce $4f^1$ valence state.
These calculations indeed showed that a warped gap of 20\,meV or larger opens for all $\vec{k}$-points in case of the nodeless \qnJz{}\,=\,1/2, but in case of the \qnJz{}\,=\,3/2 the system is metallic with several band crossings.
Without going into too much detail, this demonstrates that the microscopic real space picture provides important information also for the band structure picture.

One of the pressing open questions is whether CeRu$_4$Sn$_6$ fulfills the conditions mentioned in the theoretical studies about strongly correlated topological insulators\,\cite{Dzero2013}, i.e.\ what is the $f$ occupation and which \CF{} state forms the \CF{} ground state.

Core-level electron spectroscopy is very powerful in determining the $f$ occupation of strongly hybridized cerium compounds, that have a mixed valence ground state \(|0\rangle = c_{0} |f^{0}\rangle + c_{1}|f^{1}\underline{L}\rangle + c_{2}|f^{2}\underline{\underline{L}}\rangle\) (see \equ{eq:mixnf}).
The core hole acts differently on the different $f^n$ configurations, thus giving rise to different spectral weights.
\setcitestyle{numbers,open={},close={}}Here the absolute valence of Ce in CeRu$_4$Sn$_6$ is determined by hard x-ray photoelectron spectroscopy \HAXPES{} by analyzing the data within a full multiplet model combined with a configuration interaction model (\CIM{}) in the presence of the core hole (see \chap{subsec:CIM} and Ref.\,[\cite{Gunnarsson1983}, \hyperref[CeM2Al10_2015]{VII}]).\setcitestyle{numbers,open={[},close={]}}

The temperature dependence of the valence was determined using partial fluorescence yield x-ray absorption spectroscopy (\PFY{}-\XAS{}) at the \edge[Ce]{L}{3} edge and analyzed with empirical line shapes for the respective weights of the $f^{n+1}$ final state configurations.
The temperature dependence of the valence was then used to determine \TK{} as the Kondo scale in Ref.\,\cite[Fig.\,14]{Bickers1987}.

The dipole selection rules of linear polarized x-ray absorption spectroscopy (\LD{}-\XAS{}) give access to the ground state symmetry in tetragonal point symmetry, as does the \vecq{}-directional dependence of \NIXS{} (see \chap{sec:resvsnres}).
\LD{}-\XAS{} at the \edge[Ce]{M}{4,5} edges was performed by F.~Strigari, T.~Willers, and E.~Weschke and analyzed by F.~Strigari.
In the \LD{}-\XAS{} experiment the ground state appeared to be the \qnJz{}\,=\,1/2 with the first excited \CF{} state higher than 30\,meV, but with significant discrepancy at low \T{}\,\ref{CeRu4Sn6_2015}.
Therefore, a bulk sensitive \NIXS{} experiment at the \edge[Ce]{N}{4,5} edges was performed to show whether the missing dichroism at low \T{} is due to surface effects or arises from the bulk.

\subsection{Experimental}
\paragraph{SAMPLES} 
All experiments were performed on single crystals.
Aligned single crystals were provided by the Quantum Materials group of the Technical University of Vienna.
The single crystals were grown with the self-flux floating-zone melting method using optical heating in a four-mirror furnace\,\cite{Prokofiev_2012}.
The quality of the crystals was checked with Laue and x-ray diffraction as well as scanning electron microscopy and energy dispersive x-ray spectroscopy.
All methods show that the samples are single grain and single phase materials.
The lattice parameters are in agreement with $a$\,=\,6.8810\,\AA\ and $c$\,=\,9.7520\,\AA\ as given in Ref.\,\cite{Pottgen_1997}.
The single crystals were pre-aligned with a Laue camera.
A peculiarity of the present tetragonal structure is that $\sqrt{2}$$a$\,$\approx$\,$c$ within $\approx$\,0.2\%.
As a consequence the structure appears pseudo-cubic in Laue images and the final alignment of the single crystals had to be done via a magnetization measurement\,\cite{Paschen_2010}.

\paragraph{HAXPES}
Hard x-ray core level photoemission spectra were performed at the Taiwan beamline BL12XU at the SPring-8 synchrotron radiation facility in Japan.
The experiment was performed by S. Agrestini, K.~Chen, Y.~Utsumi, Y.-H.~Wu, and myself with technical support from K.-D. Tsuei.

High fixed incident energy of 6.47\,keV was used to ensure high bulk sensitivity.
The photo electrons were detected by a MB Scientific A-1HE analyzer with a pass energy of 200\,eV, yielding an instrumental \FWHMG{} of about 1\,eV, in vertical geometry for the Ce\,3$d$ and in horizontal geometry for the Ce\,3$p$ and Sn\,3$p$ core levels\,\cite{Weinen2016}.
The \FWHMG{} and the Fermi energy of 6470.1\,eV was determined by a Au thin film measured at the same conditions.
The samples were cooled down and cleaved in ultra high vacuum of $\approx$10$^{-8}$\,mbar and were then immediately transferred to the measurement position (pressure $<$3$\cdot$10$^{-10}$\,mbar and \T{}$<$90\,K).
In this way, a clean surface is ensured.

For the calculation of the \CIM{} and the multiplet structure the xtls9.0 program from A.~Tanaka\,\cite{Tanaka1994} has been used.

\paragraph{PFY-XAS}

Partial fluorescence yield x-ray absorption spectra were taken at the GALAXIES beamline at SOLEIL synchrotron radiation facility in France.
The experiment was performed by F.~Strigari and myself with technical support from J.-P.~Rueff and J.~M.~Ablett under supervision of A.~Severing.

The sample was cooled inside a He cryostat with polyimide windows.
The incident energy was scanned using a Si(111) double crystal monochromator and focused to the sample with a spot size of 30$\mu$m (vertical) and 90$\mu$m (horizontal) FWHM.
The incoming intensity has been normalized to scattering of some polyimide foil just before the sample.
The fluorescence was measured in backscattering geometry and the energy was set to $\approx$4839\,eV by a 1m-radius spherically bent Ge(331) analyzer at a Bragg angle of 80.8$^\circ$ and focused onto a silicon avalanche photo diode in the Rowland geometry.
Most of the beam path was filled with He to reduce air scattering.

\paragraph{NIXS}

The \NIXS{} measurements were performed at the beamline ID20 at ESRF synchrotron radiation facility in France.
The experiment was performed by F. Strigari, and myself, with technical support from M.~Moretti Sala and A.~Al-Zein under supervision of A.~Severing.

The incident energy was scanned using a Si(111) double crystal monochromator.
Two crystals with polished \xyz{100} and \xyz{001} surfaces were mounted in a He-flow cryostat with Al windows and measured in specular geometry with \qp{001} and \qp{100}, i.e.\ parallel and perpendicular to the $c$ axis.
The scattered x-rays were analyzed by three columns of three Si(660) analyzers (in total nine) at scattering angles of $2\,\Theta\,=\,140^\circ, 146^\circ$, and $153^\circ$, corresponding to an average momentum transfer of \absq{}\,=\,(9.4$\pm$0.2)\,\AA$^{-1}$.
Here the scattering plane was horizontal, i.e.\ in the plane of the polarization of the x-rays, to reach the large momentum transfer.
This configuration yields an instrumental energy resolution of \FWHMG{}\,=\,1.3\,eV.

\subsection{Hard x-ray photo electron spectroscopy results}\label{chap:CeRu4Sn6_HAXPES}

\begin{figure}
  \includegraphics[width=\textwidth]{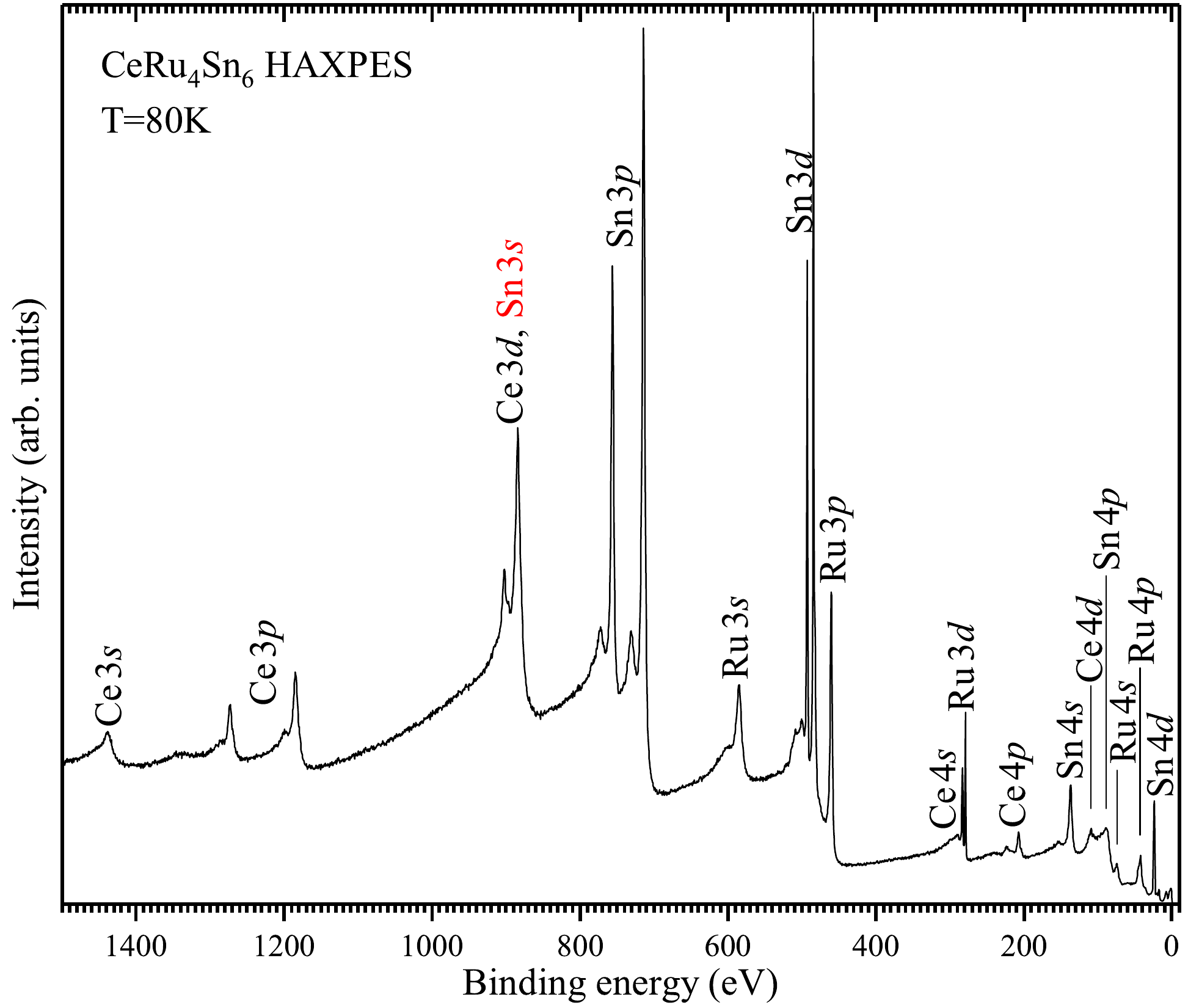}
  \caption{HAXPES spectrum of CeRu$_4$Sn$_6$ from the Fermi energy to 1.5\,keV binding energy. The observed core-level emission lines have been labeled.}
  \label{fig:CeRu4Sn6_HAXPES_wide}
\end{figure}

\Fig{fig:CeRu4Sn6_HAXPES_wide} shows the \HAXPES{} results of CeRu$_4$Sn$_6$ from the Fermi energy down to 1.5\,keV binding energy.
Zero binding energy corresponds to the Au film Fermi energy (see experimental).
All core levels have been identified and labeled.
No emission lines except the ones of Ce, Ru and Sn were observed, i.e.\ the sample is pure.
Every emission line has a satellite peak at around 16\,eV higher binding energies.
This is due to plasmon excitations upon creating the core hole or the free electron.

In the following, the emission of the Ce\,3$d$ and 3$p$ core level will be investigated to obtain information about the Ce valence in CeRu$_4$Sn$_6$.

\paragraph{Ce core level emission}

\begin{table}[t]
  \caption{Parameters of the \CIM{} (\equ{eq:Hcim} in \chap{subsec:CIM}) and the valence.}
  \label{tab:CeRu4Sn6_HAXPES_results}
  \begin{tabular*}{\columnwidth}{@{\extracolsep{\fill}}cccc||ccc}
    \hline
    $U_{ff}$ & $U_{fc}$ & $\Delta_f$ & $V_\text{eff}$ & $c_0^2$ & $c_1^2$+$c_2^2$ & valence \\
    9.1(9)\,eV & 10.6(9)\,eV & 2.5(3)\,eV & 0.26(2)\,eV & 0.08(2) & 0.92(2) & $\approx$3.04 \\
    \hline
  \end{tabular*}
\end{table}
\begin{table}[t]
  \caption{Parameters of the multiplet structure, the integrated background, and the lineshape.}
  \label{tab:CeRu4Sn6_HAXPES_lineshape}
  \begin{tabular*}{\columnwidth}{@{\extracolsep{\fill}}l||cccc||c||cccc}
    \hline
     & \redff{} & \redfc{} & \redZv{} & \redZc{} & $A$ & \FWHMG{} & \FWHML{} & $\xi$ & $\alpha$ \\
    Ce\,3$p$ & 0.6 & 0.8 & 1.0 & 0.96 & 0.0028 & 1.0\,eV & 3.8\,eV & 8\,eV & 0.4 \\
	Ce\,3$d$ & 0.6 & 0.8 & 1.0 & 0.98 & 0.009 & 1.0\,eV & 1.6\,eV & 8\,eV & 0.3 \\
    \hline
  \end{tabular*}
\end{table}
\begin{figure}
  \includegraphics[width=\textwidth]{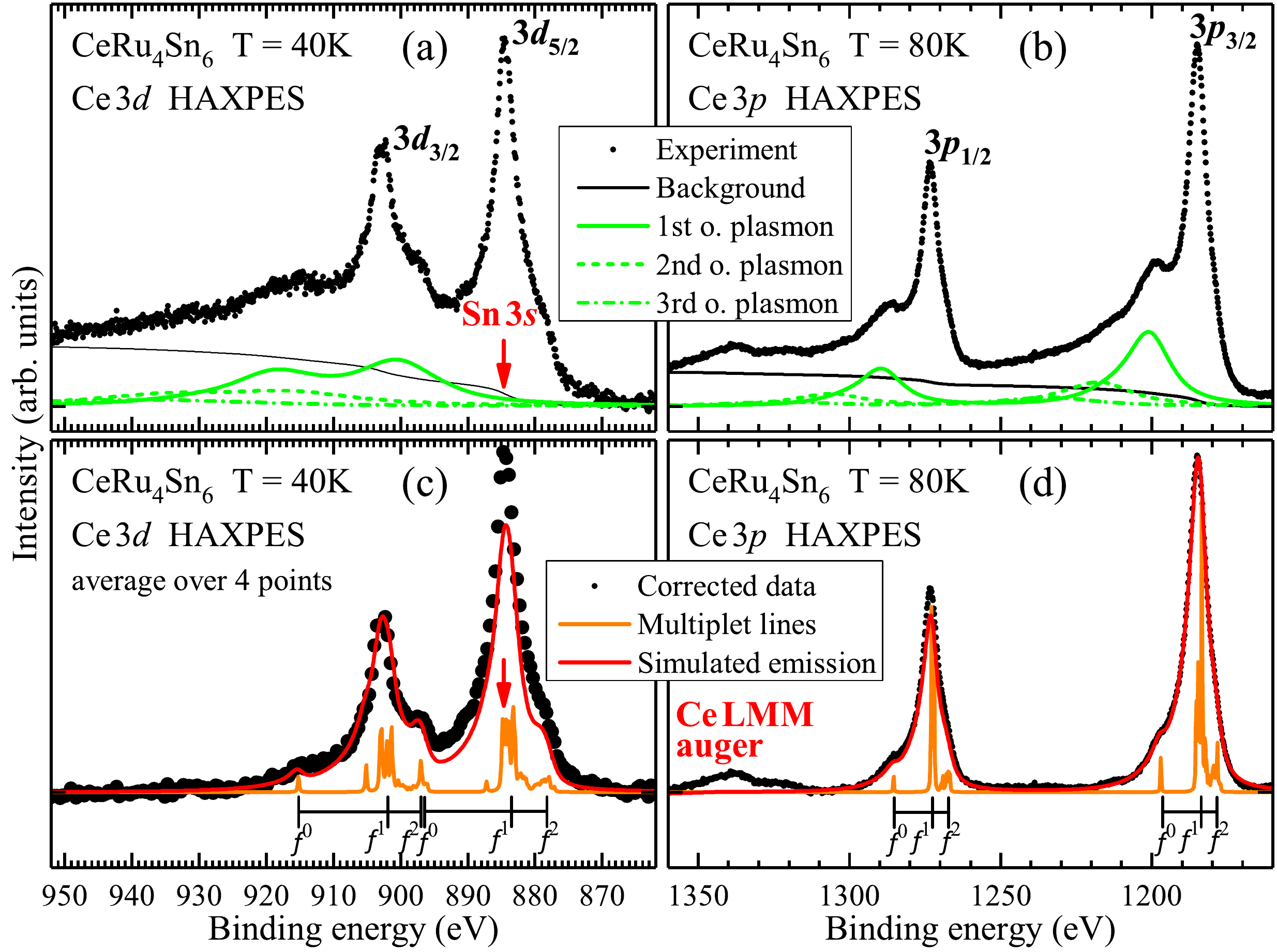}
  \caption{CeRu$_4$Sn$_6$ emission data (dots) of (a) the Ce\,3$d$ and (b) the Ce\,3$p$ core electrons. In addition the integrated background (black line) and calculated plasmon contribution (green lines) are shown. In the panels below, the corrected data (dots) after subtraction of the integrated background and the plasmon contributions of (c) the Ce\,3$d$ and (d) the Ce\,3$p$ are shown. Here, the size of the circles reflects the statistical error bar. The multiplet structure of the 3 valence contributions (bottom ruler) is also shown (orange line). The broadening of the multiplet structure yields the simulation of the core level emission (red line).}
  \label{fig:CeRu4Sn6_HAXPES_Ce3pd}
\end{figure}

\begin{figure}
\centering
  \includegraphics[width=0.8\textwidth]{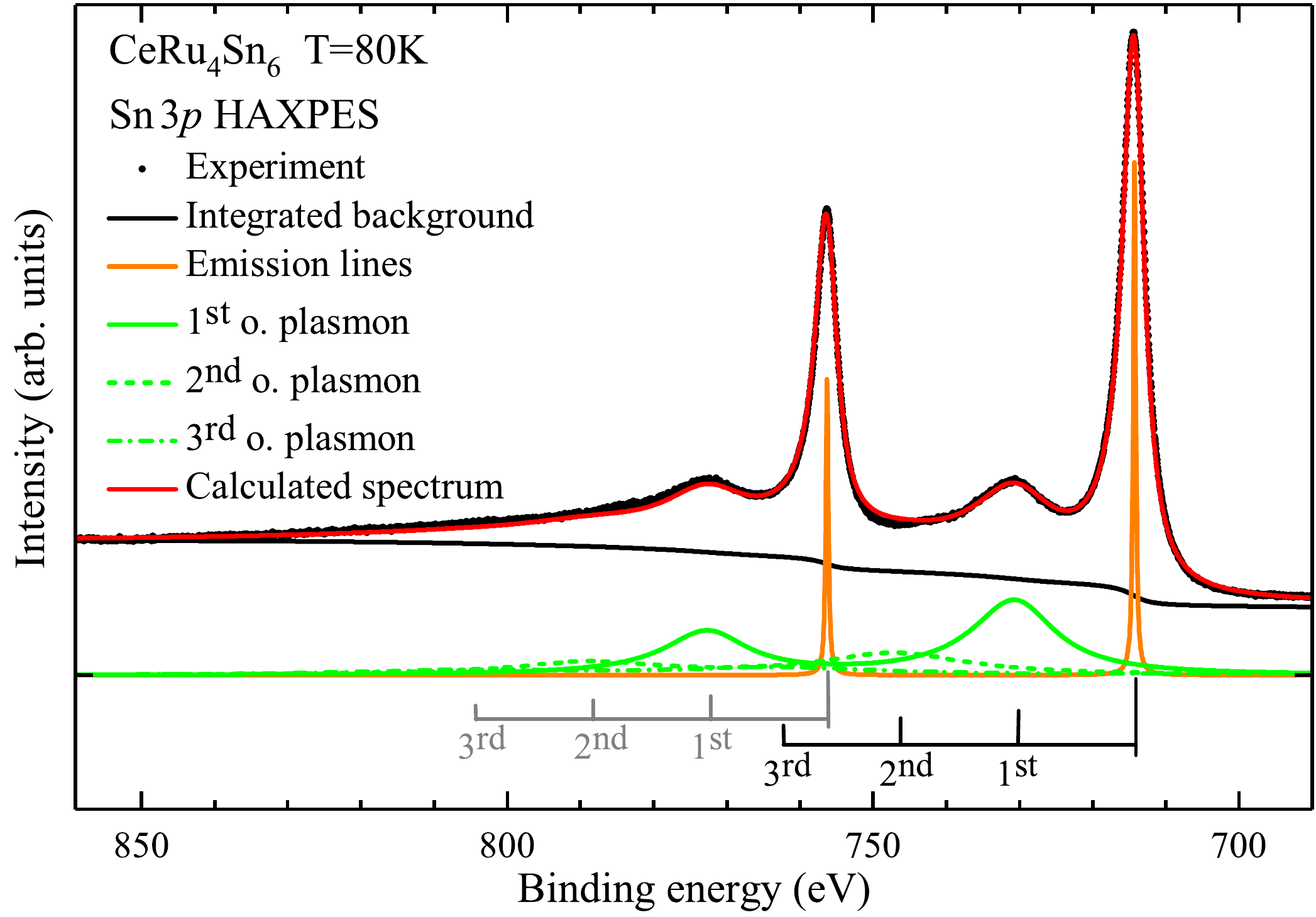}
  \caption{HAXPES data of the CeRu$_4$Sn$_6$ Sn\,3$p$ core level emission. In addition the different contributions to the spectrum are shown as lines: Sn\,3$p$ emission lines (orange), plasmon contribution (blue), integrated background (black), and their sum reflecting the total calculated spectrum (red).}
  \label{fig:CeRu4Sn6_HAXPES_Sn3p}
\end{figure}

\Fig[(a+b)]{fig:CeRu4Sn6_HAXPES_Ce3pd} shows the Ce\,3$d$ and Ce\,3$p$ core level spectra of CeRu$_4$Sn$_6$.
For both core levels two strong emission lines show up due to the strong spin-orbit splitting of the core levels.
These strong features are dominated by the $f^1$ contribution.
At binding energies about 16\,eV larger than the $f^1$ emission two smaller peaks are present.
These are the plasmons of the $f^1$ emission lines.
The lineshape of the plasmons has been identified using the Sn\,3$p$ emission line (see \fig{fig:CeRu4Sn6_HAXPES_Sn3p}) so that each emission line in the Ce core level spectra is described with a main emission line plus the 1\textsup{st} to 3\textsup{rd} order plasmons, as described in Ref.\,\ref{CeM2Al10_2015}.
\Fig[(a+b)]{fig:CeRu4Sn6_HAXPES_Ce3pd} shows, in addition to the core level data, the respective plasmon contributions and the integrated background\footnote{Integrated background: \( I_\text{IB}(E)=A \int\limits_{E_0}^{E} I(E')-I_\text{IB}(E') ~ \diff E' + I_\text{IB}(E_0) \)}.

\Fig[(c+d)]{fig:CeRu4Sn6_HAXPES_Ce3pd} shows the Ce\,3$d$ and Ce\,3$p$ core level data after subtraction of the integrated background and the plasmon contributions.
This corrected data are then expected to be solely determined by the multiplet structure and their broadening.
Both core levels show additional satellites due to the different $f$-occupancies as indicated by the bottom ruler.
In addition, the Sn\,3$s$ emission (red arrow) coincides in energy with the Ce\,3$d_{5/2}$ $f^1$ and $f^2$ emission.
This introduces some uncertainty in analyzing the spectral weights and hinders any further analysis of the Ce\,3$d$ spectra in this compound.

\Fig[d]{fig:CeRu4Sn6_HAXPES_Ce3pd} shows the background and plasmon corrected data of the Ce\,3$p$ core level.
The spin-orbit coupling of the Ce\,3$p$ electrons is stronger than for Ce\,3$d$, such that the multiplet structures (bottom ruler) of the Ce\,3$p_{1/2}$ and Ce\,3$p_{3/2}$ do not overlap.
Note that the strong intensity at the $f^0$ position in the original data is mainly due to plasmon contributions of the main $f^1$ emission line.
There are no other emission lines in this energy window.
Some extra intensity at the higher binding energy side of the Ce\,3$p_{1/2}$ is due to an Auger process of Ce, but does not overlap with the emission lines.
The distance between the different $f^n$ contributions is the same for the Ce\,3$d$ and 3$p$ core levels.
The multiplet structure of the single contributions is a bit more narrow for the Ce\,3$p$, but this advantage is canceled out by the larger lifetime broadening, which hampers the precise assignment of the $f^1$ and $f^2$ contributions.
The four \CIM{} parameters of this multiplet structure are given in \tab{tab:CeRu4Sn6_HAXPES_results}, along with the resulting valence.
For the simulation of the Ce\,3$p$ emission (red line) the multiplet line has been broadened by a Gaussian, a Lorentzian, and a Mahan\footnote{Mahan lineshape: \( I_\text{M}(E) = \frac{\Theta(E)}{\Gamma(\alpha) E} \left(\frac{E}{\xi}\right)^\alpha e^{-E/\xi} \)} lineshape.
The parameters are given in \tab{tab:CeRu4Sn6_HAXPES_lineshape}.

The \CIM{} parameters have been also applied to calculate the Ce\,3$d$ emission spectrum.
Only the lineshape parameters needed to be adjusted (see \tab{tab:CeRu4Sn6_HAXPES_lineshape}) to also reproduce the Ce\,3$d$ emission in a satisfactory manner.
Some intensity at the calculated Ce\,3$d_{5/2}$, however, is missing due to the underlying Sn\,3$s$ contribution.

The resulting initial singlet state contribution $c_0^2$ of the the \CIM{} is found to be 8$\%$.

\paragraph{Plasmon contribution}

\begin{table}
  \centering
  \caption{Properties of the Sn\,3$p$ lineshape, the integrated background, and the plasmons of the order $n\leq4$.}
  \label{tab:CeRu4Sn6_HAXPES_plasmon}
  \begin{tabular*}{\textwidth}{@{\extracolsep{\fill}}cccc||c||ccc}
    \hline
    \FWHMG{} & \FWHML{} & $\xi$ & $\alpha$ & $A$ & Amplitude & $\Delta$$E_\text{B}$ & $\Delta$\FWHML{} \\
    1\,eV & 3.2\,eV & 8\,eV & 0.1 & 0.007 & 0.5$^n$ & 16.1$\cdot$$n$\,eV & 10$\cdot$$n$\,eV \\
    \hline
  \end{tabular*}
\end{table}

The CeRu$_4$Sn$_6$ Sn\,3$p$ core level emission has been fitted to assign the plasmon contribution shown in \fig{fig:CeRu4Sn6_HAXPES_Ce3pd}.

\Fig{fig:CeRu4Sn6_HAXPES_Sn3p} shows the Sn\,3$p$ emission lines (dots) at $\approx$715\,eV and $\approx$755\,eV binding energy.
The black line shows the underlying integrated background.
At $\approx$16\,eV higher binding energies a broader feature occurs, that is also present for all other emission lines (see \fig{fig:CeRu4Sn6_HAXPES_wide}).
The intensity, which is neither described by the broadened emission lines nor the integrated background, is assigned to the plasmons.
The plasmon contributions up to 3$^{rd}$ order are also shown (green lines in \fig{fig:CeRu4Sn6_HAXPES_Sn3p}) and their properties are listed in \tab{tab:CeRu4Sn6_HAXPES_plasmon}.
The resulting overall lineshape (red line), which is given by the broadened emission lines, the integrated background, and the plasmons, reproduces the Sn\,3$p$ emission spectrum well.

\subsection{X-ray absorption spectroscopy results}\label{chap:CeRu4Sn6_PFYXAS}

\begin{figure}
  \centering
  \includegraphics[width=0.8\textwidth]{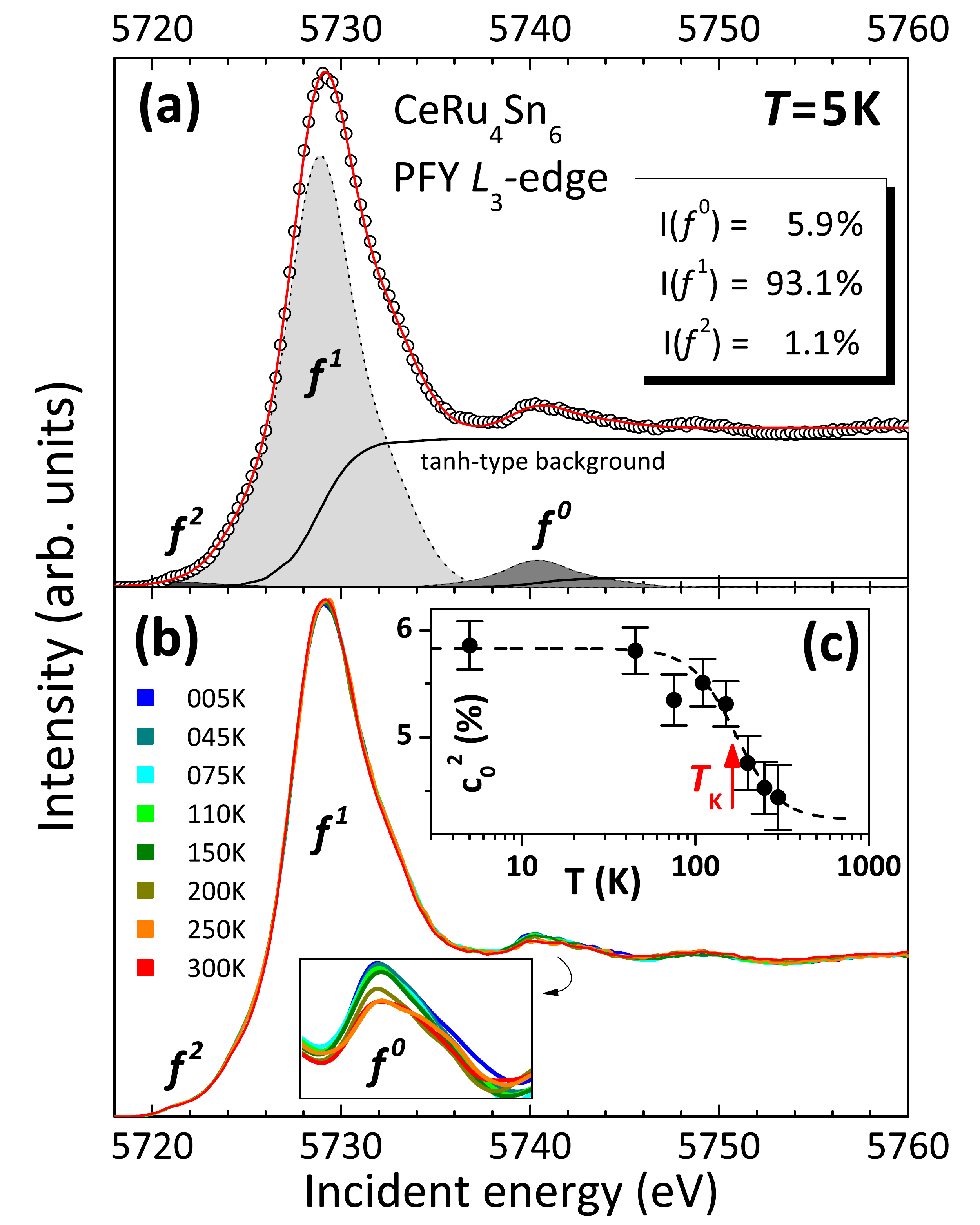}
  \caption{(a) \PFY{}-\XAS{} spectra of the \edge[Ce]{L}{3} edge at \T{}\,=\,5\,K (open circles). The size of the circles indicates the statistical error bars. The red line is a phenomenological fit, the regions of different gray refer to the $f^2$, $f^1$ and $f^0$ spectral weights, taking into account a tanh-type background (black solid lines). (b) Temperature evolution of the \edge[Ce]{L}{3} edge spectra and the $f^0$ peak region on an enlarged scale. (c) $f^0$ intensity ($\approx$c$_2^0$) as function of log(\T{}). The relative error bars amount to $\pm$0.25\%. The black, dashed line is a guide to the eye, the red arrow indicates its inflection point which is taken as Kondo temperature \TK{}.}
  \label{fig:CeRu4Sn6_PFYXAS}
\end{figure}

\Fig[a]{fig:CeRu4Sn6_PFYXAS} shows the partial fluorescence yield (\PFY{})-\XAS{} data of CeRu$_4$Sn$_6$ at the \edge[Ce]{L}{3} edge at 5\,K (open circles).
The main absorption is due to the $f^1$ contribution in the ground state, the small but well resolved hump at about 5740\,eV is due to the $f^0$ contribution, and the small shoulder on the low energy tail of the $f^1$ peak (just above 5720\,eV) is due to some $f^2$ spectral weight.
It is not straightforward to describe the \edge{L}{3} edge spectral line shape since it is determined by the unoccupied $5d$ \DOS{} in the presence of a $2p$ core hole.
Therefor the three $f^n$ spectral weights were fitted empirically\,\cite{RueffPRL106, YamaokaPRL107, Kotani_2014} by describing the spectral weight ($I$($f^0$), $I$($f^1$), and $I$($f^2$)) of each configuration with an identical line shape, consisting of three Gaussian profiles and a tanh-type step function with the maximum slope at the peak maximum to fit the background.
Only the total intensities and central energy positions were fitted.
The size of the step function was adjusted so that the modulations due to the extended absorption fine structure (EXAFS) above $\approx$5748\,eV are averaged out.
The red line in \fig[a]{fig:CeRu4Sn6_PFYXAS} is the result of this fit.

The following spectral weights were obtained:
$I(f^0$)\,=\,0.059$\pm$0.015, $I(f^1$)\,=\,0.931$\pm$ \!\!0.015 and $I(f^2$)\,=\,0.011$\pm$0.015 which corresponds to $n_f$\,$\simeq$\,0.953 and $c_0^2$\,$\simeq$\,0.059$\pm$0.015 under the assumption $I(f^n$)\,$\simeq$\,$c_n^2$.
This is an approximation since it neglects hybridization effects in the final states: in core level photo electron spectroscopy (see \chap{chap:CeRu4Sn6_HAXPES}), the spectral weights $I$($f^n$) do not relate 1:1 to the initial state $f^n$ contributions $c_n^2$.
The temperature dependence of the valence, even if the absolute values show some offset, is obtained by performing the fit for all temperatures.

\Fig[b]{fig:CeRu4Sn6_PFYXAS} shows the temperature dependence of the \PFY{}-\XAS{} data, normalized to the integrated intensity.
The bottom inset in \fig[b]{fig:CeRu4Sn6_PFYXAS} shows the $f^0$ region on an enlarged scale after smoothing the data.
The main changes occur within the temperature interval of 150 to 250\,K.

\Fig[c]{fig:CeRu4Sn6_PFYXAS} shows the result for $I(f^0)$ when applying the above fit by varying only the total intensity of the $f^1$ and $f^0$ contributions and adjusting the height of the background accordingly.
The dashed line is a broadened step function describing the temperature evolution.
From this \TK{} is determined empirically as the inflection point of $I(f^0)$\,\cite{Bickers1987, Tjeng1993, Dallera2002}, yielding \TK{}\,=\,$(170 \pm 25)$\,K.

\subsection{NIXS results}

\begin{SCfigure}
  \centering
  \includegraphics[width=0.6\textwidth]{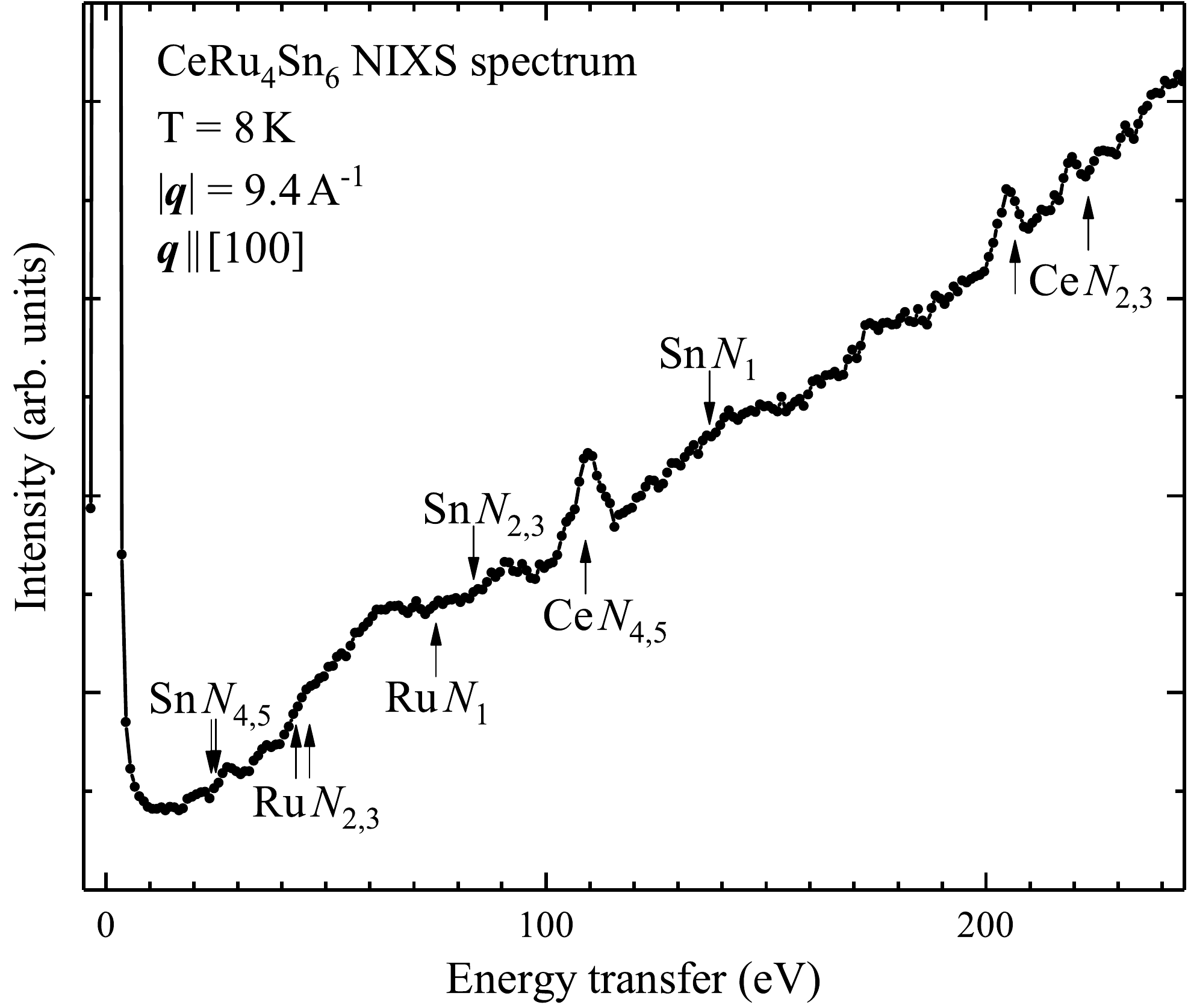}
  \caption{\NIXS{} spectrum of CeRu$_4$Sn$_6$ at \T{}\,=\,8\,K with \absq{}\,=\,9.4\,$\AA^{-1}$ and \qp{100}.}
  \label{fig:CeRu4Sn6_wide}
\end{SCfigure}

\Fig{fig:CeRu4Sn6_wide} shows the \NIXS{} spectrum of CeRu$_4$Sn$_6$ from the elastic up to 250\,eV energy transfer.
The signal to background ratio of the rare-earth \edge{N}{} edges is smaller compared to the hexaborides due to the reduced amount of Ce ions, with respect to the total number of valence electrons in this compound.
The proportion of the different Ce edges yield is still the same, i.e.\ the \edge{N}{4,5} edges yield a better signal to background ratio than the \edge{N}{2,3} edges.

\paragraph{Multiplet structure}

\begin{figure}
  \centering
  \includegraphics[width=0.7\textwidth]{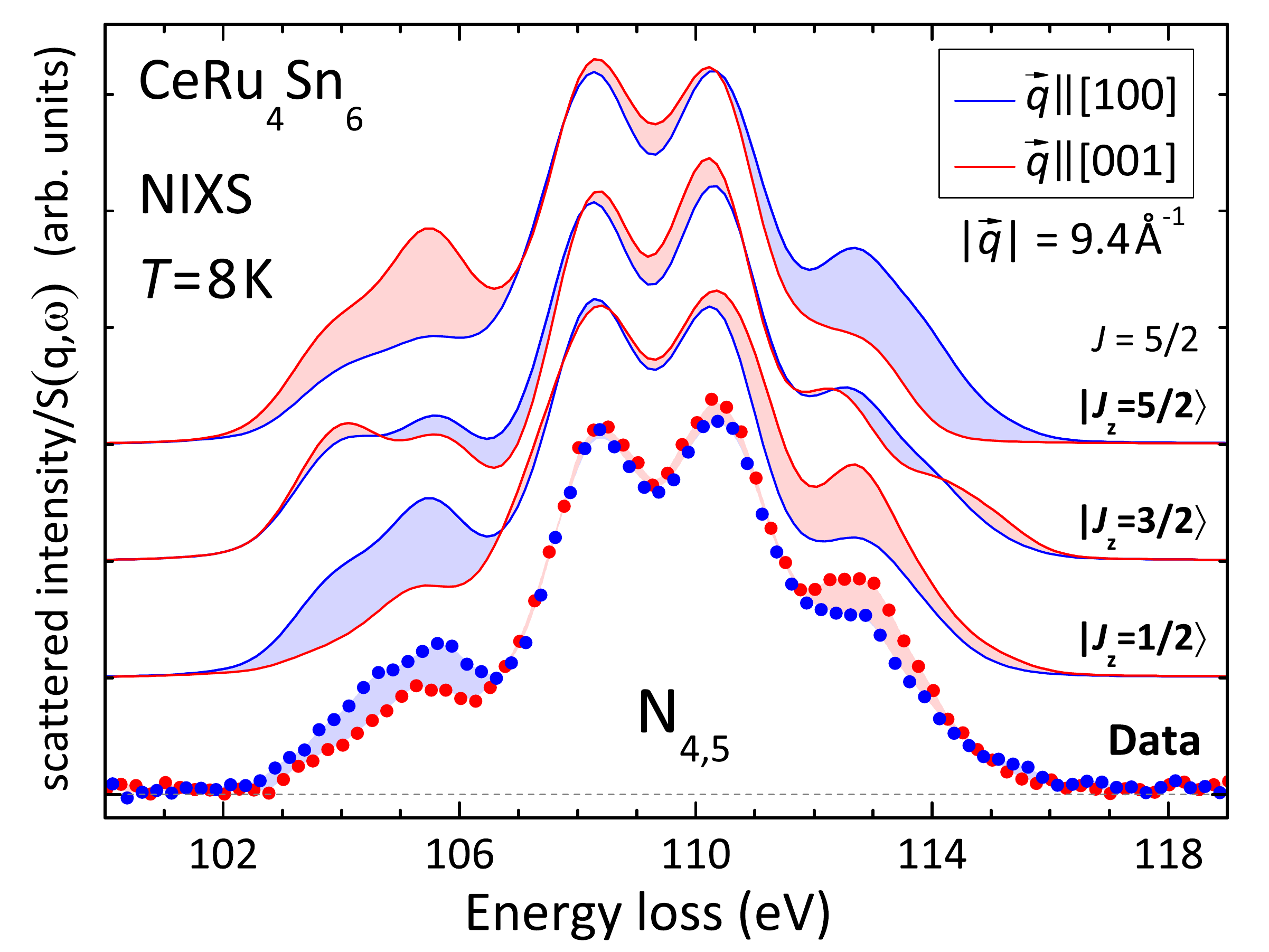}
  \caption{NIXS spectra at the \edge[Ce]{N}{4,5} edges after subtracting a linear background (dots) for \qp{100} (blue) and \qp{001} (red). The point size reflects the statistical error, the counting time was 240\,s/point. The respective calculated spectra for the pure \qnJz{} states (see \fig{fig:CeRu4Sn6scheme}) of the \qnJ{}\,=\,5/2 Hund's rule ground state are shown above.}
  \label{fig:CeRu4Sn6Fig04}
\end{figure}

The dots in \fig{fig:CeRu4Sn6Fig04} show the measured \NIXS{} spectra at the CeRu$_4$Sn$_6$ \edge[Ce]{N}{4,5} edges for the two measured directions \qp{100} and \qp{001} after subtraction of a linear background.
A pseudo-isotropic spectrum has not been used.
For this tetragonal compound the isotropic spectrum could only be realized for the dipole contribution, whereas the excitonic features mainly arise from the higher multipoles.
The reduction factors can still be adjusted as the peak positions and line width of the edge are independent of the \vecq{} direction.

\begin{table}
  \centering
  \caption{Parameters used for calculations. The values in the brackets are the variation step size of the parameters.}
  \label{tab:CeRu4Sn6parameters}
  \begin{tabular*}{\columnwidth}{@{\extracolsep{\fill}}cccc||ccc}
    \hline
	\redff{} & \redfc{} & \redZc{} & \redZv{} & \absq{} & \FWHML{} & \FWHMG{} \\
	 0.7\,(1) & 0.8\,(1) & 1.00\,(5) & 1.00\,(5) & 9.2\,(2)\AA$^{-1}$ & 0.2\,(1)\,eV & 1.3\,eV \\
	\hline
  \end{tabular*}
\end{table}

\paragraph{Directional dependence of \(\vec{\textit{q}\,}\)\!}

The experimental \NIXS{} data (dots in \fig{fig:CeRu4Sn6Fig04}) show a clear dichroic effect between the two measured direction \qp{100} (blue) and \qp{001} (red).
The strongest dichroism in the data appears at the shoulder structures at about 105 and 113\,eV.
Blue over red at 105\,eV and vice versa at 113\,eV.
In the calculations this is realized by the \qnJz{}\,=\,1/2, opposite for the \qnJz{}\,=\,5/2, and the \qnJz{}\,=\,3/2 state does not show a strong dichroism.
The \qnJz{}\,=\,3/2 and 5/2 both belong to the $\Gamma_7$ \CF{} symmetry in D\textsub{4h} and can mix, whereas the \qnJz{}\,=\,1/2 remains pure as $\Gamma_6$.
Upon mixing \qnJz{}\,=\,3/2 and 5/2 the dichroic signal changes smoothly between the two pure extremes, i.e.\ only a sizable dichroism opposite to the experiment can be realized.
Hence, the anisotropy in the CeRu$_4$Sn$_6$ \NIXS{} data must originate from the \qnJz{}\,=\,1/2 initial state.
The strength of the dichroism, however, is weaker in the experiment as compared to calculation of the \qnJz{}\,=\,1/2 initial state.

\subsection{Summary}

The \HAXPES{} experiment shows that the $f^0$ contribution to the ground state in CeRu$_4$Sn$_6$ amounts to $\approx$\,8\%.

The temperature dependence of the $f^0$ spectral weight in \PFY{}-\XAS{} indicates a large Kondo temperature of $\approx$\,170\,K.

The \NIXS{} experiment confirms the finding of the \XAS{} experiment in Ref.\,\ref{CeRu4Sn6_2015}.
The anisotropy between the in-plane and out-of-plane momentum transfer \vecq{} arises clearly from the \qnJz{}\,=\,1/2.
The experimental dichroism, however, is too small for a pure \qnJz{}\,=\,1/2 state, as it was also in the \XAS{} experiment.
This finding proves that the incomplete dichroism is a true bulk property of CeRu$_4$Sn$_6$.

\paragraph{Discussion}

The large \TK{} is in agreement with the analysis of the temperature dependence of the single crystal resistivity in Ref.\,\cite{Guritanu_2013}.

The \NIXS{} experiment proves that the missing dichroism is a bulk property.
What could be the reason for the deviation between the experiment and the simulation starting from a pure \qnJz{}\,=\,1/2?
The \XAS{} experiment did not show any change of the dichroism up to 300\,K, thus pointing towards a large \CF{} splitting and excludes any sizable Boltzmann occupation, unless two states are already equally occupied at the lowest temperature.

In the meanwhile the \CF{} potential of CeRu$_4$Sn$_6$ has been fully determined with resonant inelastic x-ray scattering (\RIXS{}) by Amorese \etal{}\,\cite{Amorese2018}.
\RIXS{} finds the two $\Gamma_7$ states at 31 and 84\,meV with $\alpha$\,$\approx$\,0 (see \fig{fig:CeRu4Sn6scheme}), i.e.\ \qnJz{} mixing is negligible.
This excludes immediately any scenario of two low lying ($\ll$5\,K) and one high lying (>300\,K) doublets.
The line shape in \RIXS{} is strongly asymmetric, which can be another indication of the strong hybridization.

Already in 1986 van der Laan \etal{} pointed out that in the presence of strong hybridization higher lying states can contribute to the \edge{M}{}-edge signal\,\cite{Laan_1986}.
In CeRu$_4$Sn$_6$ the Kondo temperature of about 170\,K is not much smaller than the \CF{} excitations ($\geq$\,30\,meV\,$\hat{=}$\,350\,K) so that a contribution of higher lying \CF{} states to the ground state is indeed to be expected.
The dichroism of these higher lying \CF{} states is smaller or even of opposite sign.
Thus, the dichroism will be reduced from that of the pure \qnJz{}\,=\,1/2 state.
In that case the reduced dichroism verifies the strong hybridization.
Nevertheless, the ground state of CeRu$_4$Sn$_6$ is predominantly of \qnJz{}\,=\,1/2 symmetry.
In fact LDA calculations\,\cite{Wissgott2016, Xu2017} support this model and show a mixing of higher states into the ground state for CeRu$_4$Sn$_6$, at most of the \qnJz{}\,=\,3/2\,\cite{Wissgott2016} or \qnJz{}\,=\,5/2\cite{Xu2017} states.
The latter also predicts the existence of Weyl-fermion states in CeRu$_4$Sn$_6$.

The present results show, that CeRu$_4$Sn$_6$ in fact fulfills the requirements to form a topological Kondo insulating state:
A strong hybridization plus the node less \qnJz{}\,=\,1/2 ground-state symmetry with some contribution of a higher lying state.
This classifies CeRu$_4$Sn$_6$ as a strongly correlated material with non-trivial topology, since the band structure calculations by Deepa Kansinathan in Ref.\,\ref{CeRu4Sn6_2015} verify that there are no non-bonding bands that otherwise could spoil the formation of a warped gap.

\clearpage
\section{UO\textsub{2}} \label{sec:UO2}
\begin{center}\textbf{
  \scalebox{1.2}{U5\textit{f} crystal-field ground state of UO\textsub{2}}\\
  \scalebox{1.2}{probed by nonresonant inelastic x-ray scattering}}\\
  \hyperref[UO2_2018]{\scalebox{1.2}{Physical Review B \textbf{98} 205108 (2018)}}
\end{center}

\FloatBarrier
\begin{SCfigure}[][h]
  \centering
  \includegraphics[width=0.48\textwidth]{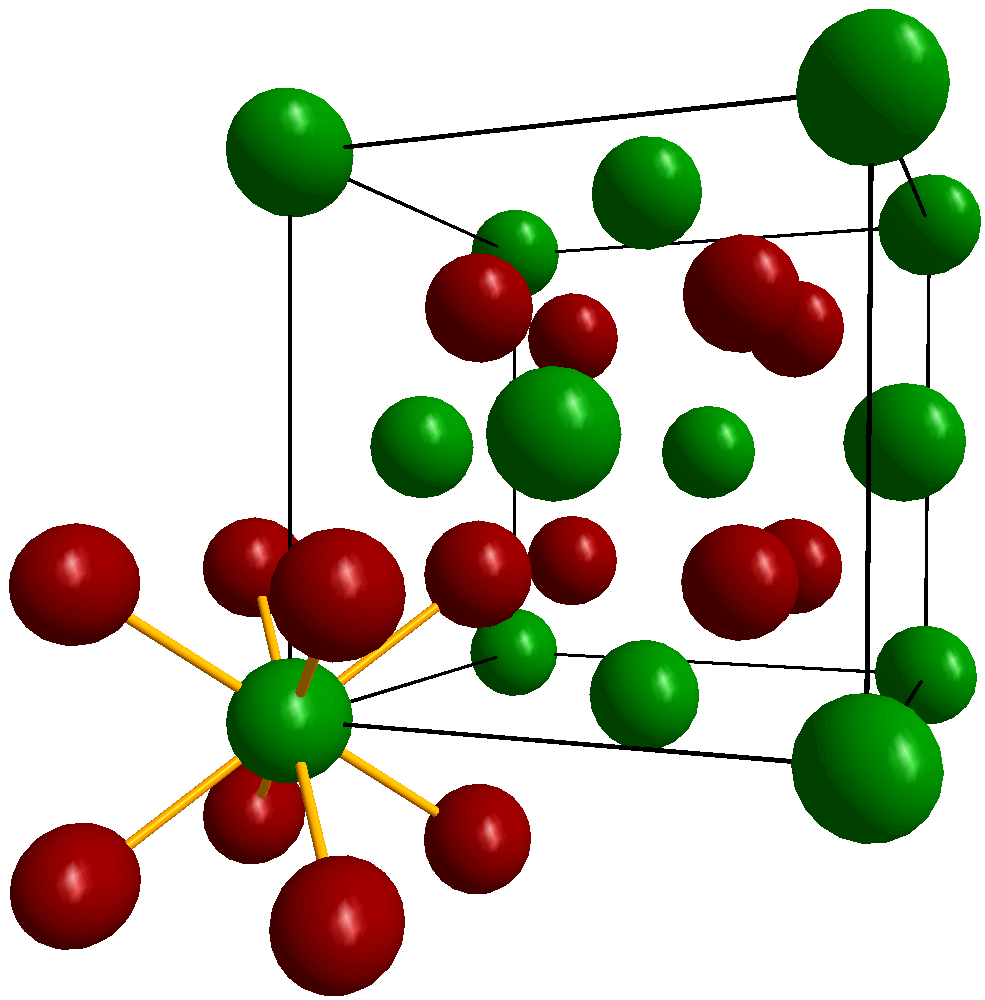}
  \caption{Unit cell (black box) and chemical environment of the U\textsup{4+} ion in cubic UO$_2$. U (green spheres), O (red spheres), and the nearest neighbor connections (yellow bars) are shown. Structure parameters from ICSD\,\cite{icsd, Szpunar2012}.}
  \label{fig:UO2structure}
\end{SCfigure}
\FloatBarrier

\subsection{Introduction}

UO$_2$ is an ionic material and its \CF{} ground state has been fairly well determined with inelastic neutron scattering.
Furthermore, it is cubic so that the number of \CF{} parameters is limited.
It is therefore very interesting to use UO$_2$ as a \textit{benchmark} material for U, just like CeB$_6$ for Ce, for testing the results obtained with \NIXS{} and \Quanty{}.

Uranium dioxide is the most studied and most cited actinide material, as it is the nuclear fuel of most of the world's reactors and a central material in the quest to understand radioactive waste and its interaction with the environment.
The ground-state symmetry of uranium in UO$_2$ has been discussed for many years.
It was first proposed theoretically over 50 years ago.
Numerous studies with neutron scattering have been interpreted with a $\Gamma_5$ triplet as the ground state.

\begin{SCfigure}
	\centering
\includegraphics[width=0.6\columnwidth]{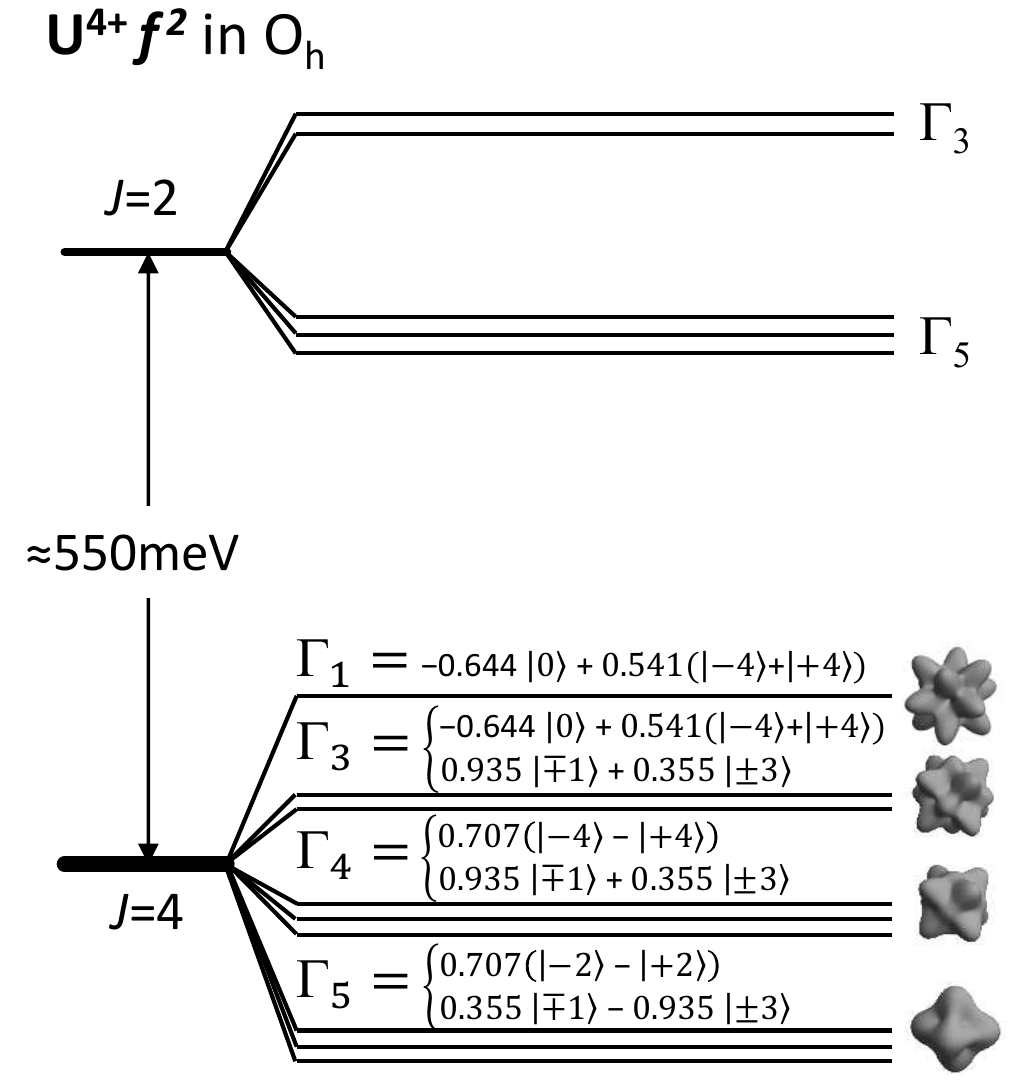}
\caption{Scheme of the low lying U\,5$f$ energy levels for U$^{4+}$ in cubic point symmetry O\textsub{h} (see \tab{tab:f2cubicrepresentations}). For the Hund's rule ground state with $J$=4 the |$J_z$$\rangle$ representation of the \CF{} eigenstates is shown. The insets show the two-electron charge densities of the \CF{} eigenstates of the $^3$H$_4$ multiplet.}
\label{fig:UO2scheme}
\end{SCfigure}

The electronic 5$f^2$ configuration has a 9-fold degenerate Hund's rule ground state with total momentum $J$\,=\,4.
In the cubic O\textsub{h} point symmetry of UO$_2$ the $J$\,=\,4 multiplet splits into one singlet $\Gamma_1$, one doublet $\Gamma_3$, and two triplets $\Gamma_4$ and $\Gamma_5$.
The $J_z$ representation of the \CF{} states is shown in \fig{fig:UO2scheme}.

The first attempt to calculate the \CF{} ground state of UO$_2$ was reported in 1966\,\cite{Rahman1966}.
By that time, the low-temperature antiferromagnetic moment was known approximately from neutron diffraction experiments\,\cite{Willis1965, Frazer1965}, and was measured more accurately later as 1.74(2)\,$\mu_B$/U\,atom\,\cite{Faber1976}.
The energy levels of the \CF{} ground-state were measured with inelastic neutron scattering\,\cite{Kern1985, Amoretti1989}.
More recently, a large number of experiments and theory have been performed on UO$_2$ and are discussed in Ref.\,\cite{Santini2009, Nakotte2010, Caciuffo2011}.
All of these studies take as a basic assumption that the \CF{} ground state is the $\Gamma_5$ triplet within the $^3H_4$ multiplet.
Moreover, this ground state, given by the \CF{} parameters $\beta V_4$ and $\gamma V_6$ in Stevens formalism, fits well with those determined for other light actinides, since $V_4$ and $V_6$ (scaled by the expectation values of the $r^{4}$ and $r^{6}$ operators over the appropriate 5$f$-electron wave function) should be approximately independent of the actinide ion considered\,\cite{Magnani2005, Zhou2012}.
$V_4$ and $V_6$ relate to the Wybourne formalism by $A_{4,0} = 8\,V_4$ and $A_{6,0} = 16\,V_6$ and the relations $A_{4,4}$\,=\,$\sqrt{5/14}\,A_{4,0}$ and $A_{6,4}$\,=\,$-\sqrt{7/2}\,A_{6,0}$ are given by the O\textsub{h} point symmetry (see \app{app_stevensformalism} and Ref.\,\cite{Amoretti1989}).

The 5$f$ electrons are more delocalized than the 4$f$ (see \fig{fig:rjrofall}).
As a result the \CF{} splittings are larger in uranium than in the rare earth and the magnetic intensities in a neutron spectrum are strongly broadened so that most uranium compounds do not exhibit sharp \CF{} excitations in neutron inelastic scattering\,\cite{Broholm1987}.
There are only a few more localized exceptions like e.g.\ the oxide UO$_2$\,\cite{Amoretti1989}, the metallic UPd$_3$\,\cite{Le2012}, or the semimetallic U$_3$Pt$_3$Sb$_4$\,\cite{Rainford1995}.
Also in \XAS{} the lines are strongly broadened and no multiplet structures have been observed in metallic uranium compounds\,\cite{Wray2015}.

\subsection{Experimental}

\NIXS{} data of UO$_2$ had been taken by R.~Caciuffo, G.~Lander and G.~van der Laan already in 2013, but had not been analyzed after several years.
Here, we joined forces and the analysis of their data became part of this thesis.

The \NIXS{} experiment was performed with L.~Simonelli at the Resonant Inelastic X-ray Spectrometer (RIXS) installed at the ID20 beamline of the European Synchrotron Radiation Facility, in Grenoble, France\,\cite{Moretti2018}.
The incident beam with an intensity of 7\,$\times$\,10$^{13}$ photons/s and a 25\,$\mu$ rad vertical divergence of the undulator radiation was generated by three consecutive undulators.
It was monochromatized by a Si(111) double-crystal monochromator and horizontally focused by a Rh-coated mirror.
Five spherically bent Si(660) analyzers were placed at scattering angles 2\thS{}\,=\,100, 110, 120, 130, and 140$^\circ$ on vertical Rowland circles.
The fixed Bragg angle of \thB{}\,=\,87.5$^\circ$ provides a resolution of $\approx$0.7\,eV at a final photon energy \Eout{} of 9.69\,keV.
The scattered intensity was recorded by a MAXIPIX fast readout, photon-counting position sensitive detector, achieving up to 1.4\,kHz frame rate with 290\,$\mu$s readout dead time, with a pixel size of 55\,$\mu$m and a detection geometry of 256\,$\times$\,256 pixels.

The samples are fully described in Ref.\,\cite{watson00}.
For the measurements two UO$_{2}$ single crystals were cut and polished by Walt Ellis at Los Alamos National Laboratory\,\cite{taylor81}.
One had the polished surface perpendicular to the \xyz{111}  direction, the other to the \xyz{100}.
The samples were aligned with the \xyz{100} and \xyz{111} surface normal for specular geometry with respect to the analyzer at 2\thS{}\,=\,120$^\circ$, i.e.\ pointing along $\phi$\,=\,60$^\circ$.
The \xyz{110} direction was realized by rotating the \xyz{100} crystal accordingly.
\NIXS{} data of the UO$_2$ \edge[U]{O}{4,5} edges ($5d\to5f$) were measured at room temperature.

For data analysis, only data collected in the highest analyzer at 2\thS{}\,=\,140$^\circ$, which also had the largest intensity in the horizontal scattering geometry, were used because here the momentum transfer is largest.
The corresponding momentum transfer \vecq{} at the elastic energy at \thS{}\,=\,70$^\circ$ amounts to \absq{}\,=\,9.1\,\AA$^{-1}$.
Note, that there is an offset of 10$^{\circ}$ between the respective crystallographic directions which were aligned at $\phi$\,=\,60$^\circ$ and \vecq{}.
This has been taken into account when analyzing the data.

\subsection{NIXS results}

\begin{table}
  \centering
  \caption{Optimized parameters used in calculations in \fig{fig:UO2iso}. The values in the brackets are the variation step size of the parameters.}
  \label{tab:UO2parameters}
  \begin{tabular*}{\textwidth}{@{\extracolsep{\fill}}cccc||ccc}
    \hline
	\redff{} & \redfc{} & \redZc{} & \redZv{} & \absq{}/$\AA^{-1}$ & \FWHML{} & \FWHMG{} \\
	 0.6\,(1) & 0.6\,(1) & 1.00\,(5) & 1.00\,(5) & 9.4\,(2) & 0.6\,(2)\,eV & 0.65\,eV \\
	\hline
  \end{tabular*}
\end{table}
\begin{figure}
  \includegraphics[width=\textwidth]{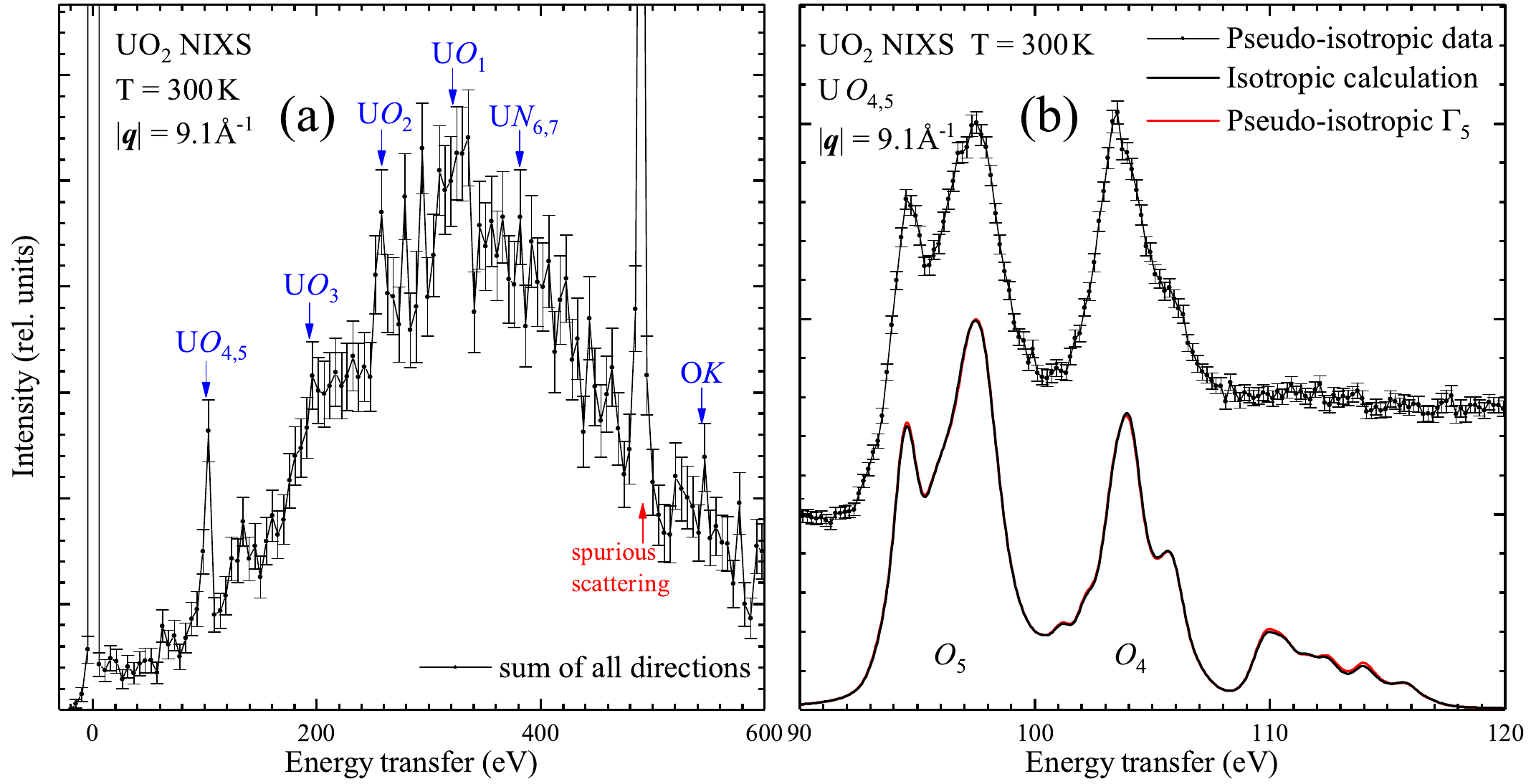}
  \caption{(a) \NIXS{} spectrum of UO\textsub{2} from zero up to 600\,eV energy transfer. For better visibility the three measured directions are summed up. The energies of the U and O core levels in this range are indicated starting from the \edge[U]{O}{4,5} edges. The strong line around 500\,eV does not belong to the \NIXS{} signal of the system. (b) (Pseudo-)Isotropic spectra of UO\textsub{2}. Data (dots) and calculation (lines) are shown.}
  \label{fig:UO2iso}
\end{figure}

\Fig[a]{fig:UO2iso} shows the UO\textsub{2} \NIXS{} spectrum for $\abs{\vec{q}}$\,$\approx$\,9.1\,$\AA^{-1}$ over a large energy range.
The three measured directions \qp{100}, \qp{110}, and \qp{111} have been added up in order to improve the statistics.
Each direction has a integration time of 10\,s per point.
The \edge[U]{O}{} edges can be seen distinctly on top of the broad Compton background that peaks at about 300\,eV.
Here, the \edge[U]{O}{4,5} edges ($5d\to5f$) is the most suited edge for studying the uranium 5$f$ wave function.
It is the strongest and has the best signal to background ratio of all visible U core levels, just as the rare-earth \edge[RE]{N}{4,5} edges ($4d\to4f$).
The huge intensity around 500\,eV is only present in the data of one direction and cannot arise form a core level, instead it is likely due to some Bragg scattering that reached the detector without passing the analyzer.
Such signals are referred to as \textit{spurious scattering}.

\begin{SCfigure}
  \centering
  \includegraphics[width=0.5\textwidth]{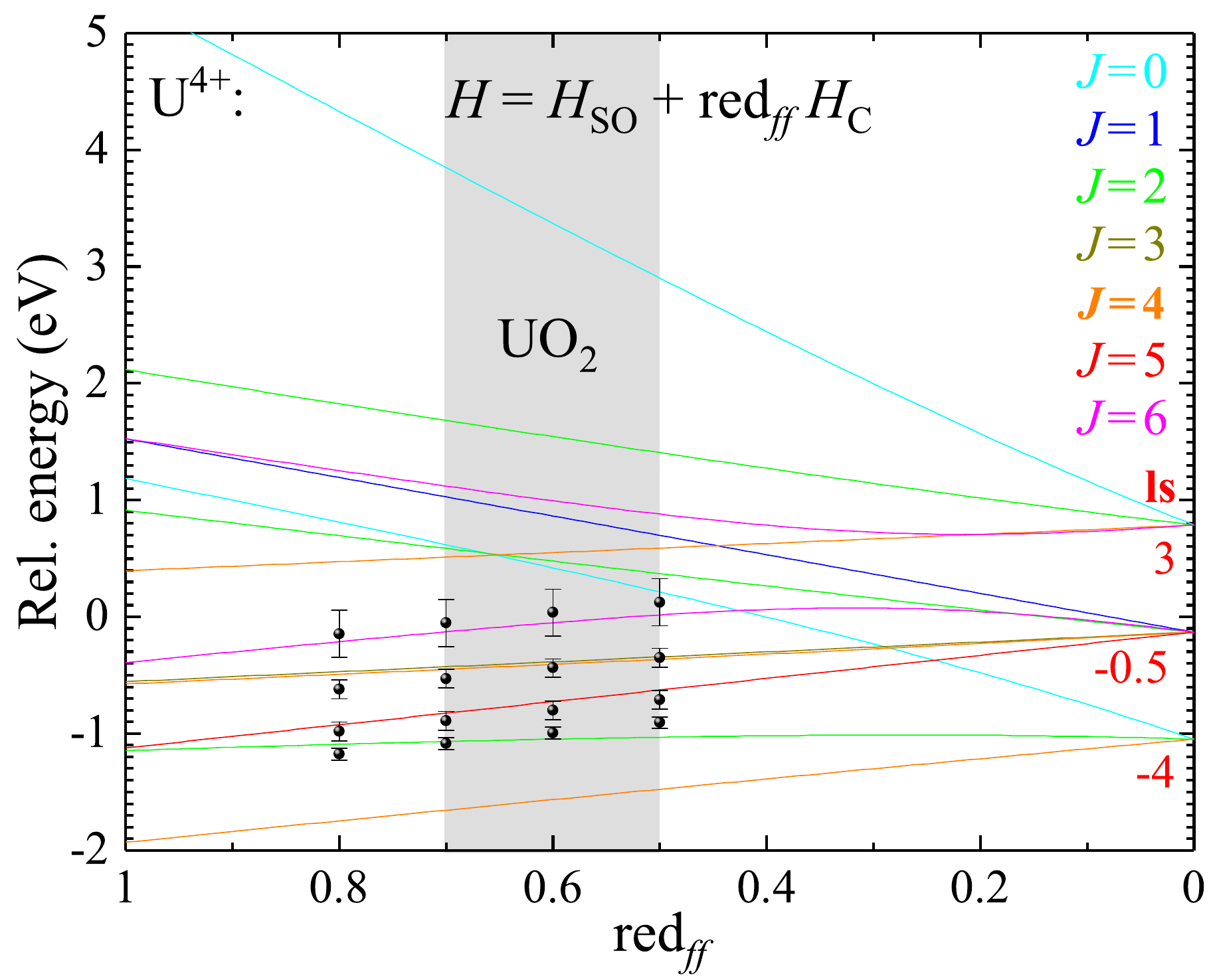}
  \caption{Energy level diagram of atomic U$^{4+}$ as a function of reduction of the 5$f$ Slater integrals. The calculations are without crystal field. The points show the excitations observed by RIXS extracted from Fig\,2 in Ref.\,\cite{Butorin2013}.}
  \label{fig:UO2ELD}
\end{SCfigure}

\paragraph{Multiplet structure}
\Fig[b]{fig:UO2iso} shows the pseudo-isotropic spectrum of the \edge[U]{O}{4,5} edges with energy steps of 0.2\,eV and an integration time of 6\,min/point per direction.
The pseudo-isotropic spectrum is the same linear combination of the three measured directions as in \equ{eq:OhQ3isotropic} for CeB$_6$, which is isotropic for the \qnQ{}\,=\,3 transition but not for \qnQ{}\,=\,5.
The black line represents the calculated isotropic spectrum.
The \CF{} splitting has been neglected at this point.
The atomic Coulomb interactions have been reduced to fit the observed peak positions.
The instrumental Gaussian line width \FWHMG{} is determined by the elastic width and the Lorentzian lifetime broadening with width \FWHML{} has been obtained by fitting the spectrum.
The momentum transfer \absq{} has been adjusted by fitting the isotropic spectrum.
The final values are given in \tab{tab:UO2parameters}.
The Hund's rule ground state of the initial state with these values can be expressed by the linear combination of the term symbols with \qnJ{}\,=\,4:
\begin{align}
86\%\,\termsymbol{3}{H}{4} + 13\%\,\termsymbol{1}{G}{4} + 1\%\,\termsymbol{3}{F}{4}.
\end{align}

The three sharp features at 94\,eV, 98\,eV, and 104\,eV energy transfer, which are important for the \CF{} analysis in the regime of higher multiplet transitions, are already described in great detail by the simple model neglecting the \CF{} splitting.
The most significant deviation between experiment and calculation is the additional intensity above 108\,eV in the calculation.
This may have two reasons as for the Ce/Sm\,$N_{4}$ edge in the hexaborides.
The dipole dominated region is shifted due to a stronger reduced G$^1$ Slater integral and/or the lines above 108\,eV have a much larger lifetime broadening due to the vicinity to the continuum states.
For UO$_2$ no extra reduction of the Coulomb dipole interaction term (G$^1$ Slater integral) was found that improves the fit significantly.
The precise description of the dipole regime, would require the inclusion of the interaction with the continuum states and make the model much more complicated, with only a minor gain for the following analysis.

For finite Coulomb and spin-orbit coupling the total momentum of \qnJ{}\,=\,4 always forms the ground state.
\Fig{fig:UO2ELD} shows the energy level diagram for the U$^{4+}$\,5$f^2$ configuration as function of the Coulomb reduction factor \redff{} without considering a finite \CF{} potential.
\redfc{} does not contribute since only the initial state eigenvalues are shown.
The gray range indicates the range that describes the UO$_2$ \NIXS{} data best.
It turns out that the calculated \qnJ{} multiplet splittings within the applicable range of \redff{} match fairly well excitations seen by high resolution ($\approx$160\,meV) \RIXS{} at the \edge[U]{O}{4,5} edges\,\cite{Butorin2013}, shown by the data points in \Fig{fig:UO2ELD}.
The best agreement with the multiplet lines is around \redff{}\,=\,0.6-0.7.
Note, this comparison not only lacks the assumption of a \CF{} splitting; it also does not take the $5d$ core hole into account, which is absent in the \RIXS{} final state.

\paragraph{\(\vec{\textit{q}\,}\)-directional dependence}

\begin{SCfigure}
  \centering
  \includegraphics[width=0.5\textwidth]{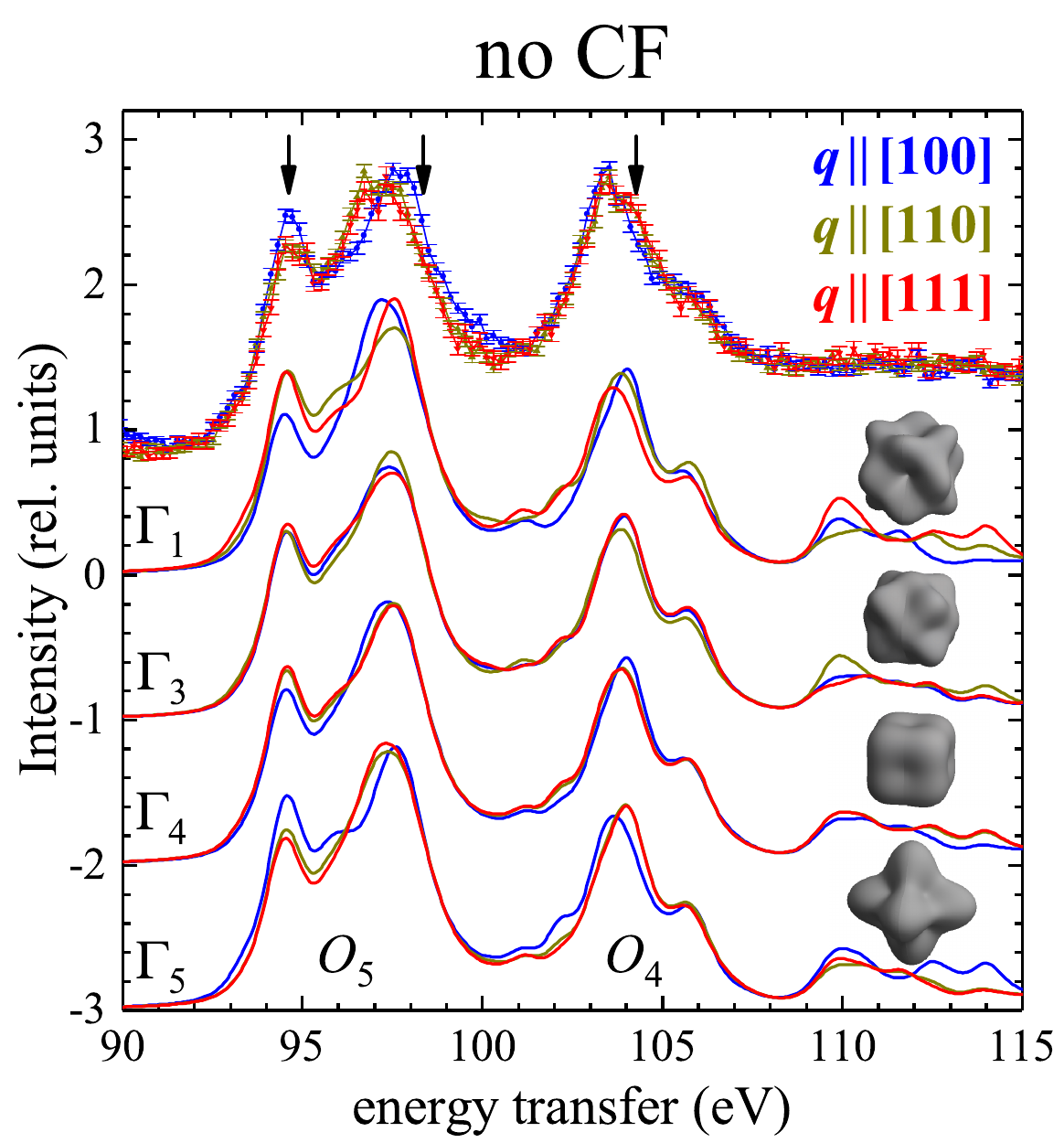}
  \caption{\NIXS{} data of the UO$_2$ \edge[U]{O}{4,5} edges for the three different directions of the momentum transfer \vecq{} at 300\,K (dots) and the calculation of the \vecq{}-directional dependence with the adjusted U$^{4+}$\,5$f^2$ parameters (\tab{tab:UO2parameters}) for the four \CF{} states. There is a 10$^\circ$ offset between \vecq{} and the labeled crystallographic directions, as described in the experimental section, which is also taken into account in the calculations. The insets show the two-electron charge densities.}
  \label{fig:UO2_noCF}
\end{SCfigure}

\Fig{fig:UO2_noCF} shows the \vecq{}-directional dependence of \sqw{} of the UO$_2$ \edge[U]{O}{4,5} edges as measured in the present \NIXS{} experiment (dots).
The arrows indicate where the dichroism is strongest, namely at 95, 98 and 104\,eV energy transfer.
The lines below represent the calculation of the spectra for the different initial \CF{} states, using the parameters in \tab{tab:UO2parameters}.
The spectra were obtained from the \CF{} eigenstates, but without using a finite \CF{} potential, in the following referred to as \textit{no \CF{}}. 
The respective two-electron charge densities are shown as insets.

As for the rare-earth hexaborides, the spectra with \qp{001} (blue) and \qp{111} (red) show the strongest directional dependence, whereas here with the 10$^\circ$ offset the \qp{110} and \qp{111} directions yield almost identical signals.
At 95\,eV only the $\Gamma_5$ reproduces the "blue-over-red" dichroism, while the other three states show an opposite behavior, which is strongest for the $\Gamma_1$ initial state, followed by $\Gamma_4$ and weakest for $\Gamma_3$.
At 104\,eV this is inverted, red over blue, and again reproduced only by the $\Gamma_5$, opposite for $\Gamma_1$ and only very little differences for $\Gamma_4$ and $\Gamma_3$.
At 98\,eV the blue line is shifted towards higher energy transfer, as compared to the red.
Such a behavior is not reproduced by any of the calculations.
In the following, the effect of a finite \CF{} is investigated.

\paragraph{Impact of a finite crystal field}

\begin{figure}
  \centering
  \includegraphics[width=\textwidth]{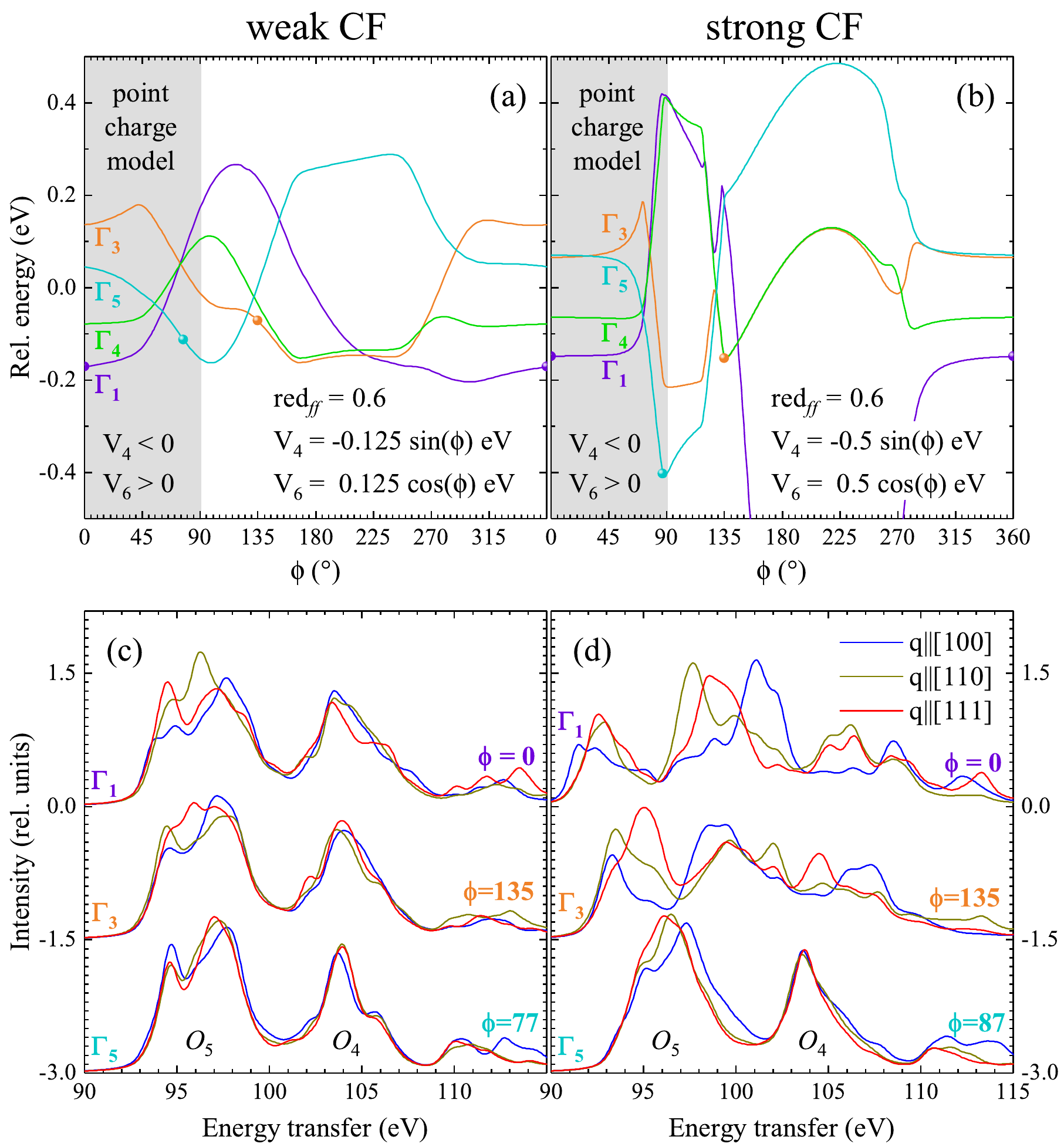}
  \caption{Realistic \CF{} calculations for UO$_2$. (a) and (b) Eigenvalues of the lowest \qnJ{}\,=\,4 multiplet calculated for red\(_{fc}\)\,=\,0.6 and \textit{weak \CF{}} and \textit{strong \CF{}} parameters V$_4$ and V$_6$. The sign of V$_4$ and V$_6$ can be determined from a point charge consideration as discussed for O\textsub{h} point symmetry in Ref.\,\cite{Lea1962}; the valid region is indicated. (c) and (d) are calculations of the \NIXS{} spectra including the \CF{} potential of (a) and (b), for 3 different V$_4$/V$_6$ ratios. The $\Gamma_4$ is missing as it can never be ground state and only the $\Gamma_1$ and $\Gamma_5$ are valid within the point charge model.}
  \label{fig:UO2_CF}
\end{figure}

\begin{SCfigure}
  \includegraphics[width=0.6\textwidth]{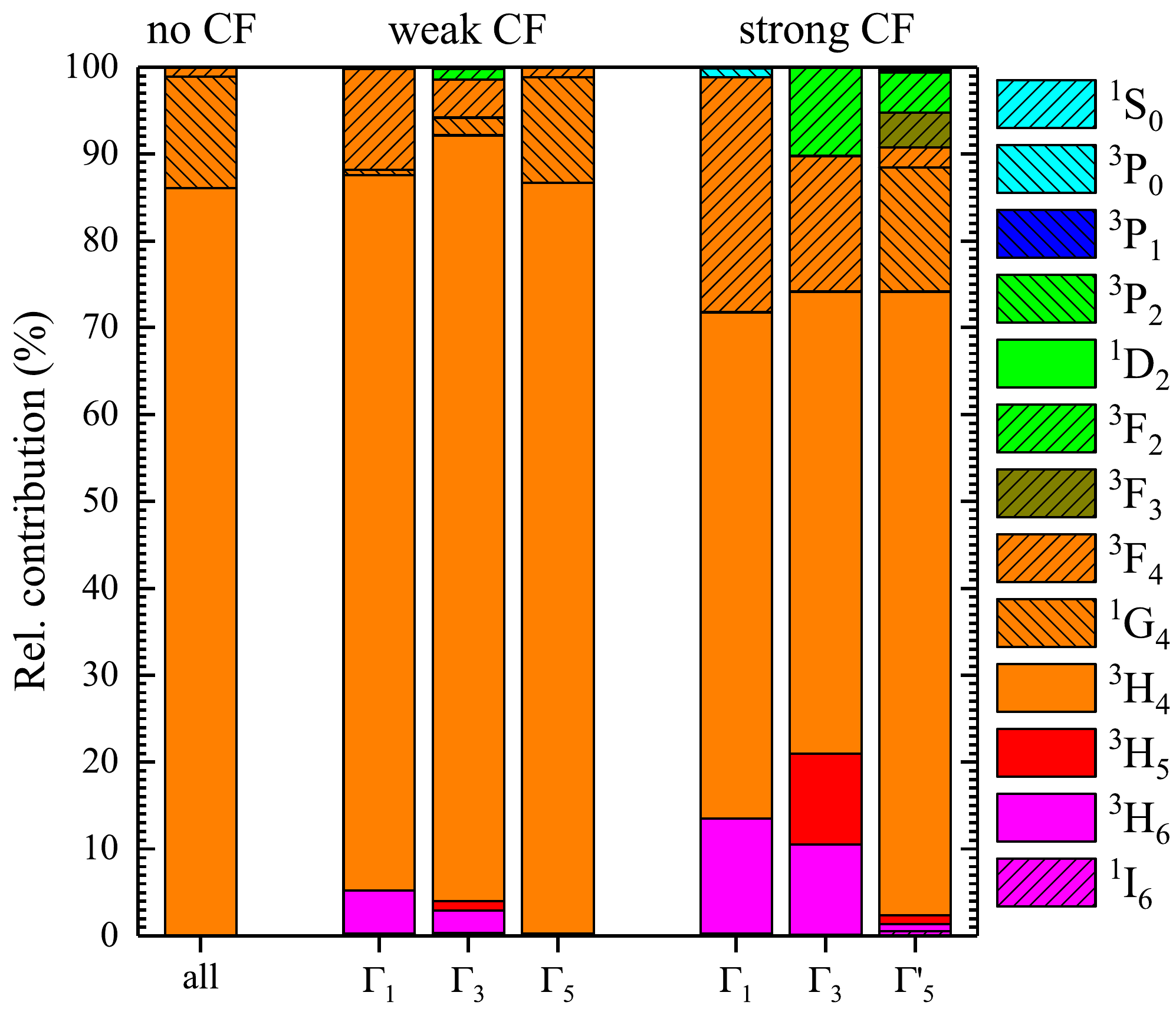}
  \caption{Mixing of the different term symbols for the different scenarios: no \CF{}, with weak \CF{}, and with strong \CF{}. Without \CF{} only \qnJ{}\,=\,4 mixes, but with finite \CF{} the mixing changes. The mixing varies for different \CF{} potentials, but not all \qnJ{} can mix with all $\Gamma_i$ (see \tab{tab:f2cubicrepresentations} for what can mix).}
  \label{fig:UO2_mixing}
\end{SCfigure}

So far the data were analyzed in the same manner as the data of the rare-earth materials in the previous chapters.
That is the atomic \SOC{} and \CI{} (with some screening) have been accounted for, but the \CF{} potential is assumed as a negligible perturbation.
The 5$f$ in the actinides, however, are significantly more extended than the 4$f$ in the rare-earth (see \fig{fig:rjrofall}), so that \CF{} effects are larger.
They may shift the multiplet lines significantly and may also lead to a mixing of different \qnJ{}.
In the following, the effect of a realistic \CF{} potential will be investigated.

\Fig[(a+b)]{fig:UO2_CF} show the \CF{} splitting of the low lying \qnJ{}\,=\,4 multiplet for a strong and a weak \CF{} potential.
The calculation uses \redff{}\,=\,0.6 as before, but now it also includes \CF{} parameters of V$_4$\,=\,-125\,sin($\phi$)\,meV and V$_6$\,=\,125\,cos($\phi$)\,meV as a function of $\phi$, referred to as \textit{weak \CF{}} and V$_4$\,=\,-500\,sin($\phi$)\,meV and V$_6$\,=\,500\,cos($\phi$)\,meV referred to as \textit{strong \CF{}}.
This is similar to the representation of Lea, Leask, and Wolf\,\cite{Lea1962}, but covers all V$_4$/V$_6$ ratios and signs of $V_4$.
The highlighted area indicates the point-charge considerations presented by Lea, Leask, and Wolf\,\cite{Lea1962}, which restricts the sign of V$_4$ and V$_6$.
In both scenarios the $\Gamma_4$ can never be ground state within the full parameter space $\phi$ of the \CF{} potential.
Moreover, within the considerations of the point charge model, the $\Gamma_3$ cannot be the ground state either.

\Fig[c]{fig:UO2_CF} shows the \NIXS{} spectra of the UO$_2$ \edge[U]{O}{4,5} edges calculated for the \textit{weak \CF{}} scenario.
Here three different values of $\phi$ are chosen, which yield the $\Gamma_1$, $\Gamma_3$, $\Gamma_5$ \CF{} ground state.
Note that the value of $\phi$ for the $\Gamma_3$ is actually no longer in the valid range of the point charge model.
The value of $\phi$\,=\,77$^\circ$ is chosen to reproduce the \CF{} splittings seen by neutron diffraction\,\cite{Amoretti1989}.
\Fig[d]{fig:UO2_CF} shows the \NIXS{} spectra of the UO$_2$ \edge[U]{O}{4,5} edges calculated for the \textit{strong \CF{}} scenario.
Here, the same \CF{} ground states are shown, but $\phi$ was adjusted for the $\Gamma_5$ to reproduce the V$_4$/V$_6$ ratio as presented in Ref.\,\cite{Rahman1966}.

Comparing the \textit{weak \CF{}} calculations in \fig[c]{fig:UO2_CF} to the calculations without \CF{} in \fig{fig:UO2_noCF} shows certain differences.
For the $\Gamma_1$ both, the \edge[U]{O}{5} and \edge{O}{4}, edges become broader with the \CF{} potential.
This has two reasons:
the \CF{} splitting of the final states multiplet lines, but also the different mixing of different \qnL{}, \qnS{}, and \qnJ{} quantum numbers.
The mixing of the different term symbols is shown in \fig{fig:UO2_mixing}.
The same happens, although weaker, for the $\Gamma_3$.
Qualitatively the dichroism remains the same:
For example for the $\Gamma_1$ the strongest signal at the \edge[U]{O}{5} edge is given by the red line, then the yellow line and finally the blue line with increasing energy transfer and the red line is weaker at the \edge{O}{4} edge.
In the case of the $\Gamma_5$ the differences are dominated by the shift of the multiplet lines due to the \CF{} potential, but the term-symbol mixing remains mostly unchanged.
The differences in the spectra are minor, but the blue line around 98\,eV in \fig{fig:UO2_noCF} gets shifted to higher energy transfer.\footnote{This shift is present also for the excited \CF{} states (not shown), thus supporting its origin from a shift of the multiplet lines.}
This is the missing piece to fully reproduce the experiment, confirming the $\Gamma_5$ initial state.

In the case of the \textit{strong \CF{}}, the changes are significant.
For the $\Gamma_1$ and $\Gamma_3$ states the multiplet lines are spread so much, that the U\,$O_5$ and $O_4$ edges are no longer separated.
Also for the $\Gamma_5$ the term-symbols mix stronger and the line shape changes further (see \fig[d]{fig:UO2_CF}), such that the experiment is no longer well described.
Hence, the \textit{strong \CF{}} scenario must be excluded as the term-symbol mixing of UO$_2$ must be close to the Hund's rule ground state.

\begin{SCfigure}
  \centering
  \includegraphics[width=0.5\textwidth]{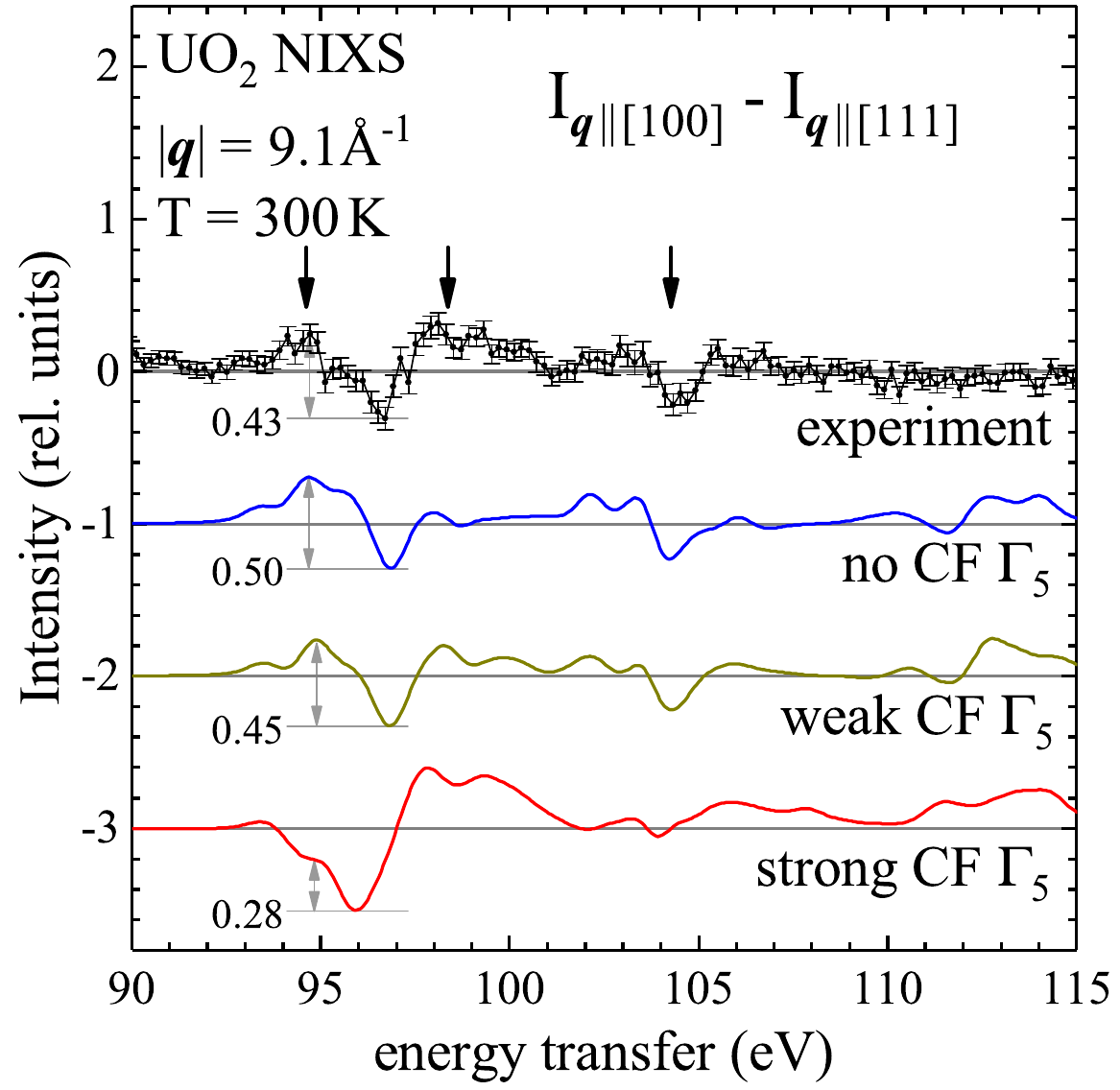}
  \caption{Difference spectrum between the NIXS data of the UO$_2$ \edge[U]{O}{4,5} edges for \qp{100} and \qp{111} at 300\,K (dots) and the $\Gamma_5$ calculations of the different \CF{} scenarios (lines); i.e.\ positive values represent blue over red and negative values red over blue in the previous \fig{fig:UO2_noCF} and \fig{fig:UO2_CF}. For better comparison, the size of the change in the difference signal around 95\,eV energy transfer is given.}
  \label{fig:UO2_dichroism}
\end{SCfigure}
When looking at the difference signals in \fig{fig:UO2_dichroism} the above considerations favoring the weak over the strong \CF{} potential are supported.
Here only the difference between the two spectra with \qp{100} and \qp{111} is shown.
For a more quantitative comparison the size of the difference spectrum between the maximum around 95\,eV and the minimum around 96\,eV, which is least affected by the continuum states, is given for each scenario.
This difference is significantly smaller than the calculated inversion without \CF{}, about the same for the weak \CF{}, and much smaller for the strong \CF{}, as the low energy blue over red anisotropy is gone.
This confirms the $\Gamma_5$ ground state in the weak \CF{} scenario.

Note, a \CF{} scenario with smaller splittings than the weak \CF{} scenario will no longer describe the shift around 98\,eV of the blue curve in the experimental data (\fig{fig:UO2_noCF}) satisfactorily, which means that the calculated maximum at 98\,eV in \fig{fig:UO2_dichroism} becomes too small to describe the experiment.
A significantly larger \CF{}, on the other hand, will at some point no longer reproduce the maximum in \fig{fig:UO2_dichroism} around 95\,eV and the minimum at 105\,eV.
Thus the values of $V_4$ and $V_6$, which are in good agreement with the experiment are limited to the values given in \tab{tab:UO2CFparameters}.
The uncertainty in the energies of the excited \CF{} states also arise from their variation with the $V_4$/$V_6$ ratio (slope relative to the ground state in \fig[a]{fig:UO2_CF}), i.e.\ due to the uncertainty of $\phi$\,=\,(81$\pm$4)$^\circ$.
\begin{table}
  \centering
  \caption{Parameters of the cubic \CF{} potential and the resulting energies of the excited \CF{} states obtained from this \NIXS{} experiment.}
  \label{tab:UO2CFparameters}
  \begin{tabular*}{\textwidth}{@{\extracolsep{\fill}}cc||ccc}
    \hline
	$V_4$/meV & $V_6$/meV & $E(\Gamma_1)$/meV & $E(\Gamma_3)$/meV & $E(\Gamma_4)$/meV \\
	 -160\,(40) & 23\,(10) & 300\,(120) & 170\,(20) & 260\,(90) \\
	\hline
  \end{tabular*}
\end{table}

\paragraph{Summary}

The \NIXS{} data of UO\textsub{2} show multiplet structures and a distinct \vecq{} dependence.
The directional \vecq{} dependence of the \NIXS{} data on the UO$_2$ \edge[U]{O}{4,5} edges unambiguously confirms the $\Gamma_5$ \CF{} ground state.
The shift of the \qp{100} (blue) peak around 98\,eV could be assigned to the finite \CF{} potential with $\Gamma_5$ ground state.
The mixing of term symbols contributing to the $\Gamma_5$ must not be changed significantly by the \CF{} potential, i.e.\ the results are in agreement with the inelastic neutron experiment\,\cite{Amoretti1989}, but not with the stronger \qnJ{} mixing as proposed in Ref.\,\cite{Rahman1966}.

As the term-symbol mixing is negligible for UO$_2$, the isotropic spectrum and directional \vecq{} dependence could be reproduced in a reasonable way in the \textit{no \CF{}} scenario.
Consequently, when the term-symbol mixing due to the \CF{} is weak ($\lesssim$10\%) the assignment of the ground state symmetry can be performed even without the knowledge of the full \CF{} scheme, i.e.\ without \CF{} parameters.

Realistic \CF{} calculations may become necessary when the term-symbol mixing is so important that the line shape of the \NIXS{} spectra is affected.
For moderate mixing, when the \edge[U]{O}{4} and \edge{O}{5} edges are still separated, but have a modified shape, a qualitative either-or question can still be answered when the dichroism is sufficiently large, as here for the $\Gamma_1$ and $\Gamma_5$.
The \textit{no \CF{}} results may simply deviate quantitatively a bit from the realistic scenario.
In conclusion, an analysis without \CF{} potential is possible as long as two distinct \edge[U]{O}{4} and \edge[U]{O}{5} edges are observed.
When the two edges are no longer distinct the term-symbol mixing due to the \CF{} becomes more important and the finite \CF{} potential must be accounted for which may be challenging in lower symmetry.

\clearpage
\section{URu\textsub{2}Si\textsub{2}} \label{sec:URu2Si2}
\begin{center}\textbf{
\scalebox{1.2}{Direct bulk sensitive probe of 5$f$ symmetry in URu\textsub{2}Si\textsub{2}}}\\
\hyperref[URu2Si2_2016]{\scalebox{1.2}{Proc.\ Natl.\ Acad.\ Sci.\ USA \textbf{113} 49, 13989\--13994 (2016)}}\end{center}
\FloatBarrier
\begin{SCfigure}[][h]
  \centering
  \includegraphics[width=0.48\textwidth]{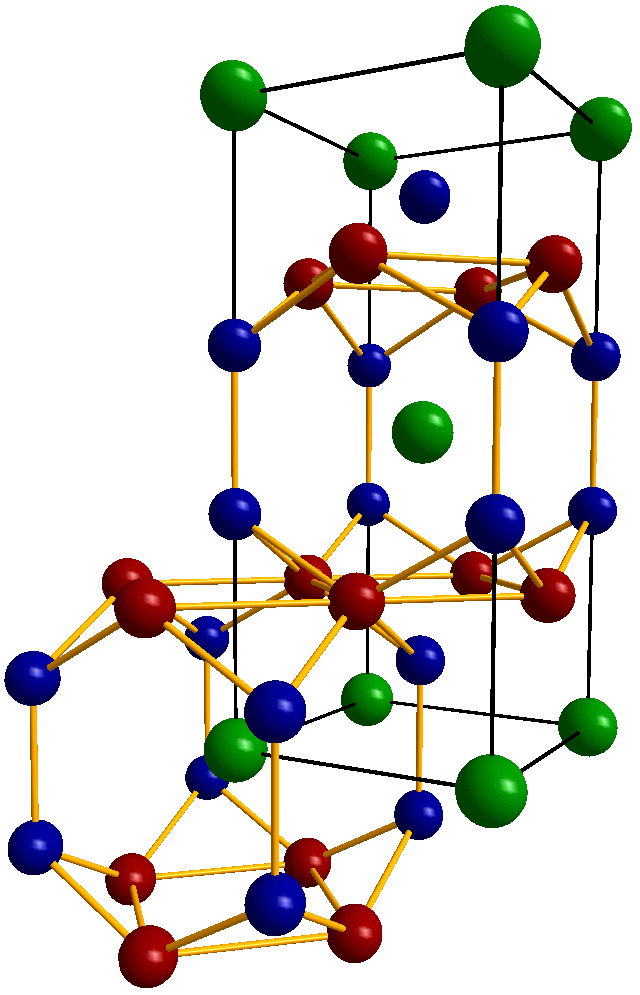}
  \caption{Unit cell (black box) and chemical environment of the U ion in tetragonal (I4/mmm) URu$_2$Si$_2$. U (green spheres), Ru (red spheres), and Si (blue spheres). The U is embedded inside a cage of Ru and Si (yellow bars) with local D\textsub{4h} point symmetry. Structure parameters from ICSD\,\cite{icsd, Ramirez1991}.}
  \label{fig:URu2Si2structure}
\end{SCfigure}
\FloatBarrier

\subsection{Introduction}
URu$_2$Si$_2$ is an intermetallic, tetragonal (see \fig{fig:URu2Si2structure}) heavy fermion compound that undergoes two phase transitions, the non-magnetic \HO{} transition at $\T{}_{\HO{}}$\,= \!17.5\,K that goes along with an appreciable loss of entropy of $\approx$0.2\,\Rigc{}\,ln\,2, and a superconducting one at about 1.5\,K\,\cite{Palstra1985, Schlabitz1986, Maple1986, Kasahara2007} (see \fig[a]{fig:URu2Si2_PD}).
With applied pressure (p\,$\ge$\,0.7\,GPa) the \HO{} order is replaced by an Large Moment AntiFerromagnetic phase (so-called \LMAF{}-phase)\,\cite{Amitsuka2007}, as shown in \fig[a]{fig:URu2Si2_PD}.
The order parameter of the \HO{} phase has been subject of intense investigations since more than 30 years, but so far it remained hidden.
This second order transition into an electronically ordered state involves a reconstruction of the Fermi surface\,\cite{Meng2013, Bareille2014} and a change of quasiparticle scattering rate\,\cite{Chatterjee2013}.
The Fermi surfaces of the \HO{} and high pressure \LMAF{} phase are very similar according to de Haas van Alphen measurements by Hassinger \textsl{et al}.\,\cite{Hassinger2010}.

\begin{figure}
  \centering
  \includegraphics[width=0.59\textwidth]{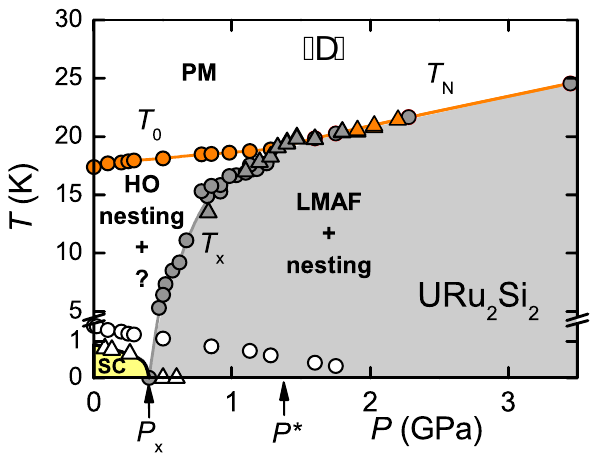}\vspace{5mm}
  \begin{minipage}[b]{0.4\textwidth}
    \includegraphics[width=\textwidth]{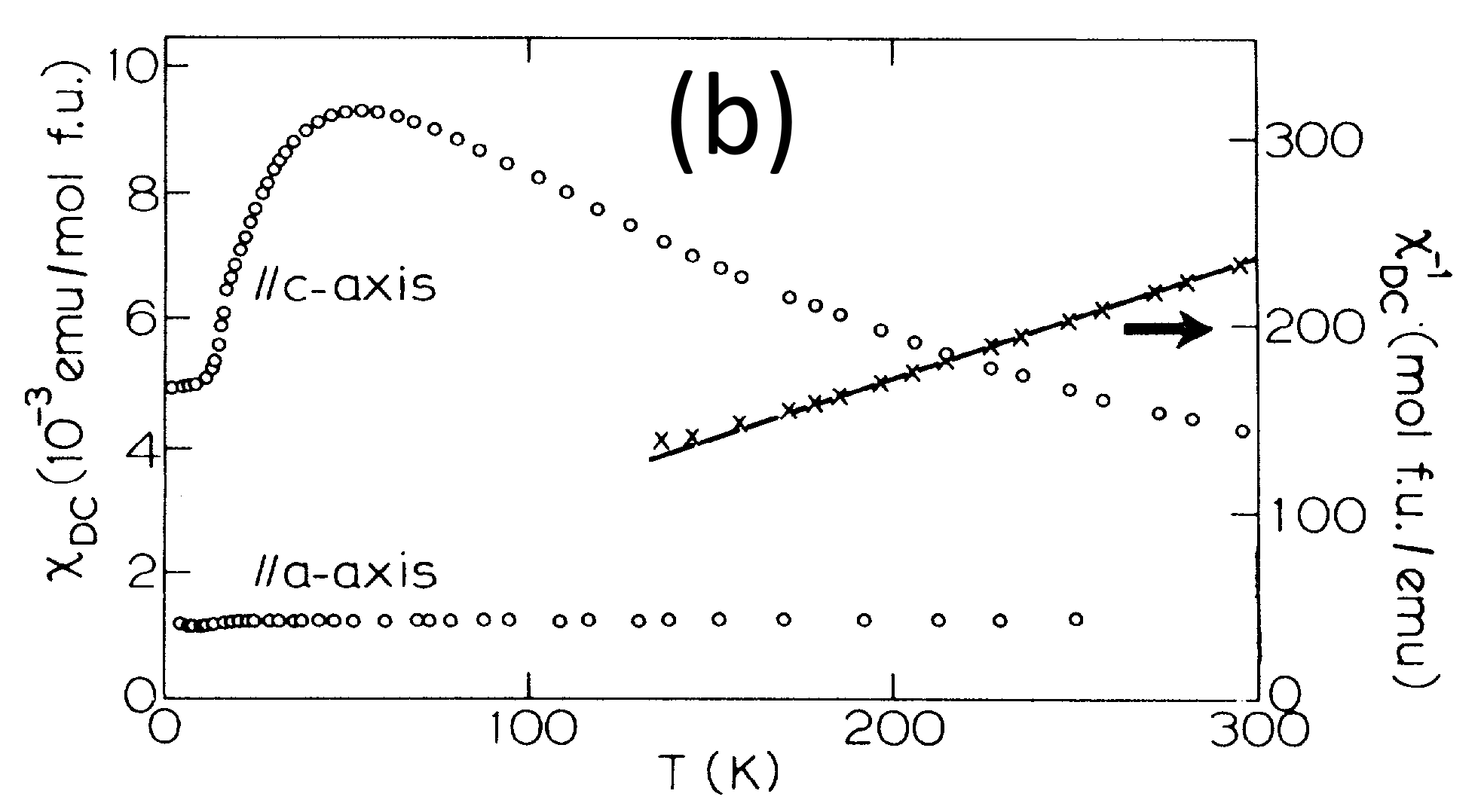}
    \includegraphics[width=0.9\textwidth]{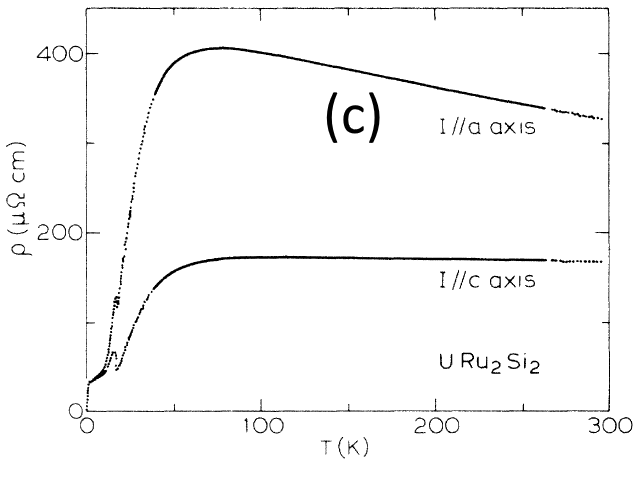}
  \end{minipage}
  \caption{(a) Low temperature phase diagram of URu$_2$Si$_2$ as a function of pressure summarizing resistivity (circles) and calorimetry (triangles) measurements from Ref.\,\cite{Hassinger2008} showing the SC phase below the HO phase and the crossover to the LMAF with pressure. (b) Susceptibility and (c) resistivity data of single crystalline URu$_2$Si$_2$ from Ref.\,\cite{Palstra1985} and Ref.\,\cite{Palstra1986}, respectively.}
  \label{fig:URu2Si2_PD}
\end{figure}

In URu$_2$Si$_2$ three energy scales have been identified: a hybridization gap of $\Delta_{hyb}$\,$\approx$ \!13\,meV (150\,K)\,\cite{Park2012} that opens below 27\,K, another gap that opens in the \HO{} phase with $\Delta_{HO}$\,$\approx$\,4.1\,meV (50\,K) in the charge\,\cite{Aynajian2010,Schmidt2010,Meng2013,Bareille2014} as well as spin channel\,\cite{Broholm1991,Wiebe2007} and a resonance mode that appears in the \HO{} gap at $\approx$1.6\,meV (18\,K), also in both channels\,\cite{Buhot2014,Kung2015,Bourdarot2010}.
A breaking of the fourfold rotational symmetry has been reported from torque experiments\,\cite{Okazaki2011} and high resolution x-ray diffraction on high quality crystals\,\cite{Tonegawa2007} but could not be verified by other groups.
A more detailed experimental and theoretical survey of the physical properties of URu$_2$Si$_2$ can be found e.g.\ in the review articles by Mydosh and Oppeneer\,\cite{Mydosh2011,Mydosh2014}.

\begin{SCfigure}
  \centering
  \includegraphics[width=0.6\textwidth]{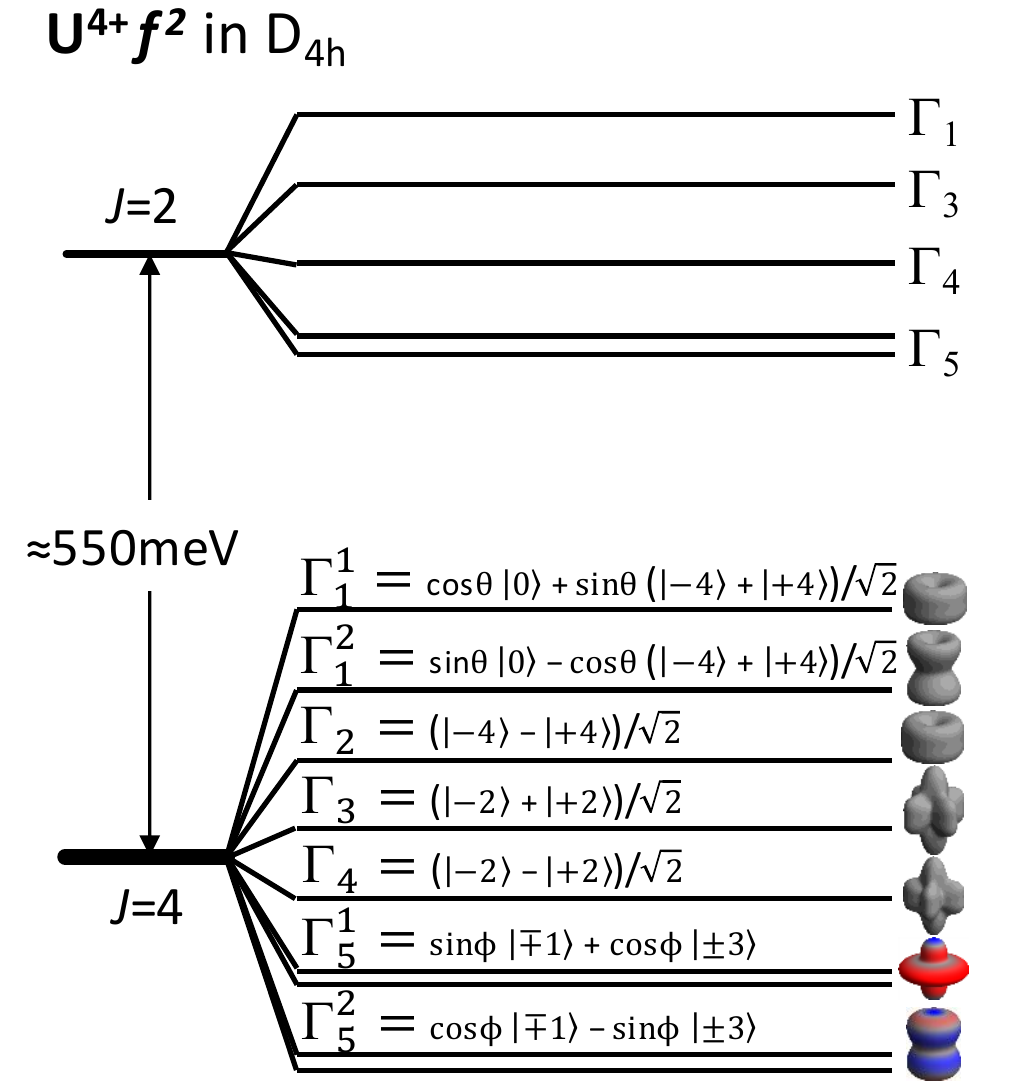}
  \caption{Scheme of the U\,5$f$ energy levels for U$^{4+}$ in D\textsub{4h} point symmetry (see \tab{tab:irD4h}). For the low laying \qnJ{}\,=\,4 multiplet the \ket{\qnJz{}} representation of the wave functions and corresponding charge density plots for $\theta$ and $\phi$\,=\,90$^\circ$ are shown.}
  \label{fig:URu2Si2scheme}
\end{SCfigure}
The transport properties of URu$_2$Si$_2$ are very anisotropic (see \fig[(b+c)]{fig:URu2Si2_PD}) due to the \CF{} acting on the U-ions.
Accordingly, many \HO{} scenarios are based on the assumption of certain ground state symmetries so that quite some efforts have been made to determine the \CF{} scheme of URu$_2$Si$_2$.
For all finite values of spin-orbit coupling and Coulomb interaction the \qnJ{}\,=\,4 multiplet forms the Hund's rule ground state.
In the tetragonal D\textsub{4h} point symmetry of U 5$f^2$ with \qnJ{}\,=\,4 the 9-fold degenerate Hund's rule ground state splits into two doublets $\Gamma_5^1$ and $\Gamma_5^2$ and five singlets $\Gamma_1^1$, $\Gamma_1^2$, $\Gamma_2$, $\Gamma_3$, and $\Gamma_4$.
The \qnJz{} representation is given in \fig{fig:URu2Si2scheme}.

The phase relations between the \qnJz{} states are defined such that the operator $\hat{J_x}$ is non-negative.
Note, that $\Gamma_1^1$(90$^\circ$)\,=\,-$\Gamma_1^2$(0$^\circ$) and $\Gamma_5^2$(90$^\circ$)\,=\,$\Gamma_5^1$(0$^\circ$) and, depending on the mixing angles $\phi$ and $\theta$, the \CF{} states correspond to "pure" $\abs{\qnJz{}}$ states 
(\,$\Gamma_1^1$(90$^\circ$)\,$\Leftrightarrow$\,\ket{4}+\ket{-4}, 
$\Gamma_2$\,$\Leftrightarrow$\,\ket{4}$-$\ket{-4}, 
$\Gamma_1^2$(90$^\circ$)\,$\Leftrightarrow$\,\ket{0}, 
$\Gamma_5^1$(90$^\circ$)\,$\Leftrightarrow$\,\ket{$\pm$3}, and\break $\Gamma_5^2$(90$^\circ$)\,$\Leftrightarrow$\,\ket{$\pm$1}\,).

In \chap{sec:UO2} it was explained that intermetallic uranium compounds do not exhibit \CF{} excitations in neutron scattering, nor multiplet lines in \XAS{}.
Determining the \CF{} potentials is therefore challenging.
This explains the disparity of results:
The anisotropy of the static susceptibility is well described with a $\Gamma_1^1$ singlet ground state, a $\Gamma_2$ as a first excited state and the next states above 15\,meV (170\,K)\,\cite{Nieuwenhuys1987}.
Analyses of elastic constant measurements find similar results\,\cite{Yanagisawa2013a, Yanagisawa2013}.
Also other models favor a $\Gamma_1^1$\,\cite{Kiss2005, Hanzawa2012, Kusunose2011} and Ref.\,\cite{Kiss2005} is also compatible with a $\Gamma_1^2$ singlet ground state, but they all propose different first excited states from their theoretical considerations.
Ref.\,\cite{Haule2009} proposes two low lying singlet states, a $\Gamma_2$ singlet ground state and a $\Gamma_1^2$ as first excited state.
This scenario is compatible with the interpretation of polarized Raman studies that find a resonance at 1.6\,meV in the $A_{2g}$ channel in the \HO{} phase\,\cite{Buhot2014, Kung2015}\footnote{Note, not only $\Gamma_1$\,$\rightleftharpoons$\,$\Gamma_2$ but also $\Gamma_3$\,$\rightleftharpoons$\,$\Gamma_4$ and $\Gamma_5$\,$\rightleftharpoons$\,$\Gamma_5$ are Raman active in the $A_{2g}$ channel (see supplementary of Ref.\,\cite{Buhot2014})}.
Thermodynamic measurements\,\cite{Santini1994} and resonant x-ray scattering data\,\cite{Nagao2005} are interpreted in terms of a $\Gamma_3$-singlet ground state with the $\Gamma_1^1$ as first excited state or alternatively with a $\Gamma_5^1$ ground state\,\cite{Nagao2005}.
Another elastic constant study\,\cite{Kuwahara1997} suggests a $\Gamma_4$ as lowest state.
$\Gamma_5^1$ and $\Gamma_5^2$ doublets as ground states are concluded by thermodynamic studies of diluted URu$_2$Si$_2$\,\cite{Amitsuka1994} and theoretical considerations\,\cite{Ohkawa1999,Chandra2013}.
Finally $O$-edge x-ray absorption measurements favor the $\Gamma_5^1$\,\cite{Wray2015} and the $\Gamma_5^2$ doublet\,\cite{Sugiyama1999} as ground state.
All these studies are considering a \CF{} ground state that arises from the U$^{4+}$ configuration, as in the case of UO$_2$.

The valence state of intermetallic actinide compounds, however, is often intermediate, and indeed, x-ray absorption and electron energy-loss spectroscopy find a valence between 3+ and 4+ for URu$_2$Si$_2$.
The different edges yield different results, namely 3.1+ for the \edge{L}{3} edge\,\cite{Booth2016}, 4+ for the \edge{M}{4,5} edges\,\cite{Kvashnina2017}, 3.3+ for the \edge{N}{4,5} edges\,\cite{Jeffries2010}, and finally 4+ for the \edge{O}{4,5} edges\,\cite{Wray2015}.
While these diverging results may origin from final state effects, as motivated in \fig{fig:AIMGS}, it is certainly correct to say that URu$_2$Si$_2$ is an intermediate valent system.
It is definitely an itinerant system based on the fact that the entropy released at $\T{}_{\HO{}}$ is not \Rigc{}\,ln\,2, but a (good) fraction of that.

The charge fluctuations can be understood to occur on top of a particular local irreducible representation of the U$^{4+}$ configuration\,\cite{Zwicknagl2003}.
The important question is now on top of which U$^{4+}$\,($5f^2$) representation the itinerant state forms and the formation of the \HO{} happens.
In this respect the dominant contribution to our \NIXS{} experiment can be assumed to be of the U$^{4+}$\,$f^2$ configuration
The observation of a sharp multiplet structure justifies the interpretation within a local approach.
Here the multiplets have a strongly asymmetric line shape which can either be taken as sign for the material being on the verge to itinerancy with a strong band dispersion in the final states or arise from a moderately strong mixing of term symbols due to \CF{} effects, as a similar modification can be seen in \chap{sec:UO2} at the example of UO$_2$, happening for the $\Gamma_5$ state in \fig[(c+d)]{fig:UO2_CF}.

\subsection{Experimental}

\paragraph{Samples}
High-quality single crystal of URu$_2$Si$_2$ were provided by Yingkai Huan, Anne de Visser, and Artem Nikitin in the group of Mark Golden at the van de Waals-Zeeman Institute of the University of Amsterdam.
One single crystal was grown with the traveling melting zone method in the two-mirror furnace under high purity argon atmosphere.
The quality of the crystal was checked and oriented with x-ray Laue diffraction for its single-crystalline nature.
The oriented crystal was cut using the spark erosion method after which the relevant \xyz{100}, \xyz{110}, and \xyz{001} surfaces were polished.
A bar-shaped piece of the single crystal was characterized by resistance measurements.
The residual resistance ratio (RRR) value of 25 is not as good as it could be, other groups have samples with RRR values greater than 200, but the \HO{} transition is robust as summarized in Ref.\,\cite{Matsuda2011}.

\paragraph{Setup}
The directional dependence of the scattering function \sqw{} was measured with \NIXS{} at the beamline ID20 at ESRF.
In contrast to UO$_2$ the experiment was preformed when the new \NIXS{} end-station was already in place.
Two monochromators, Si(111) and Si(311), were used to scan the incident energy around 9.7\,keV.
The scattered intensity was analyzed by one column of three Si(660) crystal analyzers in horizontal scattering geometry with a scattering angle 2$\theta$\,=\,153$^\circ$ and detected in a \textit{Maxipix} 2D pixel detector in the Rowland geometry.
This setting corresponds to a momentum transfer of $|\vec{q}|$\,=\,9.6\,\AA$^{-1}$ at the elastic energy of the Si(660) analyzers of $\approx$9.69\,keV and an overall energy resolution of about 0.8\,eV, measured at the elastic line at the beginning of each set of measurements.

Three crystals with \xyz{100}, \xyz{110}, and \xyz{001} surfaces allowed realizing \qp{100}, \qp{110}, and \qp{001} in specular geometry and also other directions when going off specular.
Measured specular on the \xyz{110} crystal and 45$^\circ$ off specular on the \xyz{100} crystal gave the same result, showing that the specular geometry is not necessary, as also shown for CeB$_6$.

To reach low temperatures the samples were mounted in the He-flow cryostat, first with Be windows and later with Al windows.
Initially, the \edge[Be]{K}{} edge gave a strong contribution to the scattered signal.
Clean data could be recovered, however, as the signal of the Be window and the sample were projected to different positions on the 2D-pixel detector.
Data were taken below and above the \HO{} transition at 5\,K and 25\,K and with successively rising temperature up to 300\,K.

\subsection{NIXS results}

\begin{figure}
  \centering
  \includegraphics[width=0.85\textwidth]{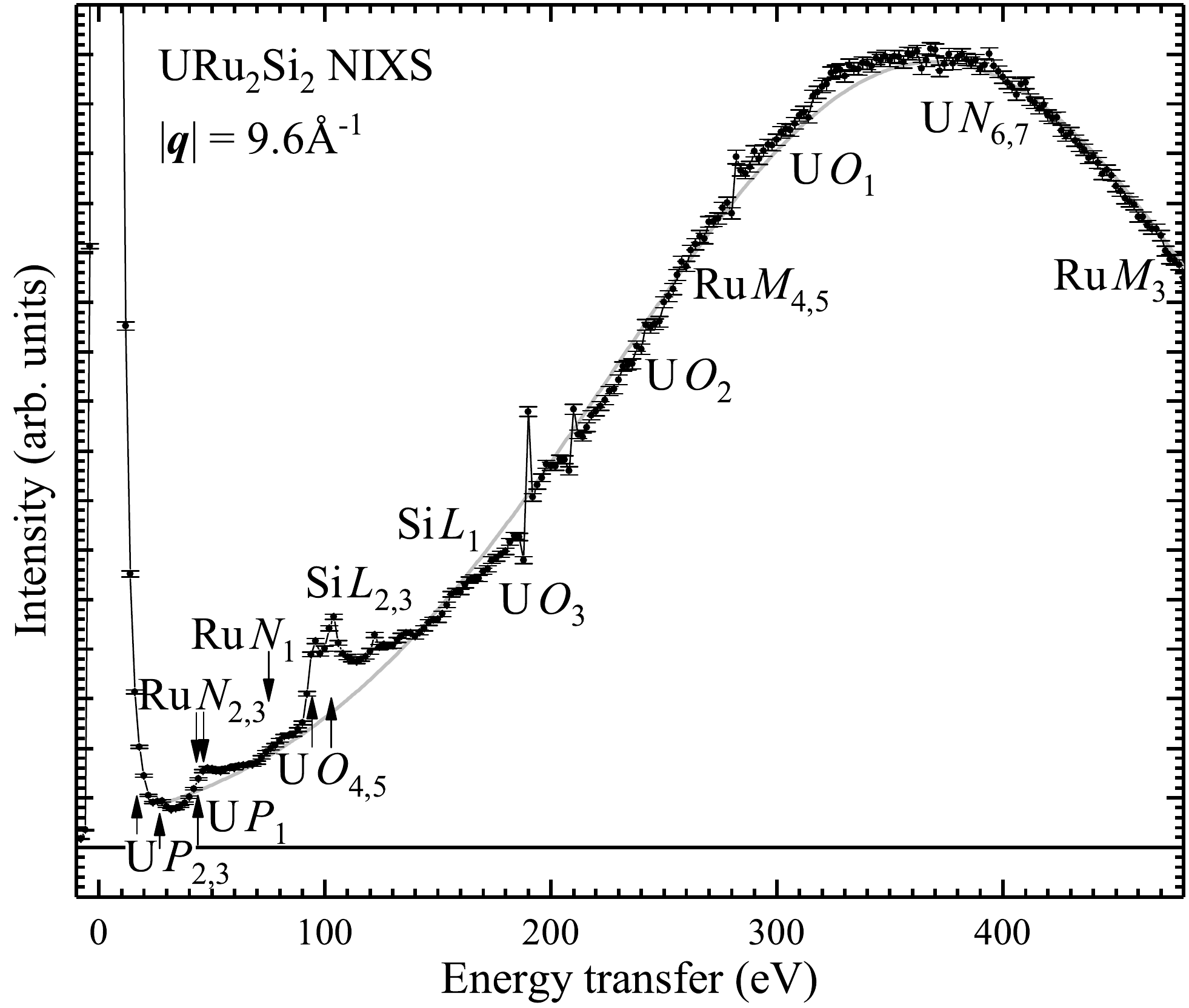}
  \caption{\NIXS{} spectrum of URu$_2$Si$_2$ over a wide energy range with a momentum transfer of \absq{}\,=\,9.6\,$\AA^{-1}$. Various \vecq{} directions and temperatures have been summed up. The electron binding energies of the core levels of the compounds elements and a Gaussian fit of the Compton (gray line) are shown.}
  \label{fig:URu2Si2_wide}
\end{figure}

\Fig{fig:URu2Si2_wide} shows the \NIXS{} spectrum of URu$_2$Si$_2$ for the first 500\,eV energy transfer.
The line shows the Gaussian fit of the Compton background.
The fits of the individual \vecq{} directions are subtracted later for the scans of the \edge[U]{O}{4,5} edges\footnote{The outstanding single points are due to glitches of the beamline optics at 100, 120, 190, 210, and 280\,eV, which are poorly corrected due to the 2\,eV step size of the wide scan.}.

As in all the investigated systems, the $d\rightarrow f$ transition at around 100\,eV energy transfer (\edge[U]{O}{4,5} edges) provides an excellent signal to background ratio.
The corresponding momentum transfer amounts to \absq{}\,$\approx$\,10\,$\AA^{-1}$.
The spectra were normalized to their pre-edge intensity (84.4-90.8\,eV) and Gaussian fits of the Compton scattering were subtracted along with some constant offset to compare of the directional dependent data of the \edge[U]{O}{4,5} edges.

The \edge[Si]{L}{2,3} edges starting from $\approx$100\,eV are in the energy range of the \edge[U]{O}{4,5} edges.
This may effect only the \edge[U]{O}{4}, but the delocalized Si $p$-valence states are not expected to show narrow excitonic features\,\cite{Johnson2016}\footnote{Compare also e.g.\ the \edge[Be]{K}{} edge of CeB$_6$ in \fig{fig:CeB6wide} or of SmB$_6$ in \fig{fig:SmB6_wide} or the \edge[Sn]{N}{} and \edge[Ru]{N}{} edges of CeRu$_4$Sn$_6$ in \fig{fig:CeRu4Sn6_wide}.}.
All other edges in \fig{fig:URu2Si2_wide}, like the \edge[Si]{L}{1} edge significantly weaker compared to the \edge[U]{O}{4,5} edges.

\paragraph{Multiplet structure}

\begin{table}
  \centering
  \caption{Parameters used for calculations. The values in the brackets are the step size used to scan the parameters.}
  \label{tab:URu2Si2parameters}
  \begin{tabular*}{\textwidth}{@{\extracolsep{\fill}}cccc||ccc}
    \hline
	\redff{} & \redfc{} & \redZc{} & \redZv{} & \absq{}/$\AA^{-1}$ & \FWHML{} & \FWHMG{} \\
	 0.5\,(1) & 0.5\,(1) & 1.00\,(5) & 1.00\,(5) & 11.5\,(5) & 1.3\,(1)\,eV & 0.8\,eV \\
	\hline
  \end{tabular*}
\end{table}

\begin{SCfigure}[][h]
  \includegraphics[width=0.5\textwidth]{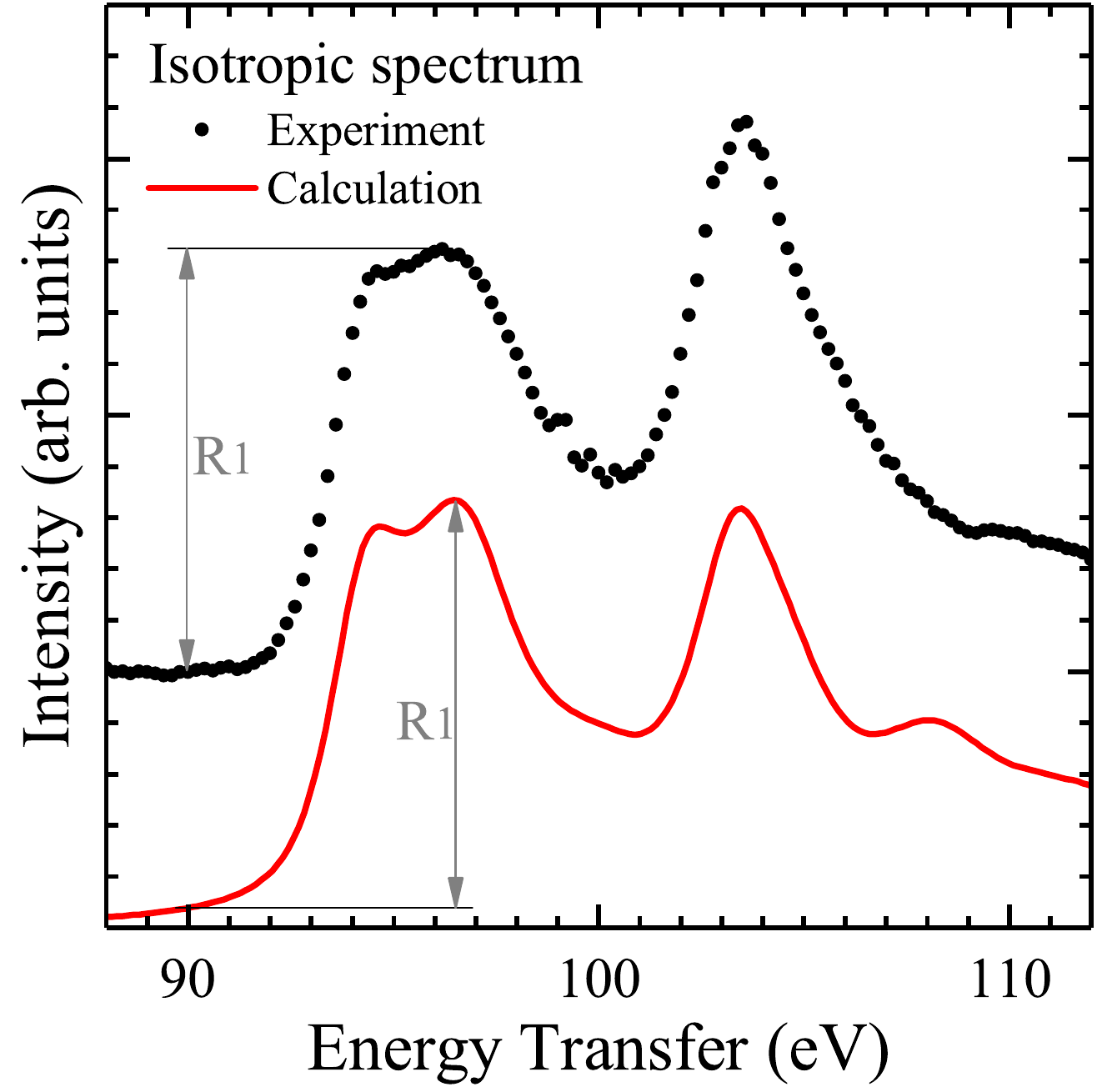}
  \caption{Experimental (black dots) and simulated (red line) isotropic spectrum of URu$_2$Si$_2$ at the \edge[U]{O}{4,5} edges for \T{}\,$\le$\,25\,K. For details see text.}
  \label{fig:URu2Si2_iso}
\end{SCfigure}

The simulations of the \NIXS{} spectra are based on an ionic model with a U$^{4+}$\,5$f^2$ configuration.
The model parameters in \tab{tab:URu2Si2parameters} have been adjusted using a pseudo-isotropic spectrum.
In D\textsub{4h} point symmetry the experimental pseudo-isotropic spectrum has been simulated by combining 10 independently measured directions, which is still not sufficient for the real isotropic spectrum of a \qnQ{}\,=\,5 transition.
The calculated isotropic spectrum is obtained by averaging over all \CF{} states, as the analysis was performed before the tensor notation (in \chap{sec:scatteringtensor}) was worked out.
The experimental and calculated spectra with the optimized parameter set in \tab{tab:URu2Si2parameters} are shown in \fig{fig:URu2Si2_iso}.

For \redff{}\,=\,0.5 the contributions of different term symbols amount to
\begin{align}
1\%\,^3\text{F}_4 + 14\%\,^1\text{G}_4+ 85\%\,^3\text{H}_4.
\end{align}
As in the previous chapters, the instrumental \FWHMG{} is adjusted to the width of the elastic line, the lifetime \FWHML{} is adjusted to reproduce the total line width at the \edge[U]{O}{4,5} edges, the reductions are chosen to fit the peak positions, and \absq{} was adjusted to reproduce the peak ratios of the lines of the isotropic spectrum.

The lineshape of the URu$_2$Si$_2$ \NIXS{} emission lines is asymmetric, which is not foreseen in the model.
It could be a \CF{} effect, which is quite similar to the effect calculated for the $\Gamma_5$ in the \textit{strong \CF{}} scenario of UO$_2$ in \fig{fig:UO2_CF}, but could as well be related to the more itinerant character of URu$_2$Si$_2$.
Because of the large number of \CF{} parameters the lineshape is reproduced in a phenomenological way:
A Mahan lineshape (as in \chap{chap:CeRu4Sn6_HAXPES}) with an asymmetry factor $\alpha$\,=\,0.18 and an energy cutoff at $\xi$\,=\,1000\,eV) has been used without physical justification, to reproduce the asymmetry.

The analysis of the \vecq{}-directional dependence of URu$_2$Si$_2$ is split into two parts due to the lower symmetry of tetragonal URu$_2$Si$_2$:
The anisotropy within the 4-fold symmetry plane and the out-of-plane.

\paragraph{Out-of-plane direction of \(\vec{\textit{q}\,}\)\! dependence}

\begin{figure}
	\includegraphics[width=\textwidth]{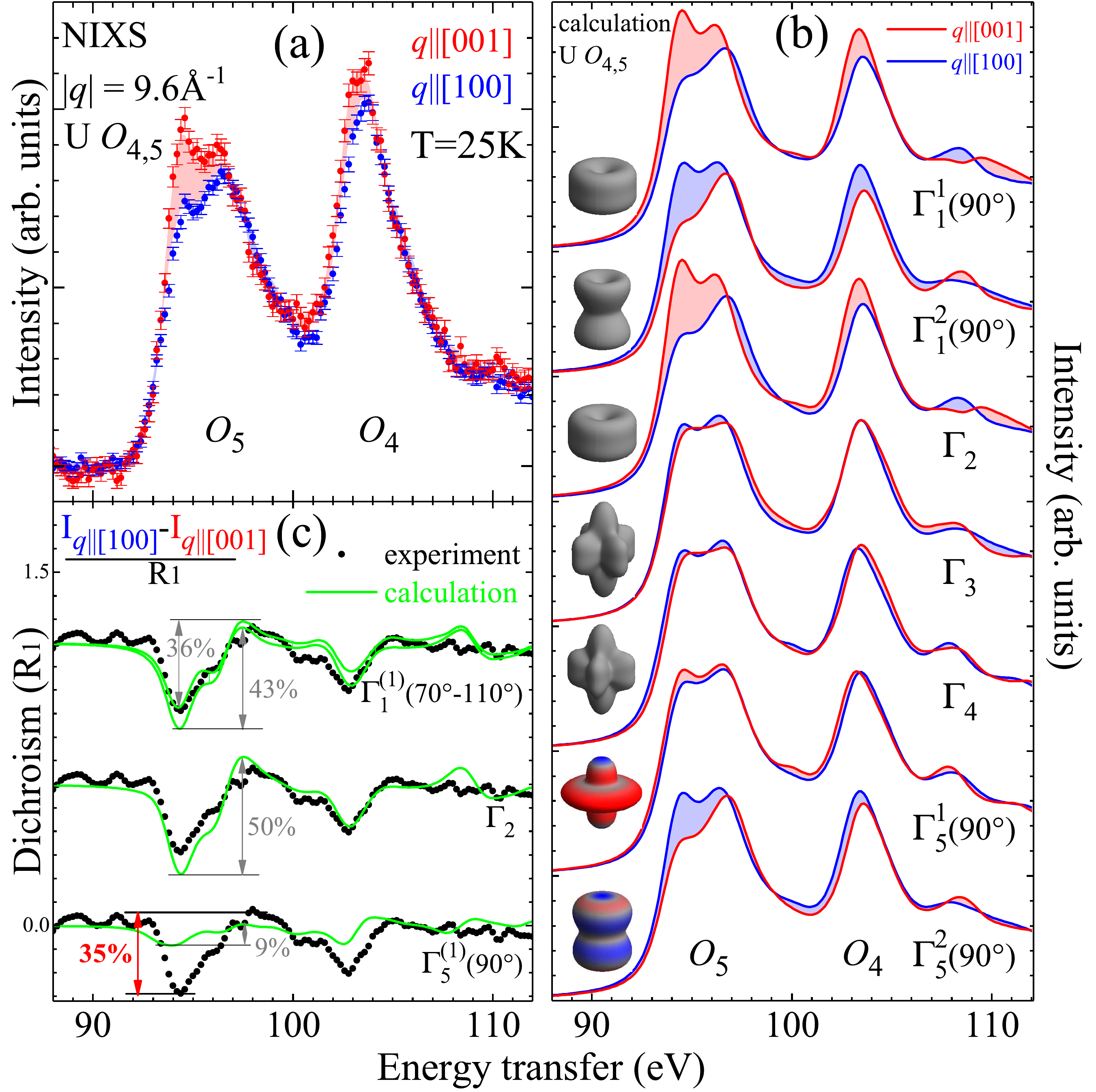}
	\caption{\NIXS{} measurements of the URu$_2$Si$_2$ \edge[U]{O}{4,5} edges for \absq{}\,=\,9.6\AA$^{-1}$ and corresponding calculations for the different \CF{} states. (a) \edge[U]{O}{4,5} \NIXS{} data for momentum transfers \qp{100} (blue) and \qp{001} (red) at \T{}\,=\,25\,K. (b) Calculations of \sqw{} of the U$^{4+}$ \CF{} states for \qnJ{}\,=\,4 in D\textsub{4h} symmetry for the two directions as in (a). The insets show the corresponding electron densities. (c) Experimental (black dots) and calculated (green lines) dichroism I(\qp{100})-I(\qp{001}) in units of the peak hight R$_1$ (defined in \fig{fig:URu2Si2_iso}). Calculations are only shown for the \CF{} states with the correct sign of the dichroism. Here the data points have been convoluted with a \FWHMG{}\,=\,0.5\,eV.}
	\label{fig:URu2Si2_ac}
\end{figure}

\begin{figure}
  \includegraphics[width=\textwidth]{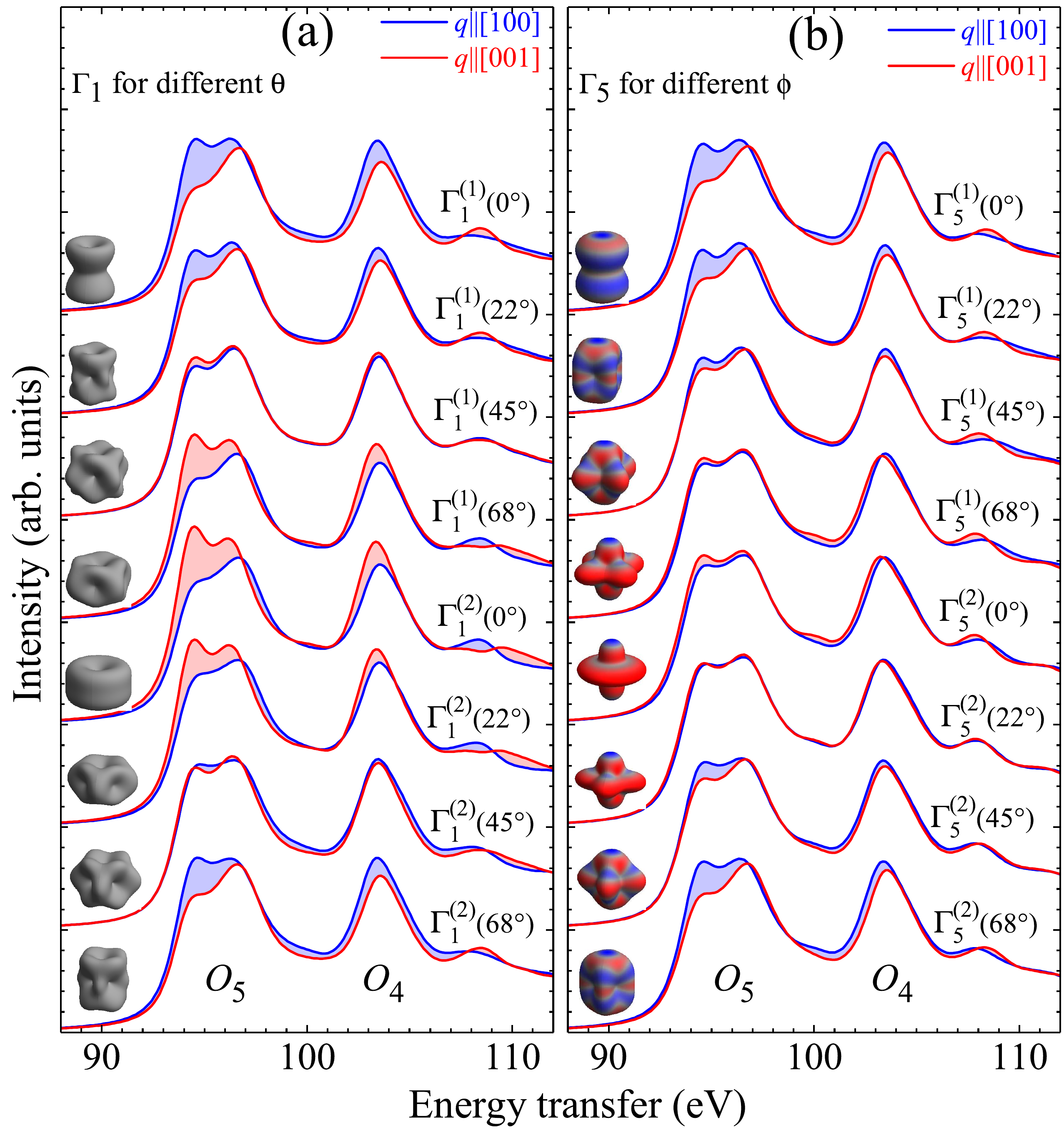}
  \caption{Calculations of the \NIXS{} \edge[U]{O}{4,5} edges as in \fig[b]{fig:URu2Si2_ac}, but for the different mixing of the two $\Gamma_1$ (a) and $\Gamma_5$ (b) states (see \fig{fig:URu2Si2scheme}).}
  \label{fig:URu2Si2_ac_mix}
\end{figure}

\Fig[a]{fig:URu2Si2_ac} shows the \NIXS{} data of URu$_2$Si$_2$ for the out-of-plane anisotropy comparing \qp{100} (blue) and \qp{001} (red).
The data shown are taken at \T{}\,=\,25\,K, i.e.\ above the \HO{} transition since they are not affected by any possible impact of the \HO{}.
There is a large anisotropy (red over blue) at both, the $O_5$ edge at $\approx$95\,eV and the $O_4$ edge at $\approx$104\,eV, that can be directly compared with the calculations.

\Fig[b]{fig:URu2Si2_ac} shows the calculations of S(\vecq{},$\omega$) of the \edge[U]{O}{4,5} edges ($5d\rightarrow5f$) transition for the nine \CF{} states of the \qnJ{}\,=\,4 ground state multiplet.
In the calculations $\theta$ and $\phi$ are chosen such that the anisotropies are maximum, i.e.\ for the extreme cases of "pure" $\abs{\qnJz{}}$ states.
The insets in \fig[b]{fig:URu2Si2_ac} show the respective two electron 5$f$ charge densities.

The out-of-plane anisotropy is very strong, so that a qualitative change due to the deviations in the asymmetric lineshape can be excluded, i.e.\ the "red-over-blue" characteristic is evident.
The $\Gamma_3$ and $\Gamma_4$ states, which show an opposite "blue-over-red" anisotropy can already be excluded at this point.
Also the $\Gamma_1^2(90^\circ)$ and $\Gamma_5^2(90^\circ)$ states can be excluded with the same argument.

\Fig[c]{fig:URu2Si2_ac} shows the experimental difference signal between the two directions I(\qp{100})-I(\qp{001}) for the experiment (black) and the remaining calculated states (green), which show the correct sign (red over blue) in \fig[b]{fig:URu2Si2_ac}.
The experimental difference amounts to 35\% of the isotropic peak hight R$_1$ (see \fig{fig:URu2Si2_iso}).
This size can be reproduced by the $\Gamma_1^1$(90$^\circ$) and the $\Gamma_2$ initial states.
The $\Gamma_5^1$(90$^\circ$) state only gives a dichroism of 9\%.
The calculations give an upper limit for the dichroism as experimental uncertainties like a misalignment or a oxidized surface layer may lower the difference.
Thus, the $\Gamma_5^1$(90$^\circ$) can be excluded as well.

\Fig{fig:URu2Si2_ac_mix} shows the out-of-plane anisotropy in the \edge[U]{O}{4,5} \NIXS{} spectra for the $\Gamma_1^{1,2}$($\theta$) (a) and $\Gamma_5^{1,2}$($\phi$) (b) initial state wave functions for values of $\theta$ and $\phi$ between 0 and 90$^\circ$.
The insets show the respective charge densities as in \fig[b]{fig:URu2Si2_ac}.
The charge densities of the "pure" states appear rotational invariant.
They do show lobes for $\theta$ and $\phi$\,$\neq$\,0 or 90$^\circ$.
All possible mixing of the states is realized between 0$^\circ$ and 90$^\circ$, because -$\Gamma_1^1$(90$^\circ$)\,=\,$\Gamma_1^2$(0$^\circ$) and $\Gamma_1^2$(90$^\circ$)\,=\,$\Gamma_1^1$(0$^\circ$) and the total sign of the wave function can not be observed.
The same holds for the $\Gamma_5^{(1,2)}$.
It is shown, that the anisotropy changes continuously and that the two scenarios at 0$^\circ$ and 90$^\circ$ yield the extreme cases.
The $\Gamma_5$ can here be excluded in general, because the small red over blue anisotropy will reduce further upon \qnJz{} mixing.
For the $\Gamma_1$ this means that a certain parameter range of the mixing around the $\Gamma_1^1(\theta\,=\,90^\circ$) is in agreement with the \NIXS{} experiment.
When taking the calculations as an upper limit once more, the \qnJz{} mixing can be limited to 70$^\circ$<$\theta$<110$^\circ$.
The range of the dichroism for this range of $\theta$ is indicated in \fig[c]{fig:URu2Si2_ac}.

The ratio of the strong anisotropy at the \edge[U]{O}{5} edge compared to the anisotropy at the \edge[U]{O}{4} edge is in good agreement with the calculation.
This supports the assumption that the anisotropy of the signal is not affected by the \edge[Si]{L}{2,3} signal.

\paragraph{In-plane direction of \vecq{} dependence}

\begin{figure}
  \centering
  \ifodd\value{page} \includegraphics[angle=90,width=0.9\textwidth]{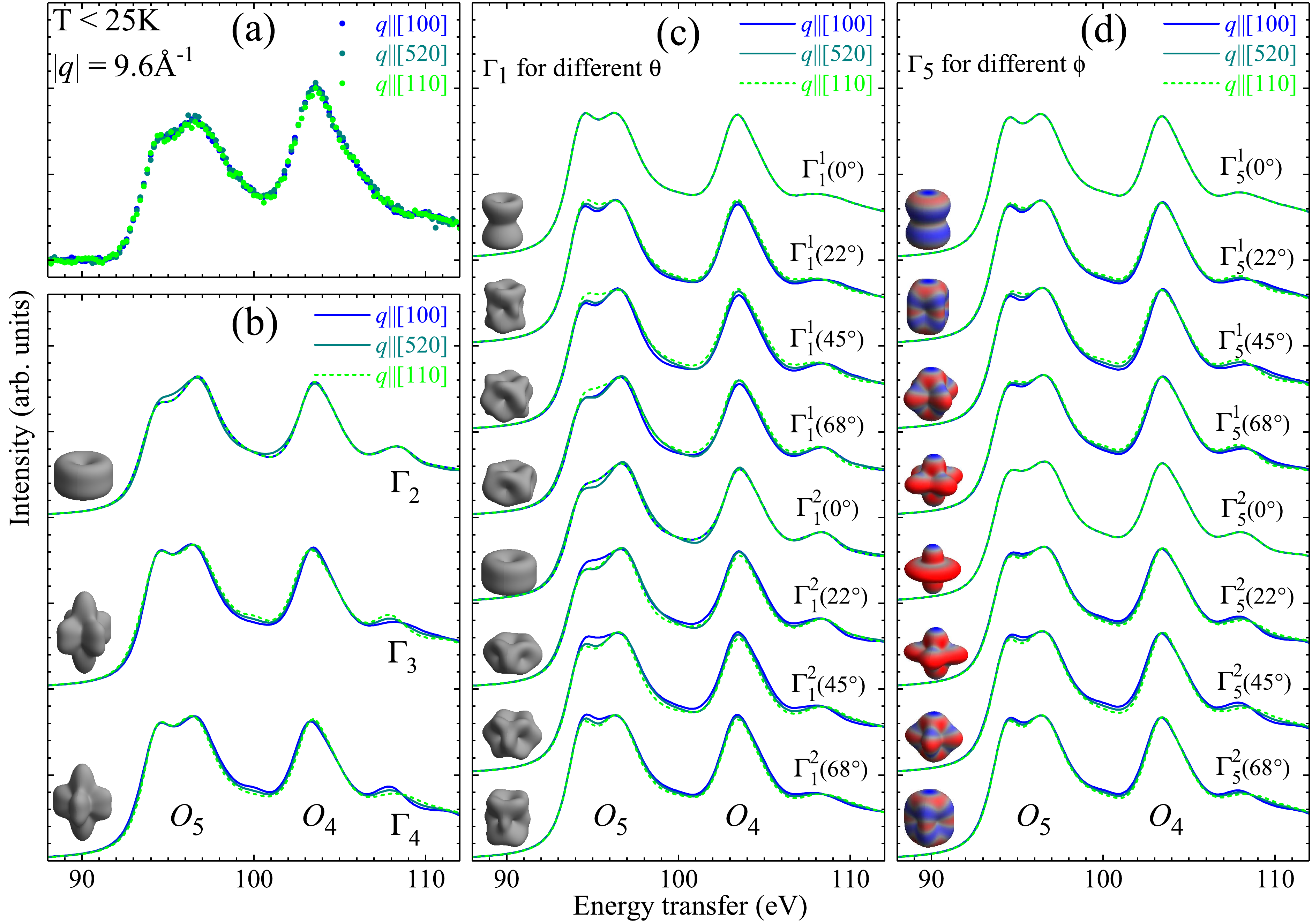} 
  \else \includegraphics[angle=-90,width=0.9\textwidth]{URu2Si2_aa.pdf} \fi
  \caption{URu$_2$Si$_2$ \NIXS{} spectra of the \edge[U]{O}{4,5} edges for momentum transfers \qp{100} (blue), \xyz{100} turned towards \xyz{010} by 22.5$^\circ$ (dark green) and by 45$^\circ$ (light green). (a) Experimental data (sum of 5\,K and 25\,K data). (b) Respective calculations of the unique singlets $\Gamma_2$, $\Gamma_3$, and $\Gamma_4$. (c+d) Calculations for the two $\Gamma_1$ singlets (c) and two $\Gamma_5$ doublets (d) for different \qnJz{} mixing parameters $\theta$ and $\phi$.}
  \label{fig:URu2Si2_aa}
\end{figure}

\Fig[a]{fig:URu2Si2_aa} shows data taken within the tetragonal 4-fold symmetry plane, for \qp{100} and for two directions rotated 22.5$^\circ$ and 45$^\circ$ towards \xyz{010}, labeled as \xyz{520} and \xyz{110}, respectively.
Neither below nor above the \HO{} order transition any anisotropy can be resolved within the statistical error bar.
Here the sum of the two identical data sets is shown.

\Fig[(b-d)]{fig:URu2Si2_aa} show the respective calculations for all \CF{} eigenstates $\Gamma_2$, $\Gamma_3$, and $\Gamma_4$ in (b), the two $\Gamma_1$ states for different \qnJz{} mixing in (c), and the two $\Gamma_5$ states for different \qnJz{} mixing in (d).
The expected directional dependencies are minor, so that the absence of any directional dependence in the data is in agreement with the calculations.

\paragraph{Temperature dependence}

\begin{SCfigure}
  \centering
  \includegraphics[width=0.55\textwidth]{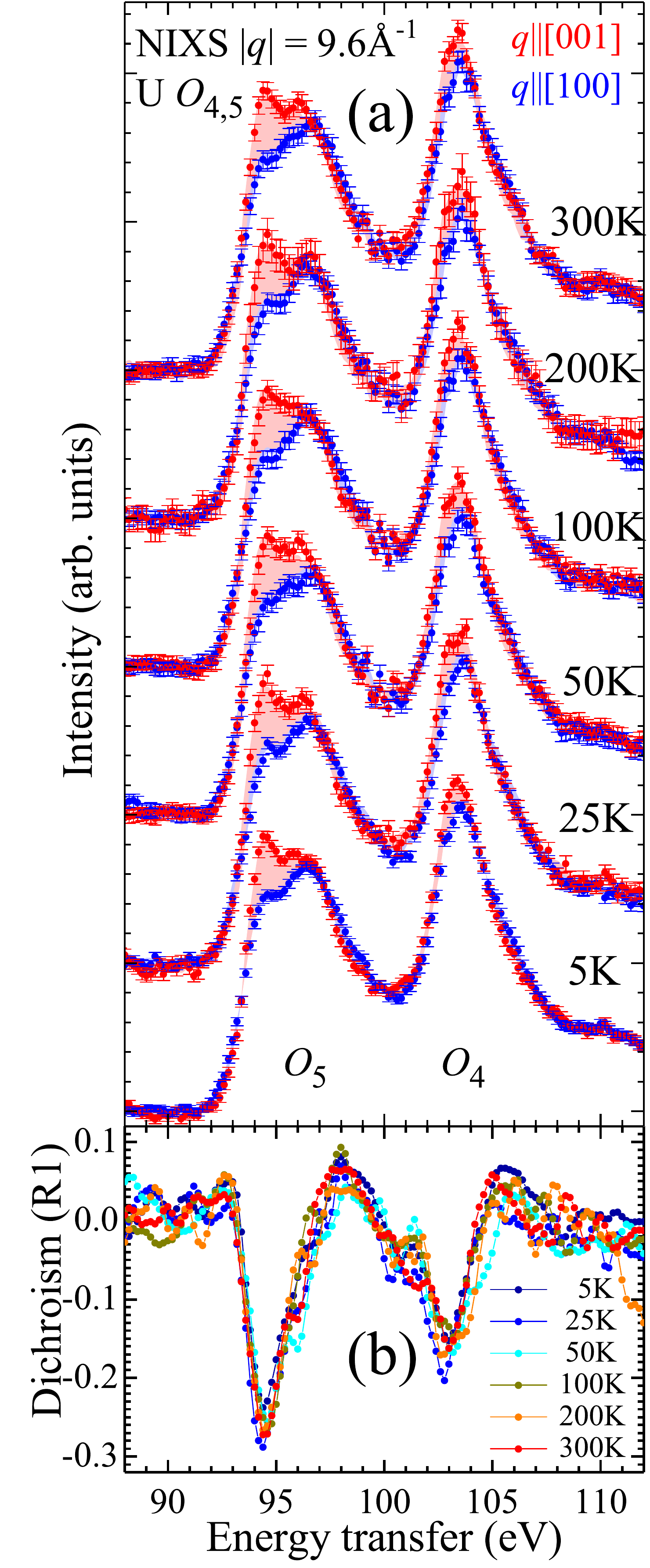}
  \caption{(a) Temperature dependence of the URu$_2$Si$_2$ \edge[U]{O}{4,5} \NIXS{} spectra for \qp{100} (blue dots) and \qp{001} (red dots). For better comparison, the \T{} dependent data are underlaid with the difference in spectral weight of the 5\,K data (light color filling). (b) Anisotropy I(\qp{100})-I(\qp{001}) for all temperatures, convoluted with a \FWHMG{}\,=\,0.5\,eV.}
  \label{fig:URu2Si2_T}
\end{SCfigure}

\Fig[a]{fig:URu2Si2_T} shows the \NIXS{} data of the \edge[U]{O}{4,5} edges for \qp{100} and \qp{001} for different temperatures between 5\,K and 300\,K.
Within the error bars, the spectra at all temperatures are identical, i.e.\ in the \HO{} state and up to room temperature.
This can be seen even better when comparing the difference signal I$_{\qp{100}}$-I$_{\qp{001}}$, as shown in \fig[b]{fig:URu2Si2_T}.

Assuming a thermal Boltzmann population of any state other than the $\Gamma_1^1$(70$^\circ$-90$^\circ$-110$^\circ$) or the $\Gamma_2$ state will change the directional dependence of the scattering due to their smaller or opposite difference signal.
From the absence of any changes in the spectra up to 300\,K it can be concluded that the ground state consists mainly of the $\Gamma_1^1$(70$^\circ$-90$^\circ$-110$^\circ$) or the $\Gamma_2$ singlet, or that one of the two singlets forms the ground state with the other state close in energy.
Within the experimental resolution, the thermal occupation of the $\Gamma_5^1$(90$^\circ$), $\Gamma_3$ and $\Gamma_4$ should be visible if the states were at less than 150\,K (13\,meV) above the ground state.
States with stronger opposite anisotropy must be even higher in energy.

\paragraph{Summary}

The \NIXS{} spectra of metallic URu$_2$Si$_2$ exhibit multiplets due to the excitonic character of the higher multipole transitions.
The \NIXS{} experiment was compared to an ionic U$^{4+}$ model with 5$f^2$ configuration and the isotropic data are well reproduced.
A huge out-of-plane anisotropy in the \NIXS{} spectra of the URu$_2$Si$_2$ \edge[U]{O}{4,5} edges has been observed.
This strong anisotropy is robust against the modification of the lineshape due to finite \CF{} effects, as the investigations in the previous chapter have shown.
The anisotropy is close to the largest theoretically possible dichroism of the $\Gamma_1^1$(90$^\circ$) and/or $\Gamma_2$ singlet states.
Other states can be excluded by this experiment and the $J_z$\,=\,0 contribution of the $\Gamma_1^1$($\theta$) could be limited to be $\leq$12\% (70$^\circ$<$\theta$<110$^\circ$).

\paragraph{Discussion}

Even if analyzed in an ionic manner, these results yield important restrictions for models which account for the intermediate valence.
\NIXS{} is sensitive to the total empty density of the $f$ states and consequently to the total $f$ occupied states.
The findings therefore show that the valence fluctuations involve mainly the U$^{4+}$ (5$f^2$) $\Gamma_1^1$(90$^\circ$) and/or $\Gamma_2$ singlet states.

The data do not exhibit any temperature dependence, neither across the \HO{} phase transition nor in the temperature interval up to 300\,K.
The latter gives constraints to the proximity of the next higher excited states.
Any state other than the $\Gamma_1^1$ or $\Gamma_2$ must be at least 13\,meV above the ground state.

These findings are in good agreement with DMFT calculations\,\cite{Haule2009}, which find the two singlet states $\Gamma_2$ and $\Gamma_1^2$($\approx$45$^\circ$) to be the two low lying states\,\cite{Haule2009}.
Linear polarized \XAS{} data at the \edge[U]{O}{4,5} edges\,\cite{Wray2015} also agree with our findings in the sense that both, the \NIXS{} and \XAS{} dichroism, rule out the  $\Gamma_1^2$(90$^\circ$), the $\Gamma_3$ and the $\Gamma_4$ as possible ground states and find no temperature dependence across the \HO{} transition.
The smaller directional dependence in the \XAS{} data that let the authors assign the $\Gamma_5^1$ doublet as ground state must be due to the higher surface sensitivity of the \XAS{} experiment.

The \NIXS{} results rule out many possibilities of different approaches to explain the \HO{} scenario in URu$_2$Si$_2$.

\clearpage

\renewcommand{\theequation}{\arabic{section}.\arabic{equation}}
\renewcommand{\thetable}{\arabic{section}.\arabic{table}}
\renewcommand{\thefigure}{\arabic{section}.\arabic{figure}}
\numberwithin{equation}{section}	
\numberwithin{table}{section}
\numberwithin{figure}{section}

\section{Summary and conclusions}
Knowledge of the wave functions that participate in ground-state formation is crucial for understanding the physics of strongly correlated $f$-electron compounds.
In this thesis non-resonant inelastic x-ray scattering (\NIXS{}) has been established as complementary method to inelastic neutron scattering and linearly polarized x-ray absorption.

\NIXS{} is a photon-in photon-out technique using hard x-rays without an intermediate state.
It has the following benefits:
\begin{itemize}
\item \NIXS{} experiments are bulk sensitive so that the samples do not need to be cleaved, and advanced sample environments (e.g.\ pressure cells) are also possible.
\item The scattering function in \NIXS{} can be calculated quantitatively in a straight forward manner.
\end{itemize}
The high incident energies allow for large momentum transfers when working at high scattering angles, and at large \absq{} higher multipole transitions give rise to new spectroscopic features.
\begin{itemize}
\item Dipole forbidden transitions can be observed.
\item These higher multipole features are more excitonic so that even metallic uranium exhibits multiplets.
\item Anisotropies with higher that twofold rotational symmetries can be measured.
\end{itemize}

Within the framework of this thesis it has been shown that self-absorption effects do not occur so that the alignment of the sample surface can be chosen freely.
The experimental conditions at the \NIXS{} beamline P01 at DESY have been optimized accordingly.
A new cryostat allows rotating the sample angle inside the cryostat at low temperatures ($\lesssim$5\,K).
New, specially prepared windows are able to withstand the hard x-ray beam, and ensure stable measurement conditions during the long counting times.
For faster scans a continuous scan has been integrated into the P01 beamline.
For data analysis a tensor notation has been introduced in analogy to the conductivity tensor (see \chap{sec:scatteringtensor}).

The \vecq{}-directional dependence of \NIXS{} spectra of 4$f$ (RE-$N$-edge, 4$d$\,$\to$\,4$f$) and 5$f$ (U-$O$-edge, 5$d$\,$\to$\,5$f$) materials has been investigated and the results show that \NIXS{} is a very useful complementary method for finding the ground-state symmetries of materials where, for example, inelastic neutron scattering or x-ray absorption yield only limited information.
Two \textit{benchmark} materials were measured, CeB$_6$ and UO$_2$, where the \NIXS{} results could be compared with previously published inelastic neutron scattering results.
The \NIXS{} data of cubic CeB$_6$ unambiguously confirm the quartet $\Gamma_8$, and the \NIXS{} data of cubic UO$_2$ confirm the triplet $\Gamma_5$ ground state.
The \NIXS{} data even support the \CF{} splittings that were reported from inelastic neutron scattering.
In the case of CeB$_6$ the temperature dependence of the directional dependent signal is conform with a 46\,meV splitting.
For UO$_2$ the spectral shape of the low temperature \NIXS{} data supports the finite \CF{} splitting reported by Caciuffo \textit{et al.} and excludes the so-called strong \CF{} scenario where a \CF{} splitting of several hundred meV would have led to a strong intermixing of multiplets and to a very different energy distribution of the multiplets.
The results of CeB$_6$ and UO$_2$ confirm that \NIXS{} is applicable to cubic materials and that the signals are truly excitonic when measured beyond the dipole limit.

Furthermore, the advantages of \NIXS{} were utilized to solve open questions in three correlated systems, SmB$_6$, CeRu$_4$Sn$_6$ and URu$_2$Si$_2$.

In the strongly intermediate valent (60\% Sm$^{3+}$, 40\% Sm$^{2+}$) cubic Kondo insulator SmB$_6$ \NIXS{} finds a $\Gamma_8$ quartet ground state for the Sm$^{3+}$ configuration -- in contradiction to many density functional theory calculations that find the hole of the \termsymbol{6}{H}{5/2} multiplet in the $\Gamma_7$ doublet.
The \NIXS{} result is convincing since with \NIXS{} the ground-state question is reduced to a simple yes/no experiment and because the signal of the Sm$^{2+}$ component does not contribute to the directional dependence -- in contrast to neutron scattering where the superposition of both configurations is always measured.
This result for the bulk is relevant for identifying the properties of the surface states of this putative strongly correlated topological insulator.

Tetragonal CeRu$_4$Sn$_6$ also exhibits a hybridization gap.
Here \NIXS{} confirms the \XAS{} results of a $\Gamma_6$ ground state.
The \XAS{} results were questionable because \XAS{} is surface sensitive and the dichroism observed in the experiment was not sufficient to reproduce the expected dichroism for a state with $J_z$\,=\,1/2.
Bulk-sensitive \NIXS{}, however, finds the same result, i.e.\ a dichroism that is smaller than the single ion expectation.
It is therefore concluded that the reduced dichroism is a bulk feature and has its origin in strong hybridization effects that mix a higher lying crystal-field state into the ground state, thus reducing the directional effect.

The isotropic \NIXS{} data of URu$_2$Si$_2$ are well described by the U\,$f^2$ configuration, and the strong directional dependence of the \edge[U]{O}{4,5}-edges of this presumably itinerant compound is well described with a crystal-field state of dominantly ($\left|-4\right\rangle$\,$\pm$\,$\left|+4\right\rangle$)/$\sqrt{2}$ character.
This means, either the singlet $\Gamma_1^1$ or $\Gamma_2$, or both close in energy form the ground state.
The lack of temperature dependence in the dichroism sets a lower limit for size of the \CF{} splitting.
The \NIXS{} result of URu$_2$Si$_2$ rules out many of the \CF{} scenarios for forming the \HO{} phase.

Based on the results of this thesis, spectroscopy of \NIXS{} core-levels has been established as a new tool to study further materials.
The results prove the validity of the technique and its analysis and in the future \NIXS{} will also be used for mapping out orbitals of transition metals.
Initial tests on NiO have already been carried out successfully.
An improvement of the resolution from 700\,meV to the lifetime limit of $\approx$\,400\,meV is in the planning by changing from a Rowland circle with a 1\,m diameter to one with a 2\,m diameter.
The motorization of the scattering angles of the sample and of the analyzers is also foreseen in the near future.
This will allow one to adjust \absq{}, e.g.\ when measuring transition metals where we expect the intensity of the higher multipoles to peak at lower \absq{}.

\clearpage
\sectioff{Appendix}
\counterwithin*{equation}{subsection}
\counterwithin*{table}{subsection}
\counterwithin*{figure}{subsection}
\renewcommand\thesection{\Alph{subsection}}
\renewcommand\thesubsection{\Alph{subsection}}
\subsection{Atomic values}\label{app_atomicvalues}

\begin{table}[htpb]
  \centering
  \caption{Atomic values in units of 1\,Rydberg ($\approx$\,13.6\,eV).\label{tab:atomicvalues}}
  \begin{tabular*}{\columnwidth}{@{\extracolsep{\fill}}|c|c|l|l|l|l|l|}
	\hline
	&			   & \multicolumn{1}{c|}{Ce$^{3+}$} & \multicolumn{1}{c|}{Sm$^{3+}$} & \multicolumn{1}{c|}{Sm$^{2+}$} & \multicolumn{1}{c|}{Eu$^{3+}$} & \multicolumn{1}{c|}{U$^{4+}$} \\
	&			   &\multicolumn{1}{c|}{$n$=4$,i$=1}&\multicolumn{1}{c|}{$n$=4$,i$=5}&\multicolumn{1}{c|}{$n$=4$,i$=6}&\multicolumn{1}{c|}{$n$=4$,i$=6}&\multicolumn{1}{c|}{$n$=5$,i$=2} \\
	\hline
	\multirow{ 4}{*}{$nf^i$}
	& $\zeta_{v}$  & 0.00639   & 0.01139   & 0.01034   & 0.01289   & 0.01926   \\
	& F$_{vv}^{2}$ & \multicolumn{1}{c|}{\---} & 1.0027339 & 0.9197937 & 1.0347590 & 0.6989795 \\
	& F$_{vv}^{4}$ & \multicolumn{1}{c|}{\---} & 0.6293212 & 0.5734748 & 0.6494242 & 0.4572877 \\
	& F$_{vv}^{6}$ & \multicolumn{1}{c|}{\---} & 0.4527923 & 0.4115580 & 0.4672594 & 0.3357035 \\
	\hline
	\multirow{ 10}{*}{$nd^9nf^{i+1}$}
	& $\zeta_{v}$  & 0.00671   & 0.01167   & 0.01068   & 0.01317   & 0.02019   \\
	& $\zeta_{c}$  & 0.09144   & 0.13476   & 0.13338   & 0.14746   & 0.23593   \\
	& F$_{vv}^{2}$ & 0.8799247 & 1.0163742 & 0.9393731 & 1.0474354 & 0.7134732 \\
	& F$_{vv}^{4}$ & 0.5525398 & 0.6384169 & 0.5864769 & 0.6578794 & 0.4675946 \\
	& F$_{vv}^{6}$ & 0.3976295 & 0.4594932 & 0.4211133 & 0.4734889 & 0.3436047 \\
	& F$_{cv}^{2}$ & 1.0002835 & 1.1421782 & 1.0849437 & 1.1746882 & 0.7826160 \\
	& F$_{cv}^{4}$ & 0.6380435 & 0.7300506 & 0.6897514 & 0.7509356 & 0.5032160 \\
	& G$_{cv}^{1}$ & 1.1873781 & 1.3503371 & 1.2808760 & 1.3872282 & 0.9223698 \\
	& G$_{cv}^{3}$ & 0.7410119 & 0.8469718 & 0.7994772 & 0.8708715 & 0.5706181 \\
	& G$_{cv}^{5}$ & 0.5228355 & 0.5987025 & 0.5640314 & 0.6157889 & 0.4072854 \\
    \hline
  \end{tabular*}
\end{table}
\clearpage

\subsection{Definitions of the bases}\label{app_basesdefinitions}
\begin{align}
\intertext{Renormalized spherical harmonics:\vspace{-3mm}}
\Clm{$l$}{$m$} &= \sqrt{\frac{(l - m)!}{(l + m)!}} ~ L_l^{\abs{m}}(\cos{\theta}) \cdot \exp{\imath m \phi}
\intertext{Renormalized tesseral harmonics:\vspace{-3mm}}
\Dlm{$l$}{$m$} &= \sqrt{\frac{(l - m)!}{(l + m)!}} ~ L_l^{\abs{m}}(\cos{\theta}) \cdot \begin{cases}
\sqrt{2} \sin{m \phi} & m<0 \\
~ 1 & m=0\\
\sqrt{2} \cos{m \phi} & m>0 \\
\end{cases}
\intertext{Spherical harmonics:\vspace{-3mm}}
\Ylm{$l$}{$m$} &= \sqrt{\frac{2l+1}{4\pi}} ~ \Clm{$l$}{$m$}
\intertext{Tesseral harmonics:\vspace{-3mm}}
\Zlm{$l$}{$m$} &= \sqrt{\frac{2l+1}{4\pi}} ~ \Dlm{$l$}{$m$}
\intertext{Here \(L_l^m(x)\) is given by the Legendre polynomials \Legendre{l}{x}:\vspace{-3mm}}
L_l^m(x) &= (-1)^m (1-x^2)^{m/2} \frac{\diff^m}{\diff x^m} \Legendre{l}{x}.
\end{align}
\begin{minipage}{\textwidth}
  \centering
  \captionof{table}{Renormalized cubic harmonic functions \Elm{$l$}{$i$} used in this thesis as defined in \Quanty{}. The cubic harmonic functions are given by \Klm{$l$}{$i$}\,=\,$\sqrt{\frac{2l+1}{4\pi}}$\,\Elm{$l$}{$i$}.\label{tab:cubicbasis}}
  \begin{tabular}{|c|c|c|c|c|}
	\hline
	$i$ & \Elm{1}{$i$} & \Elm{2}{$i$} & \Elm{3}{$i$} & \Elm{5}{$i$} \\
	\hline
	\phantom{1}1 & x & $\frac{\sqrt{3}}{2}(x^2$-$y^2)$ & $\sqrt{15}$xyz                    & $\frac{\sqrt{105}}{2}(3z^2$-$r^2)xyz$ \\
	\phantom{1}2 & y & $\frac{1}{2}(3z^2$-$r^2)$       & $\frac{1}{2}(5x^3$-$3xr^2)$       & $\frac{\sqrt{315}}{2}(x^2$-$y^2)xyz$ \\
	\phantom{1}3 & z & $\sqrt{3}yz$                    & $\frac{1}{2}(5y^3$-$3yr^2)$       & $\frac{1}{8}(63x^4$-$70x^2r^2$+$15r^4)x$ \\
	\phantom{1}4 &   & $\sqrt{3}xz$                    & $\frac{1}{2}(5z^3$-$3zr^2)$       & $\frac{1}{8}(63y^4$-$70y^2r^2$+$15r^4)y$ \\
	\phantom{1}5 &   & $\sqrt{3}xy$                    & $\frac{\sqrt{15}}{2}(y^2$-$z^2)x$ & $\frac{1}{8}(63z^4$-$70z^2r^2$+$15r^4)z$ \\
	\phantom{1}6 &   &                                 & $\frac{\sqrt{15}}{2}(z^2$-$x^2)y$ & $\frac{3}{8}(y^4$+$z^4$-$6y^2z^2)x$ \\
	\phantom{1}7 &   &                                 & $\frac{\sqrt{15}}{2}(x^2$-$y^2)z$ & $\frac{3}{8}(x^4$+$z^4$-$6x^2z^2)y$ \\
	\phantom{1}8 &   &                                 &                                   & $\frac{3}{8}(x^4$+$y^4$-$6x^2y^2)z$ \\
	\phantom{1}9 &   &                                 &                                   & $\frac{\sqrt{105}}{4}(3x^2$-$r^2)(y^2$-$z^2)x$ \\
	10 &   &                                 &                                   & $\frac{\sqrt{105}}{4}(3y^2$-$r^2)(z^2$-$x^2)y$ \\
	11 &   &                                 &                                   & $\frac{\sqrt{105}}{4}(3z^2$-$r^2)(x^2$-$y^2)z$ \\
    \hline
  \end{tabular}
  \vspace{-1cm}
\end{minipage}
\clearpage

\subsection{Wigner 3j-Symbols}\label{app_ThreeJSymbols}
The integrals over three wave functions in spherical harmonics are mathematically expressed by the 3j-Symbols\,\cite{Cowan1981}
\begin{align}
\langle l_{m \sigma}|C^{(k)}_{m_k}|l'_{m' \sigma}\rangle &= (-1)^m \sqrt{(2l+1)(2l'+1)} \left(
{\setlength\arraycolsep{2pt}\setstretch{1}\begin{array}{ccc}
 l & k & l' \\
 0 & 0 & 0 \\
\end{array}}
\right) \left(
{\setlength\arraycolsep{2pt}\setstretch{1}\begin{array}{ccc}
 l & k & l' \\
 -m & m_k & m' \\
\end{array}}
\right). \label{eq:threejsymbols} \\
\intertext{For those more familiar with the Clebsch Gordon coefficients: They can similarly be expressed by\,\cite{Mehrem2011}}
\langle l_{m \sigma}|C^{(k)}_{m_k}|l'_{m' \sigma}\rangle &= 
\sqrt{\frac{2l+1}{2l'+1}} \langle l'\,k\,0\,0\,|\,l\,0\rangle \langle l\,k\,m\,m_k\,|\,l'\,m'\rangle. \\
\intertext{For evaluation of the 3j-Symbols the Racah formula has been used\,\cite{Weisstein3jsymbols}}
\left({\setlength\arraycolsep{2pt}\setstretch{1}\begin{array}{ccc}
 l_1 & l_2 & l_3 \\
 m_1 & m_2 & m_3 \\
\end{array}}\right)
&= -1^{l_1-l_2-m_3} * \sqrt{{\frac{(l_1+l_2-l_3)!(l_1-l_2+l_3)!(-l_1+l_2+l_3)!}{(l_1+l_2+l_3+1)!}}} \nonumber\\
&* \sqrt{(l_1{+}m_1)!(l_1{-}m_1)!(l_2{+}m_2)!(l_2{-}m_2)!(l_3{+}m_3)!(l_3{-}m_3)!}\\
&* \sum\limits_t \frac{-1^t}{t!(t{-}l_2{+}m_1{+}l_3)!(t{-}l_1{-}m_2{+}l_3)!(l_1{+}l_2{-}l_3{-}t)!(l_1{-}m_1{-}t)!(l_2{+}m_2{-}t)!} \nonumber
\end{align}
summing over all integer $t$ which fulfill max(0, $l_2$-$m_1$-$l_3$, $l_1$+$m_2$-$l_3$) $\leq t \leq$ min($l_1$+$l_2$-$l_3$, $l_1$-$m_1$, $l_2$+$m_2 )$.
For this the following lua code snippet can be used, with the list f[n] containing the factorials n!:
\begin{lstlisting}[basicstyle=\scriptsize]
function ThreeJSymbol(l1, m1, l2, m2, l3, m3)
  if 0 ~= m1+m2+m3 then return 0 end
  if l2 > l1+l3 then return 0 end
  if l2 < math.abs(l1-l3) then return 0 end
  if l1 < math.abs(m1) then return 0 end
  if l2 < math.abs(m2) then return 0 end
  if l3 < math.abs(m3) then return 0 end
  local val  = 0
  local t1   = l2 - m1 - l3
  local t2   = l1 + m2 - l3
  local t3   = l1 + l2 - l3
  local t4   = l1 - m1
  local t5   = l2 + m2
  local tmin = math.max(  0, t1, t2 )
  local tmax = math.min( t3, t4, t5 )
  for t = tmin, tmax, 1 do
    val = val + (1-2*(t%2)) / ( f[t]*f[t-t1]*f[t-t2]*f[t3-t]*f[t4-t]*f[t5-t] )
  end
  val = val * (1-2*((l1-l2-m3)%2))
            * math.sqrt( f[l1+l2-l3]*f[l1-l2+l3]*f[-l1+l2+l3]/f[l1+l2+l3+1] )
            * math.sqrt( f[l1+m1]*f[l1-m1]*f[l2+m2]*f[l2-m2]*f[l3+m3]*f[l3-m3] )
  return val
end
\end{lstlisting}
\clearpage

\subsection{Stevens formalism}\label{app_stevensformalism}

Here a short connection to the Stevens formalism from Ref\,\cite{Stevens1952} will be given, summarizing Ref.\,\cite{BauerRotter2010}.
The Hamiltonian is given by
\begin{align}
H_\text{CEF} &= \sum\limits_{k=0}^{\infty} \sum\limits_{m=-k}^{k} B_{k,m} ~ O_k^m(\hat{\textbf{J}}) &\left( = \sum\limits_{k=0}^{\infty} \sum\limits_{m=-k}^{k} c \cdot B'_{k,m} ~ \langle r^k \rangle ~ O_k^m(\hat{\textbf{J}}) \right) \label{eq:CEFstevens}\\
\langle r^k \rangle &= \int_{0}^{\infty} \abs{\Rnl{\qnn{}}{\qnl{}}}^2 \, r^{k+2} \, \diff r.
\end{align}
Note, that the Stevens parameters are sometimes given with a certain prefactor $c$ or normalized to \( \langle r^k \rangle \) with the radial contribution of the wavefunction \Rnl{\qnn{}}{\qnl{}}.
The Stevens operators \(O_k^m\) are defined analog to the renormalized spherical harmonics \Clm{k}{m}, but on the basis of tesseral harmonic functions \Zlm{k}{m} in the space of $\hat{\textbf{J}}$ instead of $\hat{\textbf{r}}$
\begin{align}
Z_l^m &\equiv \begin{cases}
 \frac{\imath}{\sqrt{2}} ( Y_l^{-\abs{m}} - (-1)_{}^{\abs{m}} Y_l^{\abs{m}} ) & m<0 \\
 Y_l^0 & m=0 \\
 \frac{1}{\sqrt{2}} ( Y_l^{-\abs{m}} + (-1)_{}^{\abs{m}} Y_l^{\abs{m}} ) & m>0
\end{cases}
\end{align}
Also some unconventional prefactors \(p_{k,m}\) (see \tab{tab:pkm}) appear to transform between the two bases accounting for the \textit{missing} normalization of the Stevens operators.
\begin{align}
O_k^m(\hat{\textbf{J}}) &\equiv \left( p_{k,m} ~ \right)^{-1} ~ Z_k^m(\hat{\textbf{J}})  =  \left( p_{k,m} ~ \Theta_k \right)^{-1} \, Z_k^m(\hat{\textbf{r}})
\end{align}
\(\Theta_k\) represents the Stevens factors transforming between $\hat{\textbf{J}}$ and $\hat{\textbf{r}}$ (see e.g.\ Ref.\,\cite[Tab.\,1]{Stevens1952} for rare earth and Ref.\,\cite{Amoretti1984} for U and Np with \(\Theta_2\equiv\alpha\), \(\Theta_4\equiv\beta\), \(\Theta_6\equiv\gamma\)).

Comparing \equ{eq:CEFstevens} with \equ{eq:CEFpotential}, one finally finds the relations
\begin{align}
  A_{k,m} &= \left( \sqrt{\frac{4\pi}{2k+1}} ~ p_{k,m} ~ \Theta_k \right)^{-1} \cdot \begin{cases}
    \frac{1}{\sqrt{2}} \left( B_{k,\abs{m}} + \imath B_{k,-\abs{m}} \right) & m < 0\\
    B_{k,m} & m=0\\
    \frac{(-1)^m}{\sqrt{2}} \left( B_{k,\abs{m}} - \imath B_{k,-\abs{m}} \right) & m > 0
  \end{cases}\\
\intertext{and vise versa}
  B_{k,m} &= ~ \sqrt{\frac{4\pi}{2k+1}} ~ p_{k,m} ~ \Theta_k \, \cdot \begin{cases}
    \frac{\imath}{\sqrt{2}} \left( (-1)^{\abs{m}} A_{k,-\abs{m}} - A_{k,\abs{m}} \right) & m < 0\\
    A_{k,m} & m=0\\
    \frac{1}{\sqrt{2}} \left( A_{k,-\abs{m}} + (-1)^{\abs{m}} A_{k,\abs{m}} \right) & m > 0
  \end{cases}
\end{align}
\begin{table}
  \caption{List of the prefactors \( \sqrt{4\pi/(2k+1)} \, p_{k,m}\) obtained from comparing the definitions of \(O_k^m(J)\) and \(Z_k^m(r)\) in Ref.\,\cite[A.1 and A.2]{BauerRotter2010}.}
  \label{tab:pkm}
  \centering
  \begin{tabular}{l|ccccccc}
  \raisebox{-8pt}{k} \hspace{-20pt} \raisebox{-12pt}{\scalebox{3}{$\diagdown$}} \hspace{-24pt} m \vspace{-4pt} \hspace{-8pt} & 0 & $\pm$1 & $\pm$2 & $\pm$3 & $\pm$4 & $\pm$5 & $\pm$6 \\
   \hline
   0 &              1 &                        &                          &                          &                         &                          &                          \\
   1 &  $\frac{1}{2}$ &                      1 &                          &                          &                         &                          &                          \\
   2 &  $\frac{1}{2}$ &             $\sqrt{3}$ &     $\sqrt{\frac{3}{4}}$ &                          &                         &                          &                          \\
   3 &  $\frac{1}{2}$ &   $\sqrt{\frac{3}{8}}$ &    $\sqrt{\frac{15}{4}}$ &     $\sqrt{\frac{5}{8}}$ &                         &                          &                          \\
   4 &  $\frac{1}{8}$ &   $\sqrt{\frac{5}{8}}$ &    $\sqrt{\frac{5}{16}}$ &    $\sqrt{\frac{35}{8}}$ &  $\sqrt{\frac{35}{64}}$ &                          &                          \\
   5 &  $\frac{1}{8}$ & $\sqrt{\frac{15}{64}}$ &  $\sqrt{\frac{105}{16}}$ &  $\sqrt{\frac{35}{128}}$ & $\sqrt{\frac{315}{64}}$ &  $\sqrt{\frac{63}{128}}$ &                          \\
   6 & $\frac{1}{16}$ & $\sqrt{\frac{21}{64}}$ & $\sqrt{\frac{105}{512}}$ & $\sqrt{\frac{105}{128}}$ & $\sqrt{\frac{63}{256}}$ & $\sqrt{\frac{693}{128}}$ & $\sqrt{\frac{231}{512}}$ \\
  \end{tabular}
\end{table}

since \textit{w.l.o.g.\ }for $m>0$:
\begin{align}
  a_{k,-m} Y_k^{-m} + a_{k,m} Y_k^{m} &= b_{k,-m} Z_k^{-m} + b_{k,m} Z_k^{m} \nonumber\\
  &= b_{k,-m} \frac{\imath}{\sqrt{2}}(Y_k^{-m} - (-1)^m Y_k^{m}) + b_{k,m} \frac{1}{\sqrt{2}}(Y_k^{-m} + (-1)^m Y_k^{m}) \nonumber\\
  &= \frac{1}{\sqrt{2}} (b_{k,m} + \imath \, b_{k,-m}) Y_k^{-m} + \frac{(-1)^m}{\sqrt{2}} (b_{k,m} - \imath \, b_{k,-m}) Y_k^{m} \nonumber \\
\intertext{and}
  b_{k,-m} Z_k^{-m} + b_{k,m} Z_k^{m} &= a_{k,-m} Y_k^{-m} + a_{k,m} Y_k^{m} \nonumber\\
  &= a_{k,-m} \frac{1}{\sqrt{2}}(Z_k^{m} - \imath Z_k^{-m}) + a_{k,m} \frac{(-1)^m}{\sqrt{2}}(Z_k^{m} + \imath Z_k^{-m}) \nonumber\\
  &= \frac{\imath}{\sqrt{2}} ((-1)^m a_{k,m} - a_{k,-m}) Z_k^{-m} + \frac{1}{\sqrt{2}} ((-1)^m a_{k,m} + a_{k,-m}) Z_k^{m}. \nonumber
\end{align}

\clearpage

\subsection{D\textsub{4h} character table, double group and representations}\label{app_chartabD4h}
\begin{minipage}{\textwidth}
  \captionof{table}{Character table for D\textsub{4h} point group appended by the double group ($\Gamma_6$ and $\Gamma_7$) reproduced from Ref.\,\cite{Ballhausen1962}.\label{tab:ctD4h}}
  \begin{tabular}{ l | r r r r r r r | l }
             &  E &  R &  I & C$_4^{[001]}$ & C$_2^{[001]}$ & C$_2^{[100]}$ & C$_2^{[110]}$ & \\
	\hline
	A$_{1g}$ &  1 &  1 &  1 &     1 &     1 &      1 &       1 & $\Gamma_1$ \\
	A$_{2g}$ &  1 &  1 &  1 &     1 &     1 &     -1 &      -1 & $\Gamma_2$ \\
	B$_{1g}$ &  1 &  1 &  1 &    -1 &     1 &      1 &      -1 & $\Gamma_3$ \\
	B$_{2g}$ &  1 &  1 &  1 &    -1 &     1 &     -1 &       1 & $\Gamma_4$ \\
	E$_{1g}$ & \hphantom{+}2 & \hphantom{+}2 & \hphantom{+}2 &     0 &    -2 &      0 &       0 & $\Gamma_5$ \\
	\hline
	A$_{1u}$ &  1 &  1 & -1 &     1 &     1 &      1 &       1 & $\Gamma_1$ \\
	A$_{2u}$ &  1 &  1 & -1 &     1 &     1 &     -1 &      -1 & $\Gamma_2$ \\
	B$_{1u}$ &  1 &  1 & -1 &    -1 &     1 &      1 &      -1 & $\Gamma_3$ \\
	B$_{2u}$ &  1 &  1 & -1 &    -1 &     1 &     -1 &       1 & $\Gamma_4$ \\
	E$_{1u}$ &  2 &  2 & -2 &     0 &    -2 &      0 &       0 & $\Gamma_5$ \\
	\hline
	E$_{2}$  &  2 & -2 &  0 &$\pm \sqrt{2}$& 0 &   0 &       0 & $\Gamma_6$ \\
	E$_{3}$  &  2 & -2 &  0 &$\mp \sqrt{2}$& 0 &   0 &       0 & $\Gamma_7$ \\
  \end{tabular}
  
  \vspace{20pt}
  
  \captionof{table}{Irreducible representations in D\textsub{4h} symmetry.\label{tab:irD4h}}
  \begin{tabular}{ r | r r r r r r r | l }
 $M$ &  E &  R &  I  & C$_4^{[001]}$ & C$_2^{[001]}$ & C$_2^{[100]}$ & C$_2^{[110]}$ & irreducible representation \\
	\hline
	  0   &  1 &  1 &   1 &     1 &     1 &      1 &     1 & A$_{1g}$ \\
	  2   &  5 &  5 &   5 &    -1 &     1 &      1 &     1 & A$_{1g}$+B$_{1g}$+B$_{2g}$+E$_{g}$ \\
	 $\vdots$ &  &  &     &       &       &        &       & \\
	  +4  & +8 & +8 &  +8 &    +0 &    +0 &     +0 &    +0 & +A$_{1g}$+A$_{2g}$+B$_{1g}$+B$_{2g}$+2E$_{g}$ \\
	\hline            
	  1   &  3 &  3 &  -3 &     1 &    -1 &     -1 &    -1 & A$_{2u}$+E$_{u}$ \\
	  3   &  7 &  7 &  -7 &    -1 &    -1 &     -1 &    -1 & A$_{2u}$+B$_{1u}$+B$_{2u}$+2E$_{u}$ \\
	 $\vdots$ &  &  &     &       &       &        &       & \\
	  +4  & +8 & +8 &  -8 &    +0 &    +0 &     +0 &    +0 & +A$_{1u}$+A$_{2u}$+B$_{1u}$+B$_{2u}$+2E$_{u}$ \\
	\hline
	 1/2  &  2 & -2 &   0 &$\pm\sqrt{2}$&0&      0 &     0 & E$_{2}$ \\
	 3/2  &  4 & -4 &   0 &     0 &     0 &      0 &     0 & E$_{2}$+E$_{3}$ \\
	 5/2  &  6 & -6 &   0 &$\mp\sqrt{2}$&0&      0 &     0 & E$_{2}$+2E$_{3}$ \\
	 7/2  &  8 & -8 &   0 &     0 &     0 &      0 &     0 & 2E$_{2}$+2E$_{3}$ \\
	 $\vdots$ &  &  &     &       &       &        &       & \\
	  +4  & +8 & -8 &  +0 &    +0 &    +0 &     +0 &    +0 & +2E$_{2}$+2E$_{3}$ \\
  \end{tabular}
\end{minipage}

\vfill
\pagebreak

\subsection{D\textsub{2d} character table, double group and representations}\label{app_chartabD2d}
In D\textsub{2d} the character table is the similar to D\textsub{4h} but without the inversion I and with S$_4^{[001]}$($\equiv$I$\cdot$C$_4^{[00\bar{1}]}$) and S$_1^{[110]}$($\equiv$I$\cdot$C$_2^{[110]}$) replacing C$_4^{[001]}$ and C$_2^{[110]}$, respectively.\cite{chartabonline}
Consequently, the even and odd irreducible representations merge, but the double group is unchanged.

\begin{table}[htpb]
  \caption{Character table for D\textsub{2d} point group appended by the double group ($\Gamma_6$ and $\Gamma_7$).\label{tab:ctD2d}}
  \begin{tabular}{ l | r r r r r r | l }
\hphantom{7/2} & E &  R & S$_4^{[001]}$ & C$_2^{[001]}$ & C$_2^{[100]}$ & S$_1^{[110]}$ & \\
	\hline
	A$_{1}$ &  1 &  1 &     1 &     1 &      1 &       1 & $\Gamma_1$ \\
	A$_{2}$ &  1 &  1 &     1 &     1 &     -1 &      -1 & $\Gamma_2$ \\
	B$_{1}$ &  1 &  1 &    -1 &     1 &      1 &      -1 & $\Gamma_3$ \\
	B$_{2}$ &  1 &  1 &    -1 &     1 &     -1 &       1 & $\Gamma_4$ \\
	E$_{1}$ & \hphantom{+}2 & \hphantom{+}2 &     0 &    -2 &      0 &       0 & $\Gamma_5$ \\
	\hline
	E$_{2}$  &  2 & -2&$\pm \sqrt{2}$& 0 &   0 &       0 & $\Gamma_6$ \\
	E$_{3}$  &  2 & -2&$\mp \sqrt{2}$& 0 &   0 &       0 & $\Gamma_7$ \\
  \end{tabular}
  
  \vspace{20pt}
  
  \caption{Irreducible representations in D\textsub{2d} symmetry.\label{tab:irD2d}}
  \begin{tabular}{ r | r r r r r r | l }
 $M$ &  E &  R & S$_4^{[001]}$ & C$_2^{[001]}$ & C$_2^{[100]}$ & S$_1^{[110]}$ & irreducible representation \\
	\hline
	  0   &  1 &  1 &     1 &     1 &      1 &     1 & A$_{1}$ \\
	  1   &  3 &  3 &    -1 &    -1 &     -1 &     1 & B$_{2}$+E \\
	  2   &  5 &  5 &    -1 &     1 &      1 &     1 & A$_{1}$+B$_{1}$+B$_{2}$+E \\
	  3   &  7 &  7 &     1 &    -1 &     -1 &     1 & A$_{1}$+A$_{2}$+B$_{2}$+2E \\
	 $\vdots$ &  &  &       &       &        &       & \\
	 +4   & +8 & +8 &    +0 &    +0 &     +0 &    +0 & +A$_{1}$+A$_{2}$+B$_{1}$+B$_{2}$+2E \\
	\hline
	 1/2  &  2 & -2 &$\pm\sqrt{2}$&0&      0 &     0 & E$_{2}$ \\
	 3/2  &  4 & -4 &     0 &     0 &      0 &     0 & E$_{2}$+E$_{3}$ \\
	 5/2  &  6 & -6 &$\mp\sqrt{2}$&0&      0 &     0 & E$_{2}$+2E$_{3}$ \\
	 7/2  &  8 & -8 &     0 &     0 &      0 &     0 & 2E$_{2}$+2E$_{3}$ \\
	 $\vdots$ &  &  &       &       &        &       & \\
	 +4   & +8 & -8 &    +0 &    +0 &     +0 &    +0 & +2E$_{2}$+2E$_{3}$ \\
  \end{tabular}
\end{table}
\clearpage

\subsection{Sm\textsup{3+} energy level diagram in O\textsub{h} symmetry}\label{app_Sm3pELD}

\Fig[a]{fig:SmB6_ELD_CF} shows the energy level diagram of Sm\textsup{3+}.
The plethora of levels comes from the 2002 independent possibilities to fill the 4$f$ states with 5 electrons.
The Hund's rule ground state is always given by \qnJ{}\,=\,5/2, just like in Ce\textsup{3+}.

\begin{figure}
  \centering
  \includegraphics[width=\textwidth]{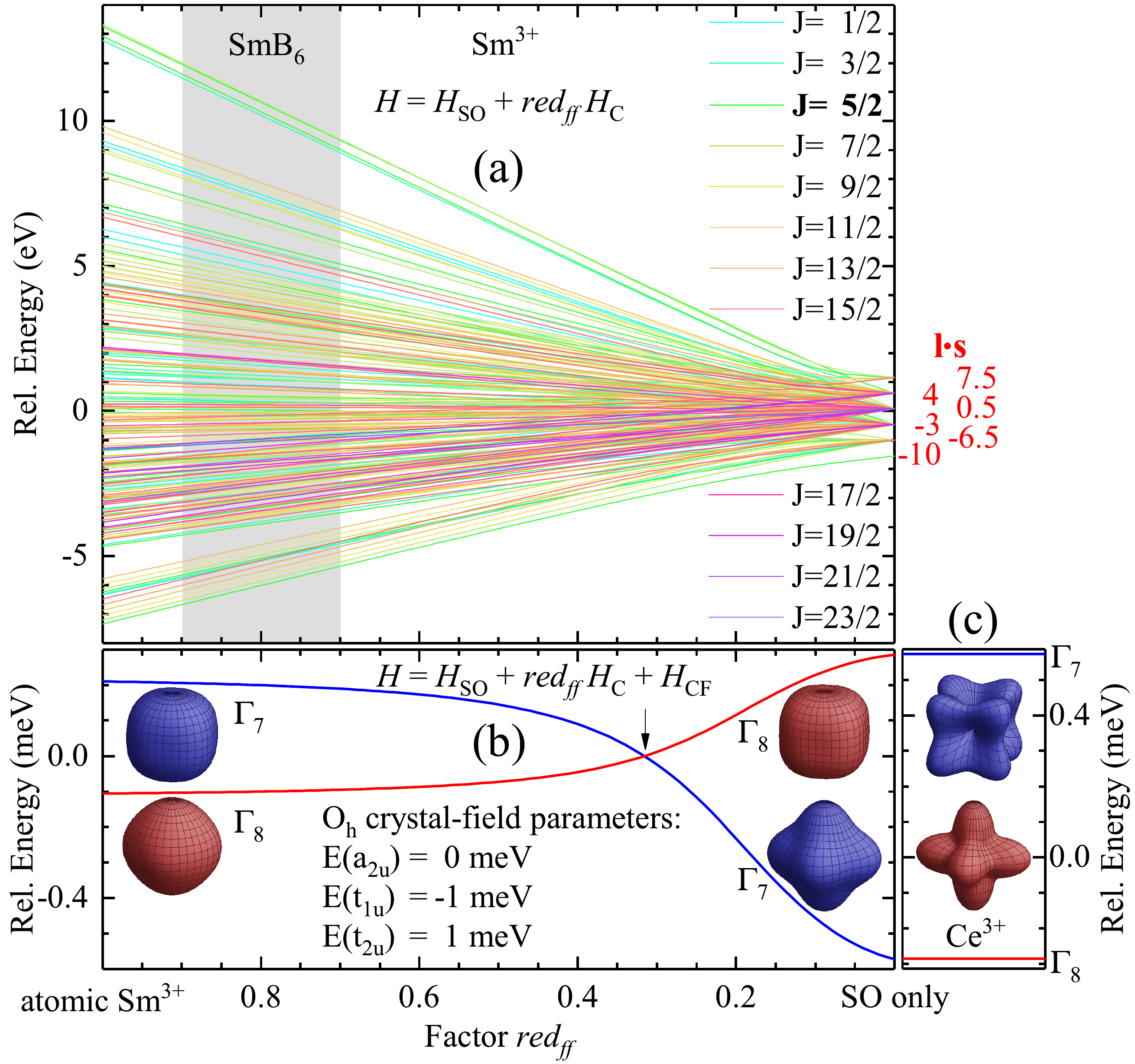}
  \caption{Energy level diagram of atomic Sm$^{3+}$ as a function of reduction of the 4$f$-4$f$ Coulomb interaction \redff{}. (a) Calculations without crystal field. The colors refer to different \qnJ{}. For the pure spin-orbit coupling case the six expectation values for $l\cdot s$ are shown. (b) Crystal field splitting of the Hund's rule ground state with \qnJ{}\,=\,5/2 in cubic O\textsub{h} symmetry into the $\Gamma_7$ (blue) and the $\Gamma_8$ (red) is shown for fixed crystal field parameters. The insets show the charge densities for the extreme cases \redff{}\,=\,0 and 1. Interestingly, there is an inversion of the crystal field ground states around \redff{}\,=\,0.3 (arrow in b), where O\textsub{h} crystal field is futile. (c) Crystal field splitting and charge densities for Ce$^{3+}$ for the same crystal field parameters.}
  \label{fig:SmB6_ELD_CF}
\end{figure}

\Fig{fig:SmB6_ELD_CF}(b) shows the \qnJ{}\,=\,5/2 Hund's rule ground state for a fixed cubic O\textsub{h} potential.
Interestingly, the two crystal-field eigenvalues of the $\Gamma_7$ (blue) and $\Gamma_8$ (red) depend on the scaling of the Coulomb interaction red$_{ff}$ and even invert around \redff{}\,=\,0.3.
The charge densities (insets of \fig{fig:SmB6_ELD_CF}(b)) however reflect these energies: The lower state always extends more towards the \xyz{100} directions and the stronger the anisotropy, the stronger the splitting; i.e.\ for \redff{}\,=\,0.8 the anisotropy is about 4 times weaker and opposite compared to \redff{}\,=\,0.
For different \redff{} the mixing of term symbols in the Hund's rule ground state changes.
This happens already in the Stevens approximation, as multiple levels of $J$\,=\,5/2 are present (2$\times$\termsymbol{4}{P}{5/2}, \termsymbol{6}{P}{5/2}, 5$\times$\termsymbol{2}{D}{5/2}, 3$\times$\termsymbol{4}{D}{5/2}, 7$\times$\termsymbol{2}{F}{5/2}, 4$\times$\termsymbol{4}{F}{5/2}, \termsymbol{6}{F}{5/2}, 4$\times$\termsymbol{4}{G}{5/2}, and \termsymbol{6}{H}{5/2}).

\Fig[c]{fig:SmB6_ELD_CF} shows the eigenstates of Ce\textsup{3+} using the same crystal-field potential as in \fig[b]{fig:SmB6_ELD_CF}.
The splitting in the Ce\textsup{3+} single-electron picture is identical to Sm\textsup{3+} (single-hole in the \qnJ{}\,=\,5/2) with pure spin-orbit coupling (\redff{}\,=\,0).
Here the charge density of the $f^5$\,$\Gamma_i$ can be obtained by the difference between the spherical $f^6$\,$\Gamma_1$ Hund's rule ground state with \qnJ{}\,=\,0 (not shown) and the $f^1$\,$\Gamma_i$.
But \fig[c]{fig:SmB6_ELD_CF} demonstrates, that this single particle picture breaks down quickly, when switching on the Coulomb interaction.

\clearpage

\subsection{Alignment of the analyzer crystals}\label{app_alignment}
\FloatBarrier
\begin{SCfigure}[][h]
  \includegraphics[width=0.45\textwidth]{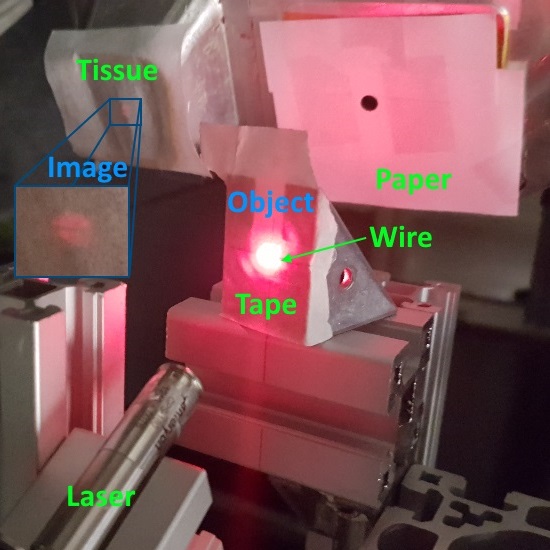}
  \caption{Photo of the pre-alignment setup. A Laser enlightens the object at the mirrors focal point \-- a tape with a wire attached to it. A sheet of paper blocks all but one analyzer. The image is projected onto a single layer of a tissue. The visibility of the wire in the image indicates the focal condition.}
  \label{fig:NIXSprealignment}
\end{SCfigure}
\FloatBarrier

Each analyzer should be aligned according to the Rowland geometry (see \fig{fig:NIXSgeometry}).
The different focal points should be aligned in a horizontal line to minimize the background due to scattering signal of the incoming beam (see \fig{fig:starmap} and text).
For this the analyzers have 3 motors for horizontal tilt, vertical tilt, and a translations for the focus.
The analyzers should be optical pre-aligned to speed up the alignment with the x-rays and save precious beamtime.

\Fig{fig:NIXSprealignment} shows the setup for the pre-alignment.
An optical Laser is targeted onto a tape, right at the later position of the sample; i.e.\ the focal point of the mirrors.
The light reflected from the analyzers can be visualized by e.g.\ a thin sheet of tissue at the later position of the detector.
This way the focal positions of the analyzers can be easily moved into the detector.
For better handling all but one hole of the collimator can be blocked to identify the projection of that analyzer.
Finally one can pre-align the focus by attaching a wire onto the targeted tape.
The focus is achieved when the wire throws a shadow on the screen with a clearly visible outline.

After the pre-alignment the focused x-ray beam should already be projected into the detector for each analyzer.
Due to the splitting of the different elastic energies the detector signal (\fig{fig:CeB6calibration}) can be seen best when the incident energy is chosen to be on top of the broad Compton, such that all analyzers are visible simultaneously.
Some adjustment may still be required to avoid an overlap of the different focal points or to optimize the focus.
\Fig{fig:starmap} shows the 2D detector signal with analyzers focused to the sample, whereas the cryostat window, signals at the top of the image, is about 1\,cm out of focus.
It appears closer on the 14\,mm large detector surface due to backscattering geometry.

\clearpage

\subsection{List of useful links and programs}\label{app_links}
NIXS beamlines:
\begin{itemize}
\item \href{} {P01 @ PETRA-III}
\item \href{} {ID20 @ ESRF}
\item \href{} {BL6-2b @ SSRL}
\item \href{} {BL12XU @ SPring-8}
\item \href{} {20-ID-B3 @ APS}
\item \href{} {GALAXIES @ SOLEIL}
\end{itemize}
Experiment preparation:
\begin{itemize}
\item \href{http://www2.fiz-karlsruhe.de/icsd_home.html} {ICSD} or \href{https://materials.springer.com} {Springer materials} for crystal structure data
\item \href{https://www.crystalimpact.de/diamond} {Diamond} to visualize crystal structures
\item \href{http://clip4.sourceforge.net} {CLIP} and/or \href{https://sourceforge.net/projects/qlaue} {QLaue} to calculate Laue patterns
\item \href{http://xdb.lbl.gov} {X-ray data booklet} for the electron binding energies
\item \href{http://henke.lbl.gov/optical_constants} {X-ray interactions with matter} to calculate x-ray transmissions
\end{itemize}
Data analysis:
\begin{itemize}
\item Python based \href{https://confluence.desy.de/display/FSP01/IXStools} {IXStools} or \href{http://ftp.esrf.fr/scisoft/XRStools} {XRStools} to process NIXS data
\end{itemize}
Calculations:
\begin{itemize}
\item \href{http://symmetry.jacobs-university.de} {Character tables} of important point groups
\item \href{https://www.tcd.ie/Physics/people/Cormac.McGuinness/Cowan} {Cowan's Atomic structure code} for atomic parameters
\item \href{http://www.quanty.org} {Quanty} or \href{} {xtls} for full multiplet calculations
\end{itemize}

\clearpage

\clearpage
\phantomsection
\renewcommand{\refname}{References}
\addcontentsline{toc}{section}{\protect\numberline{}References}

\clearpage
\sectioff{List of publications}

\subsection*{Part of this thesis:}

\begin{enumerate}[label=\text{$[$\Roman*$]$}]

\begin{minipage}{0.94\textwidth}\raggedright
\item \label{CeRu4Sn6_2015}
\textbf{CeRu\textsub{4}Sn\textsub{6}: A strongly correlated material with nontrivial topology} \\
{Martin Sundermann, Fabio Strigari, Thomas Willers, Hannes Winkler, Andrey Prokofiev, James M. Ablett, Jean-Pascal Rueff, Detlef Schmitz, Eugen Weschke, Marco Moretti Sala, Ali Al-Zein, Arata Tanaka, Maurits W. Haverkort, Deepa Kasinathan, Liu Hao Tjeng, Silke B\"uhler-Paschen \& Andrea Severing} \\
\href{http://dx.doi.org/10.1038/srep17937} {Scientific Reports \textbf{5}, 17937 (2015)}
\end{minipage}

\begin{minipage}{0.94\textwidth}\raggedright
\item \label{URu2Si2_2016}
\textbf{Direct bulk-sensitive probe of 5\textit{f} symmetry in URu\textsub{2}Si\textsub{2}} \\
{Martin Sundermann, Maurits W. Haverkort, Stefano Agrestini, Ali Al-Zein, Marco Moretti Sala, Yingkai Huang, Mark Golden, Anne de Visser, Peter Thalmeier, Liu Hao Tjeng \& Andrea Severing} \\
\href{http://dx.doi.org/10.1073/pnas.1612791113} {Proceedings of the National Academy of Sciences \textbf{113} 49, 13989\--13994 (2016)}
and
\href{http://www.pnas.org/content/suppl/2016/11/17/1612791113.DCSupplemental} {supplementary information}.
\end{minipage}

\begin{minipage}{0.94\textwidth}\raggedright
\item \label{CeRu4Sn6_2017}
\textbf{Ce\,3\textit{p} hard x-ray photoelectron spectroscopy study of the topological Kondo insulator CeRu\textsub{4}Sn\textsub{6}} \\
{Martin Sundermann, Kai Chen, Yuki Utsumi, Yu-Han Wu, Ku-Ding Tsuei, Jonathan Haenel, Andrey Prokofiev, Silke B\"uhler-Paschen, Arata Tanaka, Liu Hao Tjeng \& Andrea Severing} \\
\href{http://stacks.iop.org/1742-6596/807/i=2/a=022001} {Journal of Physics: Conference Series \textbf{807} 2, 022001 (2017)}
\end{minipage}

\begin{minipage}{0.94\textwidth}\raggedright
\item \label{CeB6_2017}
\textbf{The quartet ground state in CeB\textsub{6}: An inelastic x-ray scattering study} \\
Martin Sundermann, Kai Chen, Hasan Yava\c{s}, Hanoh Lee, Zachery Fisk, Maurits W. Haverkort, Liu Hao Tjeng \& Andrea Severing \\
\href{https://doi.org/10.1209/0295-5075/117/17003} {Europhysics Letters \textbf{117} 1, 17003 (2017)}
\end{minipage}

\begin{minipage}{0.94\textwidth}\raggedright
\item \label{SmB6_2017}
\textbf{4\textit{f} crystal field ground state of the strongly correlated topological insulator SmB\textsub{6}} \\
{Martin Sundermann, Hasan Yava\c{s}, Kai Chen, Dae-Jeong Kim, Zachery Fisk, Deepa Kasinathan, Maurits W. Haverkort, Peter Thalmeier, Andrea Severing \& Liu Hao Tjeng} \\
\href{https://doi.org/10.1103/PhysRevLett.120.016402} {Physical Review Letters \textbf{120}, 016402 (2018)}
\end{minipage}

\begin{minipage}{0.94\textwidth}\raggedright
\item \label{UO2_2018}
\textbf{U5\textit{f} crystal field ground state of UO\textsub{2} probed by nonresonant inelastic x-ray scattering} \\
{Martin Sundermann, Roberto Caciuffo, Gerry H. Lander, Maurits W. Haverkort, Andrea Severing \& Gerrit van der Laan} \\
\href{https://doi.org/10.1103/PhysRevB.98.205108} {Physical Review B \textbf{98}, 205108 (2018)}
\end{minipage}

\end{enumerate}

\begin{minipage}[t][10cm][t]{\textwidth}

\subsection*{Further publications:}

\begin{enumerate}[resume, label=\text{$[$\Roman*$]$}]

\begin{minipage}{0.94\textwidth}\raggedright
\item \label{CeM2Al10_2015}
\textbf{Quantitative study of valence and configuration interaction parameters of the Kondo semiconductors Ce\textit{M}\textsub{2}Al\textsub{10} (\textit{M}\,=\,Ru, Os and Fe) by means of bulk-sensitive hard x-ray photoelectron spectroscopy} \\
{F. Strigari, M. Sundermann, Y. Murob, K. Yutanic, T. Takabatake, K.-D. Tsuei, Y.F. Liao, A. Tanaka, P. Thalmeier, M.W. Haverkort, L.H. Tjeng \& A. Severing} \\
\href{https://doi.org/10.1016/j.elspec.2015.01.004} {Journal of Electron Spectroscopy and Related Phenomena \textbf{199}, 56\--63 (2015)}
\end{minipage}

\begin{minipage}{0.94\textwidth}\raggedright
\item \label{CeMIn5_2016}
\textbf{Quantitative study of the \textit{f} occupation in Ce\textit{M}In\textsub{5} and other cerium compounds with hard x-rays} \\
{M. Sundermann, F. Strigari, T. Willers, J. Weinen, Y.F. Liao, K.-D. Tsuei, N. Hiraoka, H. Ishii, H. Yamaoka, J. Mizuki, Y. Zekko, E.D. Bauer, J.L. Sarrao, J.D. Thompson, P. Lejay, Y. Muro, K. Yutani, T. Takabatake, A. Tanaka, N. Hollmann, L.H. Tjeng \& A. Severing} \\
\href{http://dx.doi.org/10.1016/j.elspec.2016.02.002} {Journal of Electron Spectroscopy and Related Phenomena \textbf{209}, 1\--8 (2016)}
\end{minipage}

\begin{minipage}{0.94\textwidth}\raggedright
\item \label{CeMAl4Si2_2016}
\textbf{Exchange field effect in the crystal-field ground state of Ce\textit{M}Al\textsub{4}Si\textsub{2}} \\
{K. Chen, F. Strigari, M. Sundermann, S. Agrestini, N.J. Ghimire, S.-Z. Lin, C.D. Batista, E.D. Bauer, J.D. Thompson, E. Otero, A. Tanaka, \& A. Severing} \\
\href{http://dx.doi.org/10.1103/PhysRevB.94.115111} {Physical Review B \textbf{94}, 115111 (2016)}
\end{minipage}

\begin{minipage}{0.94\textwidth}\raggedright
\item \label{CeCoIn5_2017}
\textbf{Evolution of ground state wave function in CeCoIn\textsub{5} upon Cd or Sn doping} \\
{K. Chen, F. Strigari, M. Sundermann, Z. Hu, Z. Fisk, E.D. Bauer, P.F.S. Rosa, J.L. Sarrao, J.D. Thompson, J. Herero-Martin, E. Pellegrin, D. Betto, K. Kummer, A. Tanaka, S. Wirth, \& A. Severing} \\
\href{https://link.aps.org/doi/10.1103/PhysRevB.97.045134} { Physical Review B \textbf{97}, 045134 (2018) }
\end{minipage}

\begin{minipage}{0.94\textwidth}\raggedright
\item \label{CeOs2Al10_2018}
\textbf{Tuning the hybridization and magnetic ground state of electron and hole doped CeOs\textsub{2}Al\textsub{10}: an x-ray spectroscopy study} \\
{K. Chen, M. Sundermann, F. Strigari, J. Kawabata, T. Takabatake, A. Tanaka, P. Bencok, F. Choukani \& A. Severing} \\
\href{https://link.aps.org/doi/10.1103/PhysRevB.97.155106} { Physical Review B \textbf{97}, 155106 (2018) }
\end{minipage}

\begin{minipage}{0.94\textwidth}\raggedright
\item \label{Yb2Si2Al_2018}
\textbf{Intermediate valence in single crystalline Yb\textsub{2}Si\textsub{2}Al} \\
{W.~J. Gannon, K. Chen, M. Sundermann, F. Strigari, Y. Utsumi, K.-D. Tsuei, J.-P. Rueff, P. Bencok, A. Tanaka, A. Severing \& M.~C. Aronson} \\
\href{https://doi.org/10.1103/PhysRevB.98.075101} { Physical Review B \textbf{98}, 075101 (2018) }
\end{minipage}

\begin{minipage}{0.94\textwidth}\raggedright
\item \label{NiO_2018}
\textbf{Direct imaging of orbitals in quantum materials} \\
{H. Yava\c{s}, M. Sundermann, K. Chen, A. Amorese, A. Severing, H. Gretarsson, M.~W. Haverkort \& L.~H. Tjeng} \\
\href{https://doi.org/10.1038/s41567-019-0471-2} { Nature Physics \textbf{15}, 559 (2019) }
\end{minipage}

\end{enumerate}
\end{minipage}
	
\clearpage
\sectioff{Acknowledgements\,/\,Danksagung}

At this point I would like to thank all the people, who supported me during my studies.
Only because of you, I could reach to this point.

First of all, I like to thank Dr.~Andrea Severing and Prof.~Dr.~Liu Hao Tjeng.
Andrea, I am grateful, that you gave me the opportunity to work on this project and introduced me to the fascinating field of $f$-electron physics and to the people related to this field all over the world.
Hao, thank you for welcoming me to your research group and supervising this work.
I very much appreciate your deep knowledge of the theoretical and experimental aspects of correlated electron physics and the enthusiasm you show for the pioneering research with all its important little details.

Prof.~Dr.~Maurits W.\ Haverkort, thank you very much for teaching me about the multiplet calculations and for answering my great number of questions.
I very much enjoyed our discussions and your little lectures during our coffee meetings.

Thanks to all my collaborators for the support with the various aspects of my research.
Dr.~Christoph J.\ Sahle, Dr.~Ali Al-Zein, and Dr.~Marco Moretti, thank you for the support performing my first \NIXS{} experiments.
Dr.~Jean-Pascal Rueff and Dr.~James M.\ Ablett, thank you for the support of the \XAS{} experiments.
Dr.~Hasan Yava\c{s}, Manuel Harder, and all other members of P01, thank you for your experimental support and showing how to operate a beamline.
Dr.~Fabio Strigari, Dr.~Stefano Agrestini, Dr.~Kai Chen, Dr.~Yu-Han Wu, and Dr.~Yuki Utsumi, thank you for your help running the experiments 24/7.
Dr.~Gerry H.\ Lander, Dr.~Roberto Caciuffo, and Dr.~Gerrit van der Laan, our collaboration and discussions were a great pleasure to me.
I am also grateful to the engineers and workshops, that design and manufacture the instruments for this kind of pioneering research.
A special thanks goes to the my collaborators who have been growing and providing the precious single crystals, which are the basis of this research.

I am grateful for the opportunity to participate in and learn about many different experiments, to present my work at different conferences, and to learn from so many brilliant people.
Thanks to all enthusiastic scientists, students, engineers, technicians, and staff for your kindness and your diverse contributions.

Especially, I like to thank 
Dr.~Andrea Amorese, Dr.~Fabio Strigari, Dr.~Kai Chen,
Dr.~Brett Leedahl, Christoph Becker, Dr.~Katharina Höfer, Dr.~Sahana Rößler, Dr.~Deepa Kasinathan, Dr.~Stefano Agrestini, Dr.~Yuki Utsumi,
Dr.~Hasan Yava\c{s}, Dr.~Hlynur Grettarson,
and all other group members from Cologne, Dresden, and Hamburg for the good working atmosphere and various discussions about physics and many other topics.

I further like to acknowledge the financial support of the Projects SE\,1441/2-1, SE\,1441/4-1 and SE\,1441/5-1 by the Deutsche Forschungsgemeinschaft (DFG) and the provision of synchrotron radiation by the facilities PETRA III at DESY in Germany, a member of the Helmholtz Association (HGF), the European Synchrotron Radiation Facility (ESRF) and the SOLEIL synchrotron in France, and SPring-8 in Japan with the approval of the Japan Synchrotron Radiation Research Institute (JASRI).

Prof.~Dr.~Liu Hao Tjeng, Prof.~Dr.~Jochen Geck, and Prof.~Dr.~Maurits W.\ Haverkort, thank you for the time and efforts in reviewing this work.

Ich danke auch meiner gesamten Familie für die vielfältige Unterstützung in allen Lebenslagen.
Ein besonderer Dank gilt meinen Eltern, welche mir dieses Studium ermöglicht haben.

\end{document}